\documentclass[fleqn,usenatbib]{mnras}

\usepackage{newtxtext,newtxmath}

\usepackage[T1]{fontenc}

\DeclareRobustCommand{\VAN}[3]{#2}
\let\VANthebibliography\thebibliography
\def\thebibliography{\DeclareRobustCommand{\VAN}[3]{##3}\VANthebibliography}

\usepackage{graphicx}	
\usepackage{amsmath}	
\usepackage{multicol}
\usepackage{physics}
\usepackage{bm}
\usepackage{ulem}
\usepackage{comment} 
\usepackage{xcolor}
\usepackage{mathrsfs}
\usepackage{tabularx}
\usepackage{threeparttable}
\usepackage{hyperref}
\usepackage[title]{appendix}
\usepackage{subfig}
\allowdisplaybreaks
\usepackage{CJK}

\title[Does magnetic field promote or suppress disk fragmentation?]{Does magnetic field promote or suppress fragmentation in AGN disks? Results from local shearing box simulations with simple cooling.}

\author[Tsung et al.]{Tsun Hin Navin Tsung$^{1,2}$\thanks{E-mail: tsunhinnavin.tsung@colorado.edu},
Mitchell C. Begelman$^{2,3}$,
Philip J. Armitage$^{4,5}$,
Yan-Fei Jiang (姜燕飞)$^{4}$,
\newauthor
Hannalore J. Gerling-Dunsmore$^{2,3}$
\\
$^{1}$Center for Integrated Plasma Studies, Physics Department,
390 UCB, University of Colorado, Boulder, CO 80309, USA\\
$^{2}$JILA, University of Colorado and National Institute of Standards and Technology, 440 UCB, Boulder, CO 80309-0440, USA\\
$^{3}$Department of Astrophysical and Planetary Sciences, 391 UCB, University of Colorado, Boulder, CO 80309-0391, USA\\
$^{4}$Center for Computational Astrophysics, Flatiron Institute, 162 Fifth Avenue, New York, NY 10010, USA\\
$^{5}$Department of Physics and Astronomy, Stony Brook University, Stony Brook, NY 11794, USA
}

\date{Accepted XXX. Received YYY; in original form ZZZ}

\pubyear{2015}

\begin{document}
\begin{CJK*}{UTF8}{gbsn}
\label{firstpage}
\pagerange{\pageref{firstpage}--\pageref{lastpage}}
\maketitle

\begin{abstract}
Accretion disks in Active Galactic Nuclei (AGN) are predicted to become gravitationally unstable substantially interior to the black hole's sphere of influence, at radii where the disk is simultaneously unstable to the magnetorotational instability (MRI). Using local shearing box simulations with net vertical flux and a simple cooling prescription, we investigate the effect of  magnetic fields on fragmentation in the limit of ideal magnetohydrodyamics. 
Different levels of in-disk magnetic field from the magnetorotational instability are generated by varying the initial vertical-field plasma beta $\beta_0$. We find that the disk becomes magnetically dominated when $\beta_0<10^3$, and that this transition is accompanied by a drastic drop in fragmentation (as measured by the bound mass fraction) and gravitational stress. The destabilizing influence of radial magnetic fields, which are present locally and which may promote fragmentation via magnetic tension effects, is overwhelmed by magnetic elevation, which significantly reduces the mid-plane density. The magnetic suppression of fragmentation in magnetically elevated disks has implications for the radial extent of the accretion flow in AGN disks, and for the efficiency of in situ formation of disk-embedded stars that are progenitors for single and binary compact objects.

\end{abstract}




\section{Introduction} \label{sec:intro}



Accretion disks in Active Galactic Nuclei (AGN) are predicted to become self-gravitating via the Toomre instability \citep{Toomre-1964} beyond a critical radius, which is a small fraction of the black hole's sphere of influence. The onset of disk self-gravity can lead to fragmentation and in situ star formation closer to the black hole than conventional processes allow, potentially disrupting the accretion flow and limiting black hole growth \citep{Shlosman_Begelman-1990,Goodman-2003}, but also producing compact objects and binaries that are detectable as LISA or LIGO gravitational wave sources \citep{Levin-2007,McKernan_etal-2018}.

The conditions under which gravitational instability (GI) in astrophysical disks leads to fragmentation has been studied extensively \citep{Kratter_Lodato-2016}, particularly in the context of protoplanetary disks, which can be modeled to a first approximation as non-magnetized hydrodynamic structures. In isolated disks, prompt fragmentation occurs if the cooling time is below a threshold value that is typically of the order of the local dynamical time \citep{Gammie-2001,Rice_etal-2005,Xu_etal-2024}. Rapid infall can also trigger fragmentation \citep{Kratter_etal-2010}. The applicability of these hydrodynamic criteria to AGN disks, however, is rather unclear. Estimates by \cite{Menou_Quataert-2001} suggest that, except at low accretion rates, the regions of AGN disks that are self-gravitating are also well enough ionized to be in a near-ideal magnetohydrodynamic (MHD) regime. The typical magnetic field strengths at these radii are unknown. Our main focus in this work will be on the ``magnetically elevated" regime, where moderately net strong poloidal fields lead to sustained toroidal fields whose pressure is comparable to that of the disk gas or radiation \citep{Salvesen-etal-2016,Begelman_Silk-2017}. It is also possible, however, that magnetic fields could be significantly sub-thermal, as in the case of the zero-net flux magnetorotational instability \citep{Balbus_Hawley-1998,Davis_etal-2010}, or substantially stronger as in Magnetically Arrested Disk (MAD) models \citep{Tchekhovskoy_etal-2011}. We also note recent work by \citet{Hopkins_etal-2024,Squire_etal-2025,Guo_etal-2025} that demonstrated the formation of hyper-magnetized disks distinct from the classic $\alpha$ disks and magnetically-arrested disks.

Analytically, the pressure forces associated with magnetic fields are expected to act as a stabilizing influence on disk gravitational instability. However, it has long been known that in a shearing disk magnetic tension can {\it destabilize} gravitational instability by suppressing the stabilizing effect of Coriolis forces \citep{Lynden-Bell-1966}. In a laminar background, \cite{Elmegreen-1987} and \cite{Gammie-1996} showed analytically that this process can indeed destabilize GI modes, but that the resulting growth is transient, with a brief phase of growth usually followed by an abrupt phase of damping. It is therefore unclear, from analytic work, whether magnetic suppression of the Coriolis force survives into the non-linear regime and affects the conditions for fragmentation. A different destabilizing effect may, however, arise due to the density perturbations that are generated in the saturated state of the magnetorotational instability (MRI). Although the zero-net flux MRI saturates under near incompressible conditions, the MRI in elevated disks with net flux leads to highly inhomogeneous disks \citep{Salvesen-etal-2016}.

Early simulation work confirmed that the interplay of gravitational and magnetorotational instabilities in accretion disks is complex. Gravitational stress is reduced when magnetized turbulence develops, implying suppression of GI \citep{Fromang-2005,Fromang-etal-2004b}. The MRI forms density substructures in magnetized turbulence, which can be swing-amplified to form giant clumps \citep{Kim-etal-2003}. However, the magnetic field itself reduces swing-amplication \citep{Kim_Ostriker-2001}. Thus, there was the picture that while magnetic fields tend to suppress linear development of GI, the nonlinear effects of MRI turbulence may be conducive to fragmentation through generation of denser seeds. It was unclear if the magnetic-tension-enhanced gravitational instability proposed by \citet{Lynden-Bell-1966} played a role or not.

Recent local and global simulations, including resistivity to model protoplanetary disks, have identified new physical effects. MHD turbulence can be sustained in the presence of substantial Ohmic resistivity, via a kind of spiral-wave (gravitational) dynamo that is distinct from the MRI \citep{Riols-Latter-2019,Riols_etal-2021,Lohnert_Peeters-2023}. Moreover, in high resolution global simulations, fragmented clumps form that are smaller in size than in the hydrodynamic case \citep{Deng_etal-2021,Kubli_etal-2023}. The clumps are also more numerous and long-lasting. These results have been interpreted as being due to the destabilizing effect of magnetic fields in low shear regions of the disk. As the clumps form, the magnetic field surrounding them would also be enhanced, shielding them from break-up by the shear flow, making them long-lasting. This paints a picture in which magnetic turbulence promotes fragmentation both linearly and nonlinearly.


Here, we revisit the question of whether magnetic fields stabilize or destabilize AGN disks to gravitational instability, using local shearing box simulations in ideal MHD. We vary the strength of the net vertical field to induce differing levels of saturated magnetic pressure and disk elevation. The thermodynamics is modeled with a simple ``$\beta$ cooling" law, both for comparison with prior work \citep{Gammie-2001} and to efficiently explore parameter space in preparation for forthcoming radiation hydrodynamic simulations. We investigate whether the magnetic tension enhanced gravitational instability, which we refer to as the ``Coriolis-Restricted-Magneto-Gravitational" (CRMG) instability, contributes to fragmentation. Our approach is to evaluate the physical conditions (e.g. magnetic field structure, density, etc.) of the disk when the MRI is fully developed and to identify clumps through a numerical procedure. We then explain the fragmentation behavior by examining the WKB linear growth rate. Despite changing conditions in a turbulent disk, we argue that the WKB approach  provides insight into the stability of the disk in an averaged sense.


The structure of this paper is as follows: In \S\ref{sec:setup} we describe our model and simulation setups, and describe the diagnostics we use in \S\ref{sec:diagnostics}. In \S\ref{sec:theory} we briefly outline the theory of how magnetic fields can destabilize GI. We describe our simulation results in \S\ref{sec:results} and conclude in \S\ref{sec:discussion_conclusion}.

\section{Simulation Setup} \label{sec:setup}

We utilize the 3D shearing-box setup in \texttt{Athena++} \citep{Stone-etal-2020} to solve the ideal magnetohydrodynamic (MHD) equations:
\begin{gather}
    \pdv{\rho}{t} + \nabla\cdot\qty(\rho\vb{v}) = 0, \label{eqn:continuity} \\
    \pdv{\qty(\rho\vb{v})}{t} + \nabla\cdot\qty(\rho\vb{v}\vb{v} - \vb{B}\vb{B} + P_g + B^2/2) = -2\rho\Omega\vu{z}\cross\vb{v} \label{eqn:momentum} \\ \qquad + 2q\rho\Omega^2 x\vu{x} - \rho\Omega^2 z\vu{z} - \rho\nabla\Phi, \nonumber \\
    \pdv{E}{t} + \nabla\cdot\qty[\qty(E + P_g + B^2/2)\vb{v} - \vb{B}\qty(\vb{B}\cdot\vb{v})] = -\rho\vb{v}\cdot\nabla\Phi \label{eqn:energy} \\ \qquad \rho\Omega^2\vb{v}\cdot\qty(2qx\vb{x} - z\vu{z}) - \frac{E_\mathrm{th}}{\tau_\mathrm{cool}}, \nonumber \\
    \pdv{\vb{B}}{t} = \nabla\cross\qty(\vb{v}\cross\vb{B}), \label{eqn:induction}
\end{gather}
where $E = \rho v^2/2 + P_g/(\gamma - 1) + B^2/2$ is the total (kinetic + thermal + magnetic) energy density, $E_\mathrm{th} = P_g/(\gamma - 1)$ is the thermal energy density, $\Omega$ is the angular velocity at the origin, about which the shearing-box approximation is made, $q$ is the shearing parameter ($3/2$ for Keplerian potential), and $\tau_\mathrm{cool}$ is the cooling time. We take $\gamma = 5/3$ to be the adiabatic index of the gas. The $x,y,z$-axes represent the radial, azimuthal and vertical directions, respectively. The gravitational potential $\Phi$ is calculated from Poisson's equation,
\begin{equation}
    \nabla^2\Phi = 4\pi G\rho. \label{eqn:poisson}
\end{equation}

The MHD equations are solved using a Godunov method, a second-order van Leer time integrator, the HLLD Riemann solver to calculate the hydrodynamic fluxes together with the Constrained Transport (CT) technique to ensure the magnetic field is divergence-less \citep{Stone-etal-2020}. Poisson's equation is solved using Fast Fourier Transform (FFT) \citep{Koyama_Ostriker-2009,Kim_etal-2011}.

The initial profile is characterized by 
\begin{equation}
    Q = \frac{\Omega^2}{2\pi G\rho_0}, \label{eqn:proxy_Q}
\end{equation}
which we call the proxy Toomre parameter due to its relationship to the actual Toomre parameter $Q_T$ initially (see Appendix \ref{app:initial}). Assuming a uniform, vertical magnetic field, no $x,z$ flow and that $\rho, P_g$ depend only on $z$, the initial profile can be evaluated by setting the time derivative of eq.~\ref{eqn:momentum} to zero. This gives a steady-state solution $\vb{v} = -q\Omega x\vu{y}$ for the velocity, and the following equation for $\rho, P_g$:
\begin{equation}
    \frac{1}{\rho}\dv{P_g}{z} = -\Omega^2 z - 4\pi G\int_0^{z}\rho\dd{z'}, \label{eqn:hydrostatic}
\end{equation}
which can be solved numerically assuming a polytropic EOS, provided the mid-plane density $\rho_0$ and gas pressure $P_{g0}$ are known. In practice the surface density $\Sigma$ and angular velocity $\Omega$ are more easily obtained. $(\rho_0, P_{g0})$ can be determined from $(\Sigma, \Omega)$ using the fact that
\begin{equation}
    \Sigma = 2\int_0^{\infty}\rho\dd{z}, \quad \Omega^2 = 2\pi G\rho_0 Q. \label{eqn:facts}
\end{equation}
The details of the calculation are described in Appendix \ref{app:initial}. It can be shown, given $Q,\Sigma,\Omega$, that the mid-plane density, gas pressure and temperature are given by
\begin{equation}
    \rho_0 = \frac{\Omega^2}{2\pi G Q},\quad P_{g0} = \frac{2\Sigma \Omega^2}{5Q\Psi_{3/2}},\quad T_0 = \qty(\frac{\mu m_p}{k_B})\frac{4\pi G\Sigma}{5\Psi_{3/2}}, \label{eqn:midplane}
\end{equation}
where $\Psi_{3/2}(Q)$ is a dimensionless measure of a disk's thickness dependent only on $Q$. For $Q\ll 1$, where self-gravity dominates, $\Psi_{3/2}$ will be small, while $\Psi_{3/2}$ asymptotes to a constant value for $Q\gg 1$, when the tidal potential dominates. For constant $Q$, $\rho_0\propto\Omega^2, P_{g0}\propto\Sigma\Omega^2, T_0\propto\Sigma$, while for constant $\Sigma$ and $\Omega$, $\rho_0, P_{g0}, T_0$ decrease for increasing $Q$. 

The simulations were performed in code units. We define the time unit to be $t_* = \Omega^{-1}$, length unit to be the scale-height evaluated at the mid-plane $l_* = H = c_{s0}/\Omega$, density unit to be the mid-plane density $\rho_* = \rho_0$, velocity unit to be the mid-plane sound speed $v_* = c_{s0}$, pressure unit $P_* = \rho_* v_*^2$, temperature unit $T_* = \mu m_p P_*/k_B\rho_*$ ($\mu=0.6$ is the mean molecular weight per $m_p$), and magnetic field unit $B_* = \sqrt{4\pi P_*}$. With this unit system, the unit gravitational potential is defined as $\Phi_*= v_*^2$ and `$4\pi G$' in code units by $4\pi G\rho_* t_*^2 = 2/Q$. The initial density and pressure profiles in code units are completely specified given $Q$.


We use a fiducial box domain of $20H\times20H\times24H$ for our investigation. A long vertical span is necessary to ensure that a sufficient number of nonlinear density scale heights are captured and to reduce boundary effects associated with a magnetically elevated disk. A wide horizontal span is needed for gravitational effects, which are long-range, to operate. For the strongly magnetized cases ($\beta_0\leq 10^2$, where $\beta_0$ is the initial midplane plasma beta), we set the azimuthal box length to be 2 times the radial (i.e. $20H\times 40H\times 24H$) to suppress strong zonal flows that form in the highly magnetized cases, a magnetic-wind-driven feature \citep{Riols_Lesur-2019} that can artificially increase fragmentation rates (see Appendix \ref{app:box_rationale} for more details). We adopt shearing-periodic boundary conditions in the $x$-direction and periodic boundaries in the $y$-direction. For the boundaries in the $z$-direction, we enforce outflow conditions for $\rho, P_g, v_x, v_y$ and impose outflow conditions for $v_z$ when the gas is outflowing and set the ghost zones $v_z=0$ when it is inflowing. For magnetic fields we follow \cite{Riols-Latter-2019} and \cite{Lohnert_Peeters-2023} to set $B_x = B_y = 0$ in the ghost zones and $\dv*{B_z}{z}=0$. We find that such magnetic boundary conditions generally lead to a shorter simulation runtime than outflow boundaries \citep{Bai_Stone-2013} due to the reduced Alfv\'en speed close to the boundaries. We note that while this is an often adopted recourse in shearing box simulation, setting $B_x=B_y=0$ at the vertical boundaries prevents a magnetic exchange of angular momentum between the outflow/wind and the disk, particularly if there is a magnetized wind.

We impose an initial magnetic field with a net vertical flux in the form \citep{Salvesen-etal-2016}
\begin{equation}
    B_z = B_0\qty[1 + \frac{1}{2}\sin(\frac{2\pi x}{L_x})], \label{eqn:init_b_field}
\end{equation}
where $L_x$ is the domain width in the $x$-direction. The sinusoidal component in eq.~\ref{eqn:init_b_field} is there to suppress channel flows, which could lead to rapid depletion of the disk mass. $B_0$ is specified through the mid-plane plasma beta $\beta_0 = 2P_{g0}/B_0^2$. Simulations with net vertical flux are always found to launch strong outflows that can drain the box mass in tens of orbits. To produce a steady-state solution, we therefore add mass in the form $\propto\mathrm{exp}(-z^2/H^2)$ (where $H$ is the initial scale height defined by the mid-plane $c_{s0}$) to each grid cell at the end of each iteration to keep the total mass within the box constant, mimicking mass supply from accretion. 

We implement cooling through an explicit source term using a simplistic cooling model $t_\mathrm{cool} = \tau_\mathrm{cool}\Omega^{-1}$ \citep{Gammie-2001} such that
\begin{equation}
    \dv{E_\mathrm{th}}{t} = -\frac{E_\mathrm{th} \Omega}{\tau_\mathrm{cool}} .\label{eqn:beta_prescription}
\end{equation}
This simple prescription, though not realistic, has the advantage of being freely adjustable and has been used widely in the literature \citep{Gammie-2001,Rice_etal-2005,Meru_Bate-2010}.

We trigger MRI by generating white noise on the initial density and velocities having maximum amplitude of $\abs{\delta\rho}/\rho=0.01$ and $\abs{\delta v_i/c_s}=0.01$, respectively.

We vary the initial mid-plane plasma beta $\beta_0$ through $10, 10^2, 10^3, 10^4, 10^5$ to investigate the effect of magnetic field on fragmentation and set the cooling time $\tau_\mathrm{cool}=1$. We choose $\tau_\mathrm{cool}$ to be less than 3, the fragmentation condition in the non-magnetized case \citep{Gammie-2001}, to ensure fragmentation in the weakly magnetized cases. We cannot set $\tau_\mathrm{cool}$ below 1 as the cooling time could easily be less than the signal travel time across a grid (i.e. $\Delta x/\max(v_A, c_s, v)$) for the weakly magnetized cases during run-time, leading to artificial heating and code crashing.
We set $Q=1$ initially for all our simulations. This sets the initial setup to be marginally Toomre unstable, with a Toomre parameter $Q_T=0.89$ (eq.~\ref{eqn:QT_Q_relation}). The exact choice of $Q$ is not important provided the initial magnetic field is sub-thermal and cooling is fast compared to timescales of the MRI, as the disk will be cooled below $Q_T=1$ on the cooling timescale even if $Q_T>1$ initially. 

The simulation box is resolved with a fiducial resolution of $256\times256\times512$ for the $\beta_0\geq10^3$ cases and $256\times512\times384$ for the $\beta_0\leq 10^2$ cases. We use a coarser vertical grid size for the low $\beta_0$ cases as the MRI scale is larger, but double the number of grids in the azimuthal direction due to the larger azimuthal domain width. As shown in Table \ref{tab:simulation_table}, the quality factors in the $y,z$ directions $Q_y, Q_z$ are quite large in the saturated regime, meaning that MRI scales are likely resolved. We repeat our simulations at a reduced resolution of resolution of $128\times256\times192$ (for $\beta_0\leq10^2$) and $128\times128\times192$ (for $\beta_0\geq10^3$) and compare our results in Appendix \ref{app:resolution}. To avoid excessive simulation cost associated with small time-steps in magnetically dominated regions, we impose a density floor of $\rho_\mathrm{floor} = 10^{-4}\rho_0$. We also impose a temperature floor of $T_\mathrm{floor} = 10^{-5} T_0$ and ceiling of $T_\mathrm{ceil} = 10^2 T_0$ to avoid numerical difficulties associated with excessive cooling and large sound speeds.

\section{Diagnostics} \label{sec:diagnostics}

We adopt the following notations to denote various averages used in our analysis and discussion. First, we define $\langle X\rangle_{xy}$ as horizontally averaged quantities:
\begin{equation}
    \langle X\rangle_{xy} = \frac{1}{L_x L_y}\int X\dd{x}\dd{y}, \label{eqn:xy_avg}
\end{equation}
where $L_x, L_y$ are the domain widths in the $x,y$-directions, respectively. Next, we define $\langle X\rangle_V$ as volume averaged quantities:
\begin{equation}
    \langle X\rangle_V = \frac{1}{L_x L_y L_z}\int X\dd{x}\dd{y}\dd{z}, \label{eqn:vol_avg}
\end{equation}
where $L_z$ is the domain height in the $z$-direction. We will occasionally consider mass averaged quantities as well, defined by:
\begin{equation}
    \langle X\rangle_\rho = \frac{\int\rho X\dd{x}\dd{y}\dd{z}}{\int\rho \dd{x}\dd{y}\dd{z}}. \label{eqn:mass_avg}
\end{equation}
We also define $\langle X\rangle_t$ as time averaged quantities:
\begin{equation}
    \langle X\rangle_t = \frac{1}{t_f - t_i}\int_{t_i}^{t_f}X\dd{t'}, \label{eqn:time_avg}
\end{equation}
where $t_i, t_f$ are the initial and final times the average is taken through, to be defined upon usage. 
To assess the effect of gravitation, it is useful to define the Toomre parameter $Q_T$ and the magnetic Toomre parameter $Q_{T,B}$ \citep{Kim_Ostriker-2001} as
\begin{equation}
    Q_T = \frac{\langle c_s^2\rangle_\rho^{1/2}\kappa}{\pi G\Sigma}, \quad Q_{T,B} = \frac{\langle c_s^2 + v_A^2\rangle_\rho^{1/2}\kappa}{\pi G\Sigma}, \label{eqn:Toomre}
\end{equation}
where $v^2_A = B^2/\rho$ is the Alfv\'en speed, $\kappa = \sqrt{2(2-q)}\Omega$ is the epicyclic frequency (note that for Keplerian shear $\kappa = \Omega$). $Q_T$ of the initial profile is related to $Q$ through eq.~\ref{eqn:QT_Q_relation}. For $Q=1$, which is used throughout the study in the initial setup, $Q_T=0.89$. In this study we often have to examine the stresses, as they are important to angular momentum and energy transport in accretion flows. The Reynolds, Maxwell and gravitational stress are defined respectively by
\begin{equation}
    T^R_{r\phi} = \rho v_x\delta v_y,\quad T^M_{r\phi} = -B_x B_y,\quad T^G_{r\phi} = \frac{g_x g_y}{4\pi G}, \label{eqn:stresses}
\end{equation}
where $\delta v_y = v_y + q\Omega x$ denotes the deviation of $v_y$ from the equilibrium flow. The $\alpha$ stresses, which are dimensionless measures of the stress, are defined by dividing the stresses by the gas pressure. The horizontally averaged mid-plane and volume averaged Reynolds, Maxwell and gravitational $\alpha$ parameters are defined by 
\begin{gather}
    \langle\alpha_R\rangle_V = \frac{\langle T^R_{r\phi}\rangle_V}{\langle P_g\rangle_V},\ \langle\alpha_M\rangle_V = \frac{\langle T^M_{r\phi}\rangle_V}{\langle P_g\rangle_V},\ \langle\alpha_G\rangle_V = \frac{\langle T^G_{r\phi}\rangle_V}{\langle P_g\rangle_V}.
    \label{eqn:alpha_stresses}
\end{gather}
Finally, we define $\bar{X}_w$ as the `window' average in the $x,y$-direction:
\begin{equation}
    \bar{X}_w = \frac{1}{w^2}\int^{x+w/2}_{x-w/2}\int^{y+w/2}_{y-w/2} X\dd{x'}\dd{y'}, \label{eqn:window_average}
\end{equation}
where $w$ is some specified `window' width over which this average is taken. Assuming a variable $X$ is expressed as a sum of Fourier modes 
\begin{equation}
    X = \int^{\infty}_{-\infty}\tilde{X}_{k_x,k_y}\cos(k_x x + \phi)\cos(k_y y + \phi')\dd{k_x}\dd{k_y},
\end{equation}
then its window-average
\begin{align}
    \bar{X}_w &= \frac{1}{w^2}\int_{x-w/2}^{x+w/2}\int_{y-w/2}^{y+w/2}\int^{\infty}_{-\infty}\tilde{X}_{k_x,k_y}\cos(k_x x' + \phi) \nonumber \\&\cos(k_y y' + \phi')\dd{k_x}\dd{k_y}\dd{x'}\dd{y'} \\
    &= \int^{\infty}_{-\infty}\int^{\infty}_{-\infty}\tilde{X}_{k_x,k_y}\cos(k_x x+\phi)\cos(k_y y+\phi') \nonumber \\&\frac{\sin(k_x w/2)}{(k_x w/2)}\frac{\sin(k_y w/2)}{(k_y w/2)}\dd{k_x}\dd{k_y},
\end{align}
so contributions of modes with wavenumber $k_x w, k_y w > 1$ is suppressed. This represents a smoothing procedure, or convolution, in the $x,y$-direction. When operating on real data, the upper and lower limits of the integral in eq.~\ref{eqn:window_average} are replaced by $x_\mathrm{max}$, $x_\mathrm{min}$ and $y_\mathrm{max}$, $y_\mathrm{min}$ when the domain limits are exceeded. 

\subsection{Clump identification} \label{subsec:clump_identification}

\begin{figure*}
    \centering
    \includegraphics[width=0.4\textwidth]{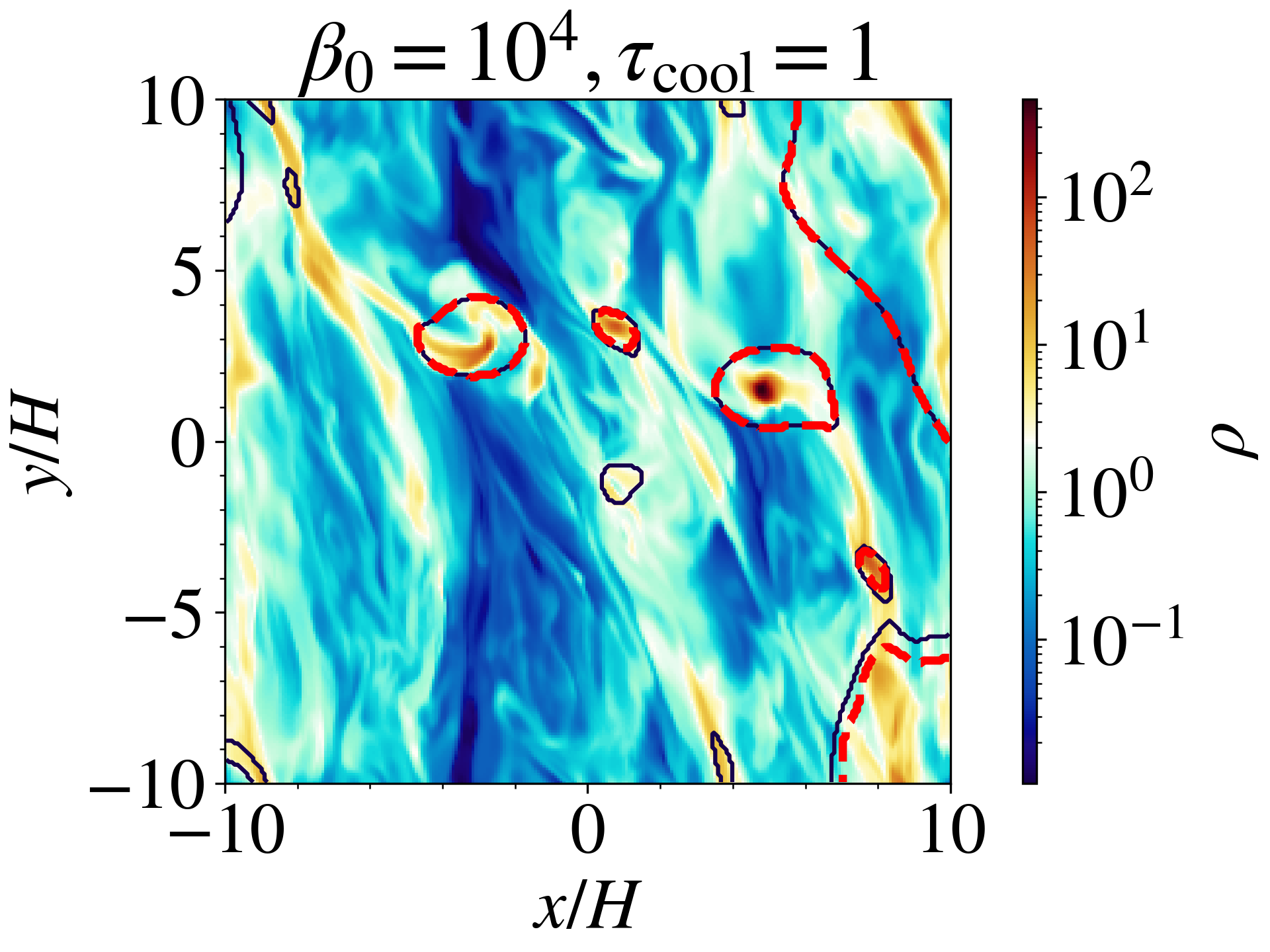}
    \includegraphics[width=0.4\textwidth]{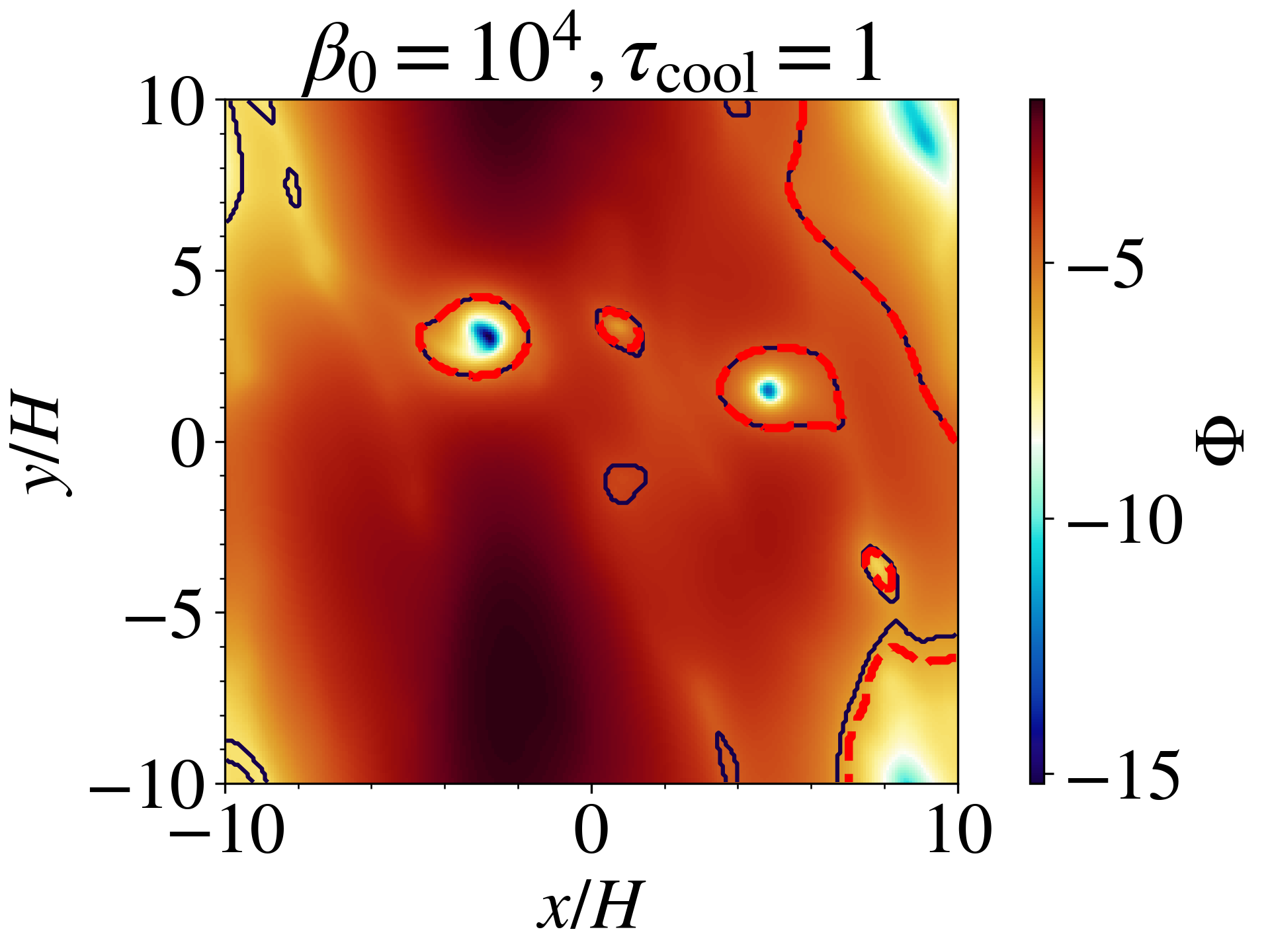}
    \caption{Clump identification using the extended GRID algorithm, with black-solid and red-dashed contours denoting GBR and TBR, respectively, overlaid on a mid-plane density (left) and gravitational potential (right) snapshot at $\Omega t=200$ for the $\beta_0=10^4,\tau_\mathrm{cool}=1$ case.}
    \label{fig:frag_example}
\end{figure*}

To assess disk fragmentation quantitatively, we follow the procedure outlined in \citet{Chen_etal-2023}, using a 3D extension of the GRID-core algorithm \citep{Gong_Ostriker-2011,Mao_etal-2020}. Briefly, for each local minimum in the gravitational potential field $\Phi$, we first identify the closed contour with the largest value of $\Phi = \Phi_\mathrm{max}$, which contains no other local minima. This region, which is referred to as a \textit{Gravitational Binding Region} (GBR), satisfies
\begin{equation}
    \int_\mathrm{GBR} \rho\qty(\Phi - \Phi_\mathrm{max})\dd{V} < 0. \label{eqn:GBR_condition}
\end{equation}
Within a GBR, we narrow the region down to cells where
\begin{equation}
    \int_\mathrm{TBR} E_\mathrm{kin} + E_\mathrm{th} + E_B + \rho\qty(\Phi - \Phi_\mathrm{max}) \dd{V} < 0 \label{eqn:frag_condition}
\end{equation}
is satisfied when integrated up to some sub-contour $\Phi'<\Phi_\mathrm{max}$, for which the gas is gravitationally bound against turbulent, thermal and magnetic pressure. Such a region is called a \textit{Total Binding Region} (TBR). Here $E_\mathrm{kin}$ is the kinetic energy density relative to the center of mass ($E_\mathrm{kin}=0.5\rho\delta \vb{v}^2_\mathrm{CM}$, where $\delta\vb{v}_\mathrm{CM}= \vb{v}-\vb{v}_\mathrm{CM}$ and $\vb{v}_\mathrm{CM}$ is the center of mass velocity of the clump), $E_\mathrm{th}$ is the thermal energy density, and $E_B$ is the magnetic energy density. The whole GBR can be a TBR if the gravitational well is deep enough. In Fig.~\ref{fig:frag_example} we show an example of clump identification using this algorithm, where the black-solid and red-dashed contours denote GBR and TBR, respectively, overlaid on the mid-plane density and gravitational potential snapshot at $\Omega t=200$. 

\section{Theoretical considerations} \label{sec:theory}

Magnetic field is often viewed as a stabilizing agent against gravitational collapse due to magnetic pressure support, counteracting gravity in linear stability analysis of the Jeans instability. On the other hand, in a rotating and shearing disk, magnetic tension can reduce the stabilizing influence of Coriolis force, leading to \textit{enhanced} gravitational instability \citep{Lynden-Bell-1966,Elmegreen-1987,Gammie-1996,Kim_Ostriker-2001}. These competing effects of magnetic field raise the  question of whether one effect would dominate the other in a realistic scenario, i.e. whether magnetic field \textit{promotes} or \textit{suppresses} gravitational instability or fragmentation. We find that this question has not been fully addressed in an MRI-turbulent disk. Past studies have focused on the mutual effects of gravitoturbulence and MRI in a mildly cooling disk \citep{Lohnert_Peeters-2023} and gravitoturbulence as a dynamo mechanism  \citep[especially in the context of a protoplanetary disk, where the gas may not be fully ionized and MRI may not be active, e.g.][]{Riols-Latter-2019,Riols_etal-2021,Bethune_Latter-2022,Deng_etal-2020}. We find it important to address the fragmentation aspect of this problem in order to better understand accretion flows and production of gravitational wave sources in AGN disks. In this section, we outline how magnetic tension could enhance gravitational instability in a rotating, shearing disk, lay out possible reasons why this effect has received little attention in the past, and offer new insights into why we reconsidered it in this study. The discussion in this section will motivate our specific focus in the results section (\S\ref{sec:results}).

\subsection{Magnetic destabilization of gravitational modes} \label{subsec:magneto_gravito}

\begin{figure}
    \centering
    \includegraphics[width=0.45\textwidth]{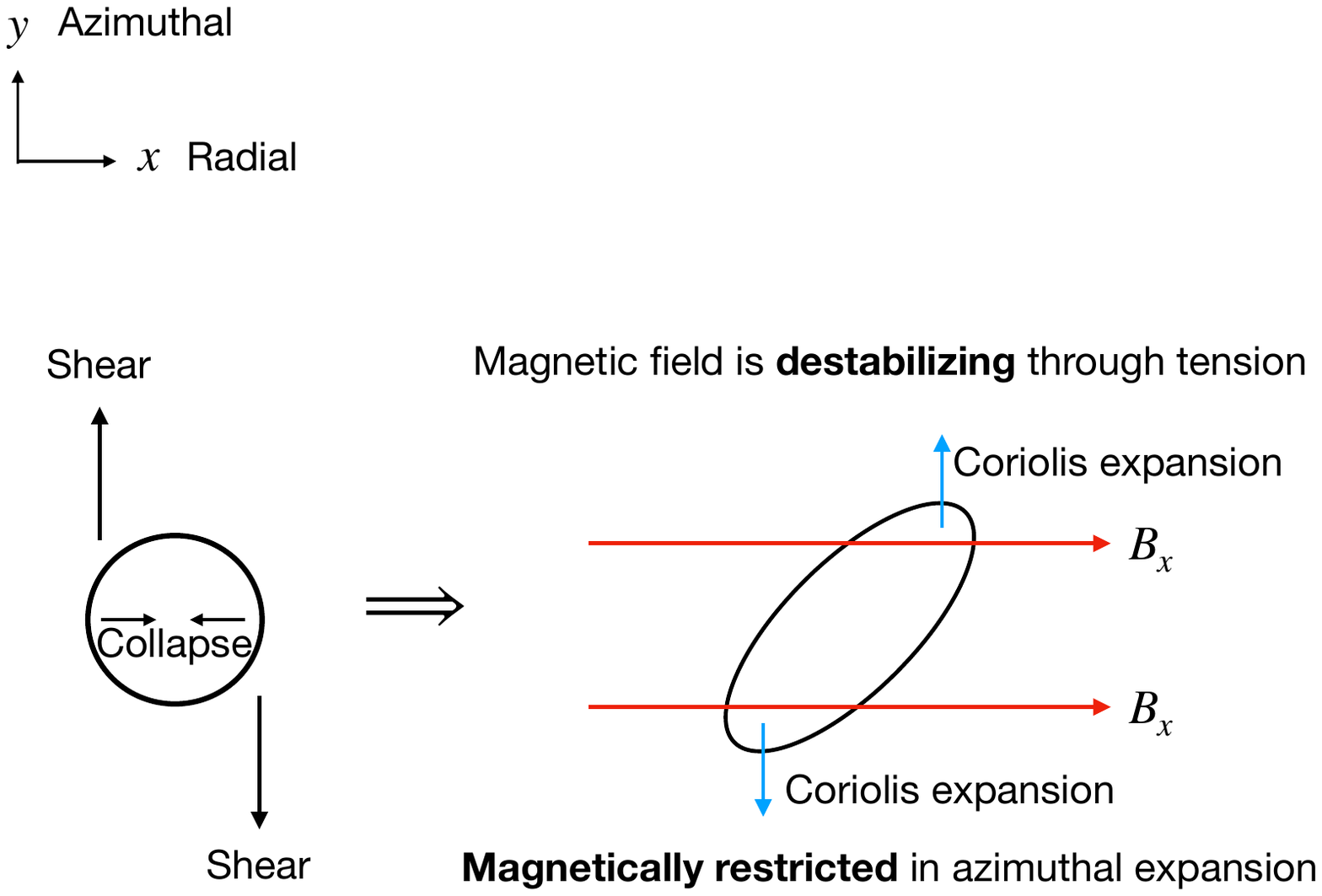}
    \caption{Intuition for the Coriolis-Restricted-Magneto-Gravitational (CRMG) instability. An overdense blob undergoing gravitational collapse experiences Coriolis forces which tend to expand it. This expansion is restricted by magnetic tension if a strong radial field is present, which promotes further collapse.}
    \label{fig:CRMG_intuition}
\end{figure}

An overdense blob generated from random motion and undergoing gravitational collapse experiences Coriolis forces which tend to expand it. This expansion is restricted by magnetic tension if a strong radial field is present, which promotes further collapse (Fig.~\ref{fig:CRMG_intuition}). We call this mechanism by which magnetic fields destabilize GI the Coriolis-Restricted-Magneto-Gravitational (CRMG) instability.

Following the local WKB stability analysis carried out in Appendix \ref{app:crmg}, axisymmetric perturbations in a rotating, razor-thin disk subjected to an in-disk magnetic field $\vb{B} = B_x\vu{x} + B_y\vu{y}$ exhibit characteristic modes with the dispersion relation \citep{Gammie-1996}
\begin{gather}
    \omega^4 + \qty[2\qty(q-2)\Omega^2 - k_x^2\qty(c_s^2 + v_A^2) + 2\pi G\Sigma_0\abs{k_x}]\omega^2 \nonumber \\\quad+ iq\Omega\omega k_x^2 v_{Ax} v_{Ay} + k_x^2 v_{Ax}^2\qty(c_s^2 - 2\pi G\Sigma_0\abs{k_x}) = 0, \label{eqn:crmg_dispersion}
\end{gather}
where $\vb{v}_A = \vb{B}/\sqrt{\Sigma_0}$ is the Alfv\'en velocity, $c_s = \sqrt{\gamma P_g/\Sigma_0}$ is the sound speed, $k_x$ is the axisymmetric wavenumber, $q$ is the shearing rate, and $\Sigma_0$ is the surface density. In this discussion, we focus on axisymmetric modes due to the simplicity of the analysis. Stability analysis can be performed on non-axisymmetric modes as well (see Appendix \ref{app:crmg}), but they generally do not admit WKB solutions due to fast variations of the non-axisymmetric wavevector. While this analysis assumes a razor-thin disk, in practice disks always have finite thickness, and in magnetically elevated disk scenarios it may be more appropriate to assume the background is approximately uniform instead of razor-thin at the mid-plane. In this case, the dispersion relation eq.~\ref{eqn:crmg_dispersion} is modified by replacing the $2\pi G\Sigma_0|k_x|$ terms by $4\pi G\rho_0$ (eq.~\ref{eqn:crmg_dispersion_uniform}). 
First consider the limit in which there is no rotation ($\Omega=0$): the dispersion relation eq.~\ref{eqn:crmg_dispersion} reduces to
\begin{equation}
    \omega^2 = k_x^2 v^2_{Ax},\quad \omega^2 = k_x^2 \qty(c_s^2 + v^2_{Ay}) - 2\pi G\abs{k_x}\Sigma_0,
\end{equation}
the former which are simply Alfv\'en waves and the latter magneto-Jeans type modes\footnote{Here we define magneto-Jeans modes as wave modes that follow the dispersion relation $\omega^2 = k_x^2(c_s^2 + v_A^2) + 2(2-q)^2\Omega^2 - 2\pi G \abs{k_x}\Sigma_0$ or close variations of this, in which magnetic field is a stabilizing influence. This definition is different from that used in \citet{Kim_Ostriker-2001}. If the background is uniform instead of razor-thin, the dispersion relation becomes $\omega^2 = k_x^2(c_s^2 + v_A^2) + 2(2-q)^2\Omega^2 - 4\pi G\rho_0$ instead, i.e. replacing the $2\pi G\abs{k_x}\Sigma_0$ term by $4\pi G\rho_0$.}. Magnetic field is a stabilizing agent or simply a medium of oscillatory wave propagation in this case. In the other limit, in which there is no radial field ($v_{Ax} = 0$), eq.~\ref{eqn:crmg_dispersion} reduces to
\begin{equation}
    \omega^2 = k_x^2 \qty(c_s^2 + v^2_{Ay}) + 2\qty(2-q)\Omega^2 - 2\pi G \abs{k_x}\Sigma_0, \label{eqn:mji_dispersion}
\end{equation}
which again is a magneto-Jeans type mode with rotation as an additional stabilizing influence, again stabilized by magnetic field. These two limits are uninteresting for the current discussion, due to the stabilizing effect of magnetic fields. It is more interesting to consider parameters away from these limits, for which magnetic fields are destabilizing. Specifically, if there is rotation and the magnetic field has a radial component ($B_x/B\neq 0$), one can solve eq.~\ref{eqn:crmg_dispersion} numerically and identify the most unstable root. The growth rate depends on 5 dimensionless parameters: the axisymmetric wavenumber $k_x V_A/\Omega$, the Toomre parameter $Q_T = c_s\kappa/\pi G \Sigma_0$, the in-disk plasma beta $\beta_\mathrm{disk} = 2P_g/(B_x^2 + B_y^2)$, orientation of the in-disk magnetic field $b_x = B_x/(B_x^2 + B_y^2)^{1/2}$, and the shearing rate $q$. 

\begin{figure}
    \centering
    \includegraphics[width=0.23\textwidth]{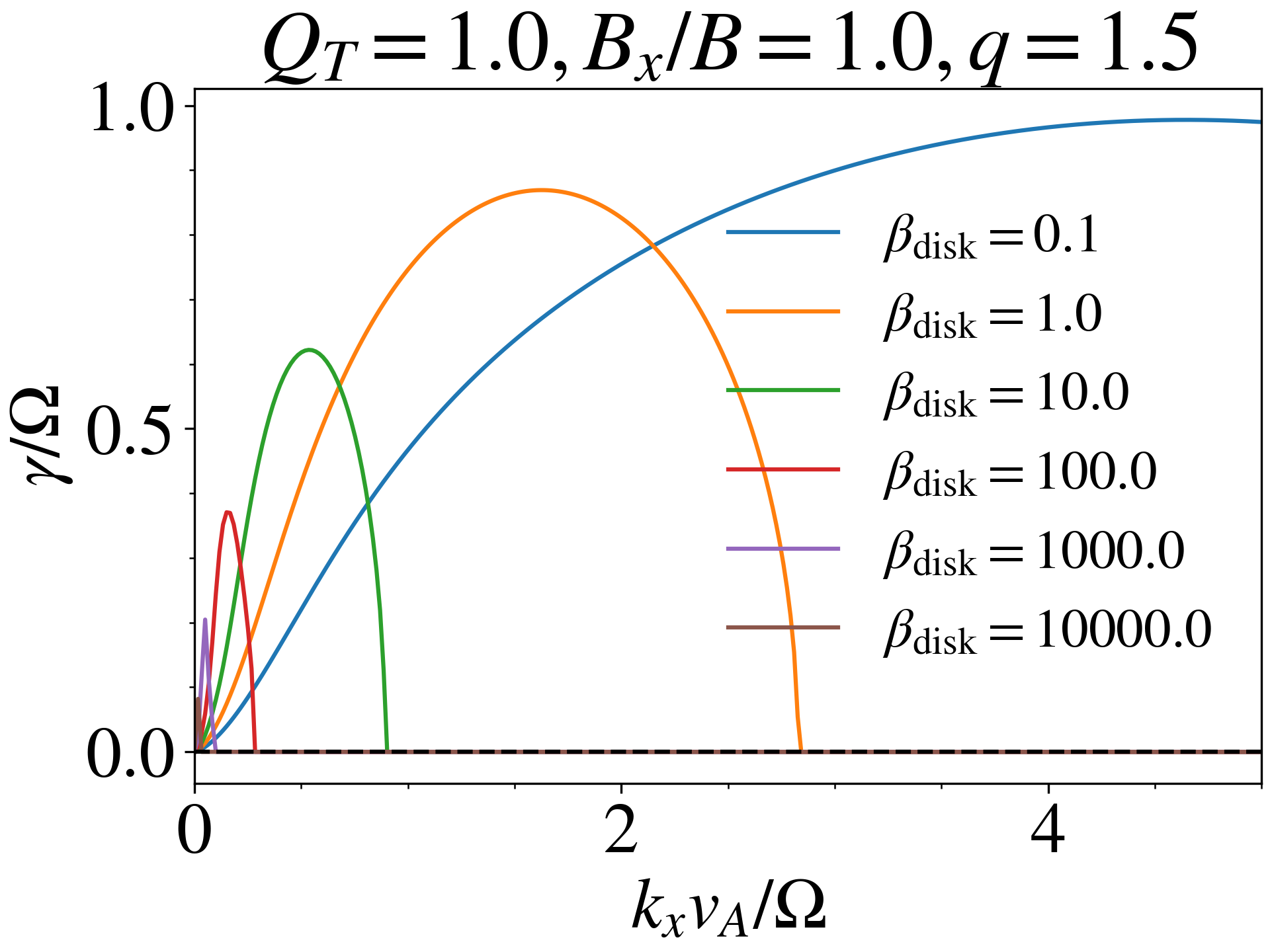}
    \includegraphics[width=0.23\textwidth]{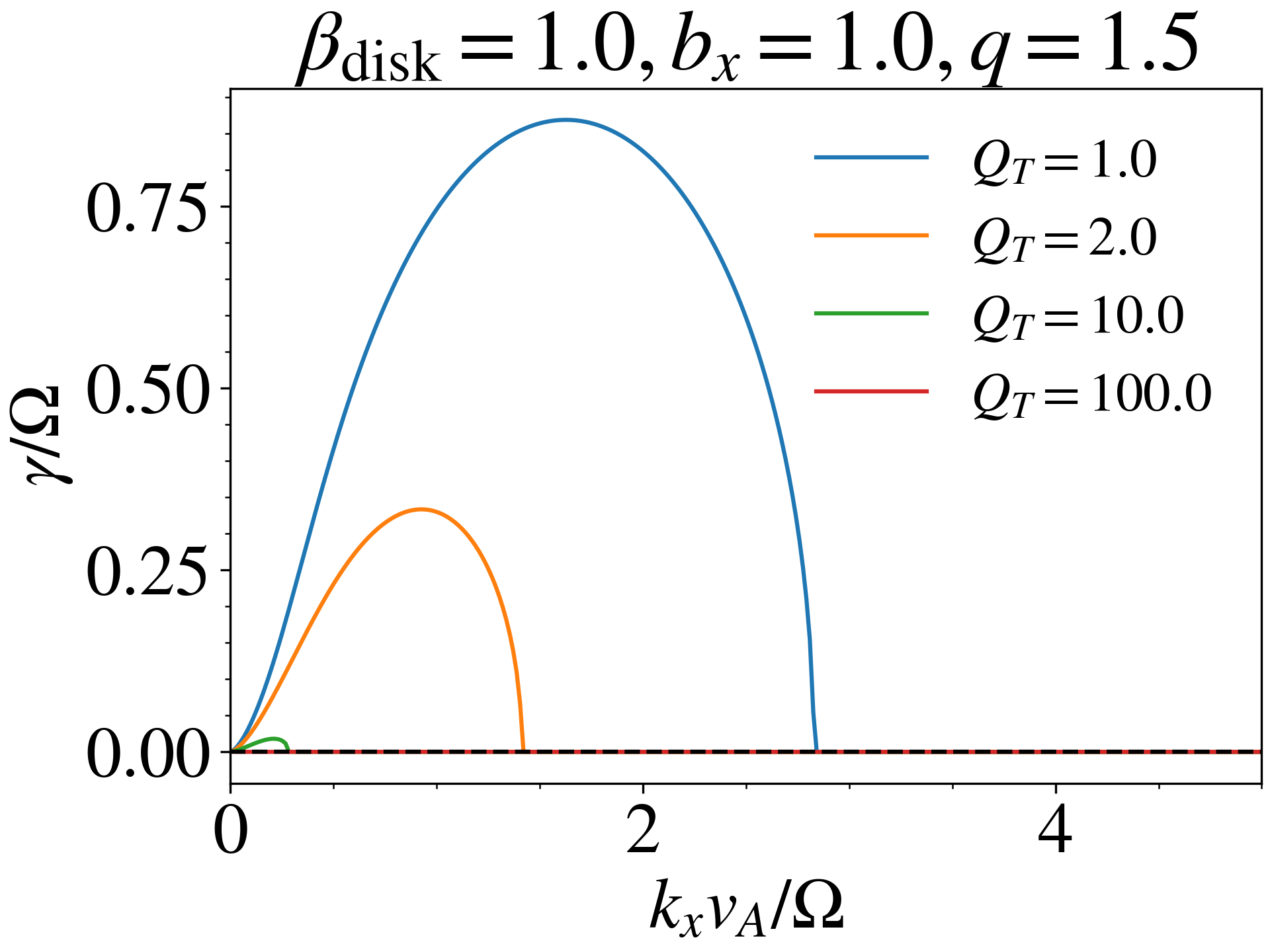} \\
    \includegraphics[width=0.23\textwidth]{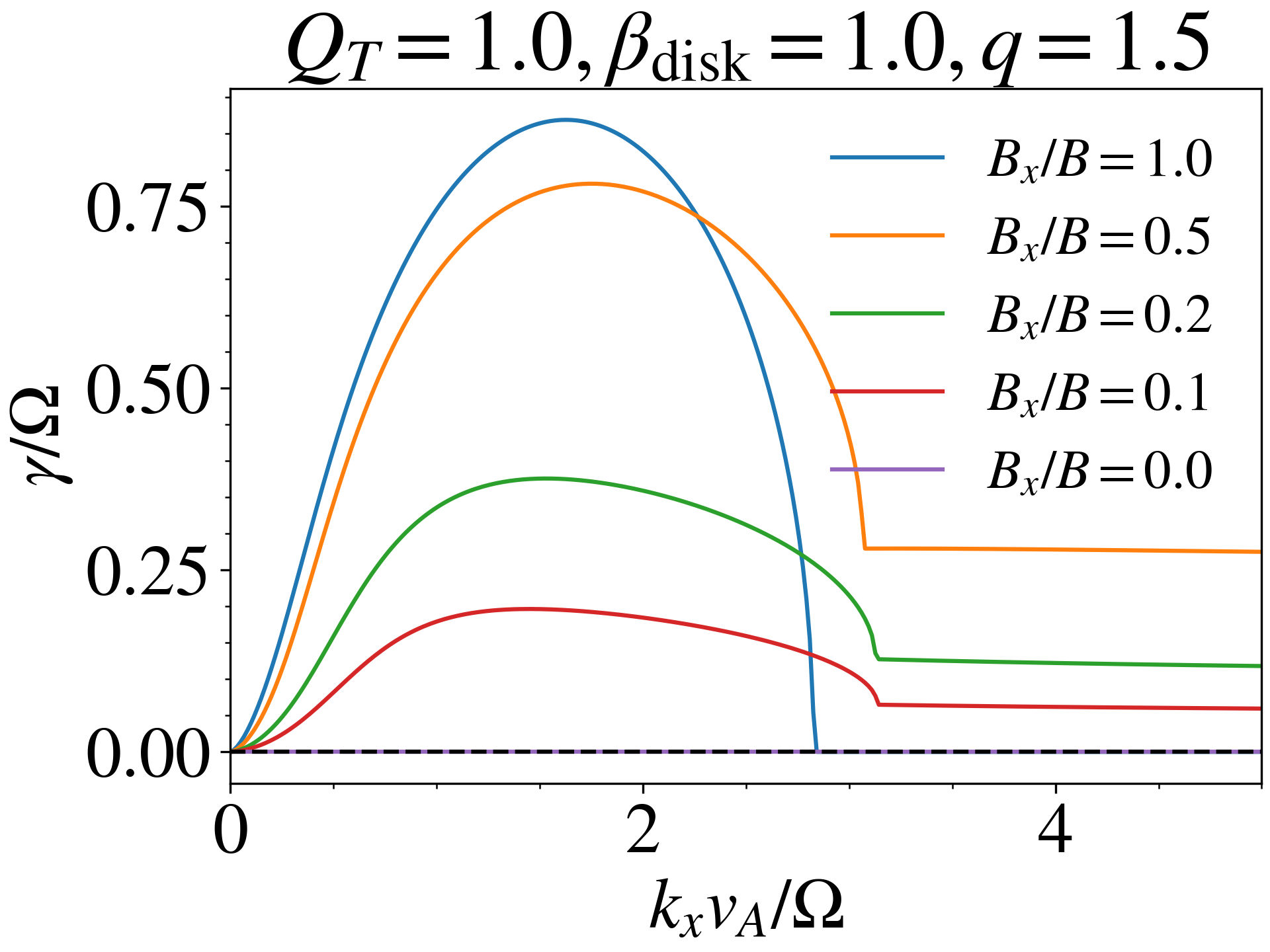}
    \includegraphics[width=0.23\textwidth]{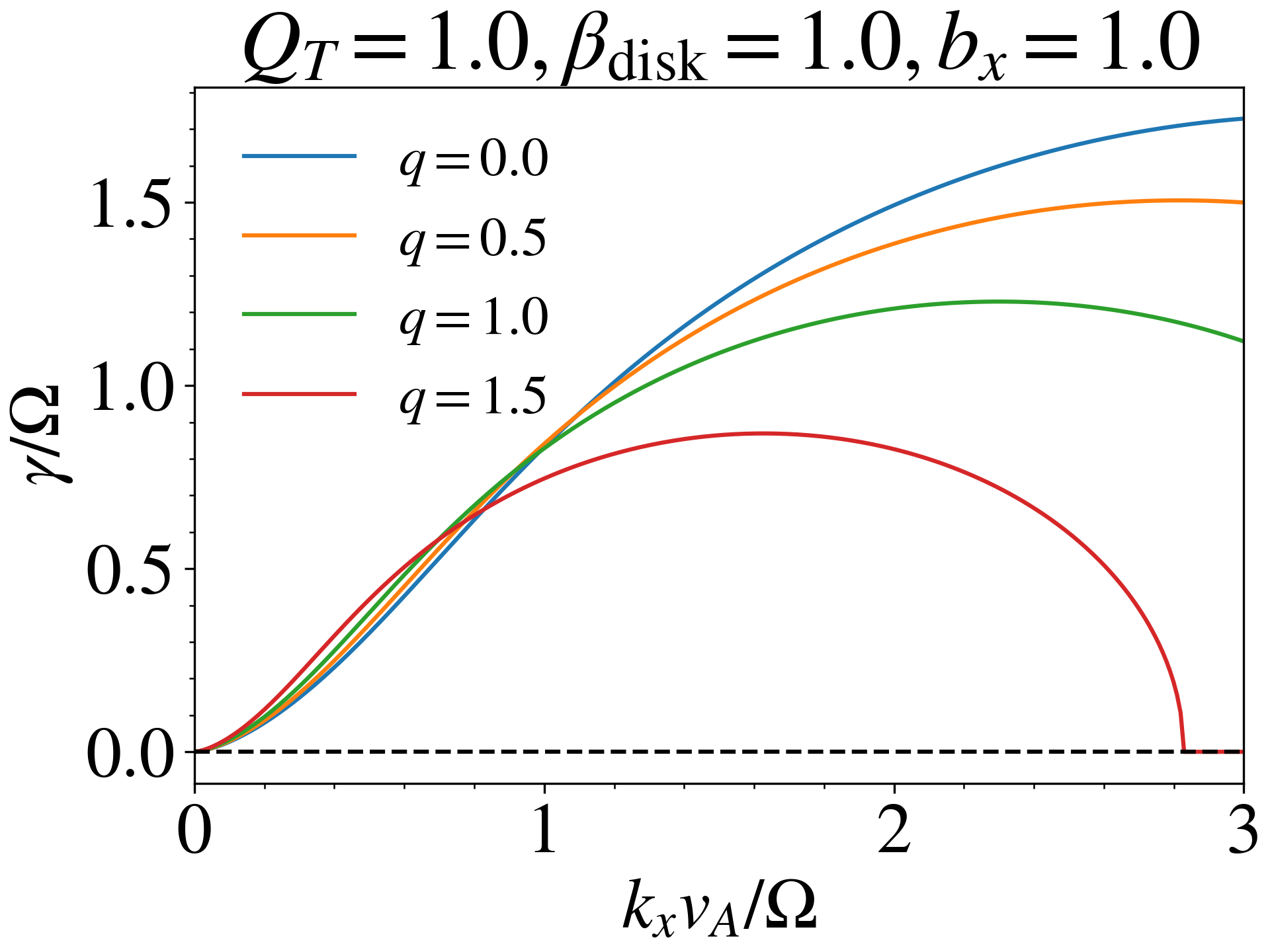}
    \caption{Growth rates $\gamma$ (normalized by $\Omega$) of axisymmetric perturbations ($k_y=0$) as a function of $k_x v_A/\Omega$ for various $\beta_\mathrm{disk} = 2P_g/(B_x^2+B_y^2)$ (top left), Toomre parameters $Q_T = c_s\Omega/\pi G\Sigma_0$ (top right), orientations of the magnetic field $b_x = B_x/(B_x^2 + B_y^2)^{1/2}$ (lower left), and shearing rates $q$ (lower right) while the other variables are kept fixed at values listed at the top of each panel.}
    \label{fig:axisymmetric_growth}
\end{figure}

In Fig.~\ref{fig:axisymmetric_growth} we show examples of the growth rates $\gamma/\Omega$ of such modes as a function of the axisymmetric wavenumber $k_x v_A/\Omega$ for various in-disk plasma betas $\beta_\mathrm{disk}$ (top left), Toomre parameters $Q_T$ (top right), magnetic field orientations $b_x$ (lower left), and shearing rates $q$ (lower right) while fixing the other parameters. We observe clearly in the top left panel that CRMG modes are more unstable in stronger magnetic fields for the same $Q_T$, both in terms of the maximum growth rate and the range of wavenumbers that are unstable. When maximally unstable, this instability acts fast, on the timescale of $\sim\Omega^{-1}$. The top right panel shows that CRMG modes are more unstable at lower $Q_T$, cementing it as a gravitational mode. However, unlike Toomre modes, which require $Q_T<1$ (or even smaller for strong magnetic fields) to be unstable, the CRMG instability criteria are more lenient on $Q_T$, as modes can be unstable even for $Q_T>1$, albeit with smaller growth rates and more restrictive ranges of unstable wavenumbers. The orientation of the magnetic field also impacts the growth rate --- as shown in the lower left panel, the instability appears weaker as the magnetic field veers off from the radial direction into the azimuthal, with the instability subdued when the field is completely azimuthal, at right angles to the axisymmetic wavevector\footnote{Technically, axisymmetric modes can be unstable even when the field is completely azimuthal. One simply recovers a Magneto-Jeans type mode (eq.~\ref{eqn:mji_dispersion}), in which the instability condition is $Q_{T,B}<1$ \citep{Kim_Ostriker-2001}. This is why in the lower left panel of Fig.~\ref{fig:axisymmetric_growth}, where $Q_T=1 < Q_{T,B}$, the system is stable when $B_x/B_0=0$. Magnetic field acts as a stabilizing agent in this case.}. This point is quite important, and we will refer to it again in the following discussion. Finally, the lower right panels shows that CRMG modes are more unstable for lower shearing rates due to reduced Coriolis stabilization.

In short, axisymmetric perturbations in a self-gravitating flow subjected to shear and an in-disk magnetic field become \textit{more} unstable with respect to a stronger field when there is a radial component in the background field, for constant Toomre parameter $Q_T$. When the field is completely azimuthal, the effect of magnetic field is \textit{stabilizing} and one recovers the magneto-Jeans type mode. We relegate more detailed discussion of the wavelength dependence of the CRMG modes, comparision between the CRMG and the magneto-Jeans type mode, and numerical verification of the growth rates to Appendix~\ref{app:crmg_wavelength}-\ref{app:numerical_crmg}. 

\begin{figure}
    \centering
    \includegraphics[width=0.45\textwidth]{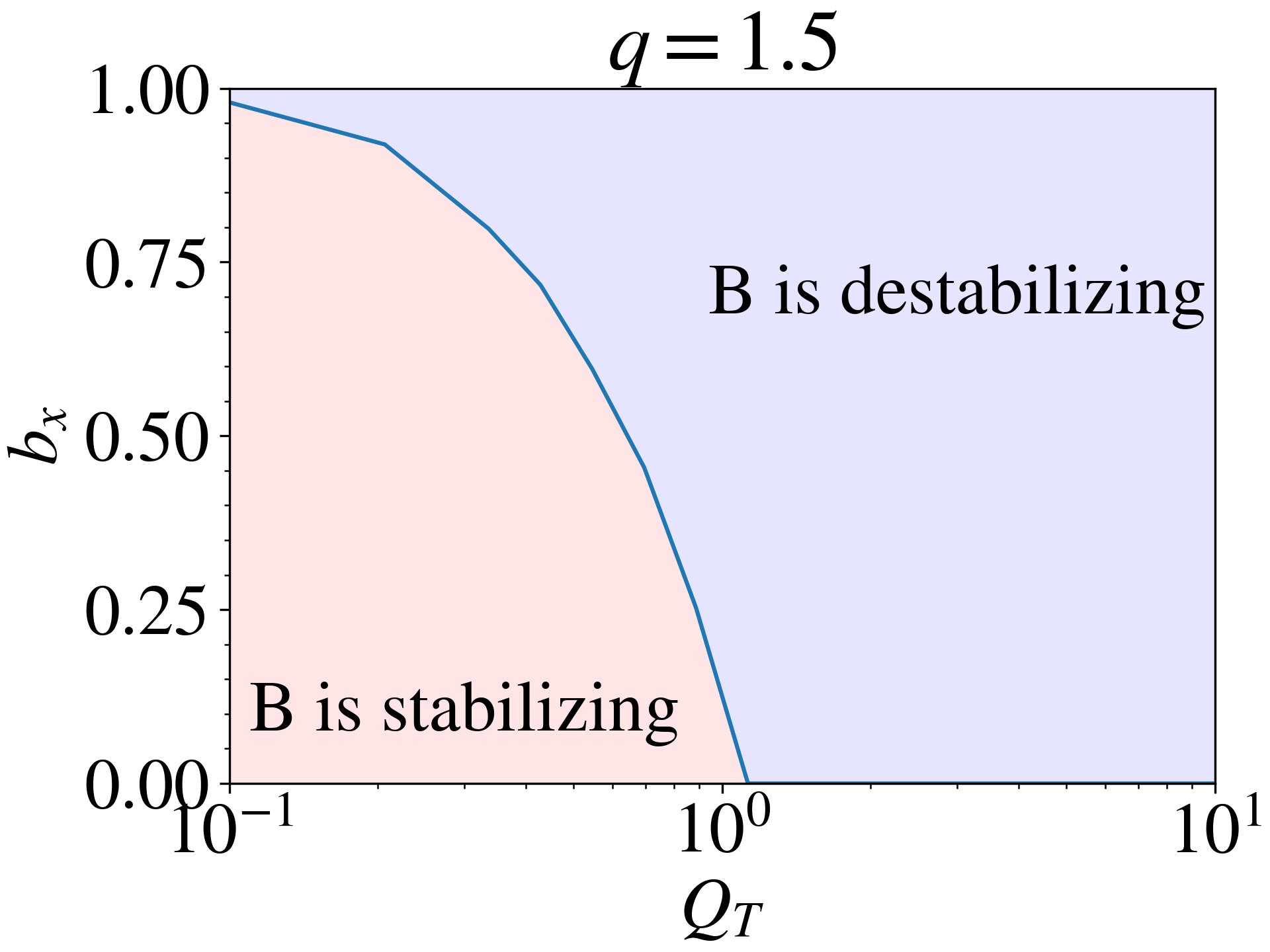}
    \caption{Plot of $b_{x,\mathrm{thres}}$ against $Q$, demarcating region where the magnetic field is destabilizing (above the curve, shaded in blue) and vice versa. }
    \label{fig:B_transition}
\end{figure}

We pointed out that magnetic field destabilizes axisymmetric CRMG modes when the field is radial, becoming stabilizing when the field is toroidal. Mathematically, this means that when $b_x=B_x/(B_x^2+B_y^2)^{1/2}$, the relative contribution of the radial component to the in-disk field, is close to unity, the growth rate increases when $\beta_\mathrm{disk}$ decreases. In contrast, if the field is largely toroidal, $b_x\approx 0$, the growth rate decreases when $\beta_\mathrm{disk}$ decreases. The transition from destabilization to stabilization occurs at some intermediate value of $b_x$, which we call $b_{x,\mathrm{thres}}$. In Fig.~\ref{fig:B_transition} we plot $b_{x,\mathrm{thres}}$ as a function of $Q_T$ for $q=1.5$. The area under the curve denotes values of $b_x$ for which the magnetic field is stabilizing, and vice versa for the area above the curve. 
Studies have found that $Q_T$ maintains a value slightly above 1 in the nonlinear saturation stage when GI/fragmentation occurs \citep{Gammie-2001}. On the other hand, $b_x$ can take on an appreciable value in MRI simulations with strong vertical fluxes. These observations suggest magnetic destabilization of gravitational modes is entirely possible.

Despite the promising signs shown by this WKB analysis that strong magnetic fields can enhance gravitational instability, \citet{Gammie-1996} argued against its validity based on the fact that radial magnetic field will be toroidally directed by shear on the timescale $(q\Omega)^{-1}$. Thus, magnetic fields will destabilize axisymmetric gravitational modes only transiently, and become stabilizing long after $(q\Omega)^{-1}$, thus bringing this WKB analysis into question. An extension of this analysis shows that it is possible for non-axisymmetric modes to be destabilized by a toroidal magnetic field\footnote{The stability analysis involving non-axisymmetric modes is more complicated as the wavevector is time-dependent. Nonetheless, one can derive a set of equations \citep[called the `shearing sheet equations' by][]{Goldreich_Lynden-Bell-1965,Julian_Toomre-1966,Elmegreen-1987,Gammie-1996,Kim_Ostriker-2001} that governs the amplitude of the perturbation, which can be solved in the same manner as one would solve an initial value problem. To characterize the strength of the instability, one then considers the maximum amplitude the density perturbations can grow to (a.k.a the `responsiveness').}. However, as a non-axisymmetric wavevector will eventually tend towards axisymmetry due to shear on the timescale $(q\Omega)^{-1}$ --- $\vb{k} =  (k_x + q\Omega t k_y)\vu{x} + k_y\vu{y}$, where $k_x,k_y$ are the initial axisymmetric and non-axisymmetric wavenumbers (See Appendix \ref{app:crmg} for a visual explanation.) --- non-axisymmetric modes are only transiently unstable to a toroidal field\footnote{Similar to axisymmetric CRMG modes, transient growth of non-axisymmetric modes in the presence of toroidal fields is faster for stronger fields \citep{Elmegreen-1987}. The problem is that this period of growth may be too short for clumps to deepen their gravitational potential well enough to become bound.}. Thus we have a scenario where the wavevector becomes axisymmetric on the same (fast) timescale as the field becoming toroidal, which is the configuration for which the magnetic field is a stabilizing influence. 

We argue that this WKB analysis of axisymmetric modes can still, to a good extent, capture the growth structure in realistic  scenarios based on the fact that radial magnetic fields lost to the toroidal direction due to shear can be continuously replenished by the MRI-dynamo on the dynamical timescale, thus providing, in the mean-field sense, a roughly time-independent background field. In the weak vertical field case, the radial field is generally much weaker than the toroidal field. However, in the strong ($\beta \lesssim 10^2$) vertical field case, it has been shown that a relatively strong radial field with sustained polarity can be generated \citep{Salvesen-etal-2016}. We will characterize the strength and time-steadiness of this field in our simulations. The question is, can the MRI-dynamo produce a radial field strong and time-steady enough to trigger the CRMG instability? In the strong field limit it is likely that nonlinear effects such as magnetic elevation will be active --- would such effects inhibit fragmentation? We will clarify these issues in the following discussion. We note that a recent study by \citet{Kubli_etal-2023} found that significantly smaller clumps were formed when a magnetic field, amplified and maintained by a gravitoturbulent dynamo, was present. They noticed that most of the clumps were formed in low shear (small $q$) regions, where the wavevector and background magnetic field change in direction over a much longer time and the problem suggested by \citet{Gammie-1996} is assuaged. However, their focus was on protoplanetary disks, where MRI is less active and resistivity effects are strong. The gas in AGN disks tends to be more ionized within $\sim 10^3r_g$, where MRI plays an important role in a dynamo. The shear parameter is most likely Keplerian ($q=1.5$) at radii where self-gravity is important. Even if $q$ were small at some distinct locations, it is unclear how the gas can be magnetized as MRI is likely suppressed for small $q$. Therefore, conditions in AGN disks could be completely different from protoplanetary disks, and our conclusions regarding the role of magnetic field could be different from \citet{Kubli_etal-2023}.

\section{Results} \label{sec:results}

\begin{table*}
    \centering
    \begin{tabular}{c|c|c|c|c|c|c|c}
        $\beta_0$ & domain & grid & $\langle Q_y\rangle_t$ & $\langle Q_z\rangle_t$ & $\langle\langle\alpha_M\rangle_V\rangle_t$ & $\langle\langle\alpha_R\rangle_V\rangle_t$ & $\langle\langle\alpha_G\rangle_V\rangle_t$ \\
        \hline
        $10$ & $20 H\times 40H\times 24H$ & $256\times512\times384$ & 482 & 146 & 3.28 & 0.788 & 0.00125 \\
        $10^2$ & $20 H\times 40H\times 24H$ & $256\times512\times384$ & 391 & 88.2 & 1.34 & 0.305 & 0.00111 \\
        $10^3$ & $20 H\times 20H\times 24H$ & $256\times256\times512$ & 207 & 35.8 & 1.15 & 0.224 & 0.0124 \\
        $10^4$ & $20 H\times 20H\times 24H$ & $256\times256\times512$ & 78.6 & 24.8 & 0.461 & 0.302 & 0.132 \\
        $10^5$ & $20 H\times 20H\times 24H$ & $256\times256\times512$ & 82.5 & 28.7 & 0.380 & 0.320 & 0.132 
    \end{tabular}
    \caption{Simulation parameters and transport quantities of the different simulations. $\langle\cdot\rangle_t$ denotes time-averaged quantities taken over $\Omega t=100-300$, except for the $\beta_0=10^5$ case, which is taken over $\Omega t=250-450$. $Q_y, Q_z$ are the quality factors in the $y,z$-direction, $Q_y=2\pi v_{A,y}/\Omega\Delta y$, $Q_z=2\pi v_{A,z}/\Omega\Delta z$.}
    \label{tab:simulation_table}
\end{table*}

In this section, we present results of our disk simulations. First, we outline general properties of the disk structure and discuss the $\alpha$ stresses (Reynolds, Maxwell, gravitational) found for simulations with different values of $\beta_0$. To clarify the role of magnetic fields in the gravitational stability of the system, we isolate various contributions to the magnetic budget and discuss the magnetic field structure of the disk, focusing on the in-disk field strength, orientation and time-steadiness as motivated by our discussion of the CRMG instability in \S\ref{sec:theory}. Having laid out the magnetic properties of the disk, we then discuss fragmentation in relation to the field structure and other disk parameters (e.g. the Toomre parameters). Using the dispersion relation eq.~\ref{eqn:crmg_dispersion}, we show quantitatively how different aspects of the magnetic field conspire to promote or suppress fragmentation. In Table \ref{tab:simulation_table} we list the simulation parameters and transport quantities of the cases explored in this study.

\subsection{Disk structure} \label{subsec:sg_disk_structure}

\begin{figure*}
    \centering
    \includegraphics[width=0.31\textwidth]{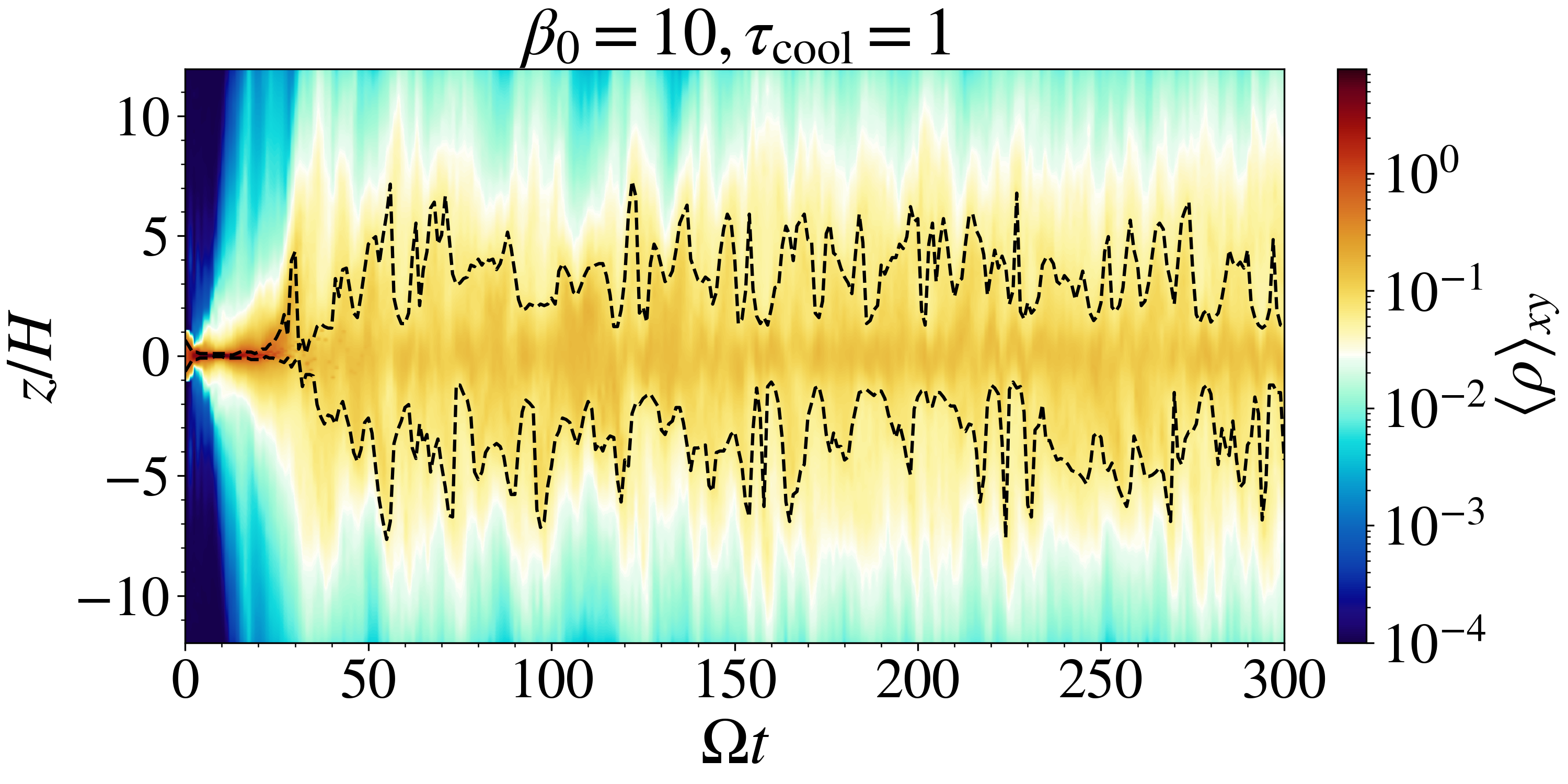}
    \includegraphics[width=0.31\textwidth]{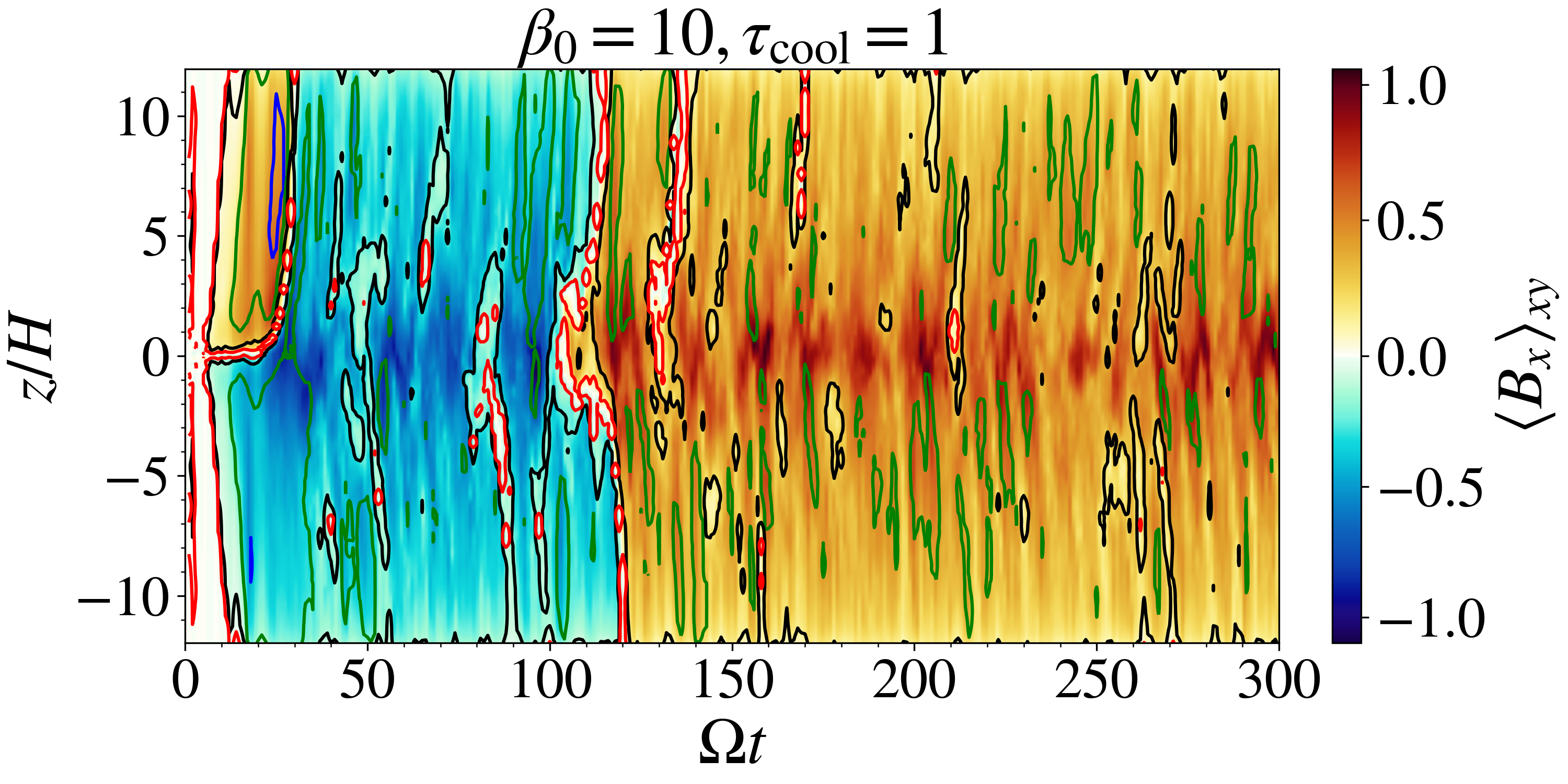}
    \includegraphics[width=0.31\textwidth]{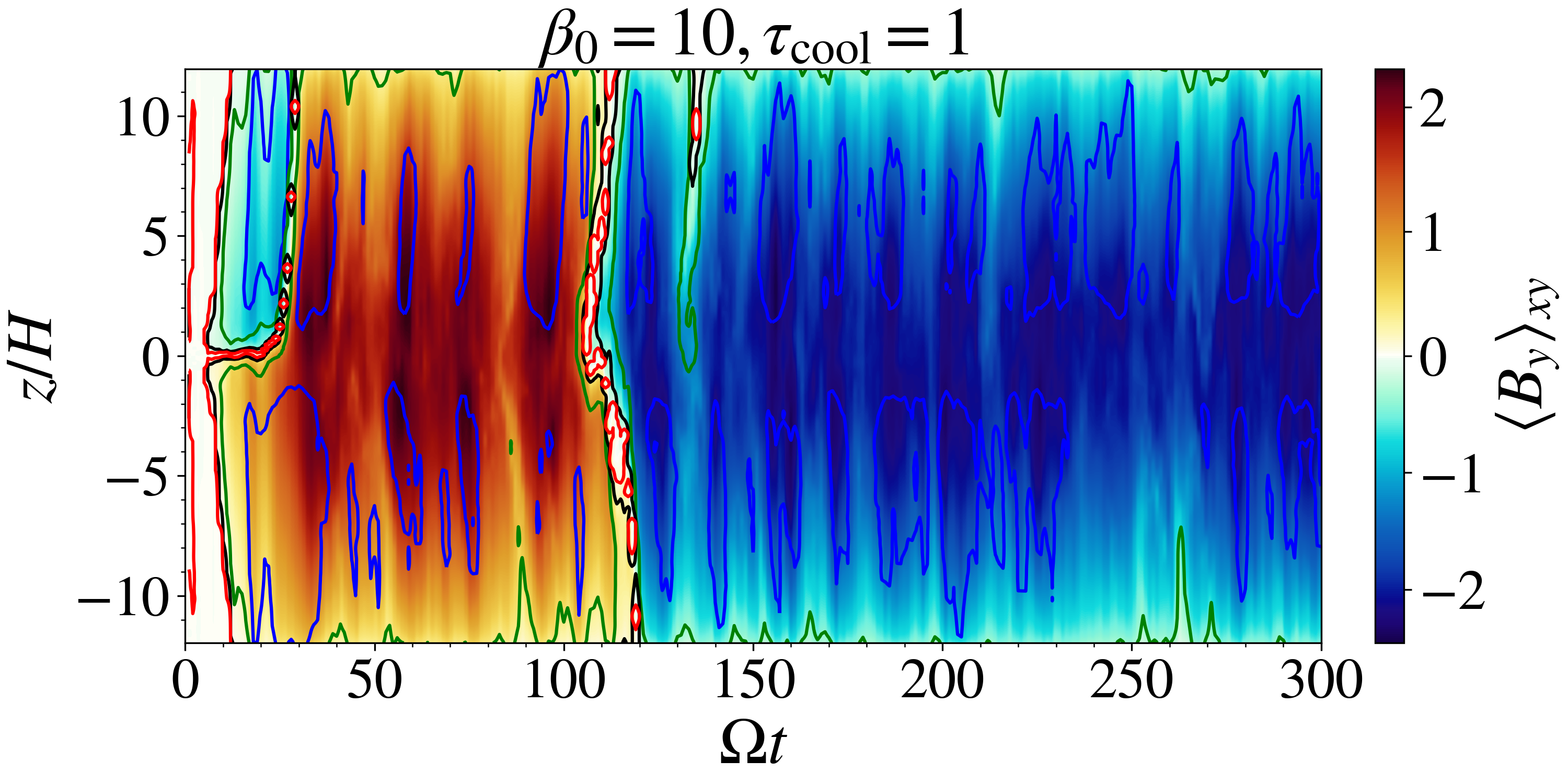} \\
    \includegraphics[width=0.31\textwidth]{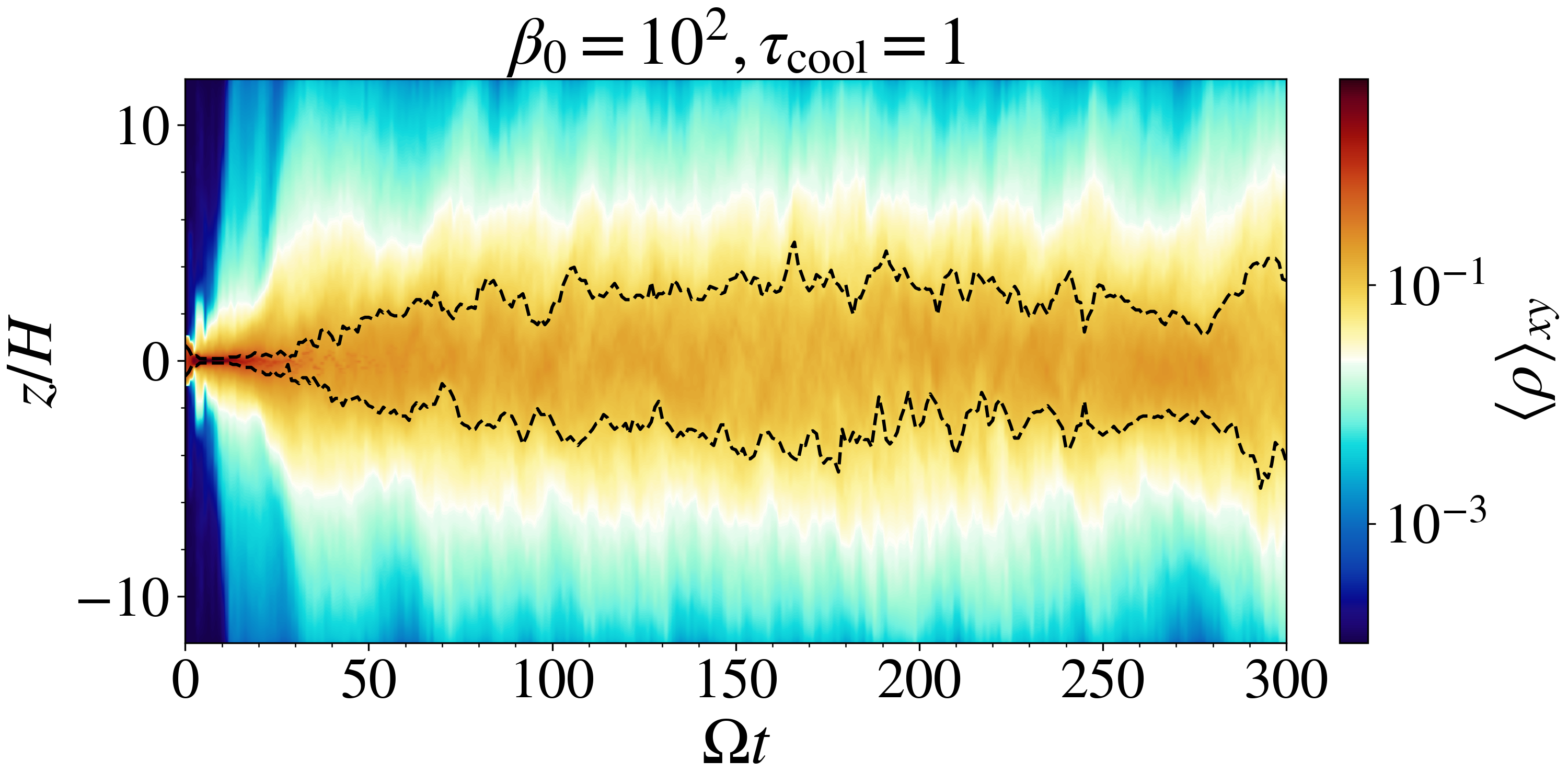}
    \includegraphics[width=0.31\textwidth]{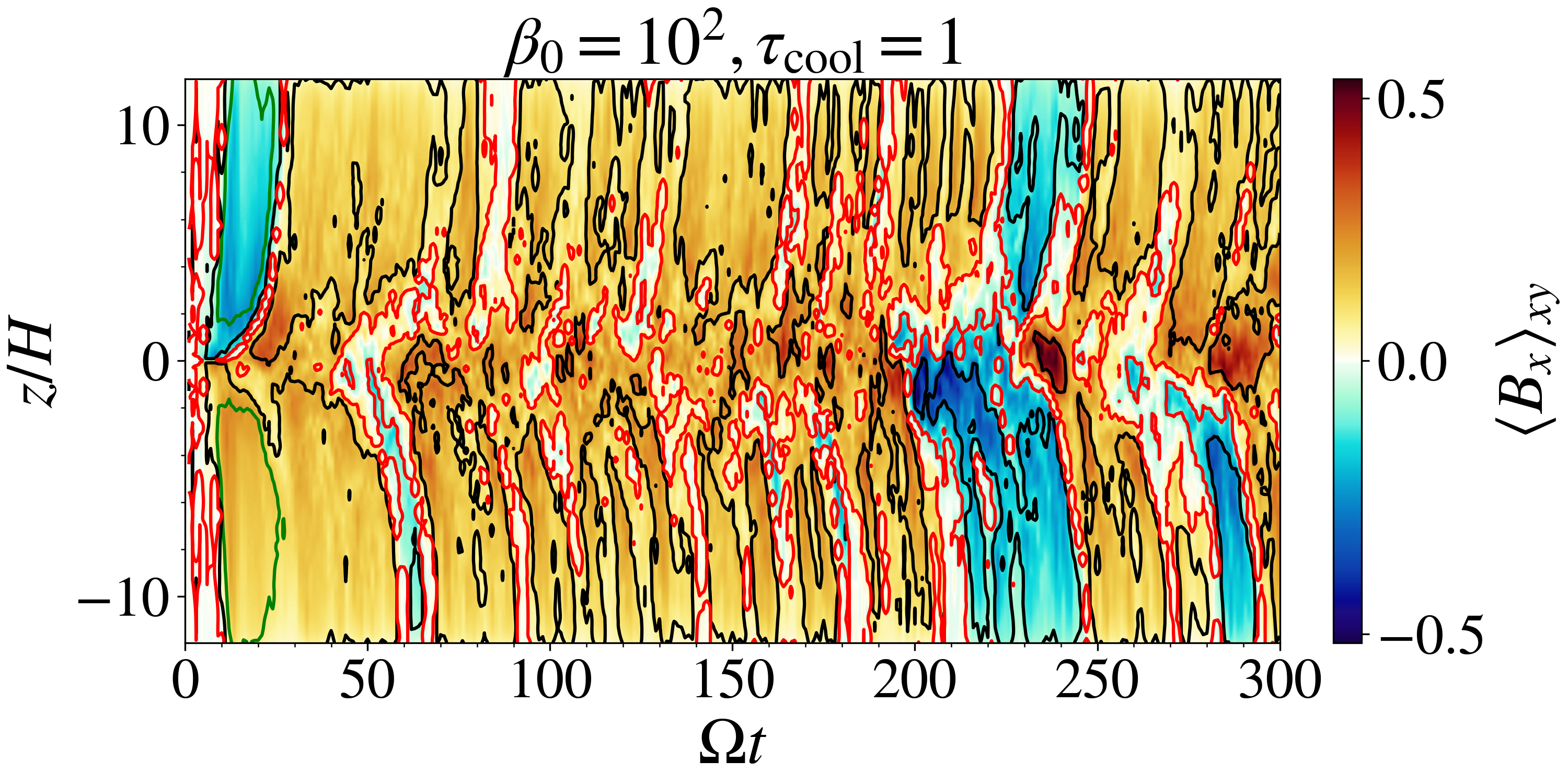}
    \includegraphics[width=0.31\textwidth]{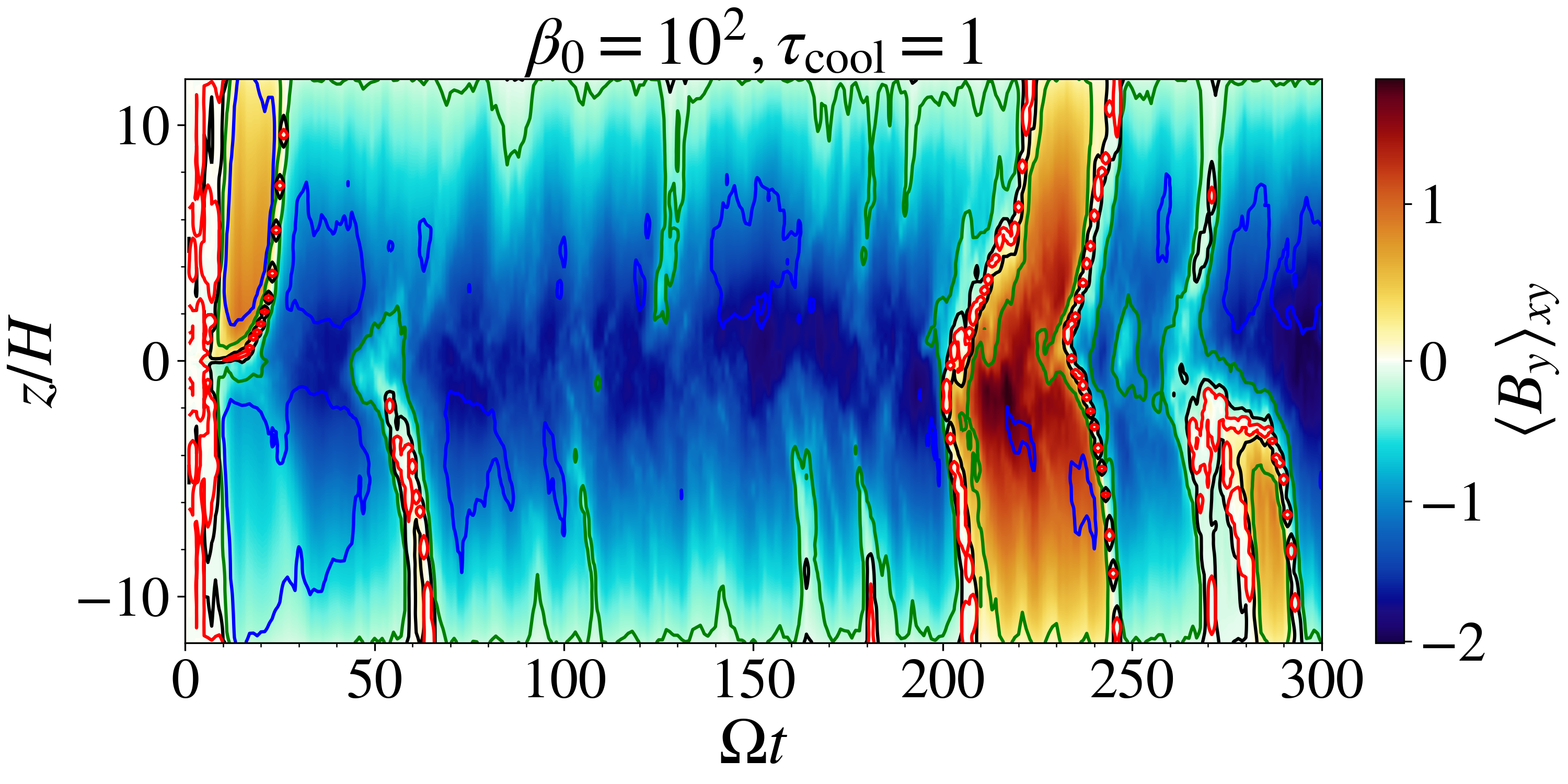} \\
    \includegraphics[width=0.31\textwidth]{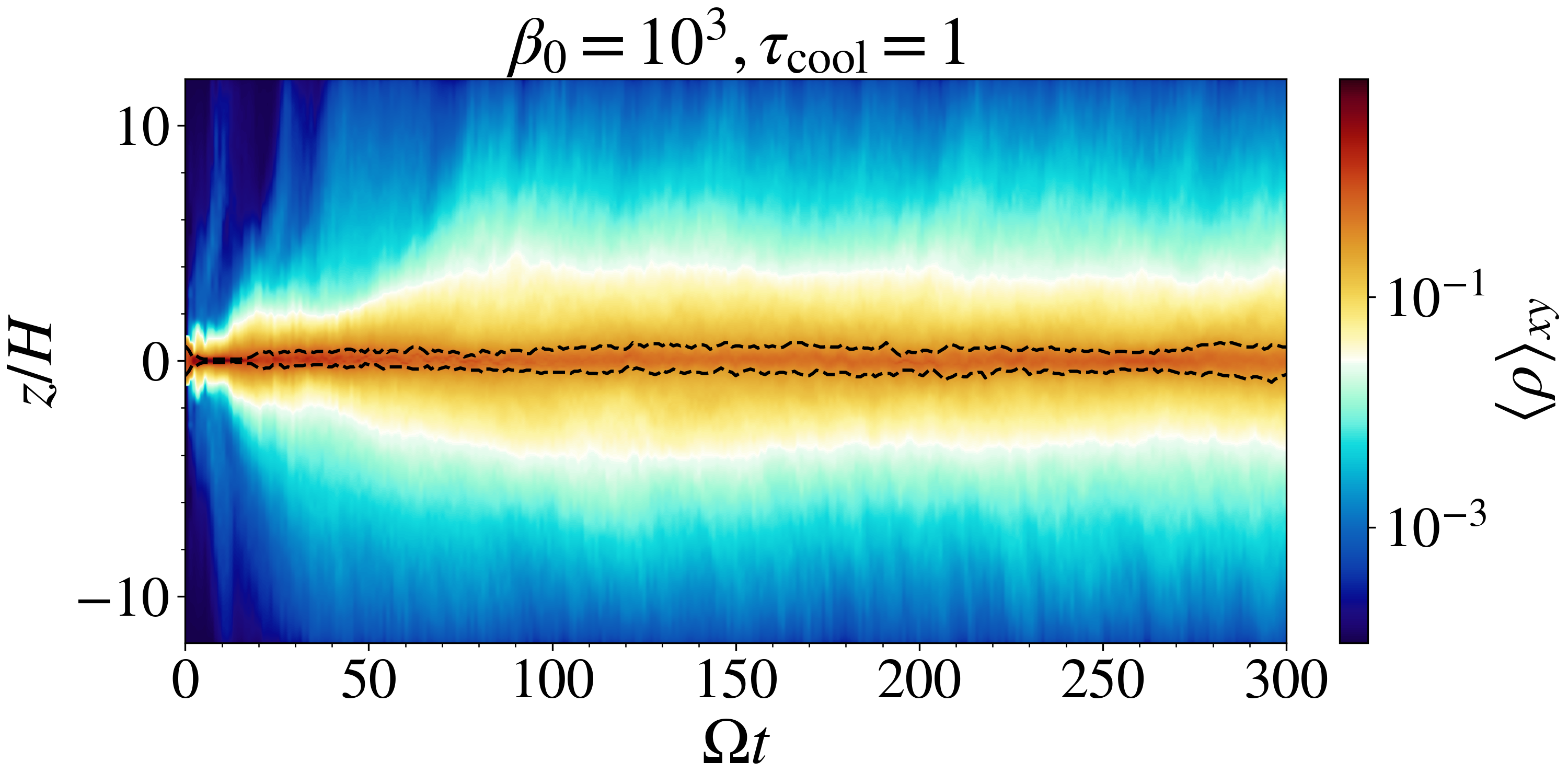}
    \includegraphics[width=0.31\textwidth]{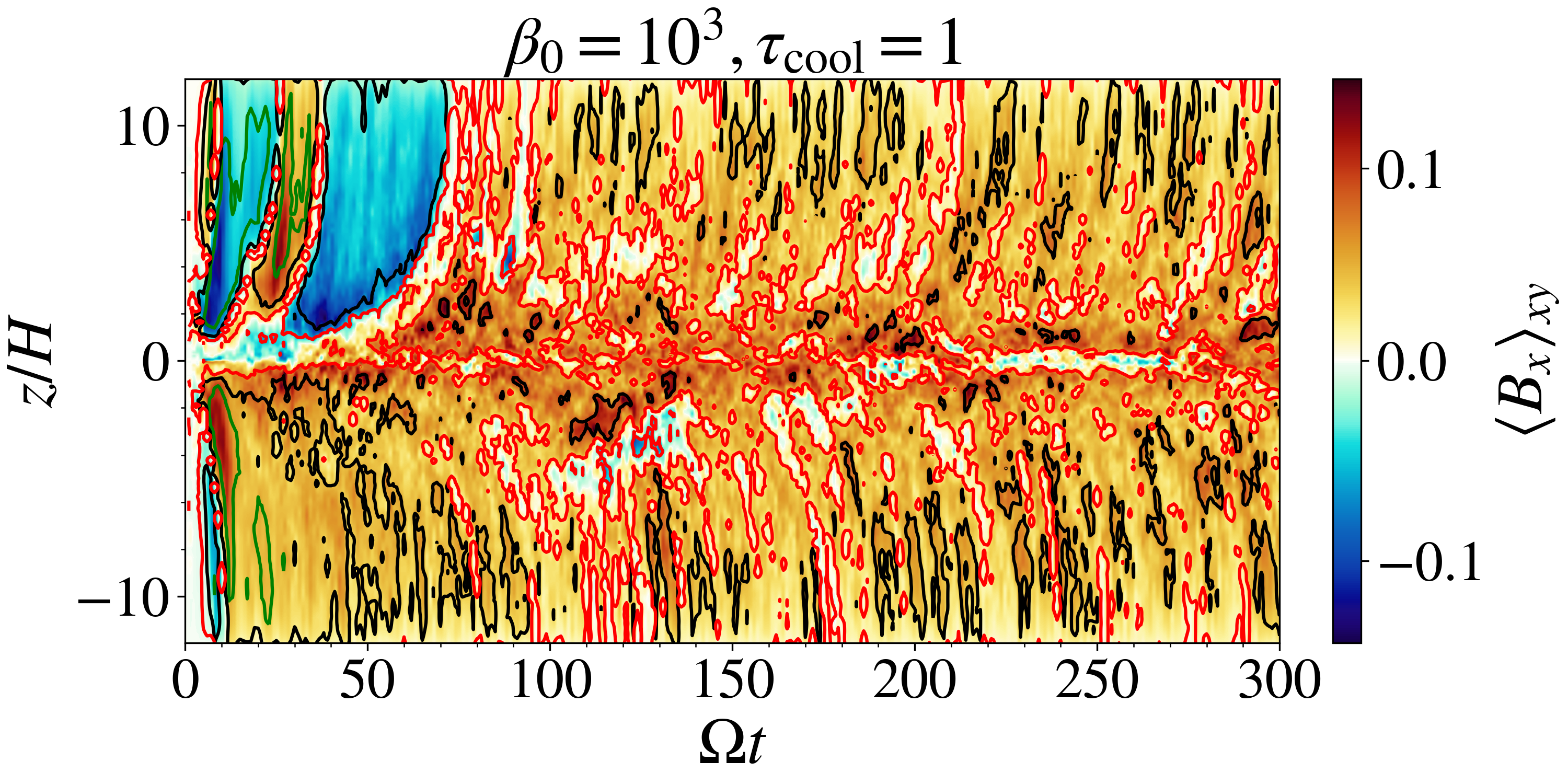}
    \includegraphics[width=0.31\textwidth]{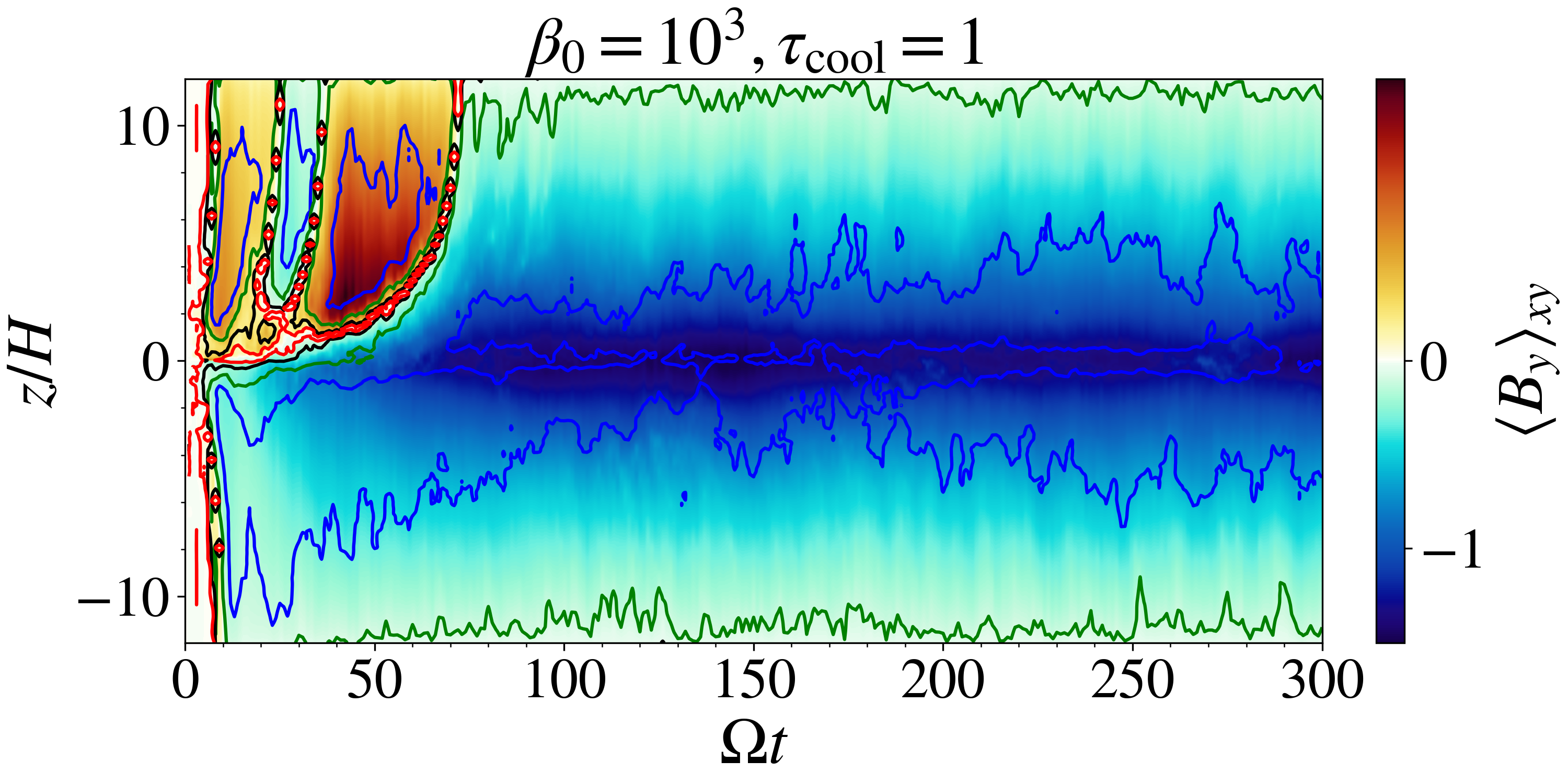} \\
    \includegraphics[width=0.31\textwidth]{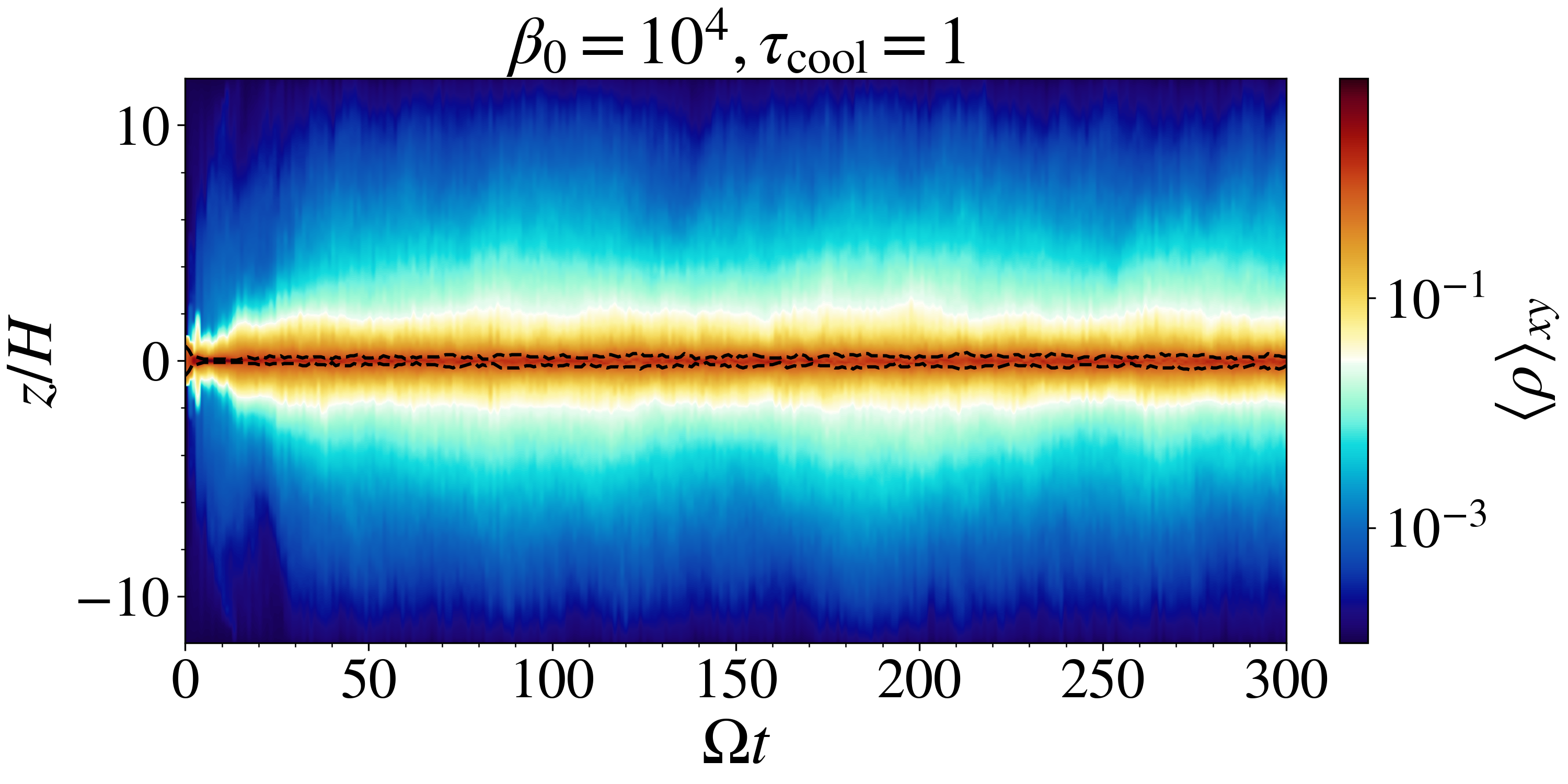}
    \includegraphics[width=0.31\textwidth]{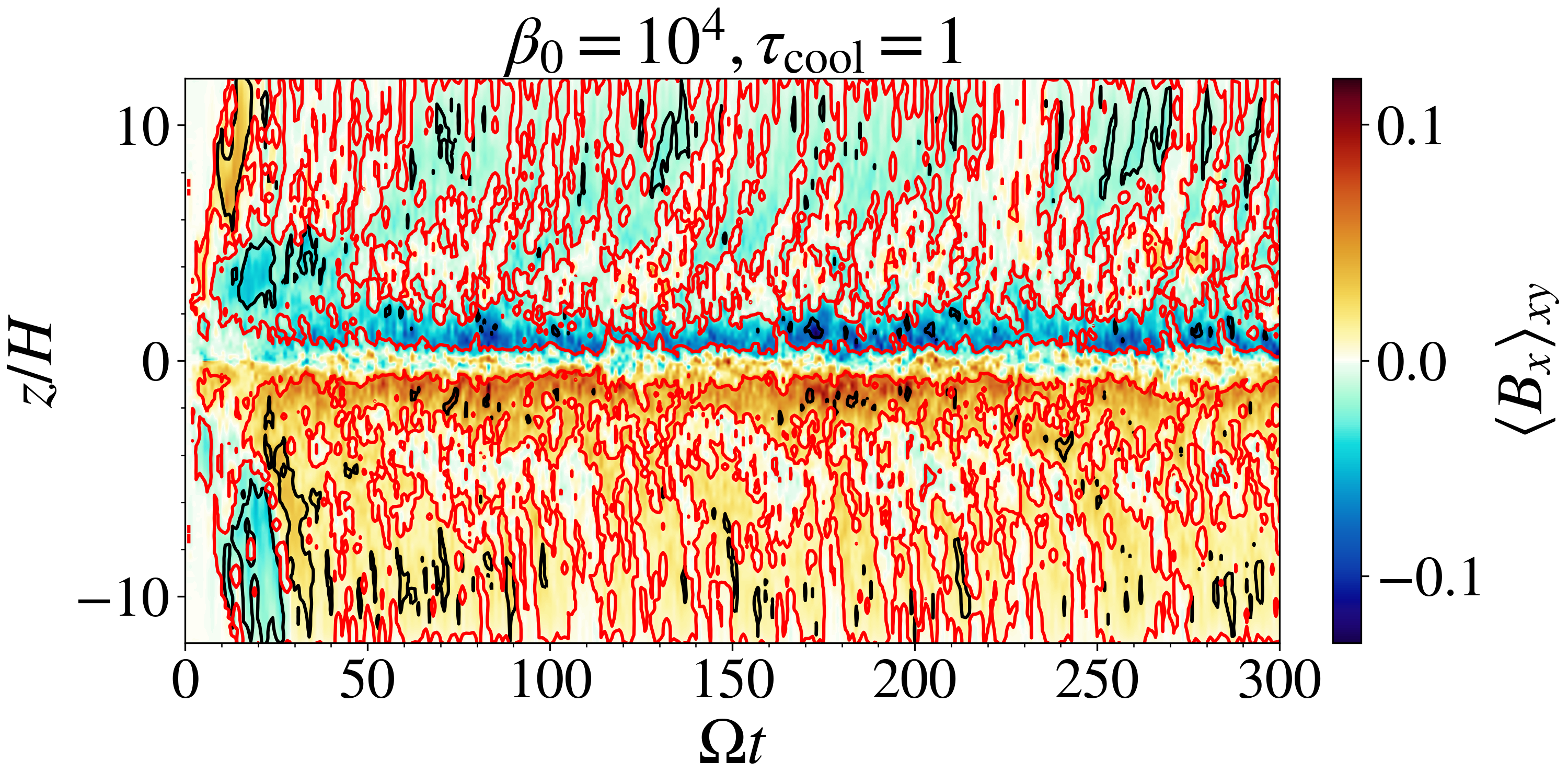}
    \includegraphics[width=0.31\textwidth]{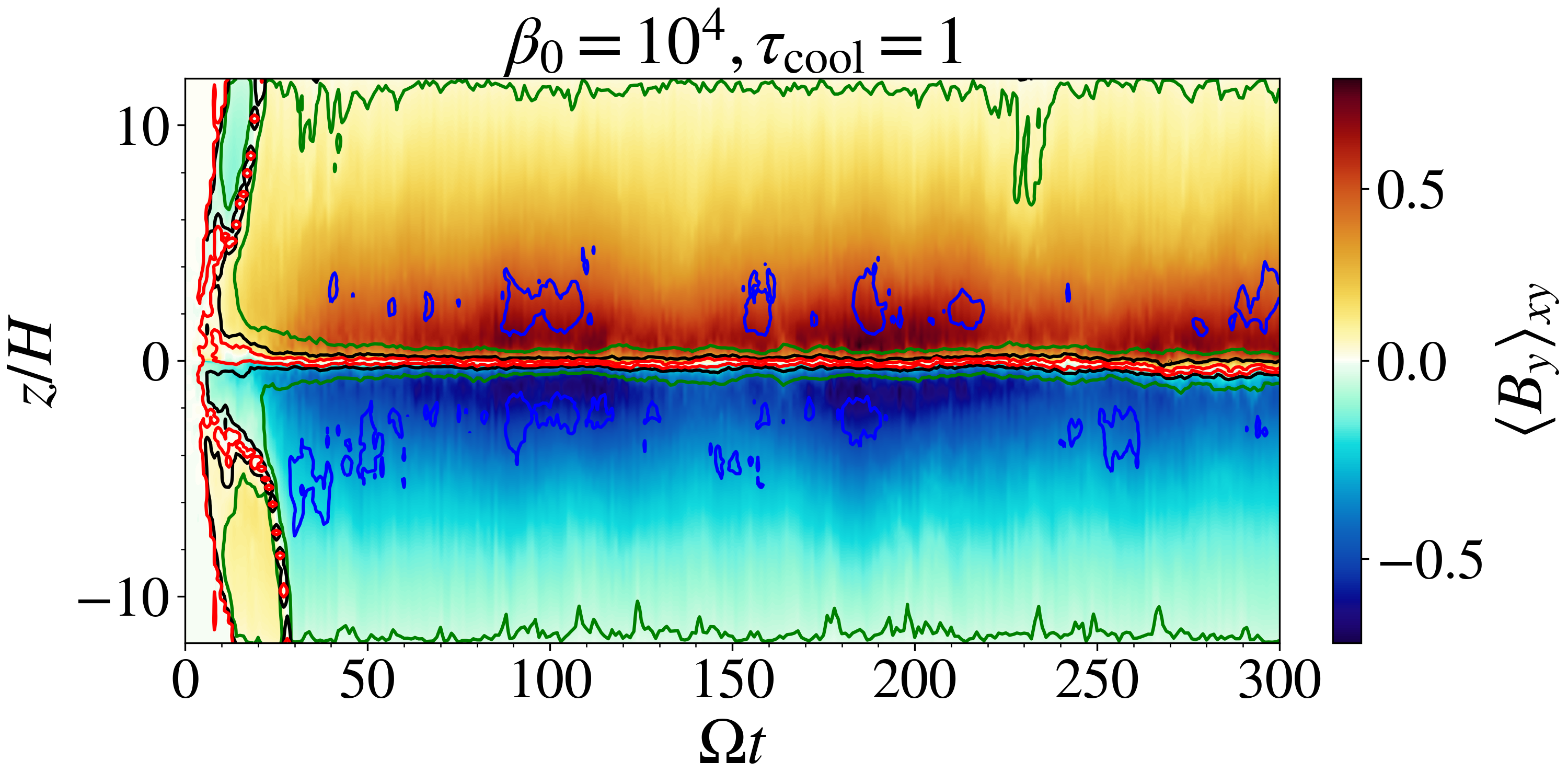} \\
    \includegraphics[width=0.31\textwidth]{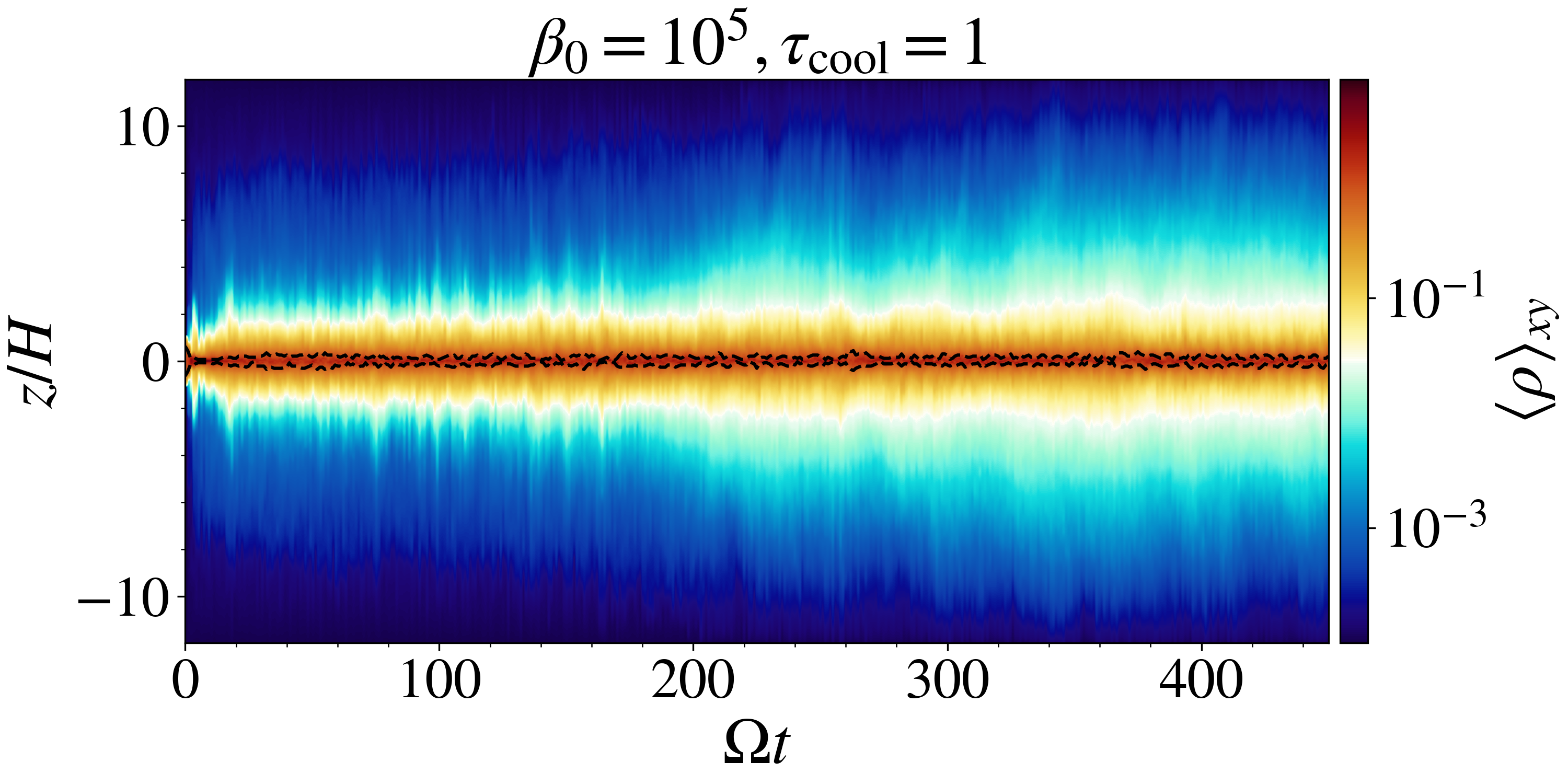}
    \includegraphics[width=0.31\textwidth]{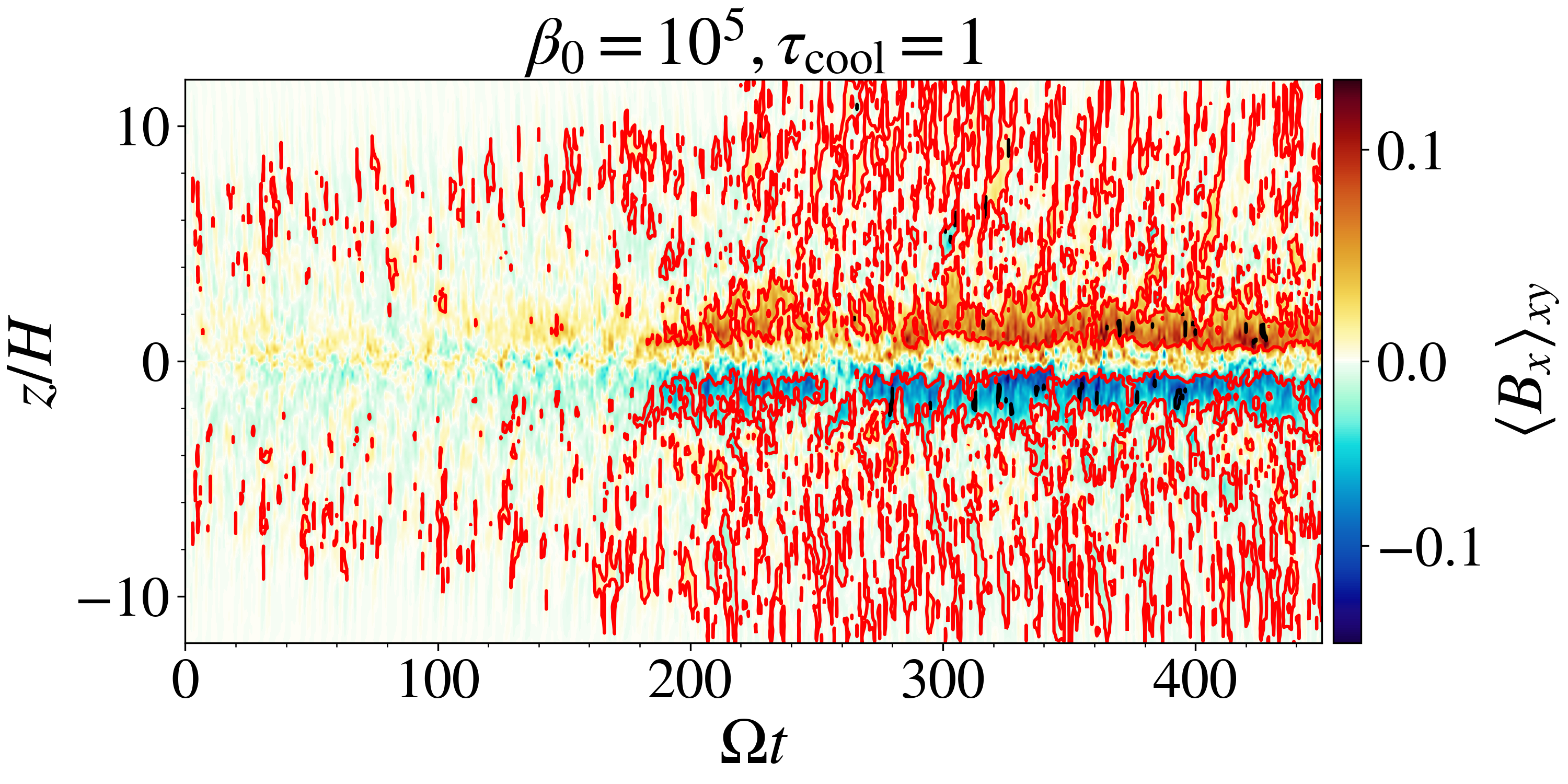}
    \includegraphics[width=0.31\textwidth]{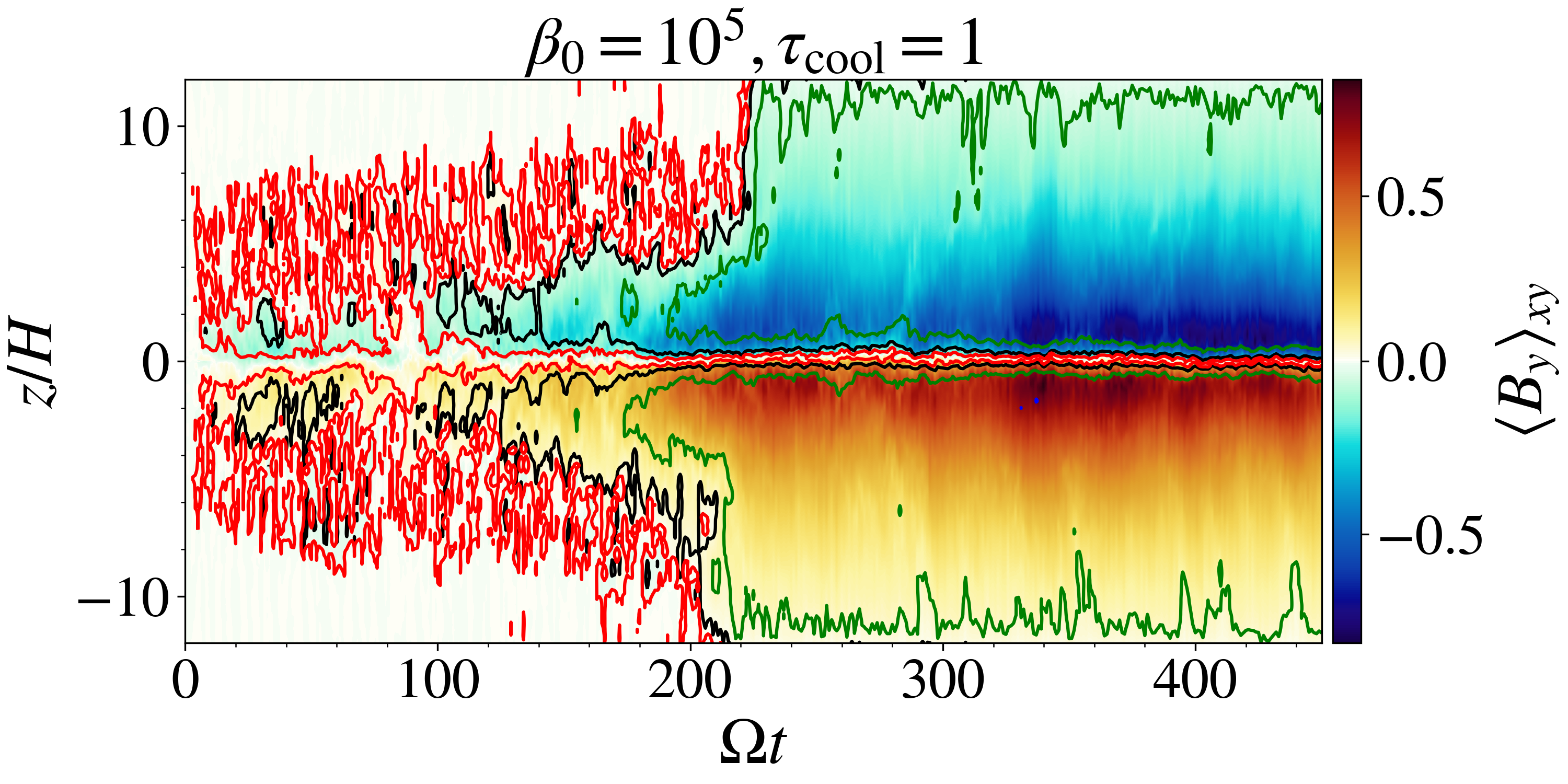} 
    \caption{Time-series plot of $\langle\rho\rangle_{xy}$ (left column), $\langle B_x\rangle_{xy}$ (middle column), $\langle B_y\rangle_{xy}$ (right column), in code units. The initial mid-plane plasma beta $\beta_0$ increases through $10, 10^2, 10^3, 10^4, 10^5$ from top to bottom, with the cooling time $\tau_\mathrm{cool}=1$. In the left column, the black contour denotes the height for which the density is half the maximum at that particular time, i.e. the thickness of the disk. In the middle column, the contours denotes various levels of $\beta_x = 2\langle P_g\rangle_{xy}/\langle B_x\rangle^2_{xy}$, blue: $\beta_x=0.1$, green: $\beta_x=1$, black: $\beta_x=10$, red: $\beta_x=100$. The contours in the right column have the same meaning as in the middle column except now they denotes various levels of $\beta_y = 2\langle P_g\rangle_{xy}/\langle B_y\rangle^2_{xy}$.}
    \label{fig:timeseries}
\end{figure*}

\begin{figure}
    \centering
    \includegraphics[width=0.23\textwidth]{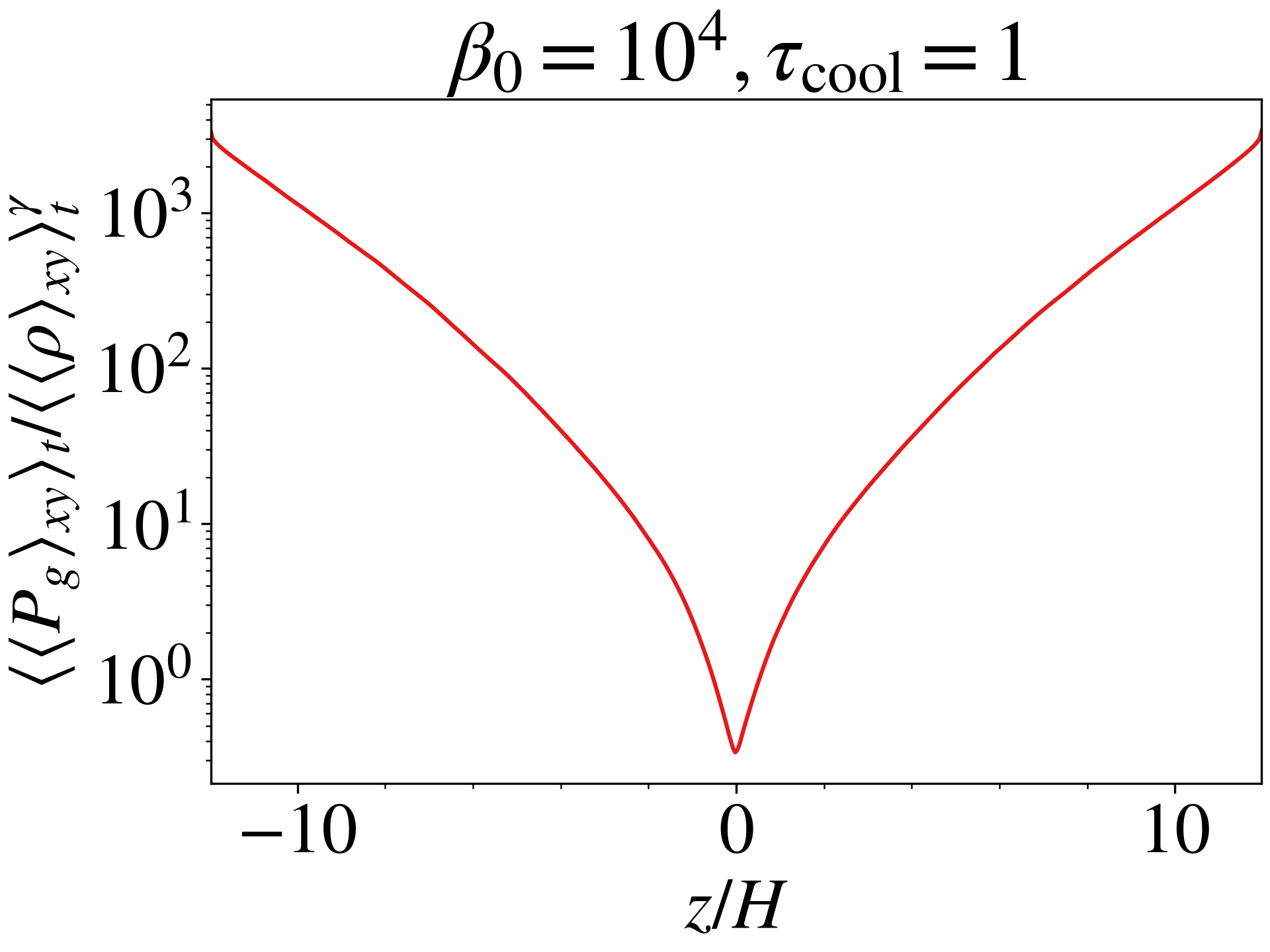}
    \includegraphics[width=0.23\textwidth]{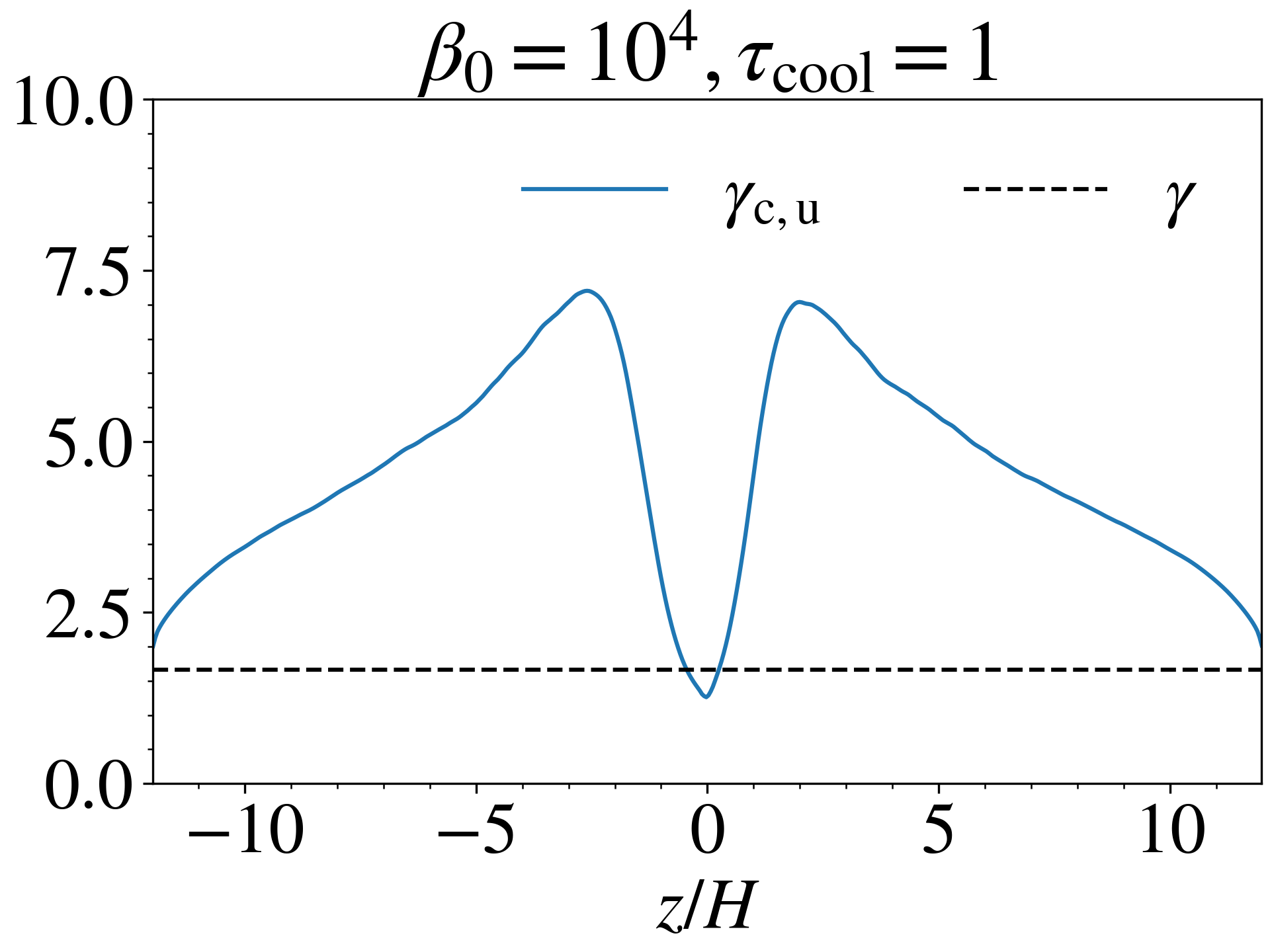}
    \caption{Left: Time-averaged, horizontally averaged entropy profile of the $\beta_0=10^4,\tau_\mathrm{cool}=1$ case. Entropy here is measured by $\langle\langle P_g\rangle_{xy}\rangle_t/\langle\langle\rho\rangle_{xy}\rangle_t^\gamma$. Time-average is taken from $\Omega t = 100-300$. Right: Comparison of the adiabatic index used in the simulation ($\gamma = 5/3$, black dashed) against the critical $\gamma$ for undular instability (blue solid line). $\gamma < \gamma_\mathrm{c,u}$ implies undular instability.}
    \label{fig:entropy}
\end{figure}

In Fig.~\ref{fig:timeseries} we present time-series plots of the horizontally averaged density $\langle\rho\rangle_{xy}$ (left column), and the horizontally averaged radial field $\langle B_x\rangle_{xy}$ (middle column) and toroidal field $\langle B_y\rangle_{xy}$ (right column) for $\tau_\mathrm{cool}=1$ and initial mid-plane plasma beta $\beta_0=10, 10^2, 10^3, 10^4, 10^5$ (increasing from top to bottom). Most of the test cases were run until $\Omega t=300$, except the $\beta_0=10^5$ case, which was run until $\Omega t=450$ due to the MRI saturating later. The black dashed lines in the left column indicate locations above and below the disk where the density is half the maximum, as a measure of the disk thickness. The colored contours in the middle and right columns indicate different levels of $\beta_x = 2\langle P_g\rangle_{xy}/\langle B_x\rangle^2_{xy}$ and $\beta_y = 2\langle P_g\rangle_{xy}/\langle B_y\rangle^2_{xy}$,  respectively. As can be seen in the figure, stronger vertical magnetic field tends to puff up the disk, inducing strong outflows. Observe also that $\langle B_y\rangle_{xy}$ is generally stronger than $\langle B_x\rangle_{xy}$. This is consistent with the fact that any radial field generated from the MRI-dynamo is subjected to shear which winds it up in the toroidal direction\footnote{An initially radial field $\vb{B}_0 = B_{x0}\vu{x}$ subjected to shear $\vb{v} = -q\Omega x\vu{y}$ does not remain radial. From the induction equation, $\partial\vb{B}/\partial t = -q\Omega B_{x0}\vu{y}$, which gives $B_x = B_{x0}, B_y = -q\Omega B_{x0} t$, i.e. the field becomes toroidal with time.}. Additionally, the polarities of the radial and toroidal magnetic field tend to be the same above and below the disk for the lower $\beta_0$ cases ($\beta_0 = 10, 10^2, 10^3$) in the saturated stage ($\Omega t \gtrsim 100$), while they reverse in polarity at the mid-plane for the high $\beta_0$ cases ($\beta_0 = 10^4, 10^5$). We do not observe quasi-periodic flip-flops of the magnetic field (butterfly structure) for the high $\beta_0$ cases ($\beta_0=10^4,10^5$), as was observed in \citet{Salvesen-etal-2016,Bai_Stone-2013}. A possible reason is that since cooling is proportional to the local thermal energy density in the cooling prescription (eq.~\ref{eqn:beta_prescription}), cooling is weaker further from the disk mid-plane and the temperature rises with vertical height. This causes the entropy to increase with $\abs{z}$ (left panel of Fig.~\ref{fig:entropy}), which by the Newcomb condition is stable against buoyant rise and interchange instabilities \citep{Newcomb-1961,Riols_Latter-2018}. This may prevent radial or toroidal flux from escaping vertically, which is key to the butterfly structure seen in \citet{Salvesen-etal-2016,Bai_Stone-2013}. The magnetic field profiles found in our simulations suggest the disk may be unstable to Parker (undular) modes except close to the mid-plane, according to the condition derived by \citet{Parker-1966} (right panel of Fig.~\ref{fig:entropy})\footnote{The condition for undular modes to be unstable in an isothermal equilibrium background is when the adiabatic index $\gamma$ is less than some critical value $\gamma_\mathrm{c,u}=(1+\beta^{-1})^2/(1 + 1.5\beta^{-1})$, where $\beta = P_g/P_B$ is the plasma beta \citep{Parker-1966}.}. However, the instability condition derived by \citet{Parker-1966} comes from a global stability analysis and assumes an isothermal equilibrium profile whereas our simulated profile is not isothermal. \citet{Kim_Ryu-2001} also suggested that magnetic tension due to field randomness (which our turbulent disks exhibit) could reduce or completely suppress Parker modes. Thus it is unclear if Parker instability would be active. Finally, we note that the in-disk field $B_x, B_y$ tapers off near the vertical boundaries as a result of the boundary conditions we use ($B_x = B_y = 0$). 

In short, the disk thickness varies with the magnetic field in an expected way, and the magnetic fields in the low $\beta_0$ cases also agree with previous studies. However, the butterfly structure is absent in the high $\beta_0$ cases. This could be a result of the cooling prescription used, which seems to suppress buoyant rise of magnetic flux.

\subsection{$\alpha$ stresses} \label{subsec:alpha_stress}

\begin{figure}
    \centering
    \includegraphics[width=0.23\textwidth]{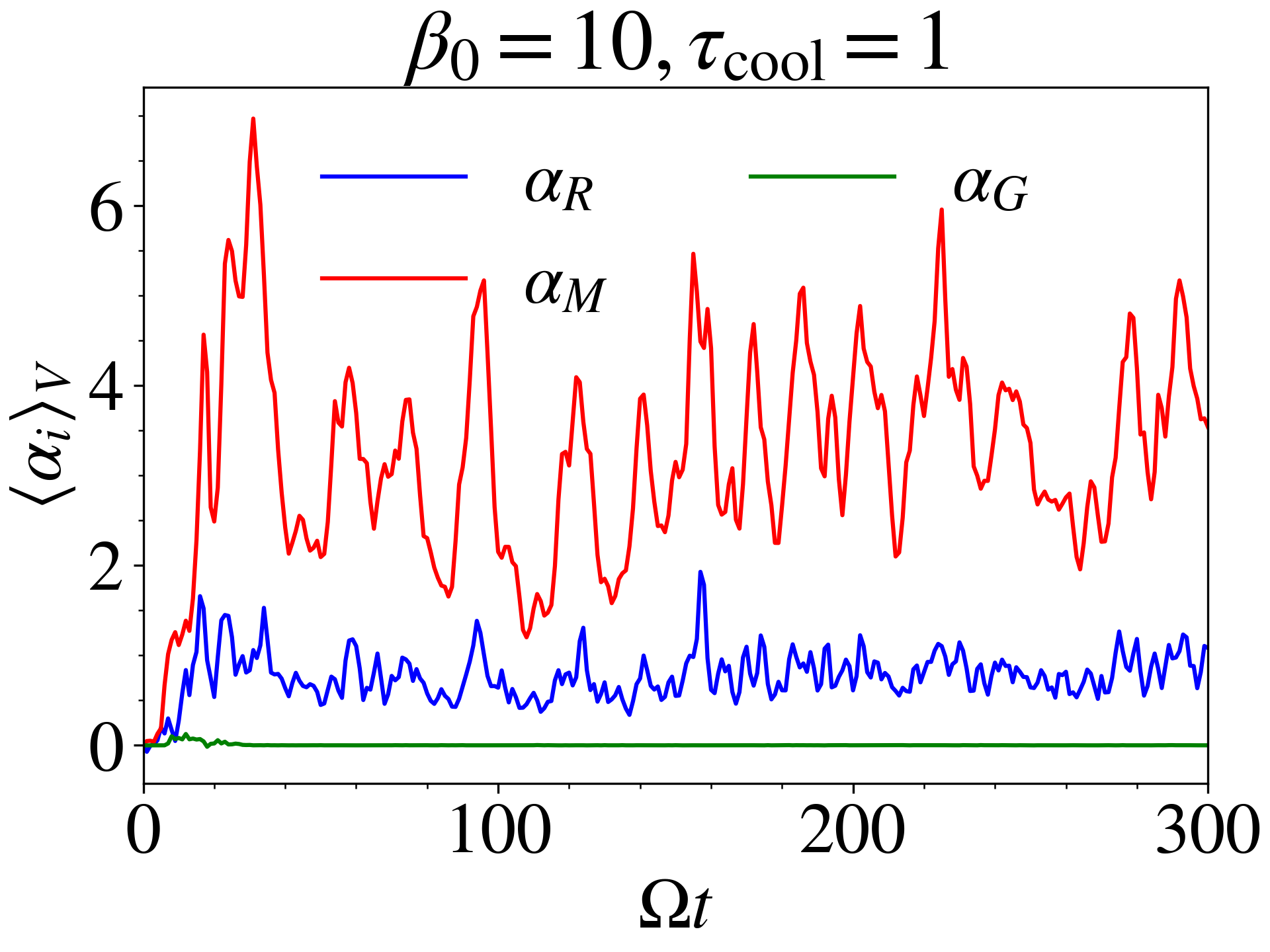}
    \includegraphics[width=0.23\textwidth]{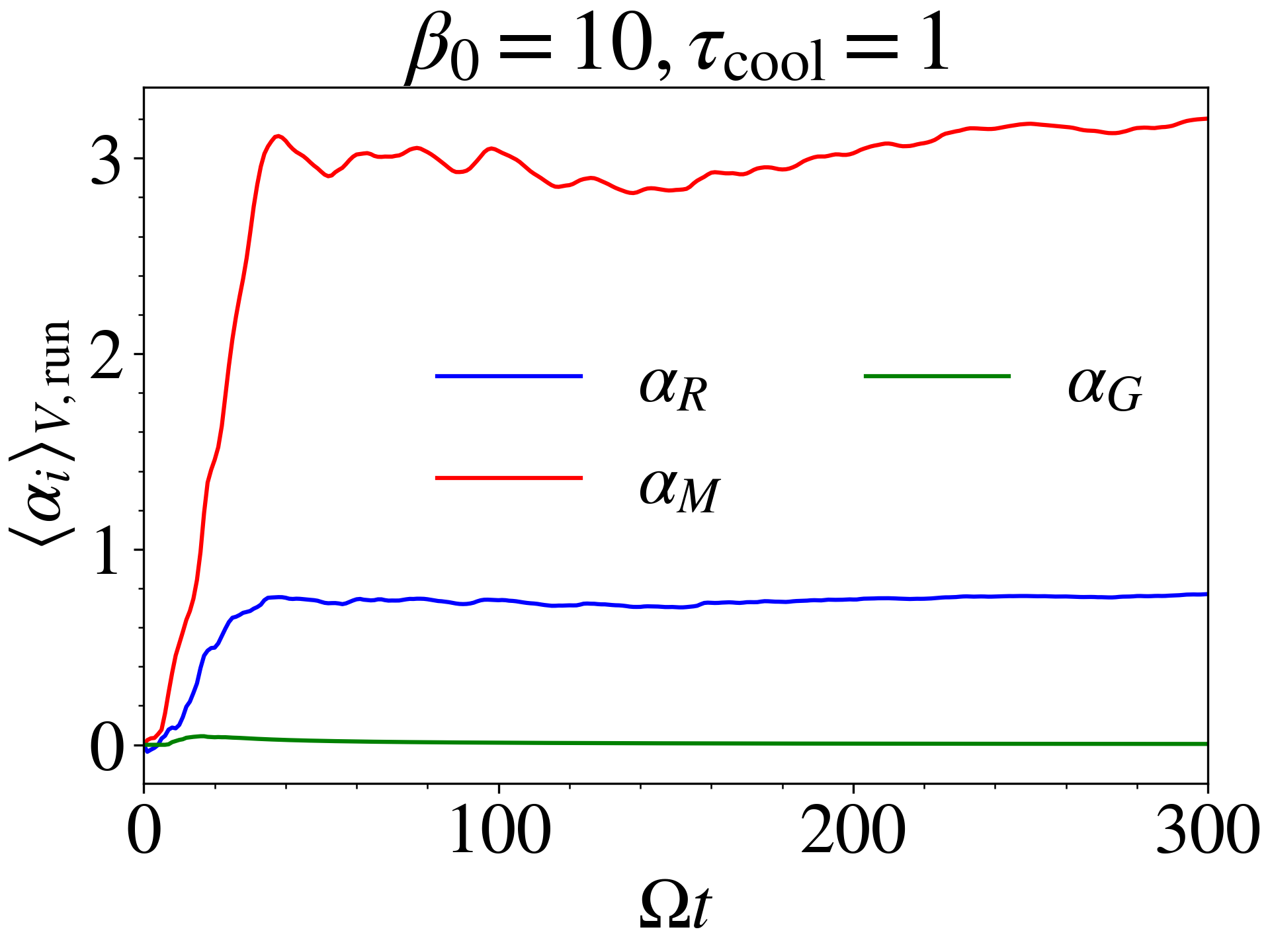} \\
    \includegraphics[width=0.23\textwidth]{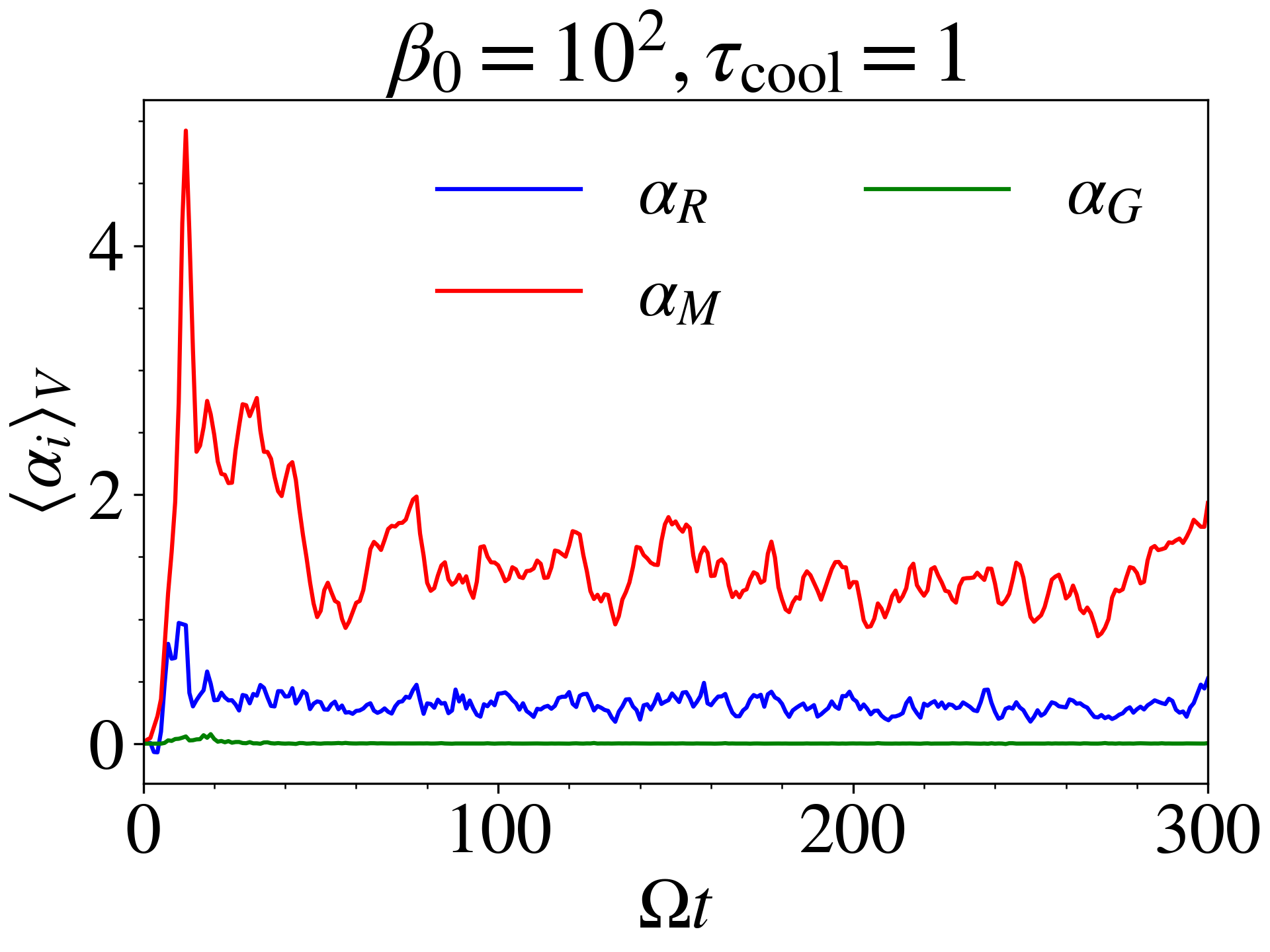}
    \includegraphics[width=0.23\textwidth]{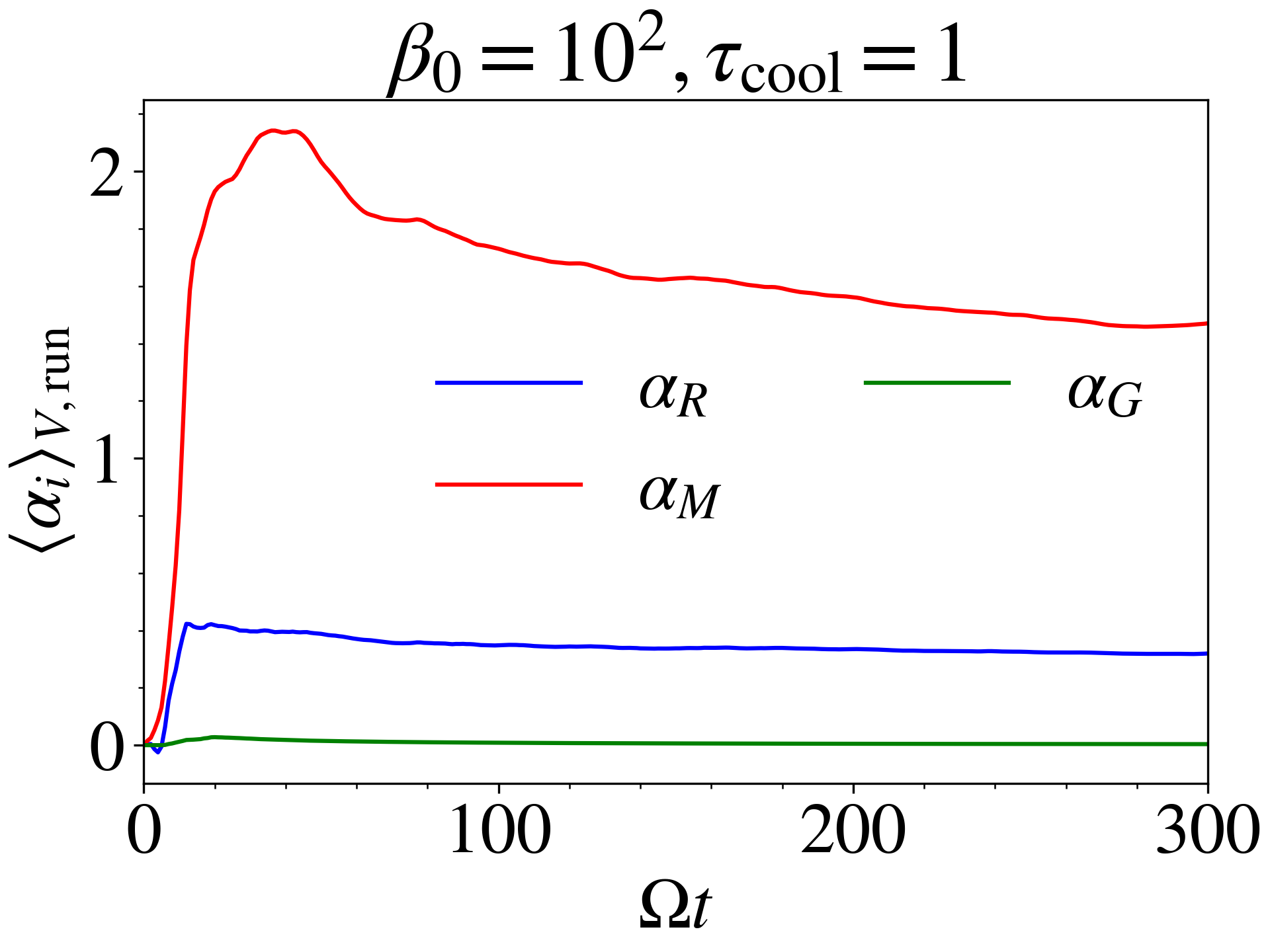} \\
    \includegraphics[width=0.23\textwidth]{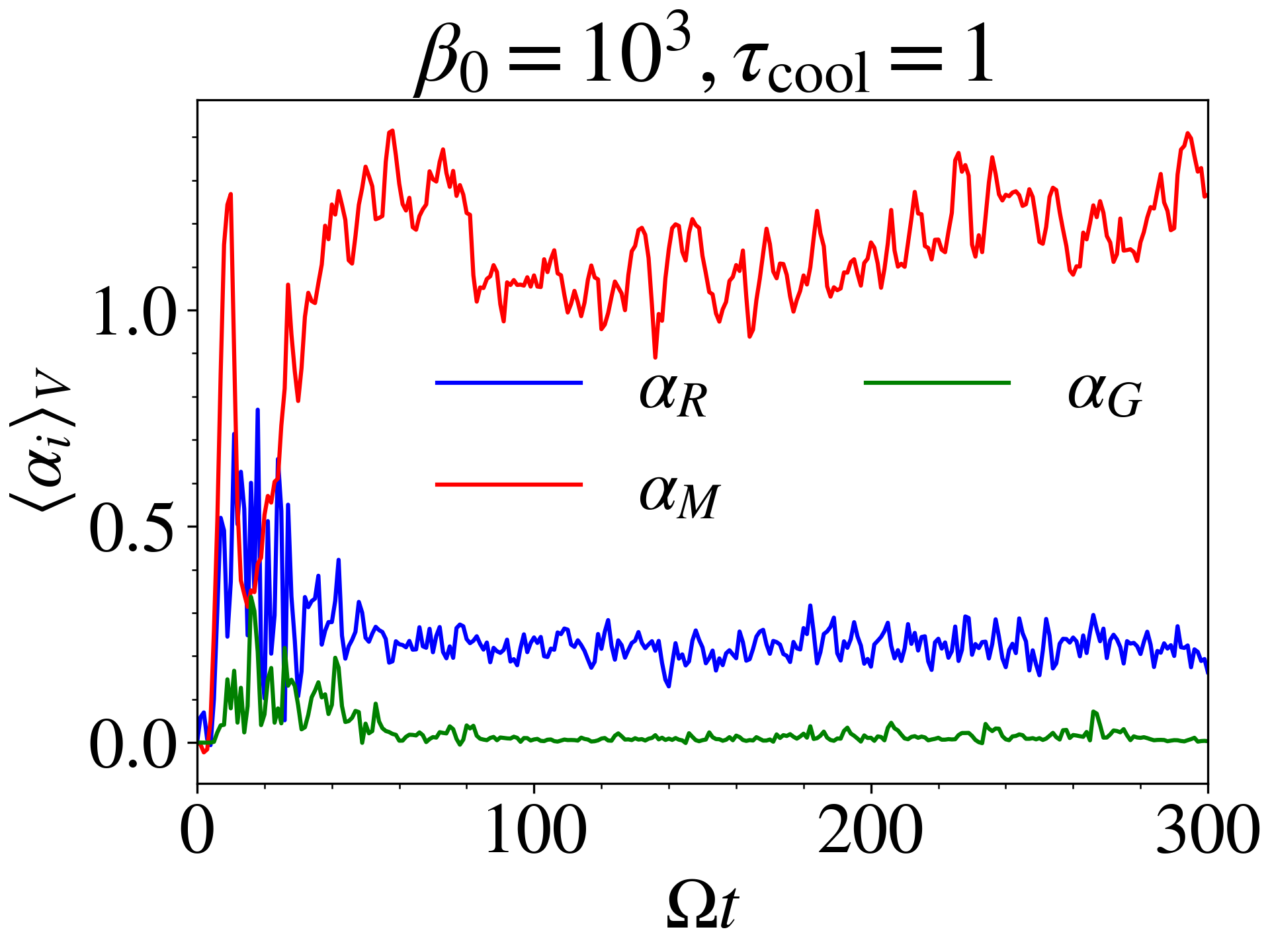}
    \includegraphics[width=0.23\textwidth]{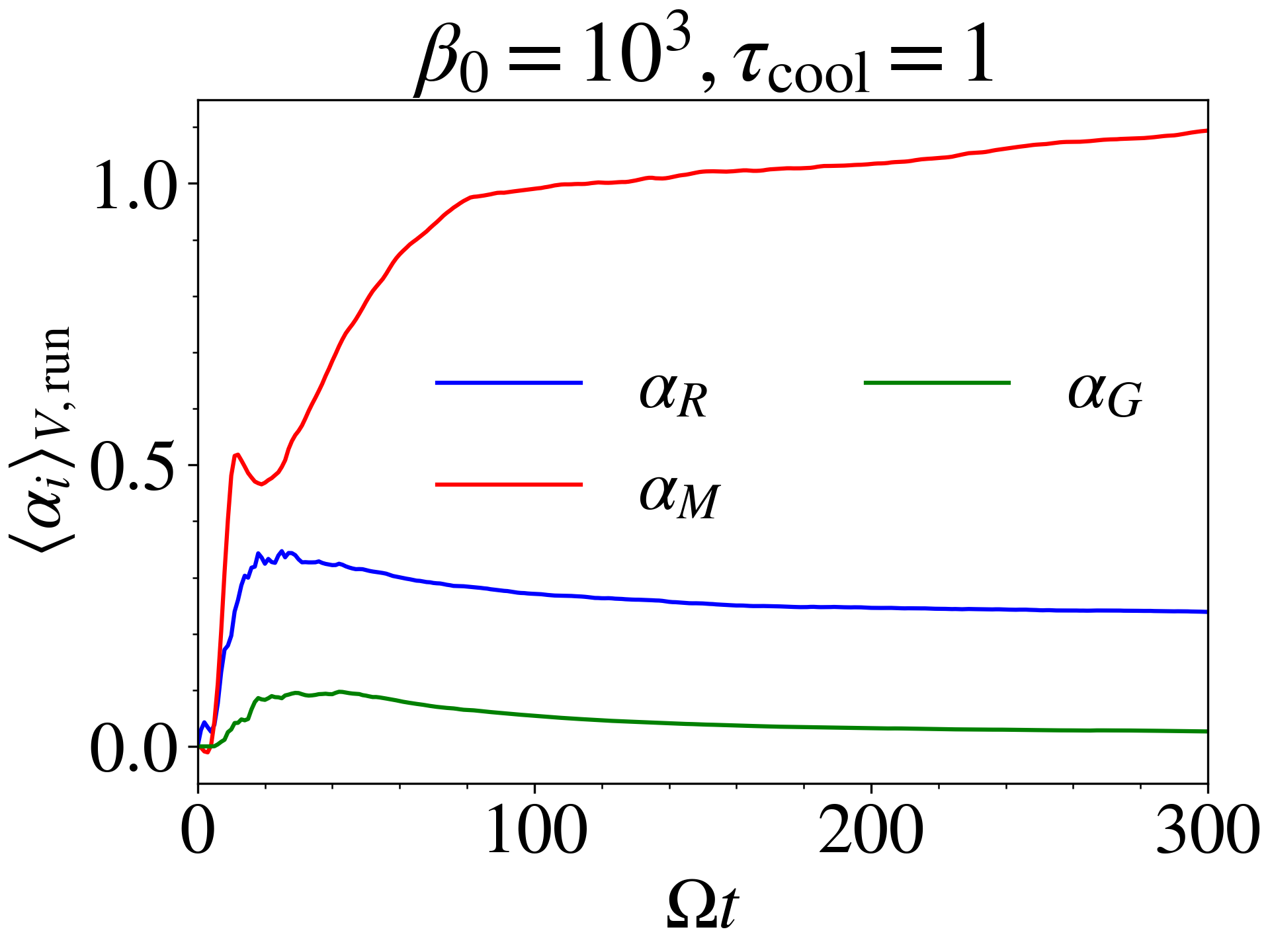} \\
    \includegraphics[width=0.23\textwidth]{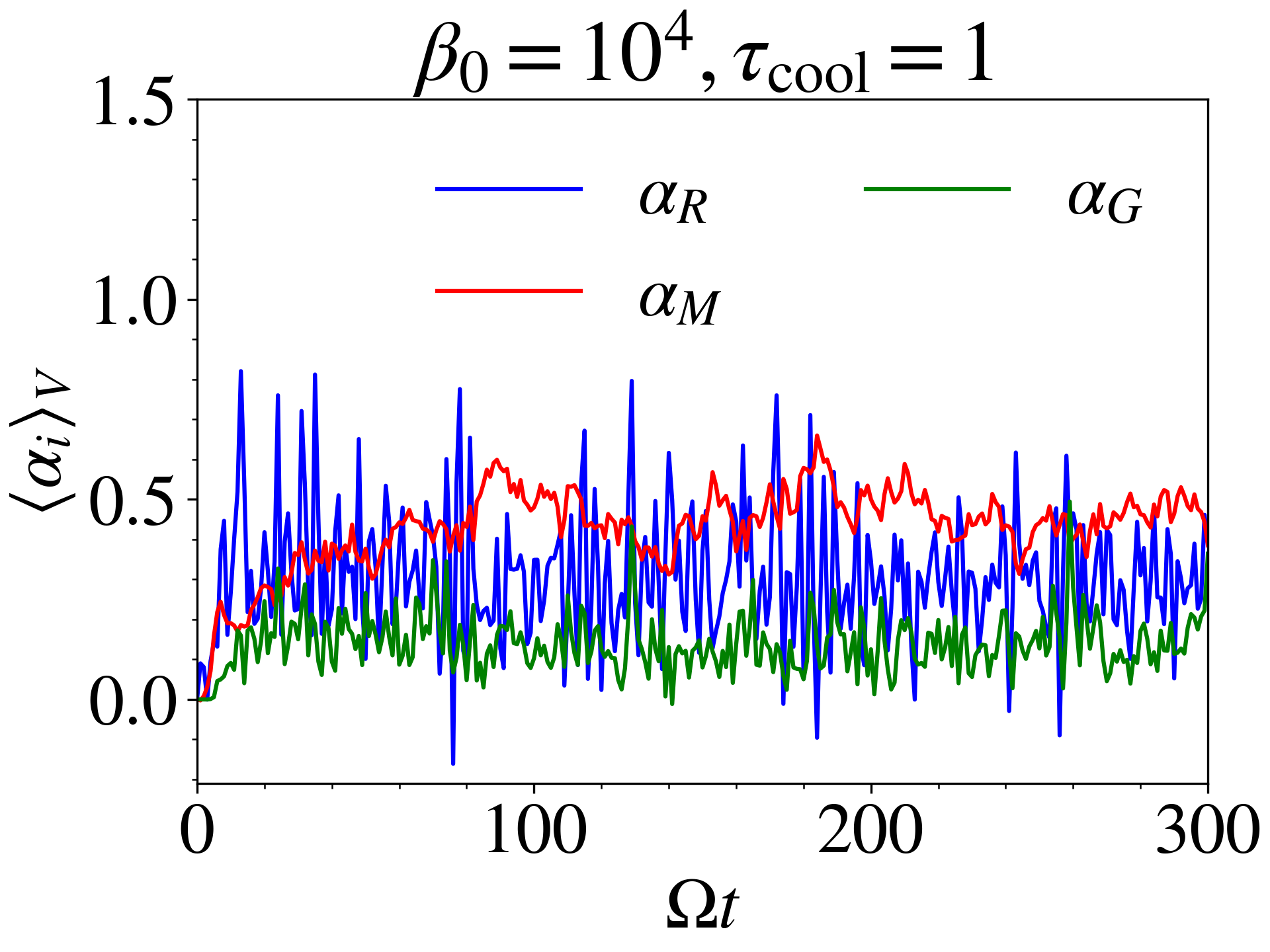}
    \includegraphics[width=0.23\textwidth]{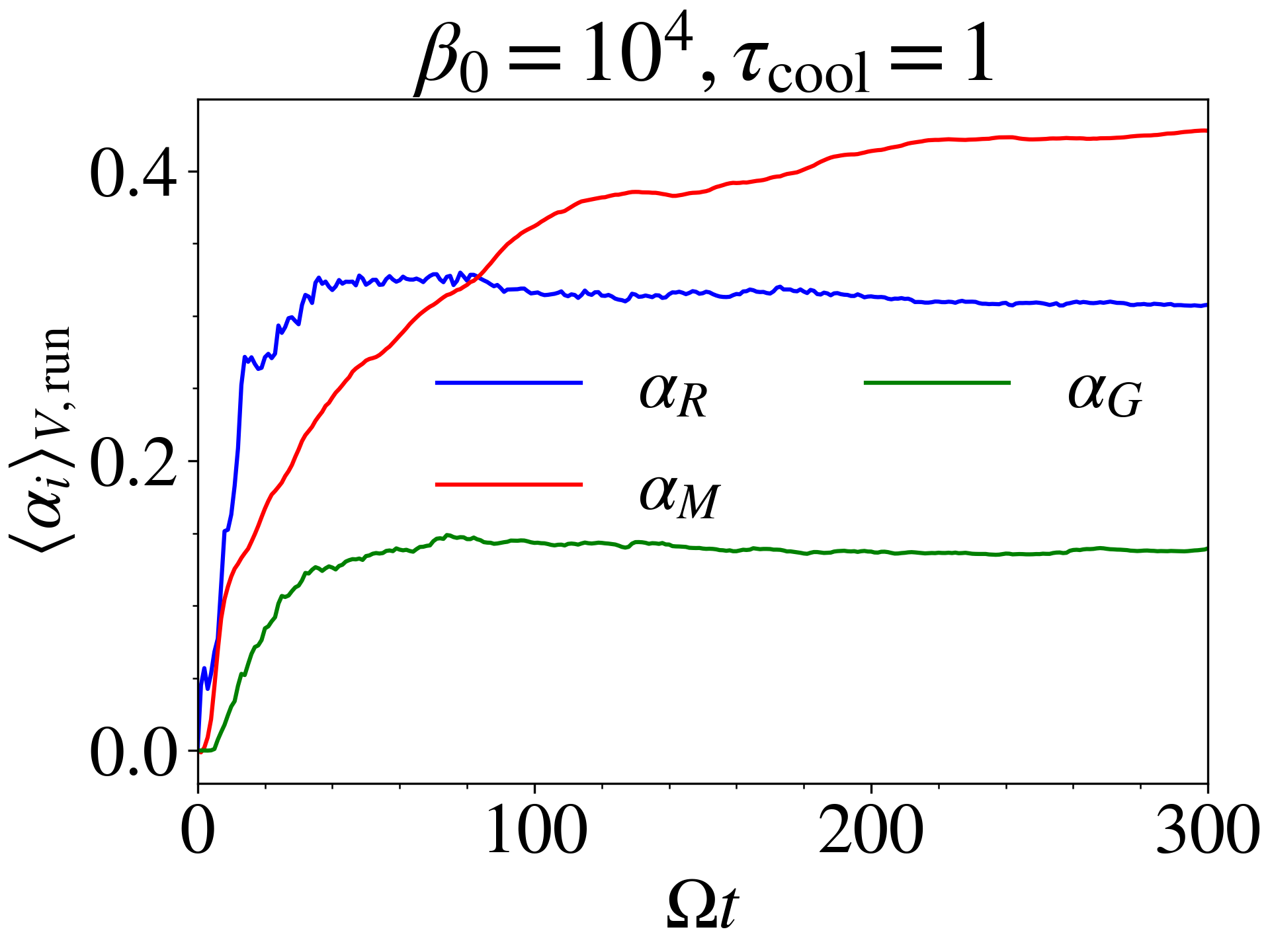} \\
    \includegraphics[width=0.23\textwidth]{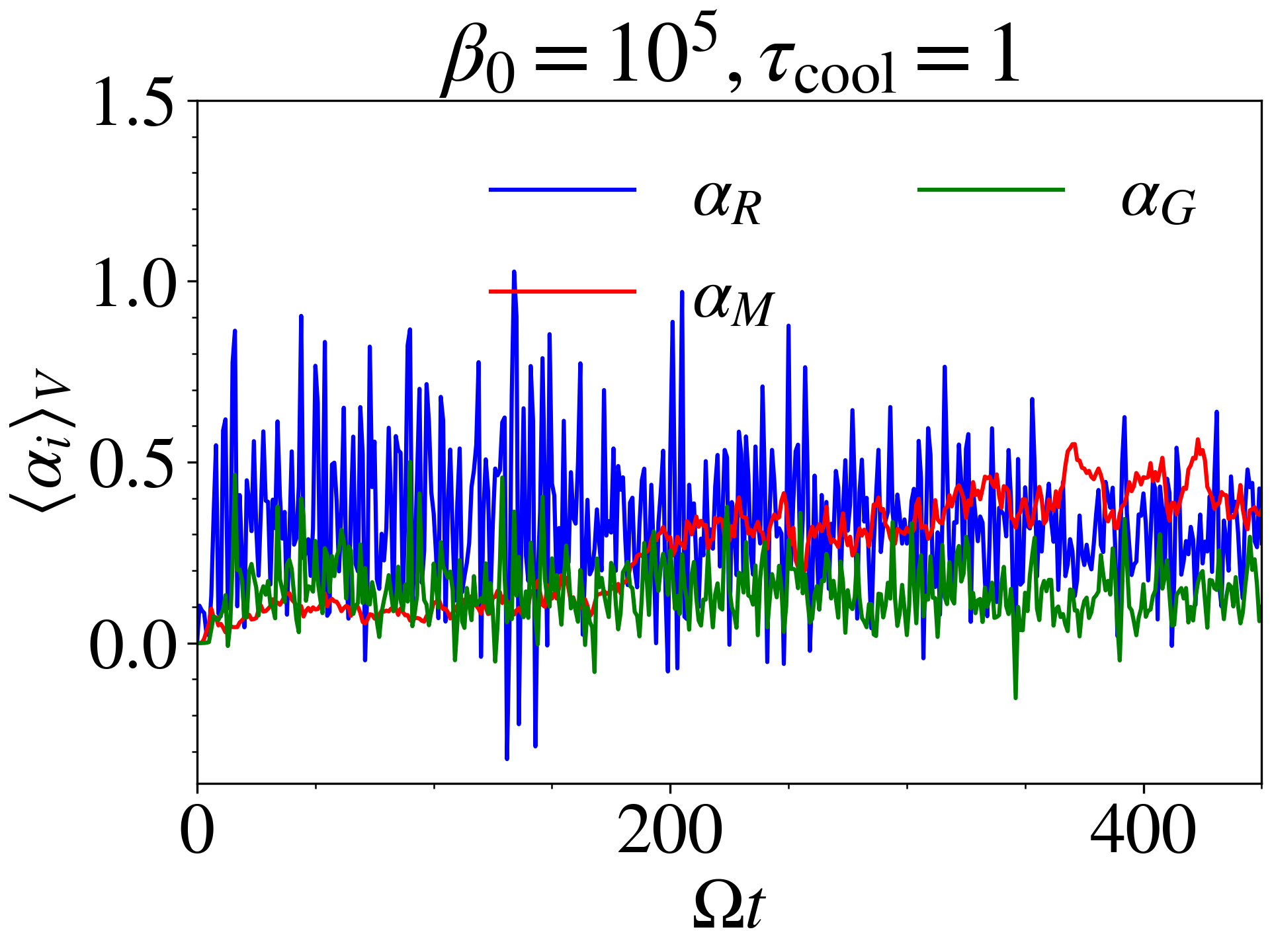}
    \includegraphics[width=0.23\textwidth]{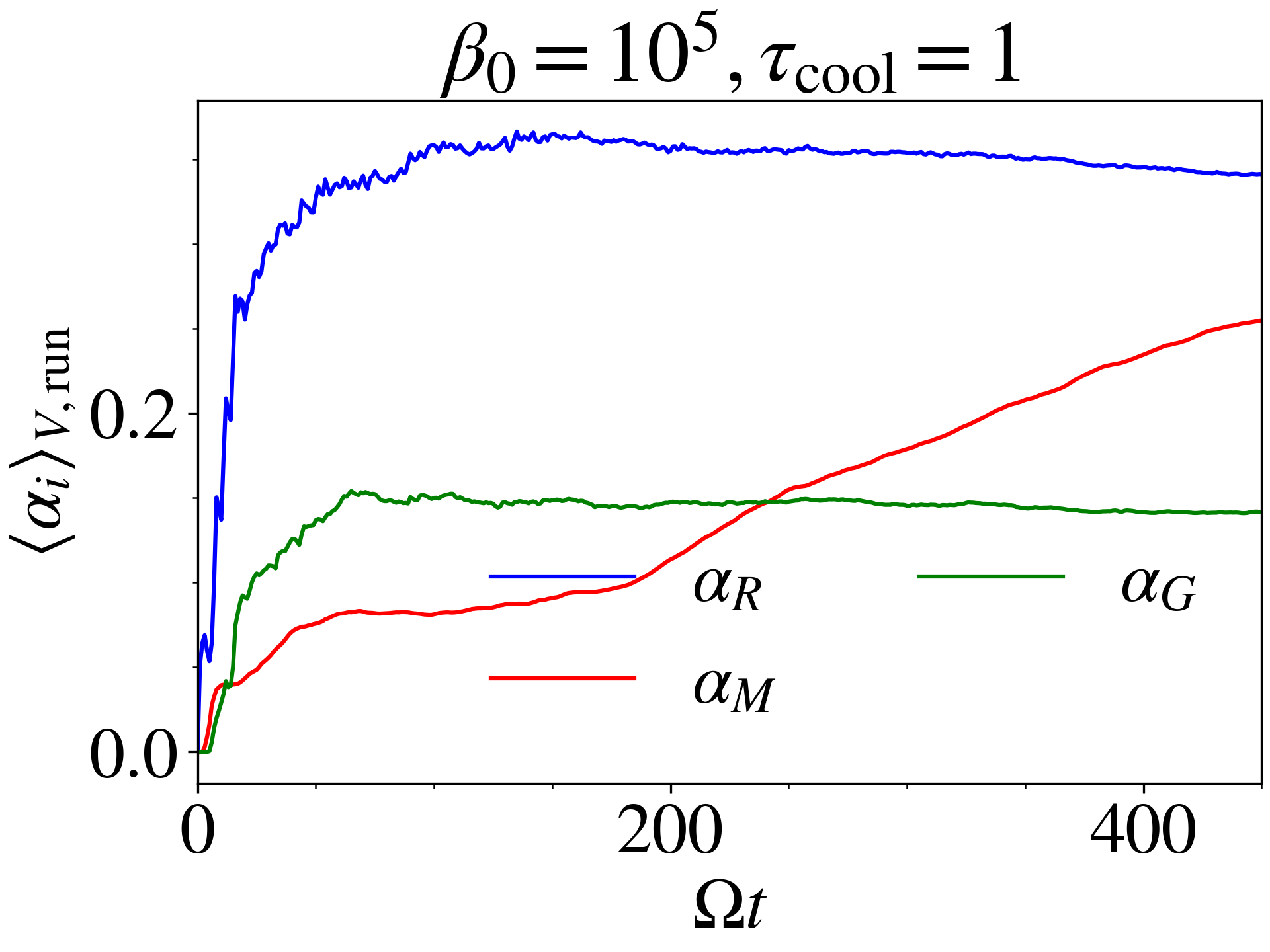} 
    \caption{Volume averaged Reynolds (blue), Maxwell (red) and gravitational (green) stresses as a function of time for various $\beta_0$ cases. Left column: The instantaneous $\langle\alpha_i\rangle_V$ at each time. Right column: The running-average of $\langle\alpha_i\rangle_V$.}
    \label{fig:stresses}
\end{figure}

We examine the various kinds of stresses in this section. In Fig.~\ref{fig:stresses} we plot the volume averaged Reynolds (blue), Maxwell (red), and gravitational (green) stress against time. We plot the instantaneous $\langle\alpha_i\rangle_V$ values on the left column and the running-average $\langle\alpha_i\rangle_{V,\mathrm{run}}$ on the right for clarity. The running-average of a quantity at time $t$ is the cumulative time-average of it up to $t$. In all cases explored, the Maxwell stress is the greatest, followed by the Reynolds stress and then the gravitational stress. In the saturated stage, the ratio of the Maxwell to Reynolds stress for the low $\beta_0$ cases ($10,10^2,10^3$) is roughly 3-4, consistent with previous stratified MRI simulations. For the high $\beta_0$ cases ($\beta_0=10^4,10^5$), the onset of MRI is delayed (to $\Omega t\sim 100$ and $250$ for the $\beta_0=10^4,10^5$ cases, respectively). Taking the average over the last third of the simulation time span, we find that this ratio drops to roughly 1.5. The ratio of the gravitational to Reynolds stress exhibits an opposite trend, being $\ll 1$ in the low $\beta_0$ cases, but rising to $\sim 0.5$ for the high $\beta_0$ cases, hinting at strong suppression of GI in the low $\beta_0$ cases. This ordering of the stresses is observed in \citet{Riols_Latter-2018} as well, in their simulation with net vertical flux. The Maxwell stress $\alpha_M$ obtained in our simulations is slightly on the high side, but generally consistent with existing studies in the strong cooling regime for low $\beta_0$ cases \citep[e.g.][]{Gammie-2001,Salvesen-etal-2016}. However, the stresses in the high $\beta_0$ cases are orders of magnitude higher than standard MRI calculations. $\alpha_M$ reaches up to $\sim 3$ for $\beta_0=10$, and $\sim 0.3$ for $\beta_0=10^5$, showing a decreasing trend with respect to $\beta_0$. This decreasing trend is very weak, with a dependence $\sim \beta_0^{-1/4}$, compared with the $\sim\beta^{-1/2}$ dependence obtained in previous simulations \citep{Salvesen-etal-2016}. Note that there is no self-gravity in \citet{Salvesen-etal-2016}; This is suggestive of gravitoturbulence at play boosting the stresses in the high $\beta_0$ cases. 

\begin{figure}
    \centering
    \includegraphics[width=0.23\textwidth]{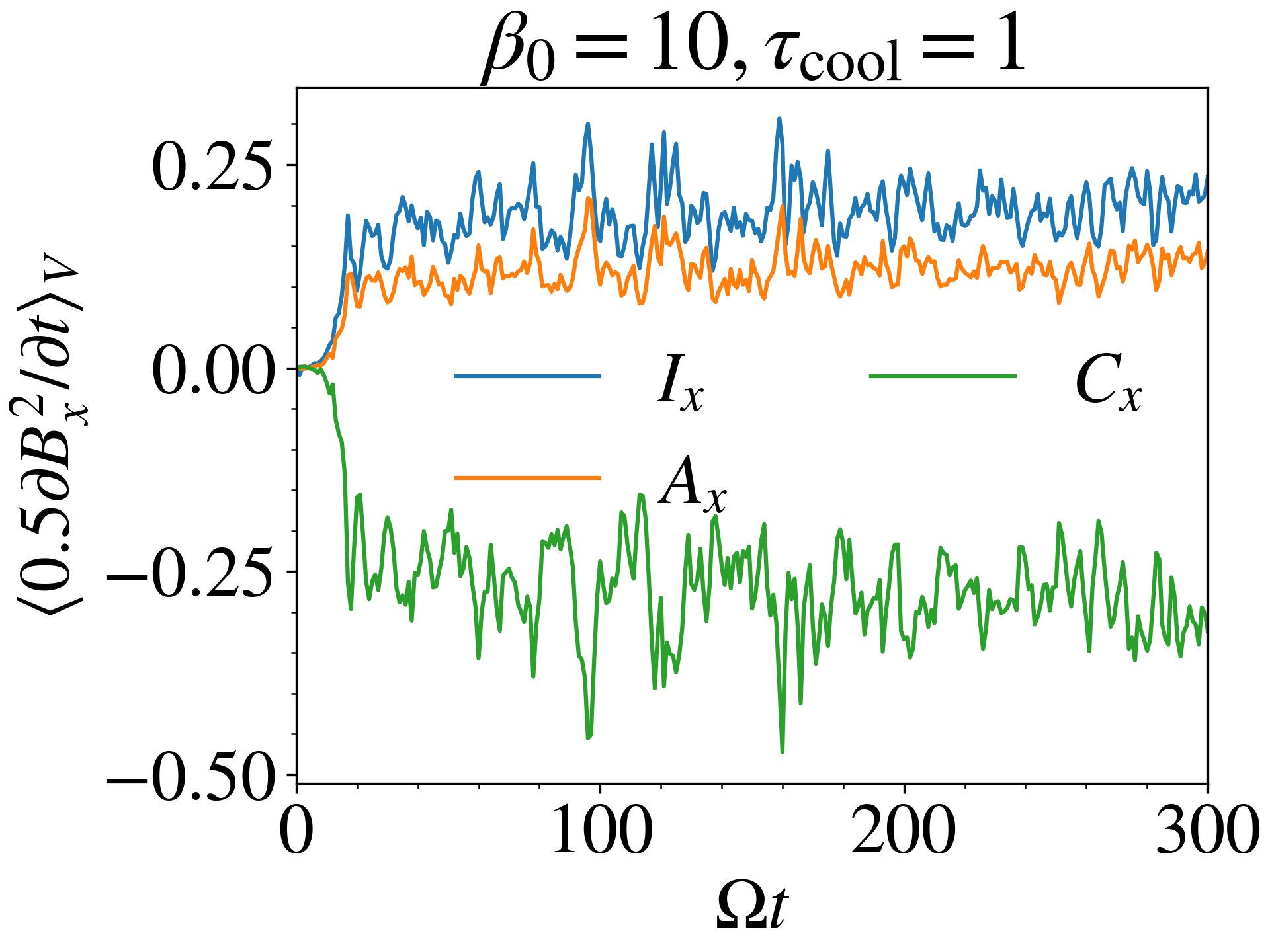}
    \includegraphics[width=0.23\textwidth]{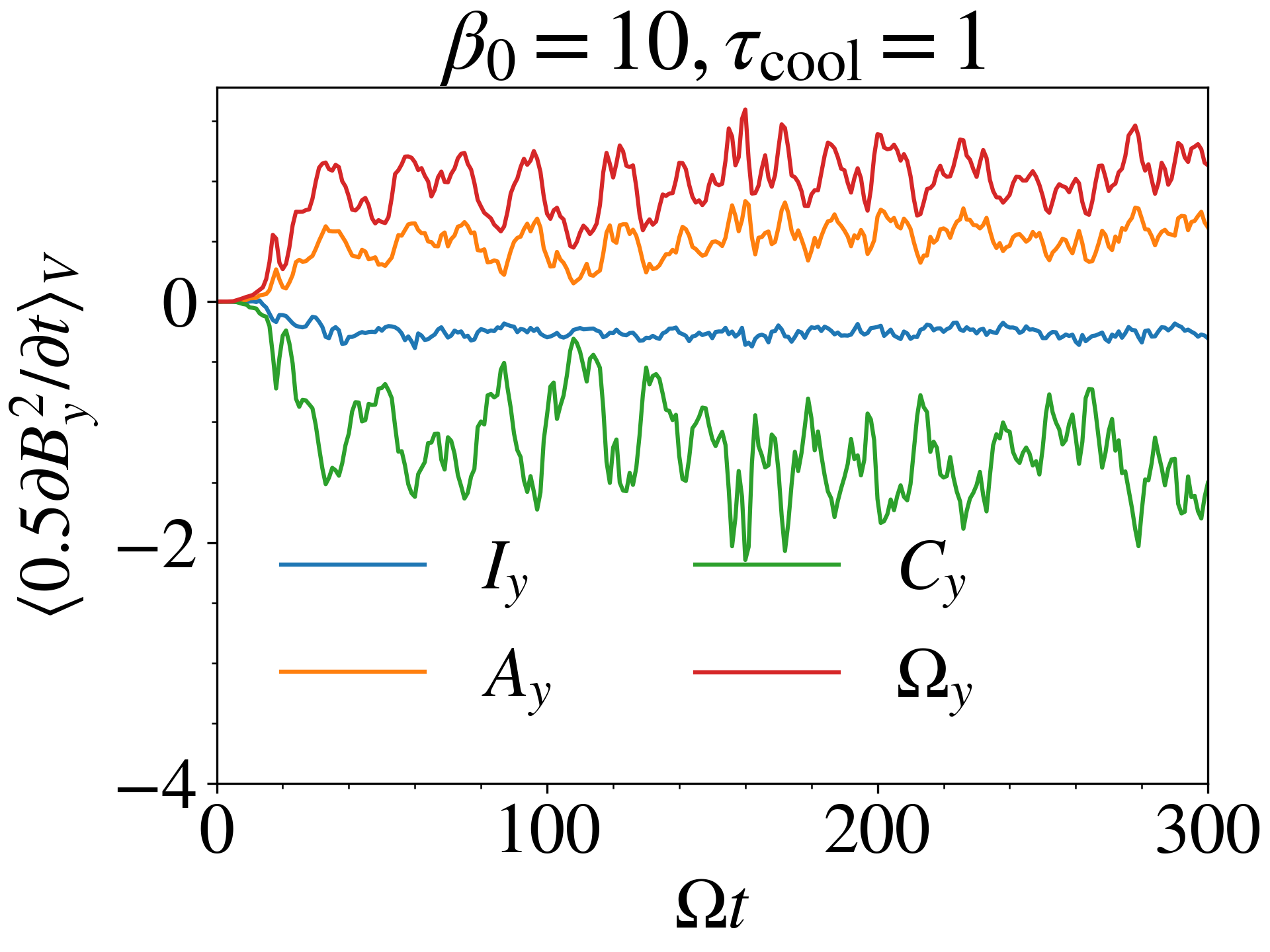} \\
    \includegraphics[width=0.23\textwidth]{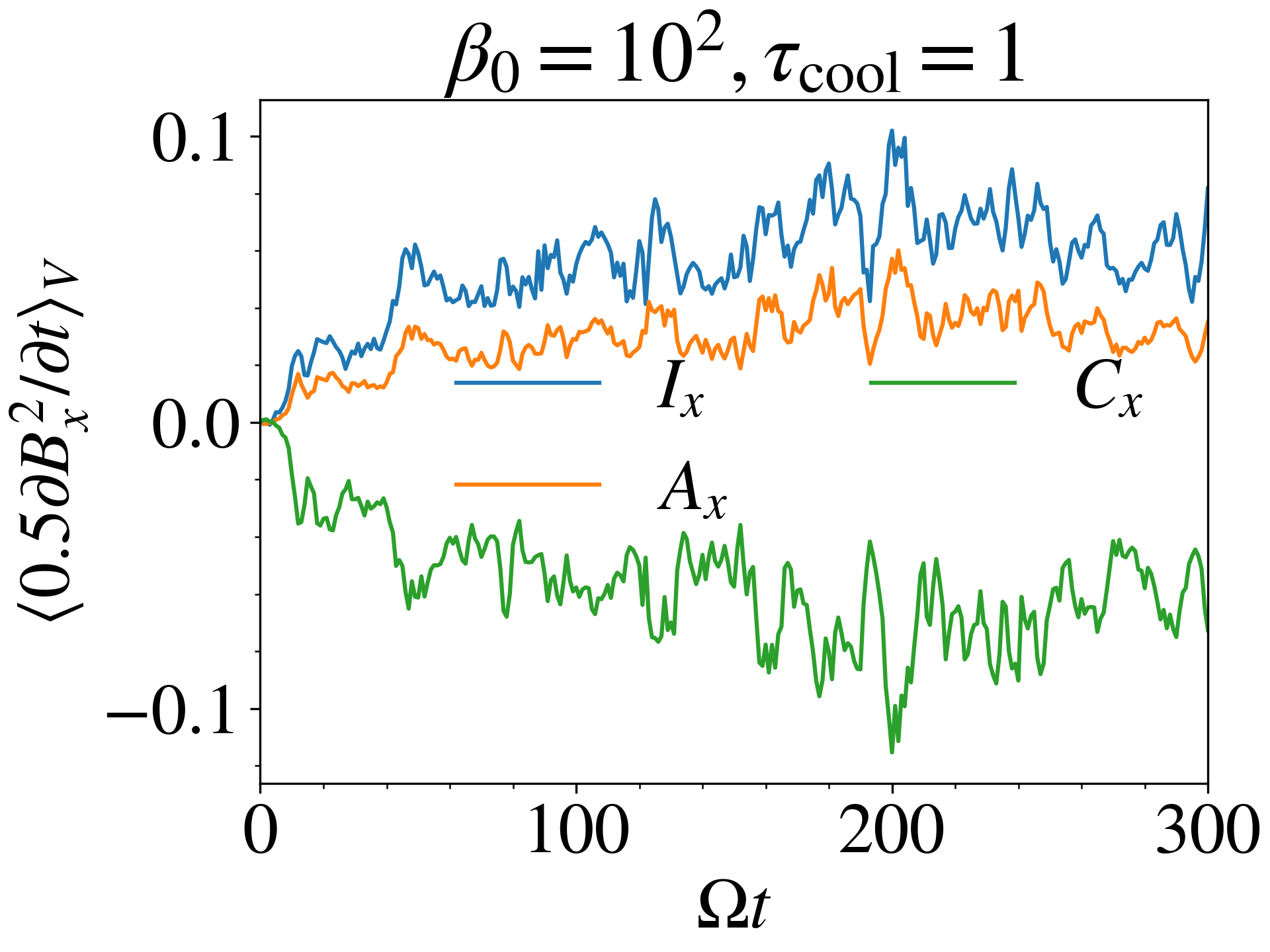}
    \includegraphics[width=0.23\textwidth]{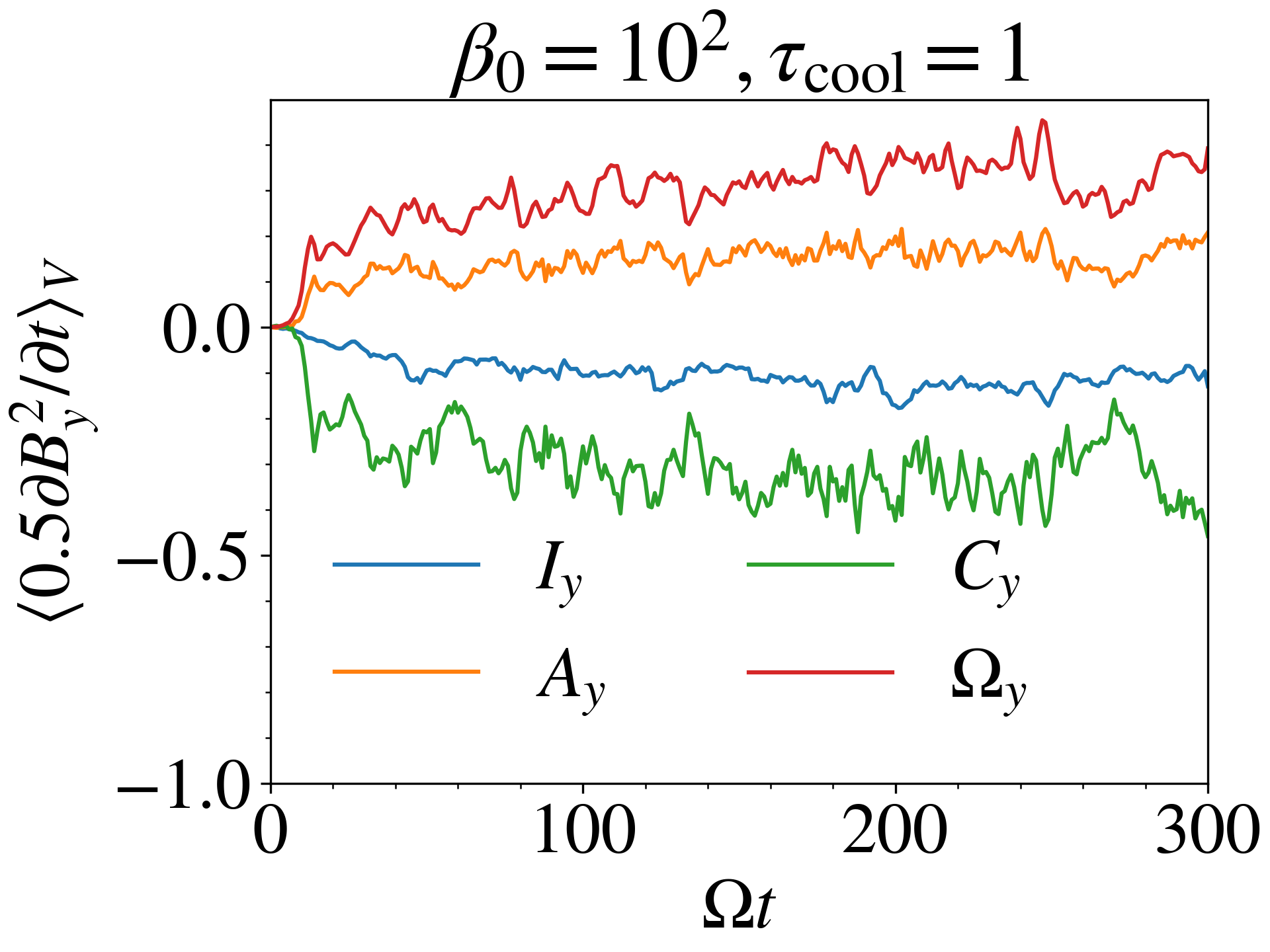} \\
    \includegraphics[width=0.23\textwidth]{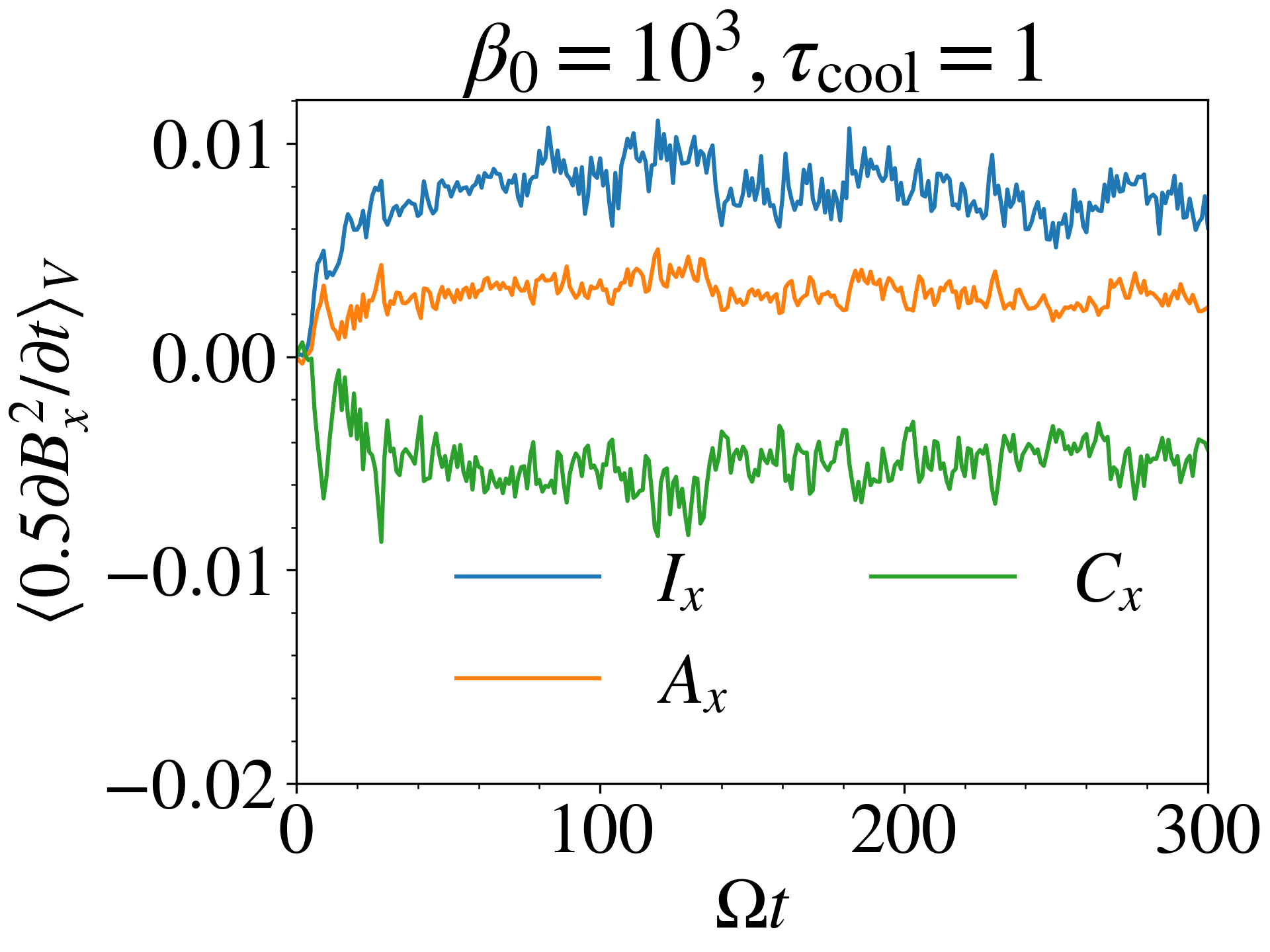}
    \includegraphics[width=0.23\textwidth]{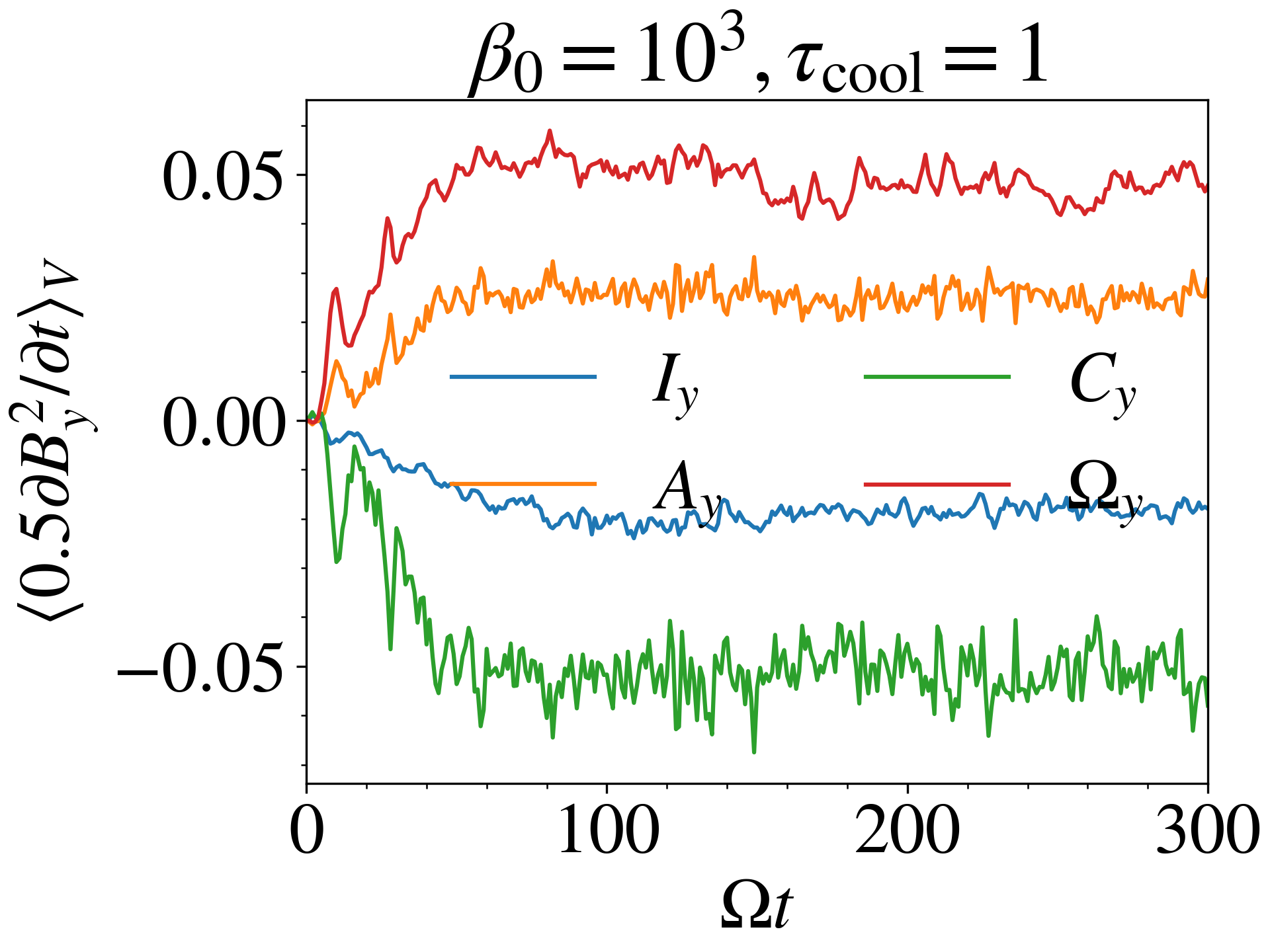} \\
    \includegraphics[width=0.23\textwidth]{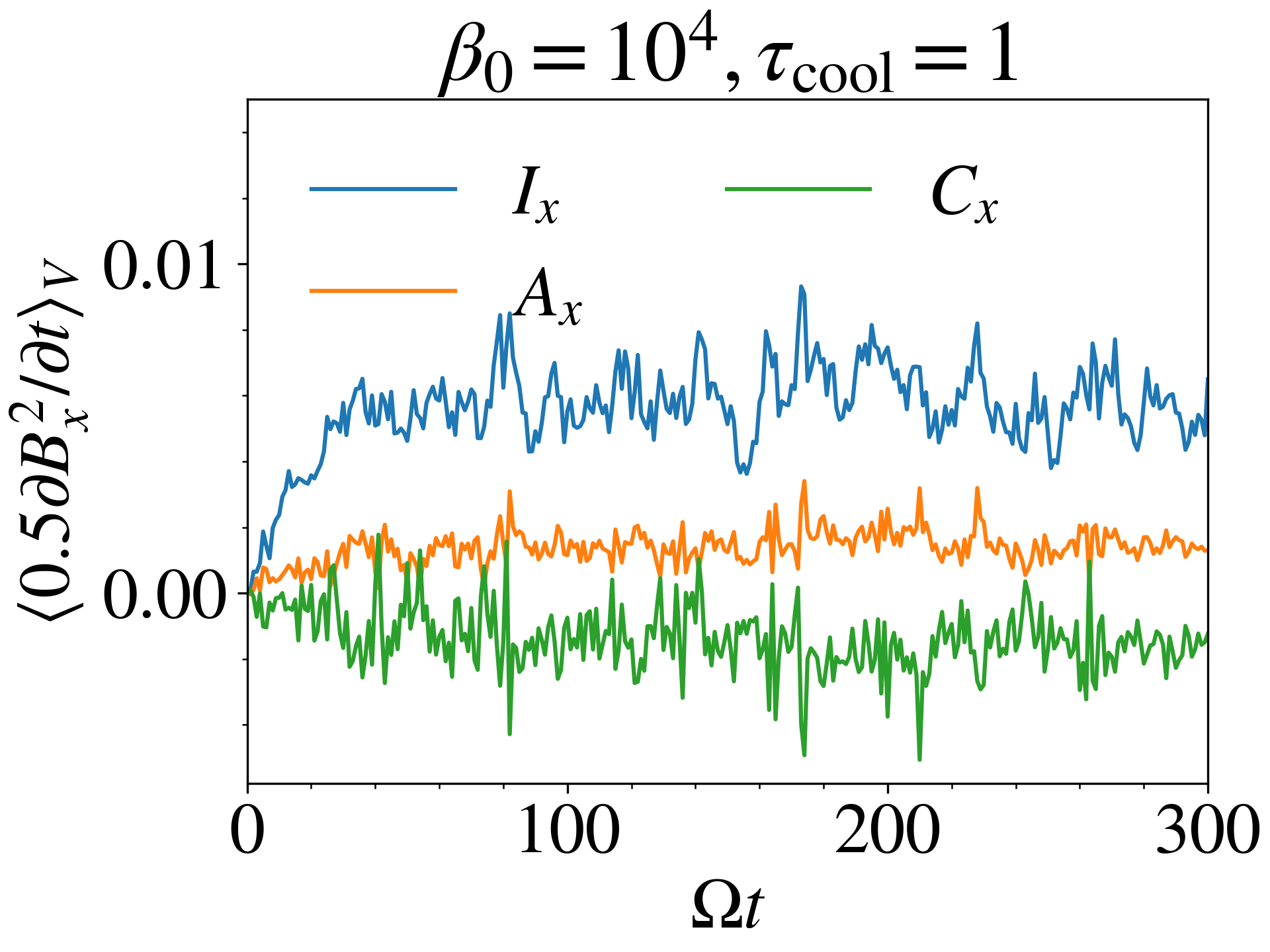}
    \includegraphics[width=0.23\textwidth]{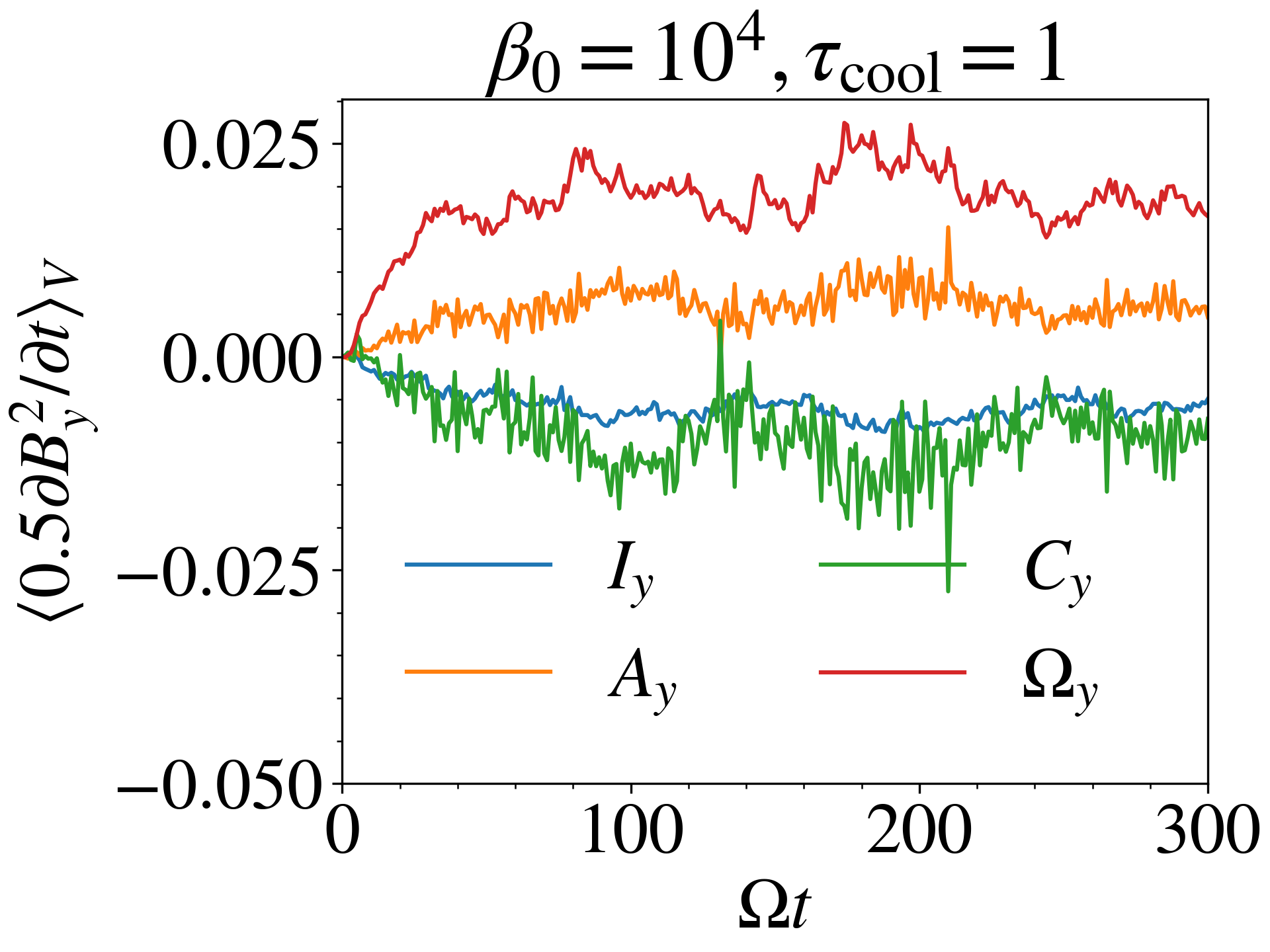} \\
    \includegraphics[width=0.23\textwidth]{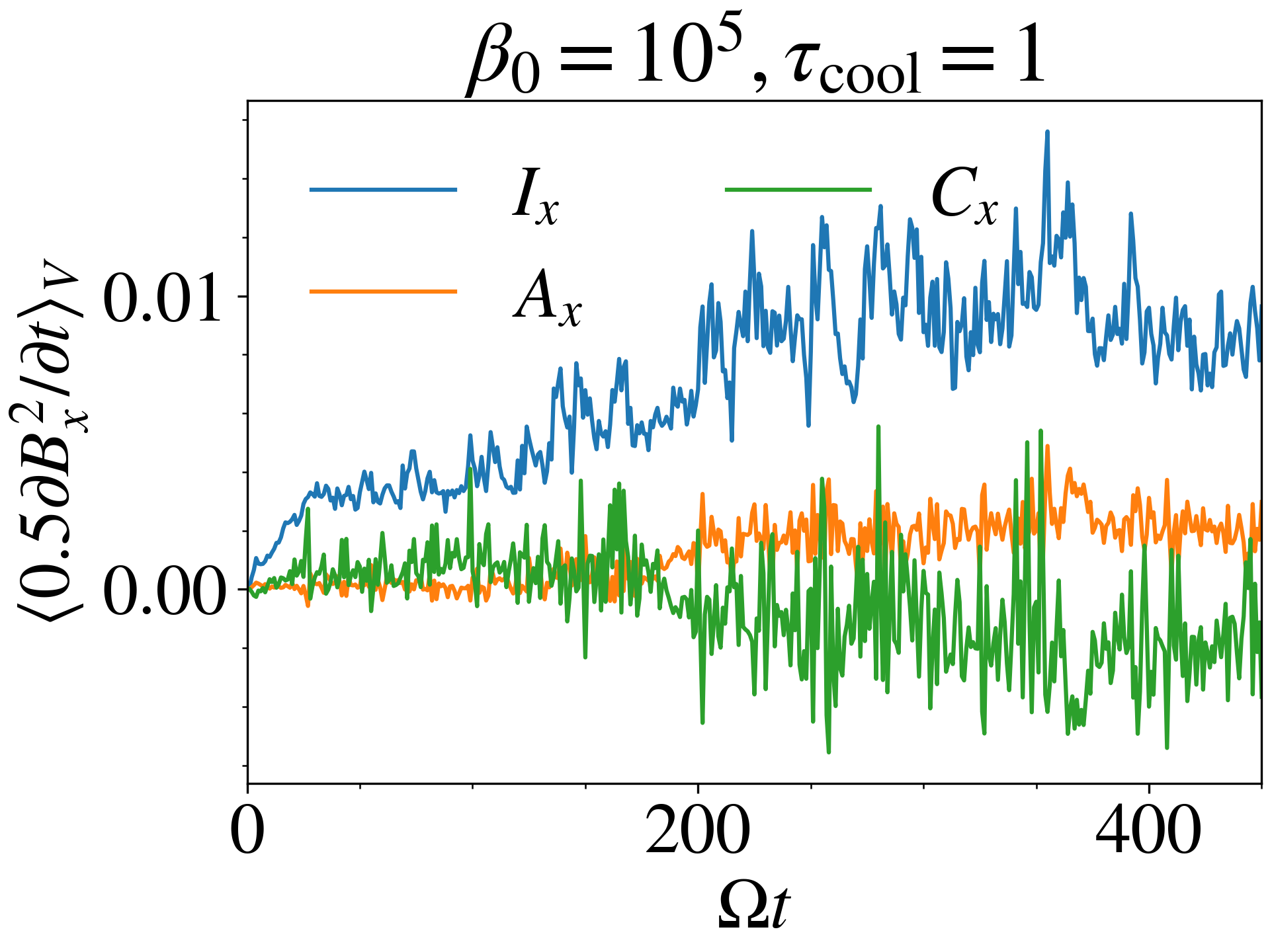}
    \includegraphics[width=0.23\textwidth]{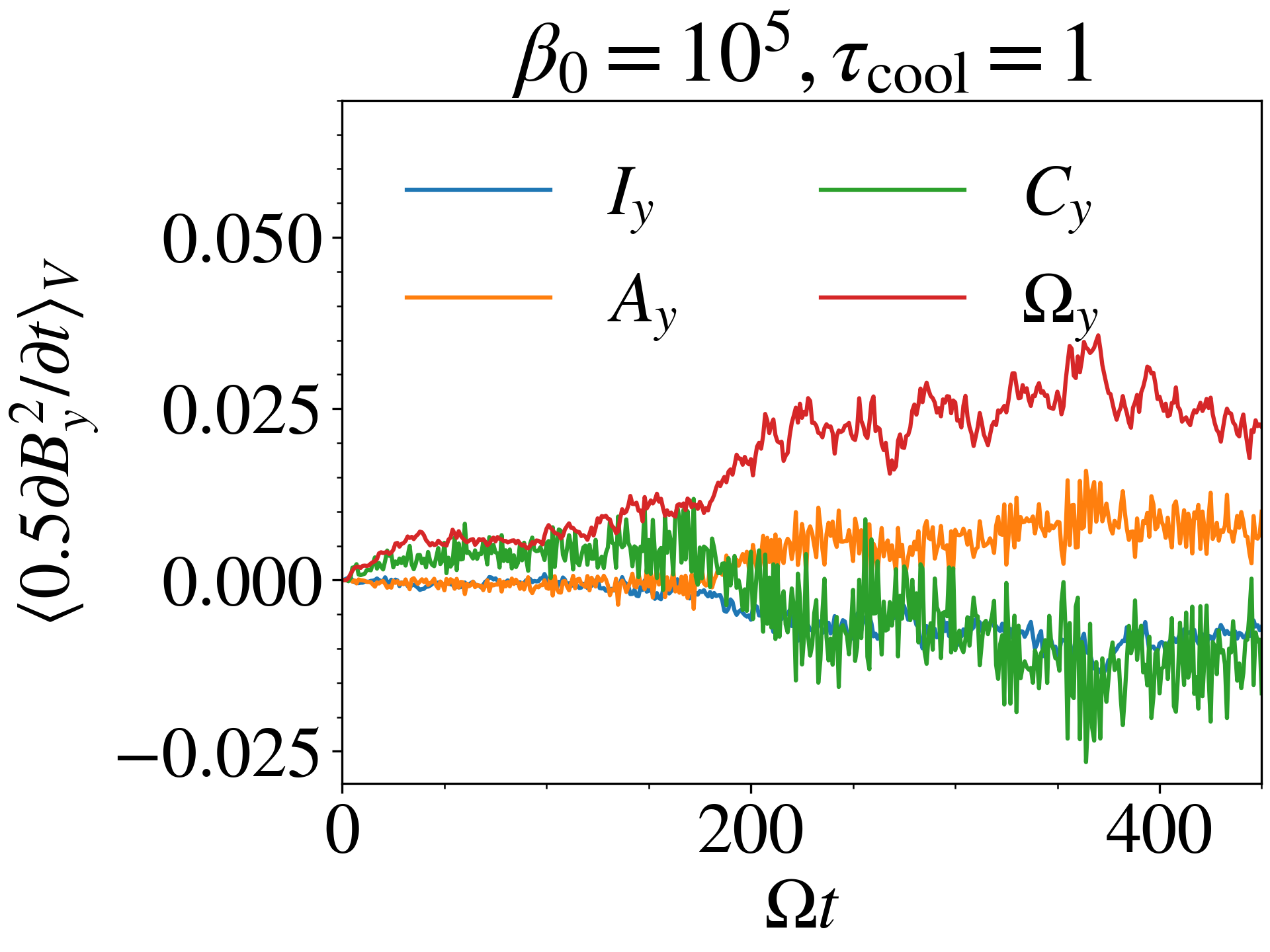}
    \caption{Decomposition for $\partial B_x^2/\partial t$ (left column) and $\partial B_y^2/\partial t$ (right column), each curve denoting the volume average of a term in the decomposition (eq.~\ref{eqn:helmholtz}), against time for various $\beta_0$ cases.}
    \label{fig:helmholtz_decom}
\end{figure}

To aid our understanding of the growth and sustenance of the magnetic field, we decompose the magnetic budget into various contributions, as in \citet{Riols_Latter-2018}, and plot them in Fig.~\ref{fig:helmholtz_decom}. From the induction equation, we obtain
\begin{equation}
    \frac{1}{2}\pdv{B_i^2}{t} = I_i + A_i + C_i + \Omega_i, \label{eqn:helmholtz}
\end{equation}
where $I_i=B_i B_j\pdv*{\delta v_i}{x_j}$ denotes the `stretching' term, $A_i = -B_i\delta v_j\pdv*{B_i}{x_j}$ is the advection term, $C_i = -B_i^2\nabla\cdot\delta\vb{v}$ is the compression term, and $\Omega_i = q\Omega (x B_i\pdv*{B_i}{y} - B_i B_x)\delta_{iy}$ is the $\Omega$-effect (linear stretching by shear). Note that summation is implied for $j$ but not $i$ in eq.~\ref{eqn:helmholtz}, and $\delta\vb{v} = \vb{v} + q\Omega x\vu{y}$ is the fluctuation velocity. In practice, the first term of the $\Omega$ contribution ($q\Omega x B_i\pdv*{B_i}{y}$) is negligible, so $\Omega_i\approx -q\Omega B_i B_x$ effectively. We note that the stretching term $I_x$ is the dominant contribution to the radial field, while the $\Omega$ term (stretching by shear) is the dominant contribution to the toroidal field. These results are consistent with \citet{Riols_Latter-2018}. Unlike their findings, advection ($A_i$) and compression ($C_i$) terms are not negligible in our simulations, especially for low $\beta_0$ cases, where $A_i, C_i$ can be comparable in magnitude to the dominant term. Advection effects ($A_i$) generally supplement field generation while compression effects ($C_i$) are the dominant counter-balance terms for the magnetic field budget in all cases. An examination of the constituent terms for $A_i, C_i$ (not shown here) shows that at low $\beta_0$, the contributions from terms involving $v_z$ dominate, leading to exceptionally large magnitudes of $A_i, C_i$. In high $\beta_0$ cases where $v_z$ is small, $A_i,C_i$ are weaker. While this suggests vertical expansion and advection by a magnetically driven wind at low $\beta_0$ could be important to field generation, note that $A_i$ and $C_i$ are anticorrelated. We know for a fact that the vertical wind speed generally increases with height $\abs{z}$ while the magnetic field generally decreases with $\abs{z}$. Suppose
\begin{equation}
    v_z\qty(z)\propto B_i^{-p}\qty(z), \label{eqn:wind_power_law}
\end{equation}
where $p>0$ generally. After some manipulation, we find that
\begin{equation}
    -B_i^2\pdv{v_z}{z} = p v_z\pdv{z}\qty(\frac{B_i^2}{2}). \label{eqn:manipulate}
\end{equation}
Since the terms involving $v_z$ dominate the contributions to $A_i, C_i$ when the wind is strong, eq.~\ref{eqn:manipulate} implies
\begin{equation}
    C_i \approx -p A_i.
\end{equation}
That is, if $p=1$, then the advective term cancels the compression term. If $p<1$ and $A_i$ remains positive, the combined advective and compression contributions from the wind then contribute positively to magnetic field generation. 


\citet{Riols_Latter-2018} found in several zero-net-flux simulations with self-gravity and strong cooling ($\tau_\mathrm{cool}\lesssim 5$) that magnetic field is generated mostly by a gravitoturbulent dynamo, as demonstrated by high levels of gravitational stress that are comparable to or exceed the Reynolds and Maxwell stresses. They also noticed a change in the location of field generation from $\abs{z}\sim H$ (above and below the disk) in pure MRI to $\abs{z}\sim 0$ (within the disk) in gravitoturbulent dynamos. The subdominant levels of gravitational stress observed in our low $\beta_0$ simulations seem to suggest against a gravitoturbulent dynamo at play in those cases. However, in the high $\beta_0$ cases, the relatively high levels of gravitational stresses together with the higher stresses compared to standard MRI calculations is suggestive of a gravitoturbulent dynamo at play, consistent with \citet{Riols_Latter-2018}. Unlike in pure MRI (no cooling or self-gravity), field generation occurs mostly at $z\sim 0$ in our simulations. We attribute this to cooling suppressing the buoyant rise of disk gas. We conclude that MRI-generated turbulence is still responsible for the dynamo process needed for field generation in our low $\beta_0$ simulations, while there could be mixed contributions from MRI and gravitoturbulence in the high $\beta_0$ cases. In a simulation performed by \citet{Riols_Latter-2018} with net flux and a moderately strong initial field ($\beta_0\approx 225$), it was found that the gravitational stress is the most subdominant stress, consistent with our low $\beta_0$ test cases.

Magnetic field generation in setups with self-gravity and cooling is a profound subject warranting a separate study. For our purposes, we simply note that evidence does not seem to support self-gravity being responsible for the dynamo process in our low $\beta_0$ simulations, but could be conducive to such due to fragmentation in the high $\beta_0$ cases.  We also observe that flux-tube-stretching $I_x$ is the dominant contribution to the radial field which, when fed into flows with shear, contributes to most of the toroidal field generated through the shearing term $\Omega_i$.



\subsection{Magnetic field structure} \label{subsec:magnetic_field}

\begin{figure}
    \centering
    \includegraphics[width=0.23\textwidth]{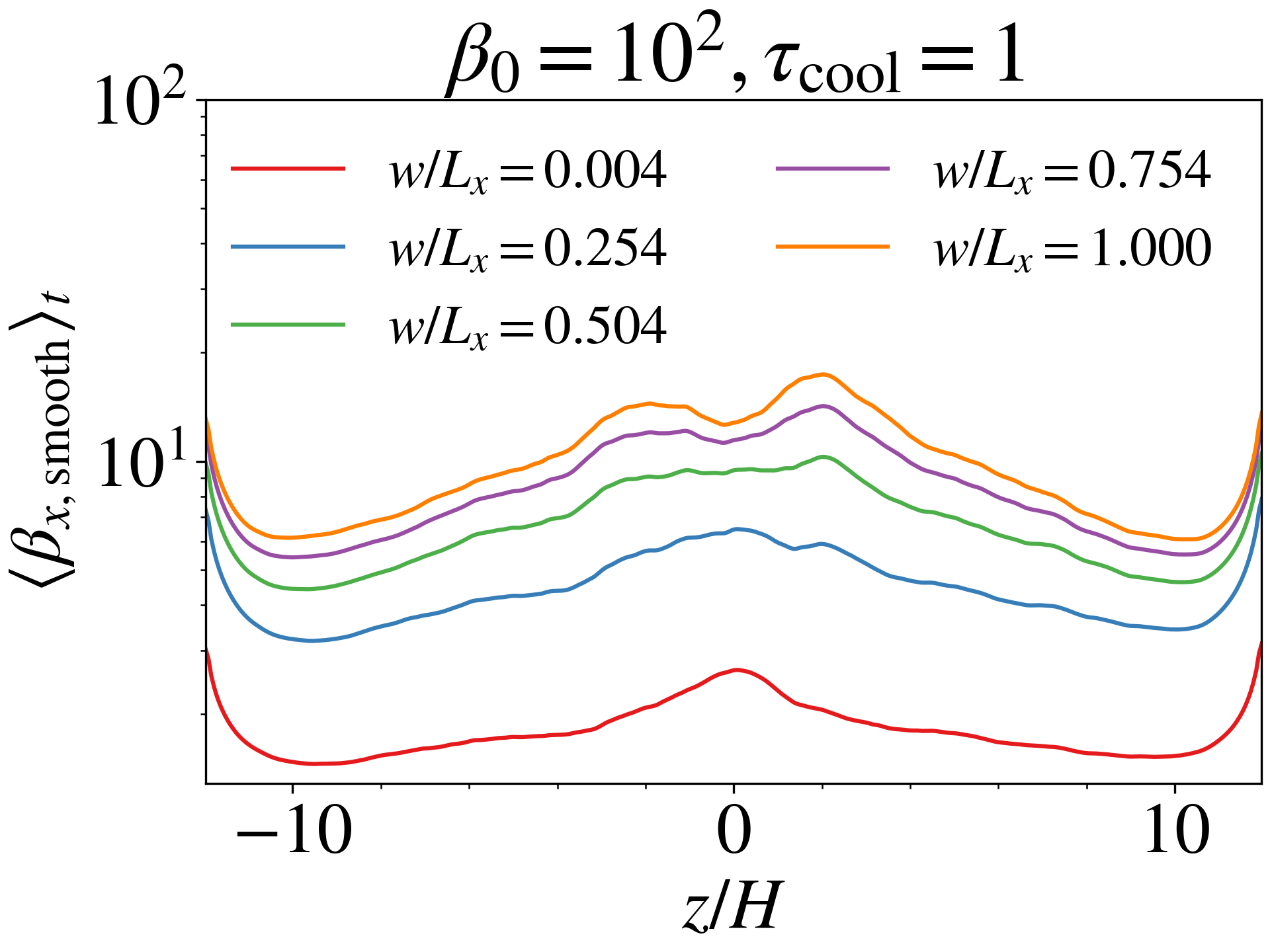}
    \includegraphics[width=0.23\textwidth]{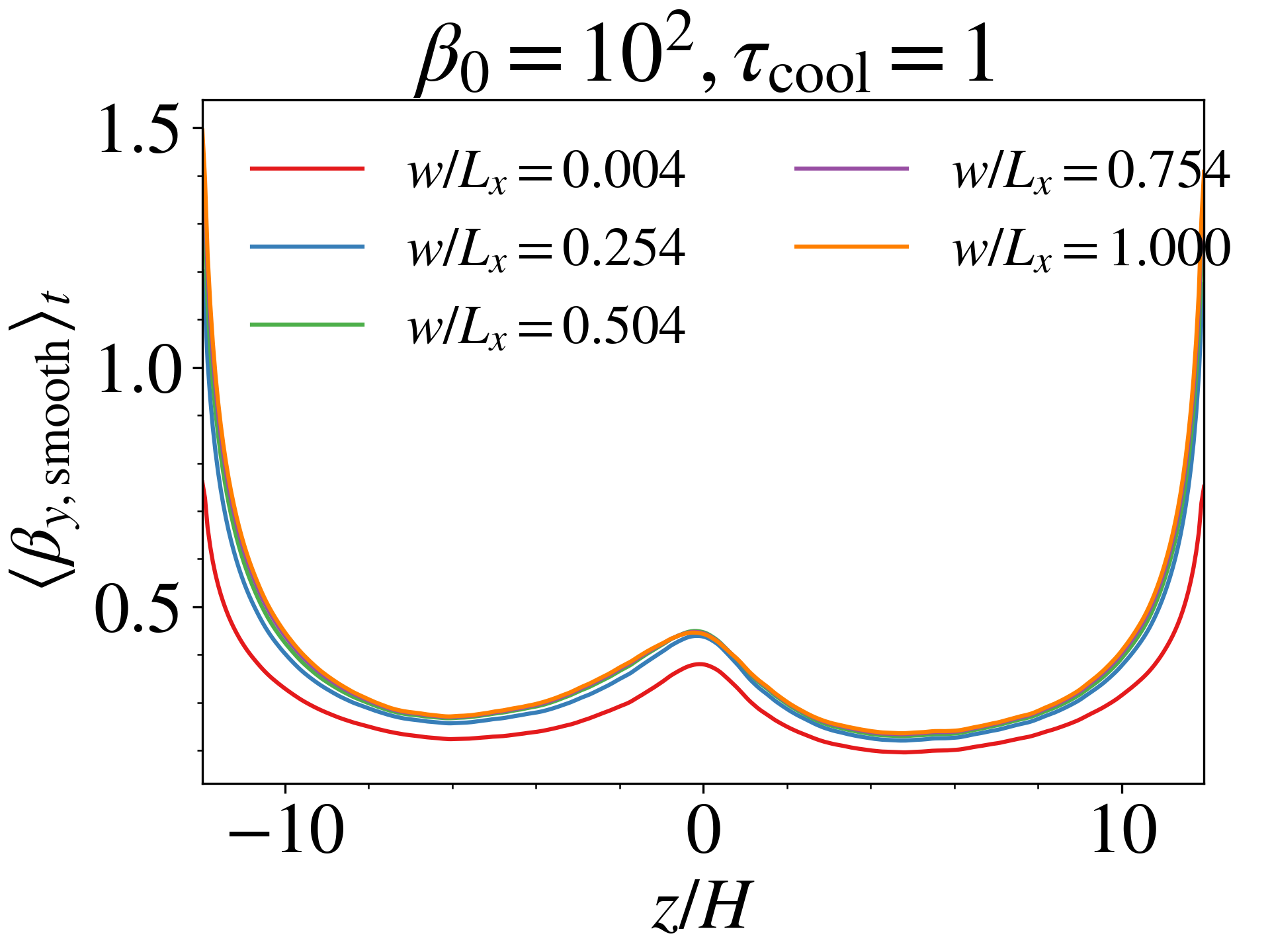} 
    \includegraphics[width=0.23\textwidth]{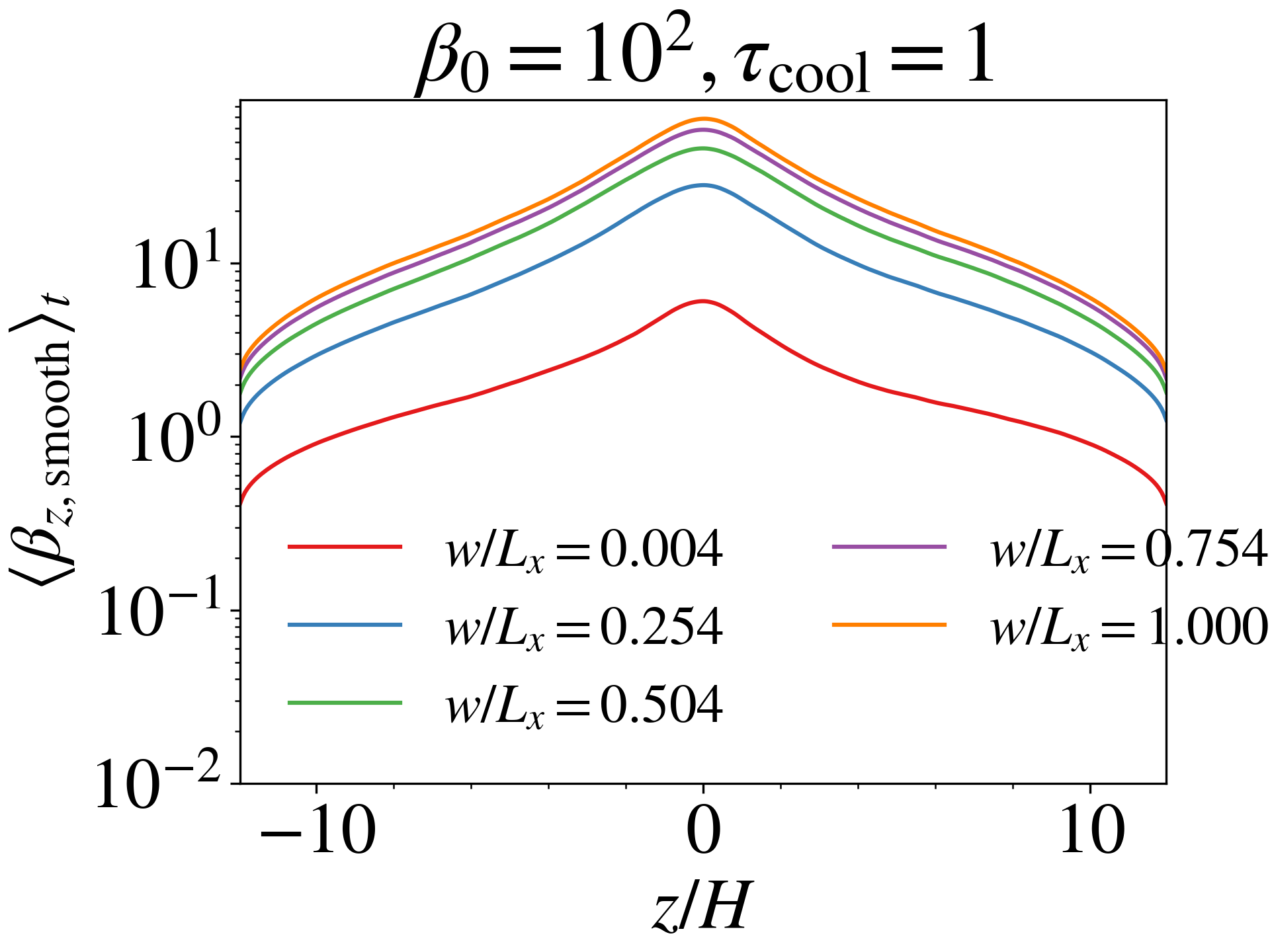} 
    \caption{Top left: Window-averaged, time-averaged plasma beta of the radial field defined by $\langle\beta_{x,\mathrm{smooth}}\rangle_t = 2\langle\langle\bar{P}_{g,w}\rangle_{xy}\rangle_t/\langle\langle\bar{B}^2_{x,w}\rangle_{xy}\rangle_t$. Top right: Window-averaged, time-averaged plasma beta of the toroidal field $\langle\beta_{x,\mathrm{smooth}}\rangle_t = 2\langle\langle\bar{P}_{g,w}\rangle_{xy}\rangle_t/\langle\langle\bar{B}^2_{y,w}\rangle_{xy}\rangle_t$. Bottom: Window-averaged, time-averaged plasma beta of the vertical field $\langle\beta_{z,\mathrm{smooth}}\rangle_t = 2\langle\langle\bar{P}_{g,w}\rangle_{xy}\rangle_t/\langle\langle\bar{B}^2_{z,w}\rangle_{xy}\rangle_t$.  Various window lengths $w$ are displayed. The case displayed is $\beta_0=10^2,\beta_\mathrm{cool}=1$. Time-average is taken from $\Omega t=100-300$.}
    \label{fig:beta_100_cool1_field}
\end{figure}

In this section, we examine the magnetic fields in our simulations. The first thing we would like to point out is that the strength of the radial and vertical fields varies with scale. The meaning of this is illustrated in Fig.~\ref{fig:beta_100_cool1_field}, which shows the window-averaged, time-averaged plasma beta parameter corresponding to the radial field $\langle\beta_{x,\mathrm{smooth}}\rangle_t = 2\langle\langle\bar{P}_{g,w}\rangle_{xy}\rangle_t/\langle\langle\bar{B}^2_{x,w}\rangle_{xy}\rangle_t$ (top left), toroidal field $\langle\beta_{y,\mathrm{smooth}}\rangle_t = 2\langle\langle\bar{P}_{g,w}\rangle_{xy}\rangle_t/\langle\langle\bar{B}^2_{y,w}\rangle_{xy}\rangle_t$ (top right), and vertical field $\langle\beta_{z,\mathrm{smooth}}\rangle_t = 2\langle\langle\bar{P}_{g,w}\rangle_{xy}\rangle_t/\langle\langle\bar{B}^2_{z,w}\rangle_{xy}\rangle_t$ (bottom) for different window lengths, where the time average is taken from $\Omega t=100-300$. The displayed case is $\beta_0=10^2,\tau_\mathrm{cool}=1$. The meaning of window-average, and the notation $\bar{X}_w$, is explained in \S\ref{sec:diagnostics}. Briefly, it is a smoothing operation in the $x,y$-direction, with $w$ being the smoothing window. By performing such smoothing operations, we are essentially removing contributions from modes with wavenumbers $k_x w, k_y w > 1$. A clear difference between the top right panel and the rest of Fig.~\ref{fig:beta_100_cool1_field} is that $\beta_{x,\mathrm{smooth}}, \beta_{z,\mathrm{smooth}}$ are sensitive to the window length chosen, while $\beta_{y,\mathrm{smooth}}$ is insensitive to it. This means there is a lot of small scale structure in the radial and vertical field while the azimuthal field structure consists mostly of a single, large scale field. The difference in $\beta_{x,\mathrm{smooth}}, \beta_{z,\mathrm{smooth}}$ when one includes all small scale modes\footnote{$\beta$ depends inversely on field strength so a low value for the small window means a lot more mangetic energy density.} (the smallest window width $w/L_x=0.004$) compared to when one includes only the domain-sized mode (the largest window width $w/L_x=1$) can be an order of magnitude for the case shown. In this section, when discussing quantities related to the field, we will display only window-averaged results with window lengths $w/L_x=0.004$ and $w/L_x=1$ (or as specified) to reduce clutter, representing results including the full spectrum and just the large-scale, respectively.

\begin{figure}
    \centering
    \includegraphics[width=0.23\textwidth]{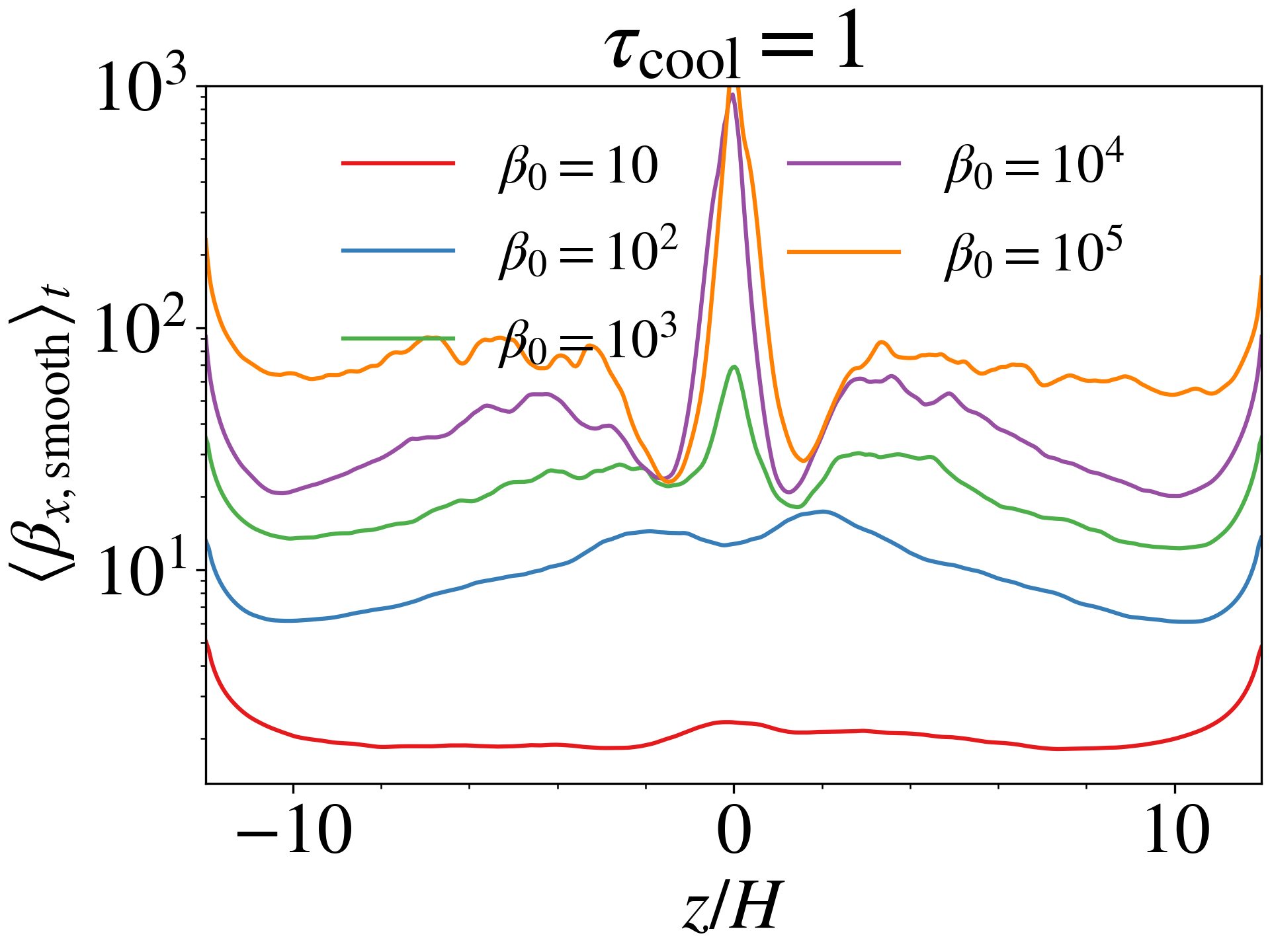}
    \includegraphics[width=0.23\textwidth]{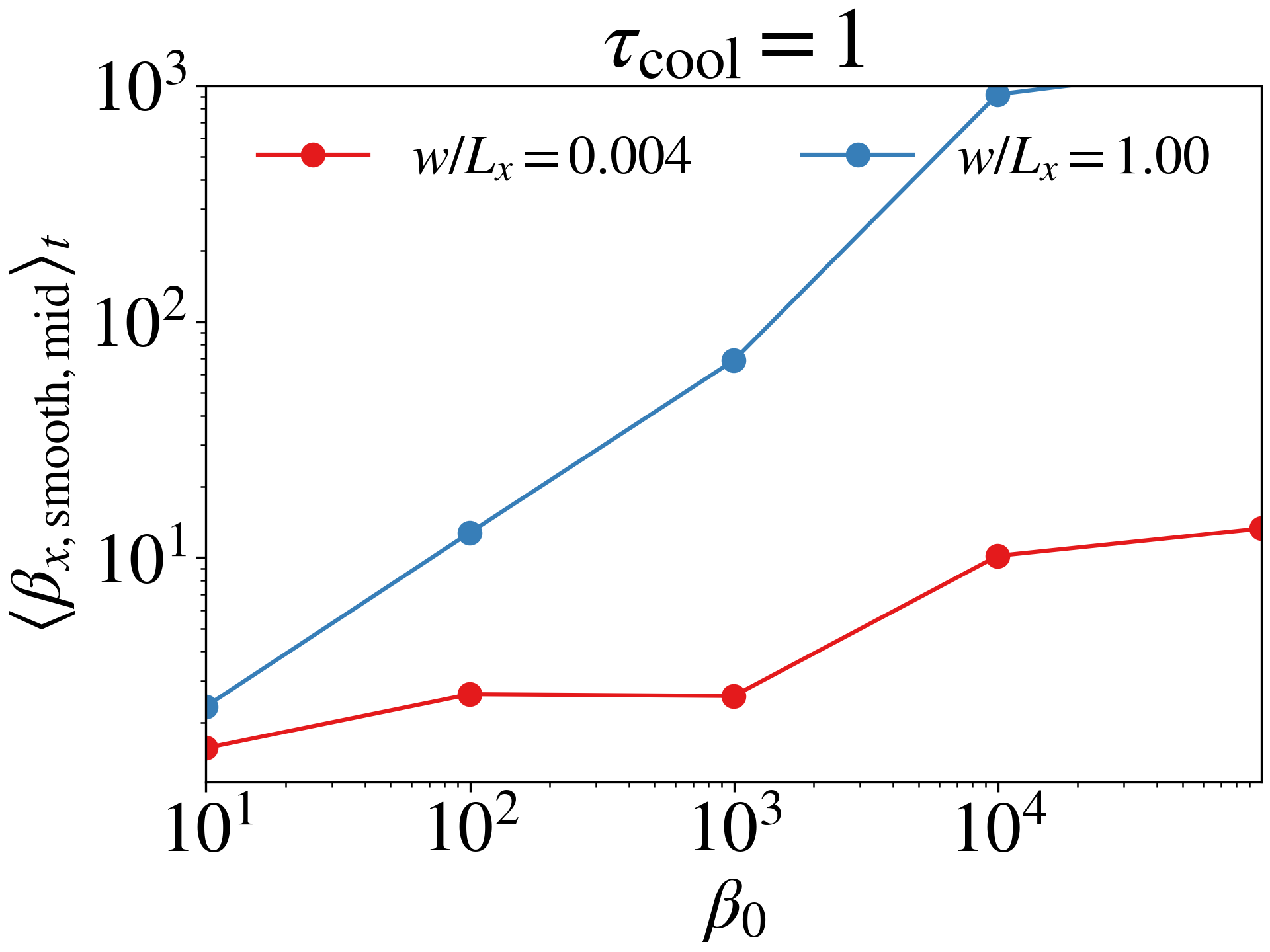} \\
    \includegraphics[width=0.23\textwidth]{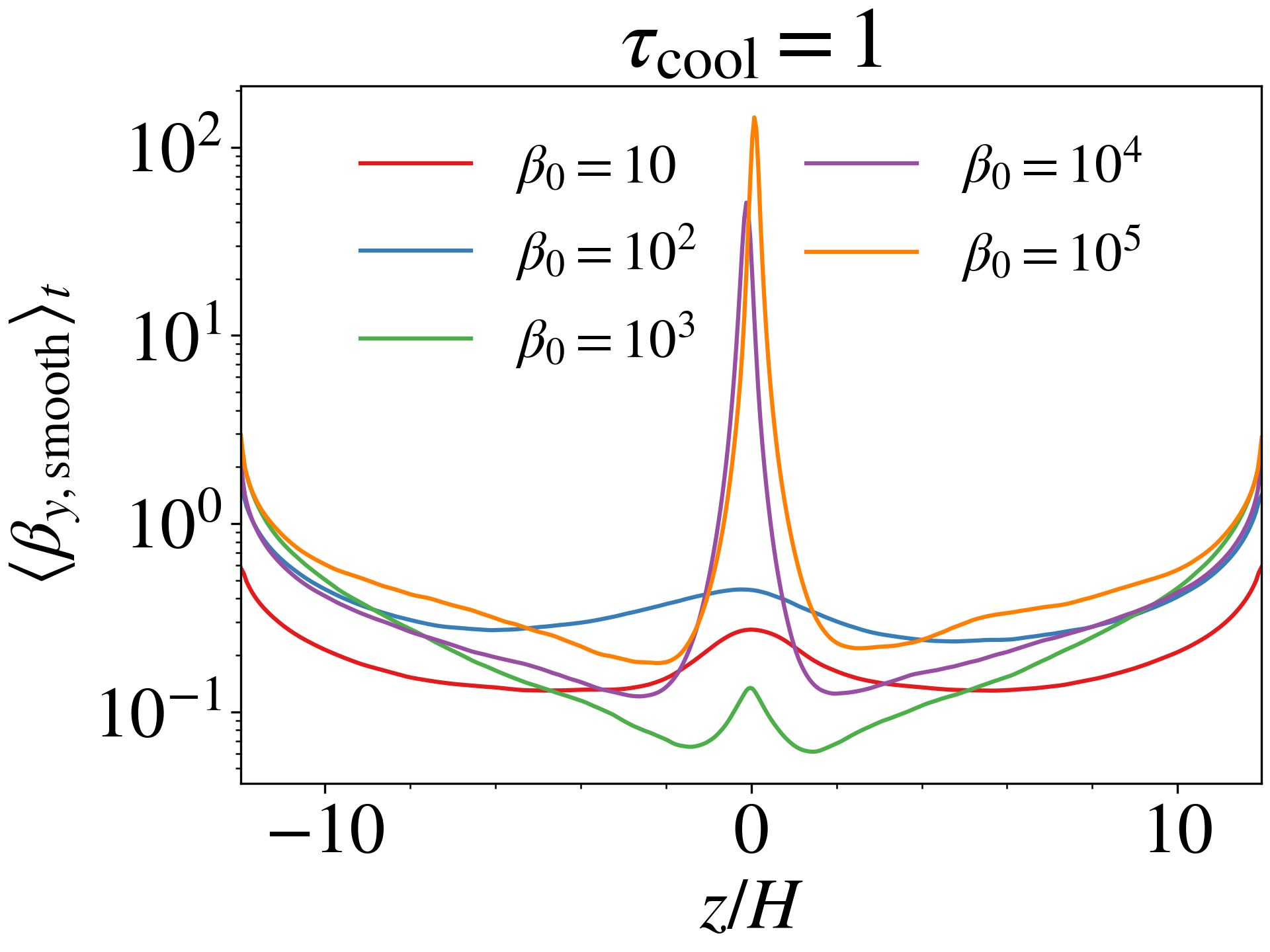}
    \includegraphics[width=0.23\textwidth]{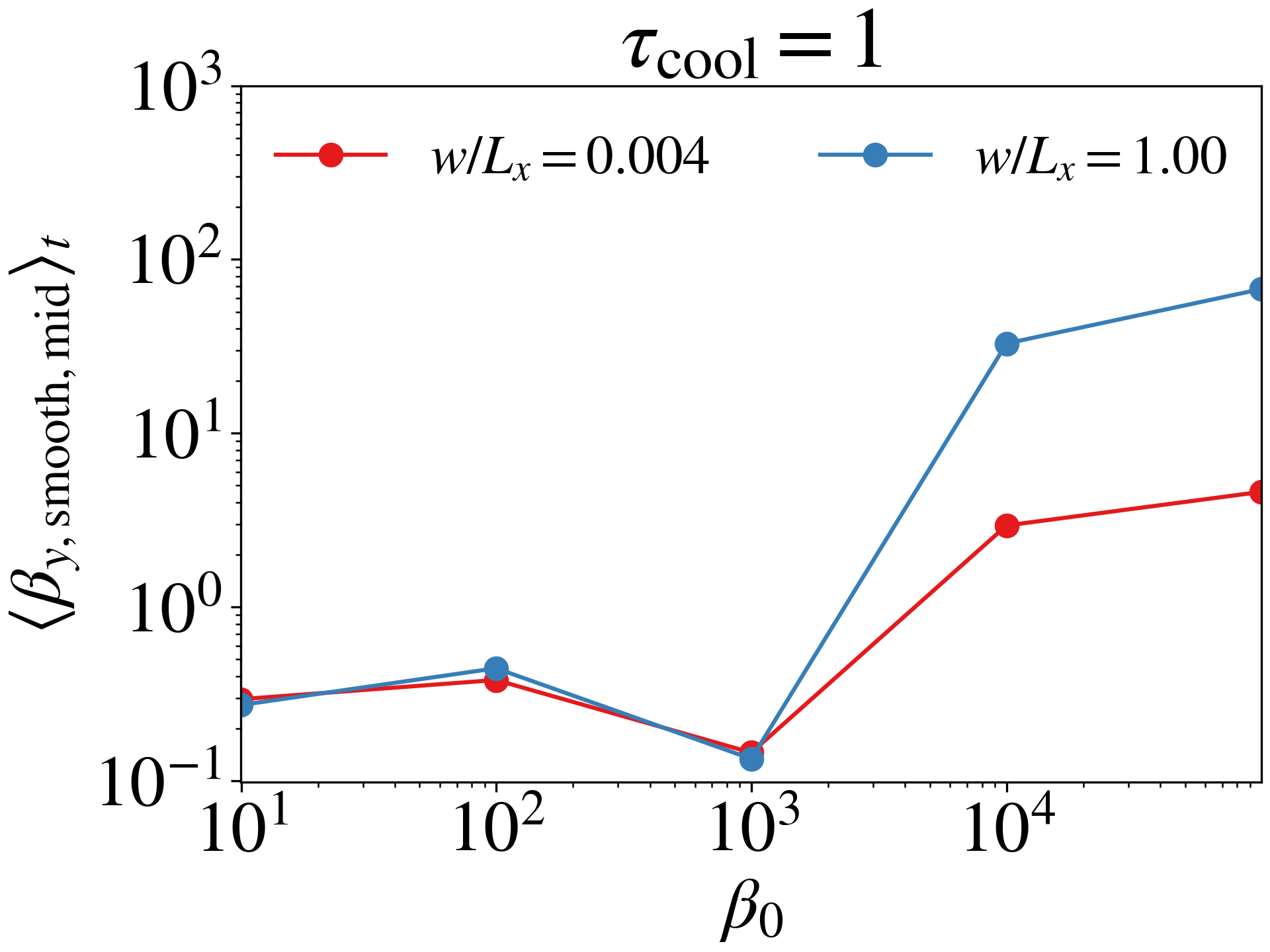} \\
    \includegraphics[width=0.23\textwidth]{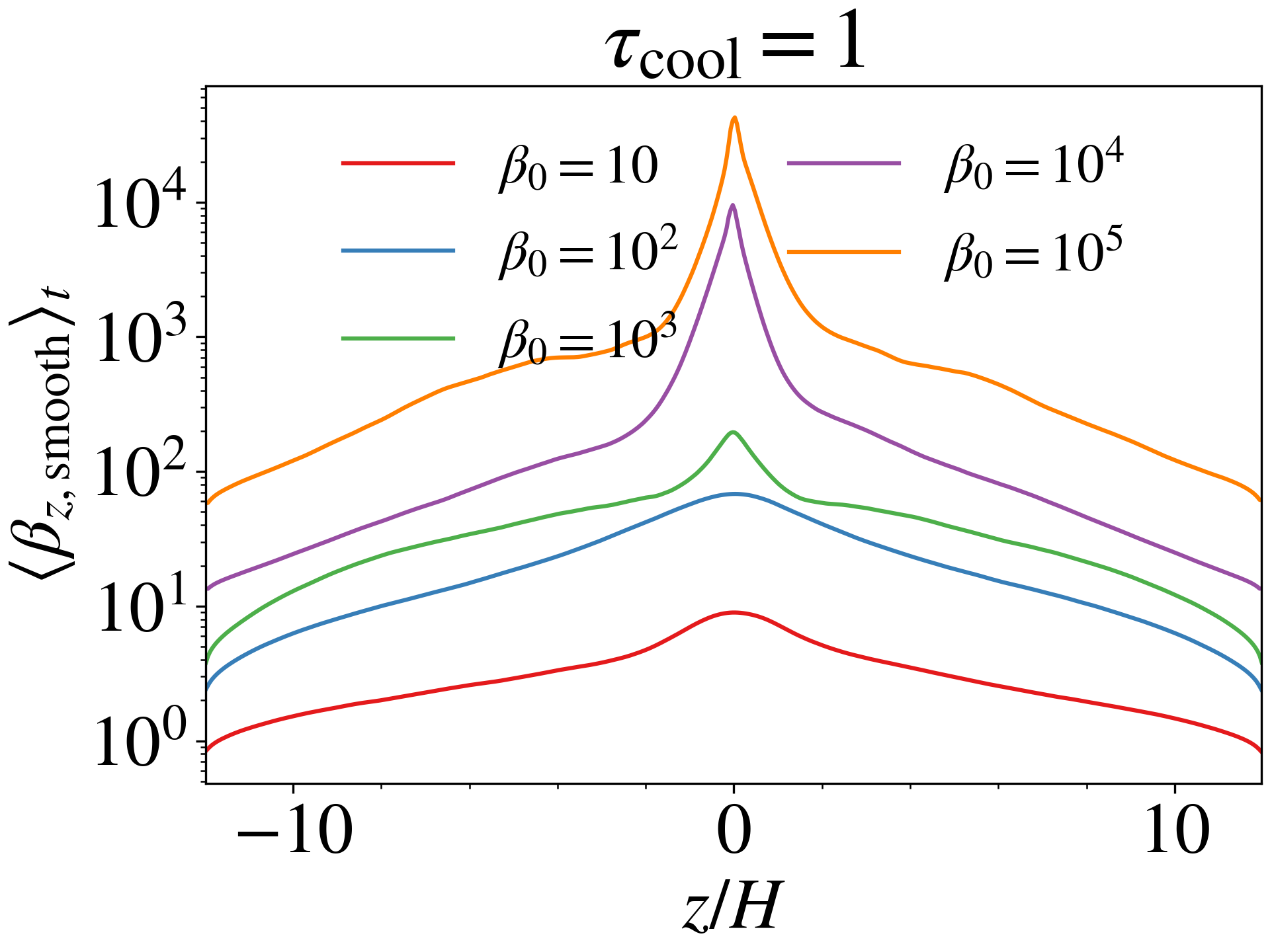}
    \includegraphics[width=0.23\textwidth]{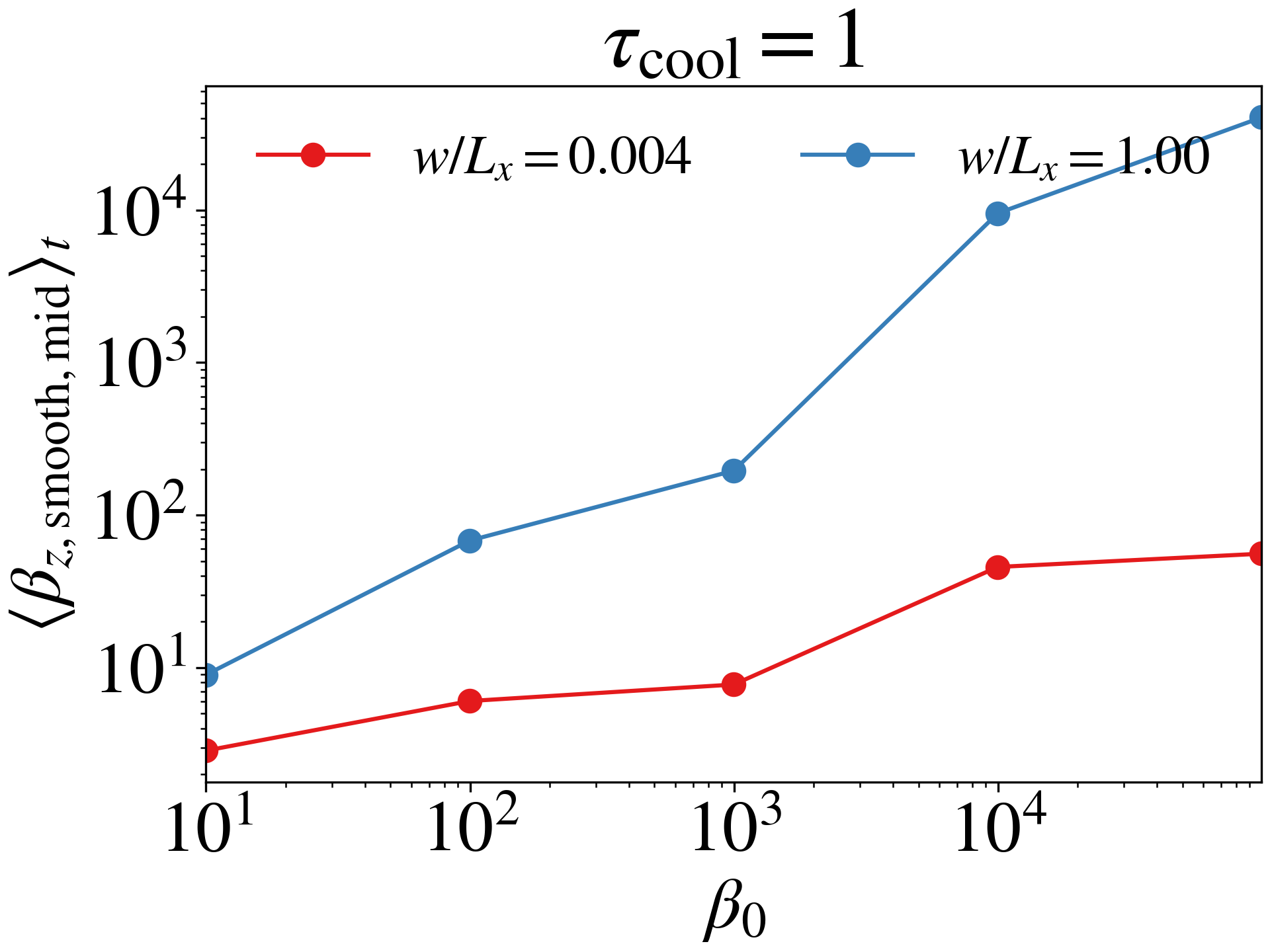} \\
    \includegraphics[width=0.23\textwidth]{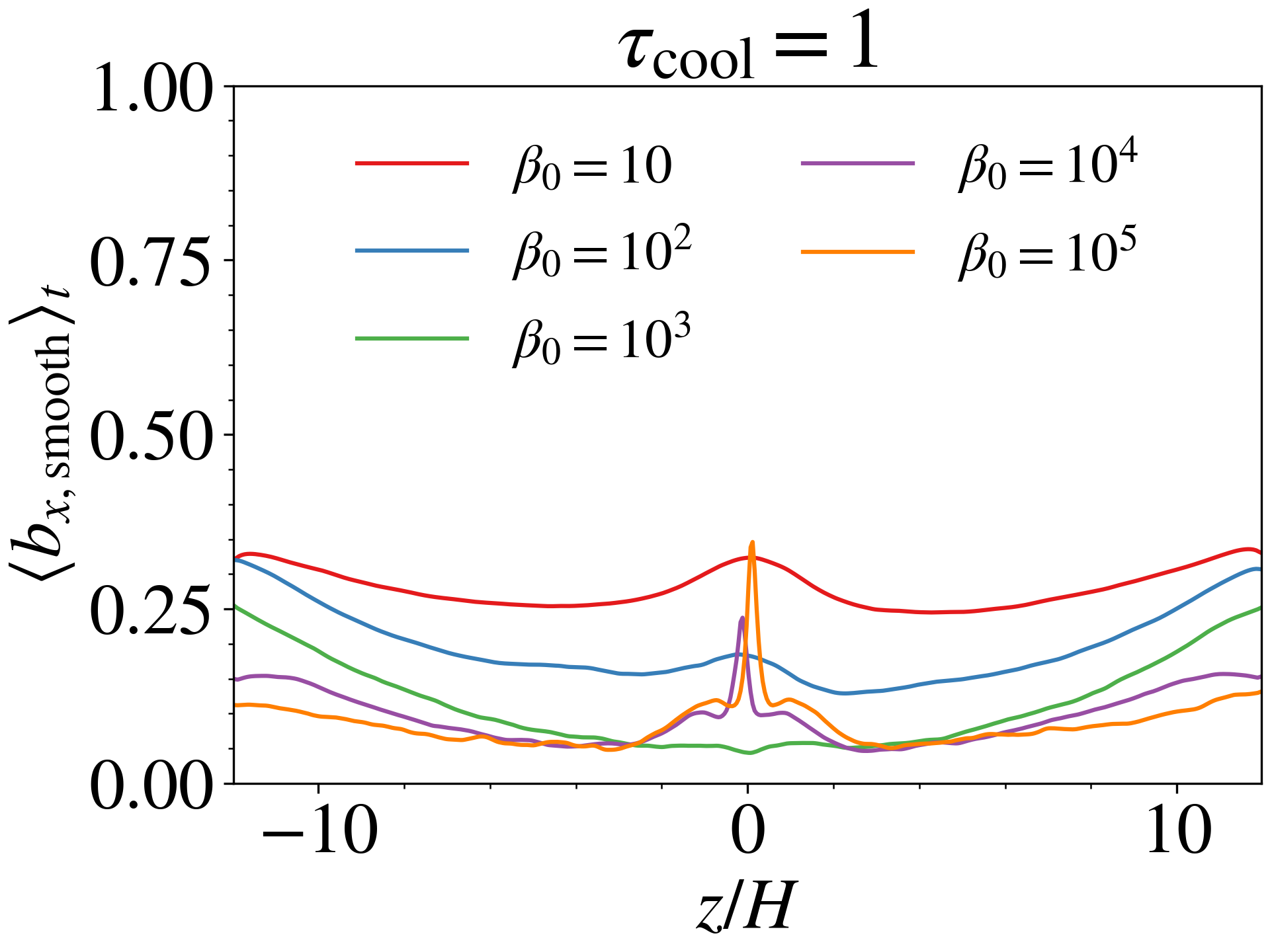}
    \includegraphics[width=0.23\textwidth]{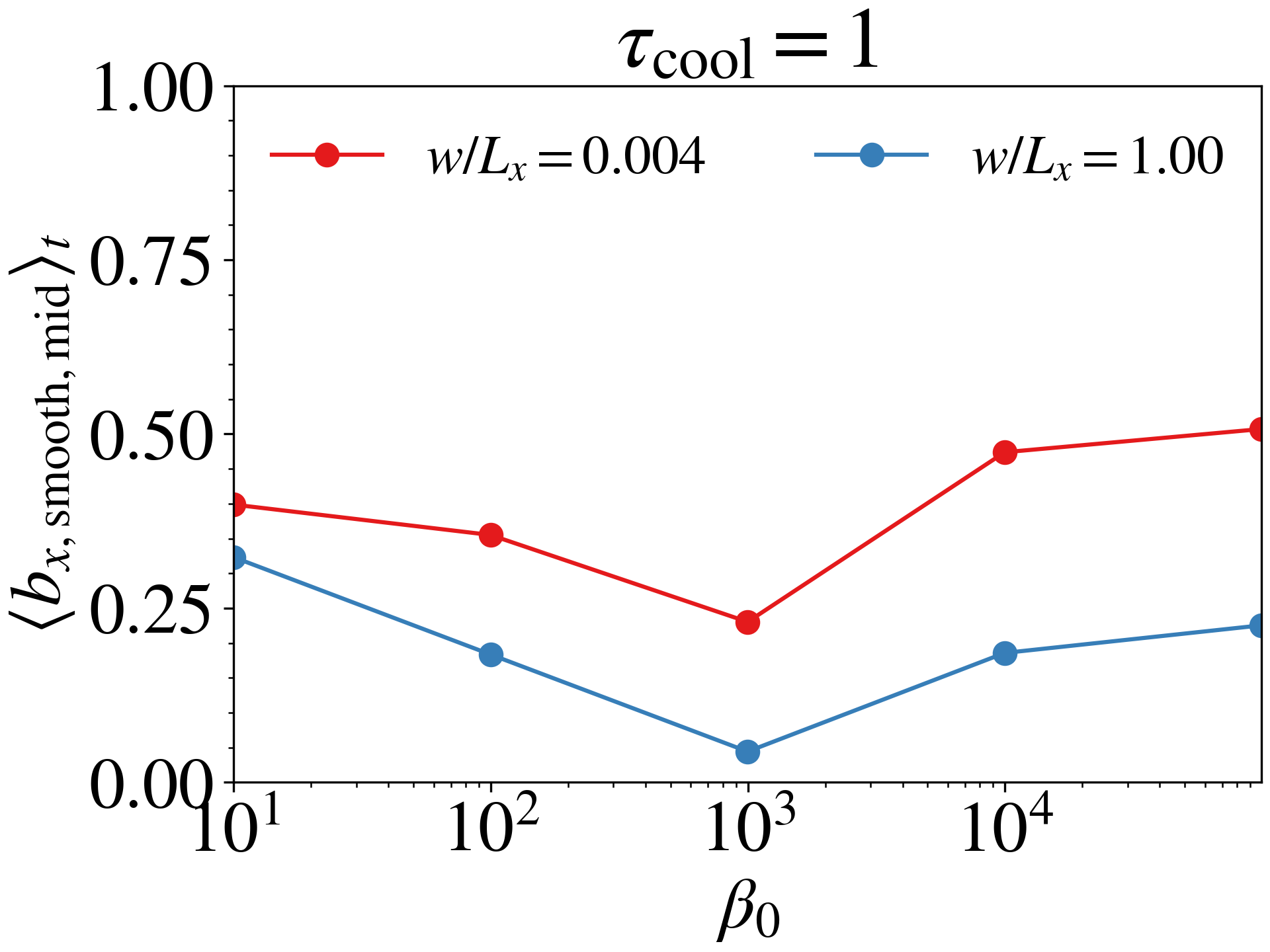}
    \caption{Left column: Window-averaged, time-averaged profiles of the radial field plasma beta $\langle\beta_{x,\mathrm{smooth}}\rangle_t=2\langle\langle\bar{P}_{g,w}\rangle_{xy}\rangle_t/\langle\langle\bar{B}^2_{x,w}\rangle_{xy}\rangle_t$ (top left), the toroidal field plasma beta $\langle\beta_{y,\mathrm{smooth}}\rangle_t=2\langle\langle\bar{P}_{g,w}\rangle_{xy}\rangle_t/\langle\langle\bar{B}^2_{y,w}\rangle_{xy}\rangle_t$ (second row left), the vertical field plasma beta $\langle\beta_{z,\mathrm{smooth}}\rangle_t=2\langle\langle\bar{P}_{g,w}\rangle_{xy}\rangle_t/\langle\langle\bar{B}^2_{z,w}\rangle_{xy}\rangle_t$ (third row left), and the relative radial field strength $\langle b_{x,\mathrm{smooth}}\rangle_t=[\langle\langle\bar{B}^2_{x,w}\rangle_{xy}\rangle_t/(\langle\langle\bar{B}^2_{x,w}\rangle_{xy}\rangle_t+\langle\langle\bar{B}^2_{y,w}\rangle_{xy}\rangle_t)]^{1/2}$ (bottom left) for various $\beta_0$, with $w/L_x=1$ as the window length. Right column: Plot of $\langle\beta_{x,\mathrm{smooth},\mathrm{mid}}\rangle_t$ (top right), $\langle\beta_{y,\mathrm{smooth},\mathrm{mid}}\rangle_t$ (second row right), $\langle\beta_{z,\mathrm{smooth},\mathrm{mid}}\rangle_t$ (third row right) and $\langle b_{x,\mathrm{smooth},\mathrm{mid}}\rangle_t$ (bottom right), where the subscript `mid' denotes values taken at $z=0$, as a function of $\beta_0$, for two window lengths: $w/L_x=0.004$ (red line), $w/L_x=1$ (blue line). The cases displayed have $\tau_\mathrm{cool}=1$. Time-average is taken from $\Omega t=100-300$ for the $\beta_0\leq 10^4$ cases and from $\Omega t=250-450$ for the $\beta_0=10^5$ case.}
    \label{fig:cool0.5_field_cases}
\end{figure}

The second thing we would like to point out is that the magnetization in the nonlinear stage generally increases with decreasing $\beta_0$, but not necessarily in a monotonic manner for some field components. In Fig.~\ref{fig:cool0.5_field_cases} we plot in the left column the window-averaged, time-averaged profiles of the radial field plasma beta $\langle\beta_{x,\mathrm{smooth}}\rangle_t$ (top left), the toroidal field plasma beta $\langle\beta_{y,\mathrm{smooth}}\rangle_t$ (second row left), the vertical field plasma beta $\langle\beta_{z,\mathrm{smooth}}\rangle_t$ (third row left), and the relative radial field strength $\langle b_{x,\mathrm{smooth}}\rangle_t=[\langle\langle\bar{B}^2_{x,w}\rangle_{xy}\rangle_t/(\langle\langle\bar{B}^2_{x,w}\rangle_{xy}\rangle_t+\langle\langle\bar{B}^2_{y,w}\rangle_{xy}\rangle_t)]^{1/2}$ (bottom left) for various $\beta_0$, with $w/L_x=1$ as the window length. We note that the magnetic field is typically stronger above and below the disk, with the disk mid-plane being the least magnetically dominant\footnote{The upward drift in $\beta_x,\beta_y$ at large $|z|$ is due to the boundary conditions used, for which $B_x, B_y=0$ at the $z$-boundaries.}. The mid-plane magnetic field strength is particularly weak for the high $\beta_0$ cases ($\beta_0=10^4,10^5$), reflecting field reversals seen in Fig.~\ref{fig:timeseries} at $z=0$. The low $\beta_0$ cases $\beta_0=10,10^2$ do not suffer such a flip in polarity at the mid-plane, and are able to maintain field strength at the mid-plane that is comparable to the corona. In the right column of Fig.~\ref{fig:cool0.5_field_cases} we pick out the values of $\langle\beta_{x,\mathrm{smooth}}\rangle_t$, $\langle\beta_{y,\mathrm{smooth}}\rangle_t$, $\langle\beta_{z,\mathrm{smooth}}\rangle_t$ and $\langle b_{x,\mathrm{smooth}}\rangle_t$ at $z=0$ (hence the subscript `mid') and plot them against $\beta_0$. The mid-plane is particularly relevant as fragmentation happens mostly there. We observe that the mid-plane radial plasma beta $\langle\beta_{x,\mathrm{smooth},\mathrm{mid}}\rangle_t$ increases monotonically with respect to $\beta_0$ regardless of the window length chosen (i.e. scale), with the large-scale beta dropping from $\sim 10^3$ for $\beta_0=10^5$ to $\sim 1$ for $\beta_0=10$. The toroidal plasma beta $\langle\beta_{y,\mathrm{smooth},\mathrm{mid}}\rangle_t$ does not drop monotonically with respect to $\beta_0$. It exhibits an asymmetric U-shape pattern, with the minimum located at $\beta_0=10^3$, at which the value is $\sim 0.1$. The two ends of the U-shaped pattern have values of $\sim 10^2$ and $\sim 0.5$ at $\beta_0=10^5, 10$ respectively. Similar to the radial plasma beta, the vertical plasma beta $\langle\beta_{z,\mathrm{smooth},\mathrm{mid}}\rangle_t$ again decreases monotonically with respect to $\beta_0$, with the large-scale field dropping from $\sim 10^5$ for $\beta_0=10^5$ to $\sim 10$ for $\beta_0=10$, just as in the initial setup, while the small-scale components are significantly stronger, hovering between $5-15$. We know that MRI does not modify the mean vertical field much, so this implies that the averaged mid-plane gas pressure remains more or less the same as the initial value despite the thermodynamical changes throughout the simulation. The relative radial field strength $\langle b_{x,\mathrm{smooth},\mathrm{mid}}\rangle_t$ exhibits the same U-shaped trend as $\beta_{y,\mathrm{smooth},\mathrm{mid}}$, with a minimum located at $\beta_0=10^3$ in which the value is close to zero, implying the in-disk field is mostly toroidal. The two ends of the U-shaped pattern have $\langle b_{x,\mathrm{smooth},\mathrm{mid}}\rangle_t$ values between $0.25$ to $0.5$ depending on the scale, indicating the toroidal field component is still dominant, but the radial field component is non-negligible. 


\begin{figure*}
    \centering
    \includegraphics[width=0.3\textwidth]{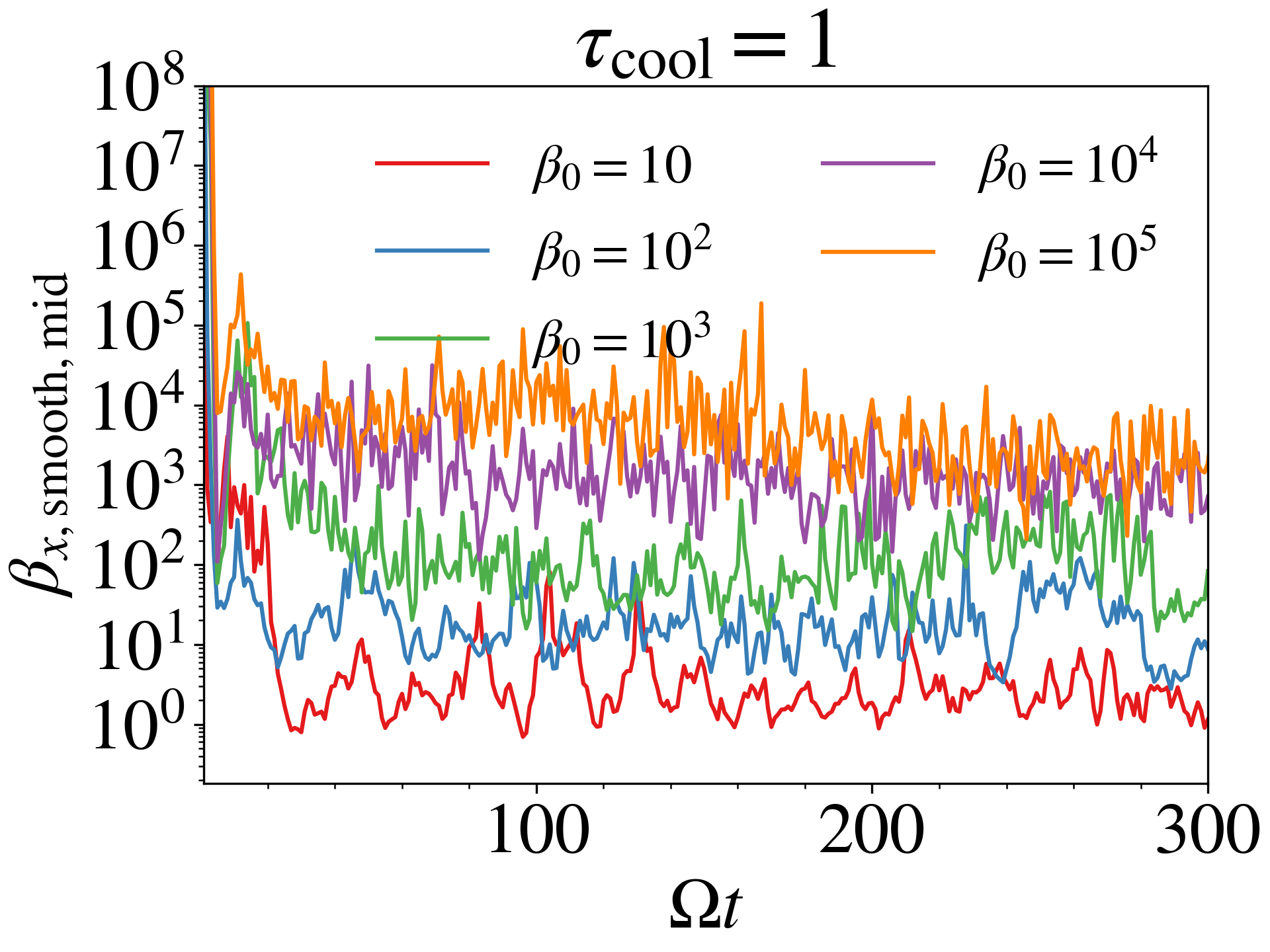} 
    \includegraphics[width=0.3\textwidth]{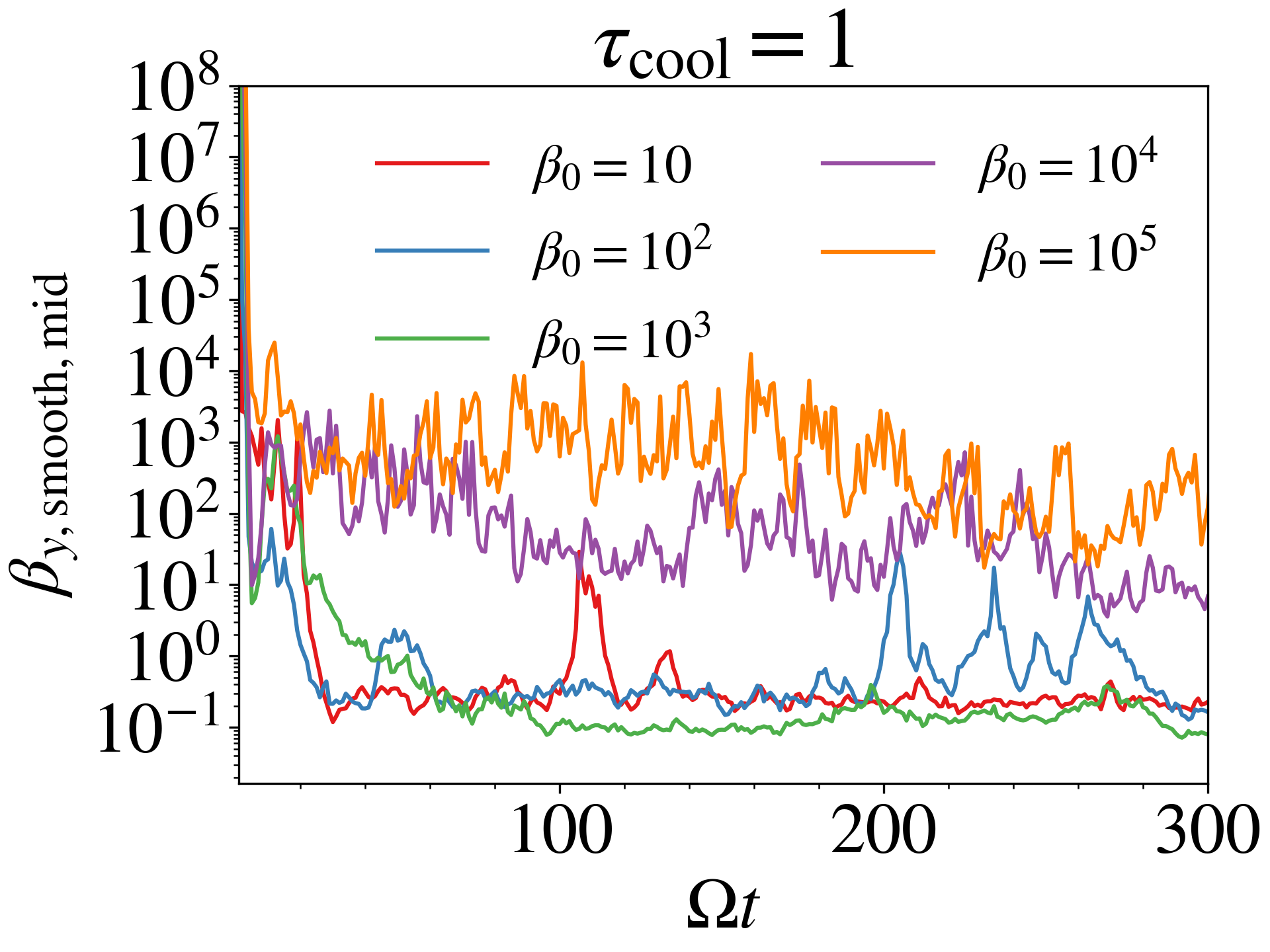} 
    \includegraphics[width=0.3\textwidth]{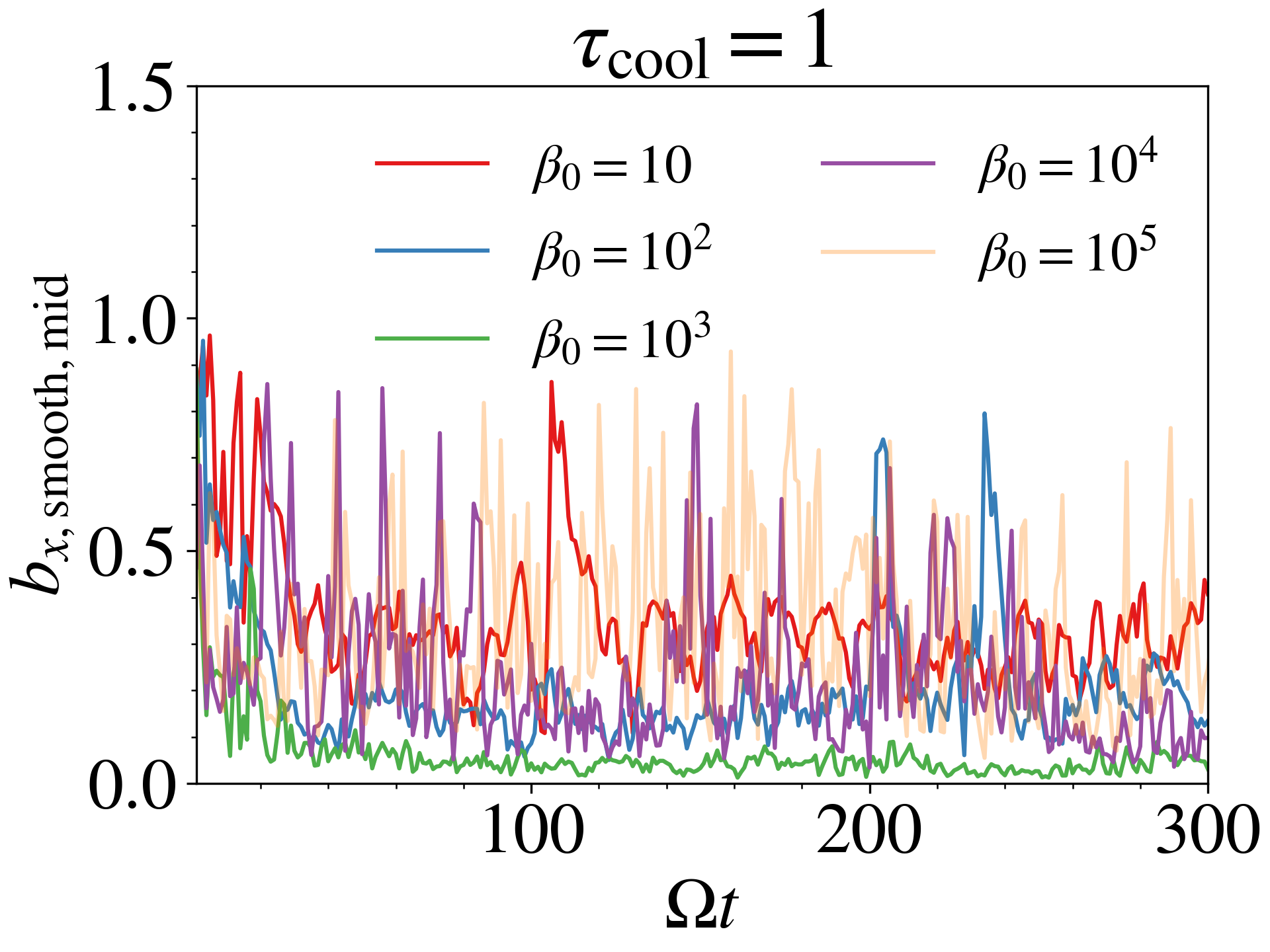} 
    \caption{Window-averaged radial field plasma beta $\beta_{x,\mathrm{smooth},\mathrm{mid}} = 2\langle\bar{P}_{g,w,\mathrm{mid}}\rangle_{xy}/\langle\hat{B}^2_{x,w,\mathrm{mid}}\rangle_{xy}$ (left), toroidal field plasma beta $\beta_{y,\mathrm{smooth},\mathrm{mid}} = 2\langle\bar{P}_{g,w,\mathrm{mid}}\rangle_{xy}/\langle\hat{B}^2_{y,w,\mathrm{mid}}\rangle_{xy}$ (middle), and relative radial field strength $b_{x,\mathrm{smooth},\mathrm{mid}} = [\langle\bar{B}^2_{x,w,\mathrm{mid}}\rangle_{xy}/(\langle\bar{B}^2_{x,w,\mathrm{mid}}\rangle_{xy}+\langle\bar{B}^2_{y,w,\mathrm{mid}}\rangle_{xy})]^{1/2}$ at the mid-plane as a function of time for various initial mid-plane plasma beta $\beta_0$. The cooling strength for the displayed cases is $\tau_\mathrm{cool}=1$. The window-length used for the window average is $w/L_x=1$.}
    \label{fig:timeseries_Bfield_mid}
\end{figure*}

\begin{figure}
    \centering
    \includegraphics[width=0.23\textwidth]{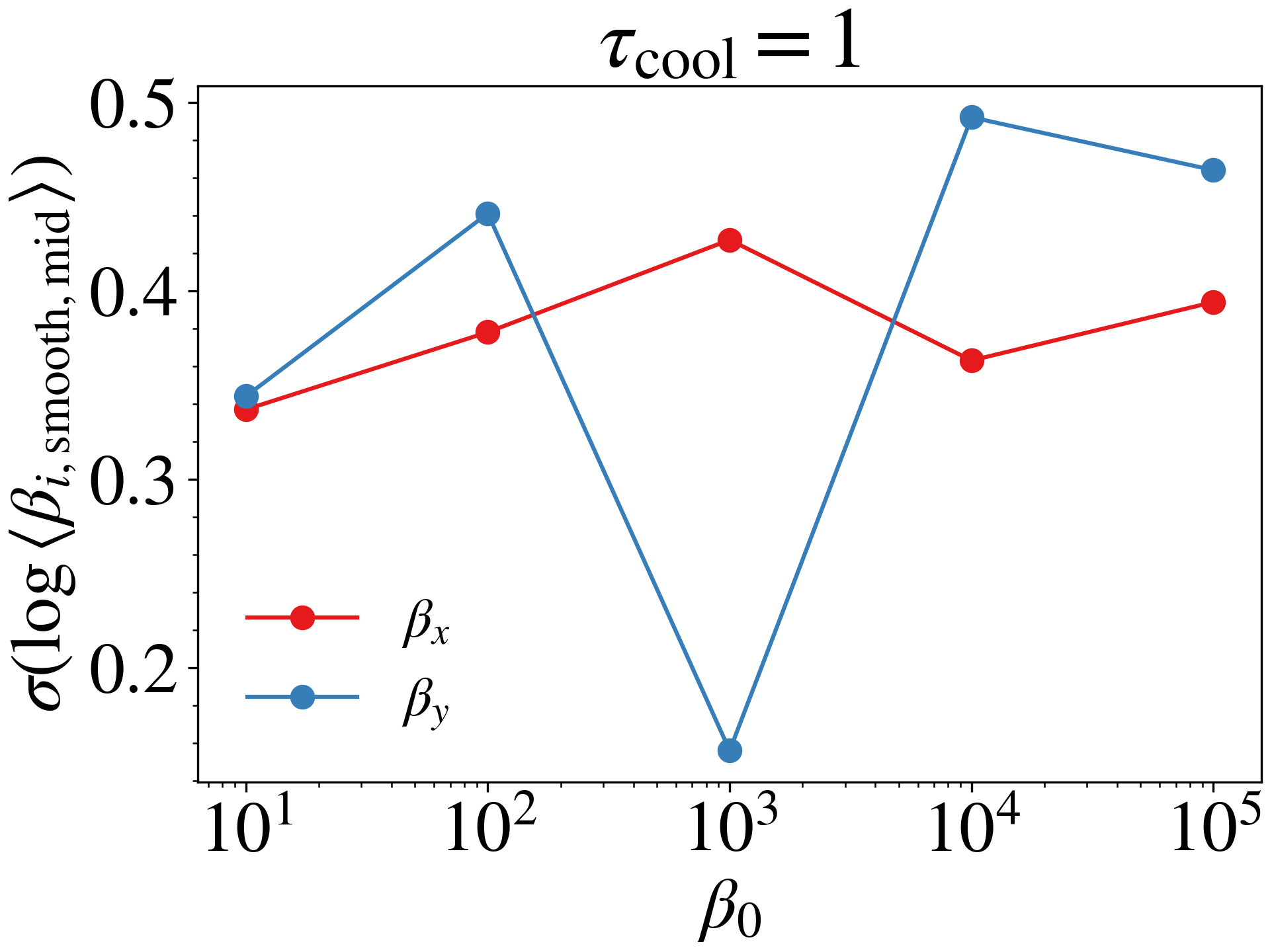}
    \includegraphics[width=0.23\textwidth]{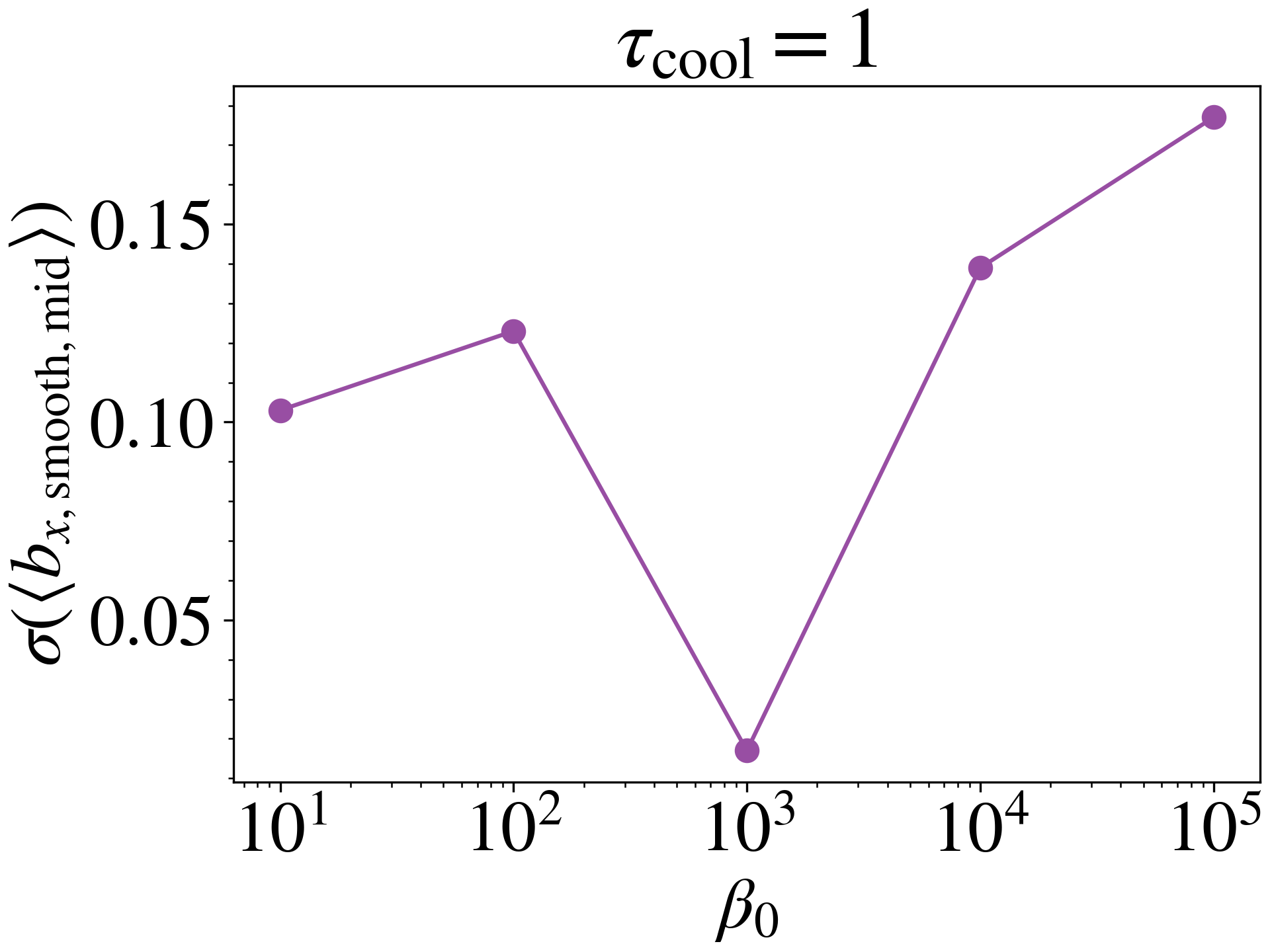} \\
    \includegraphics[width=0.23\textwidth]{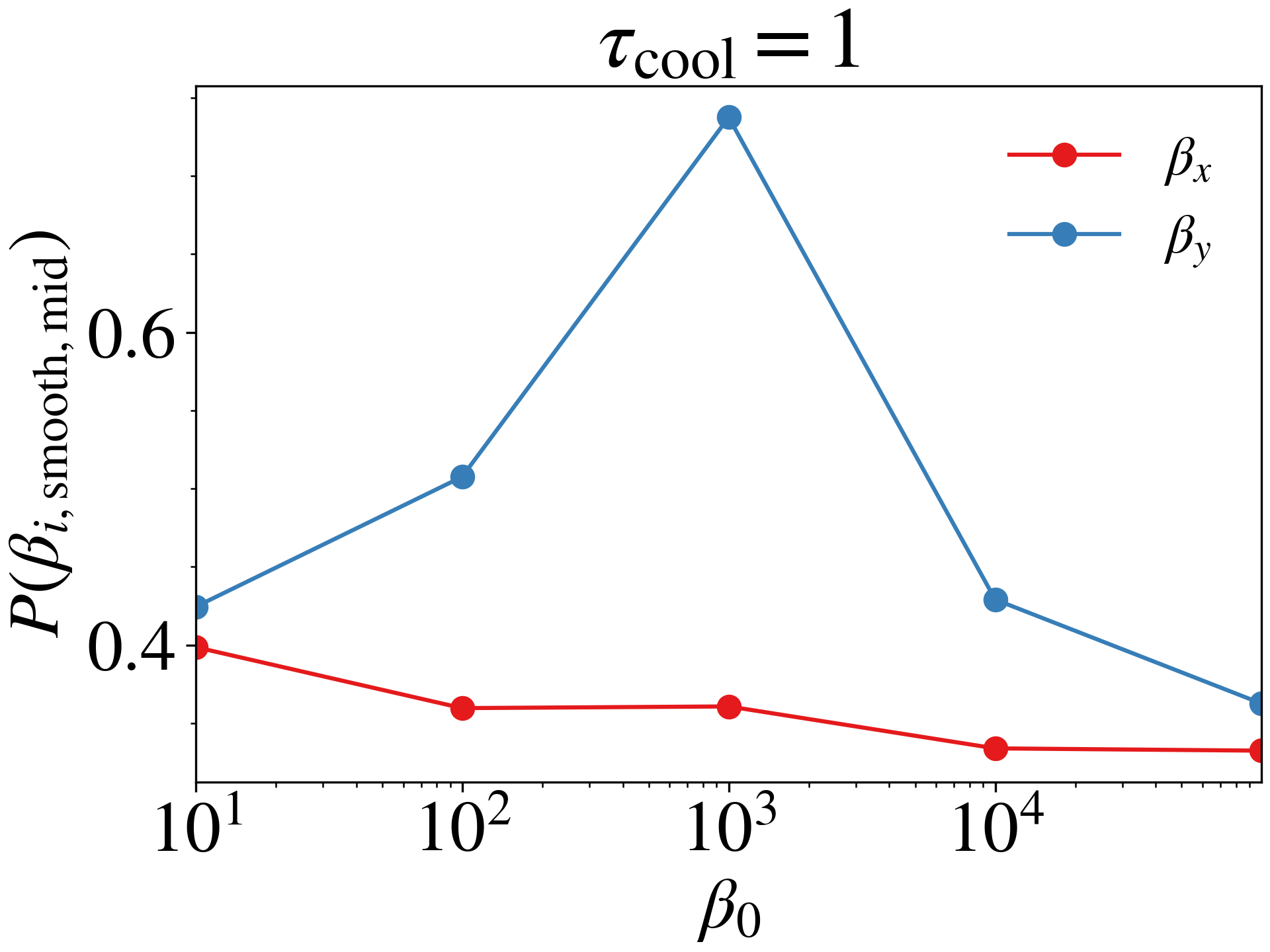}
    \includegraphics[width=0.23\textwidth]{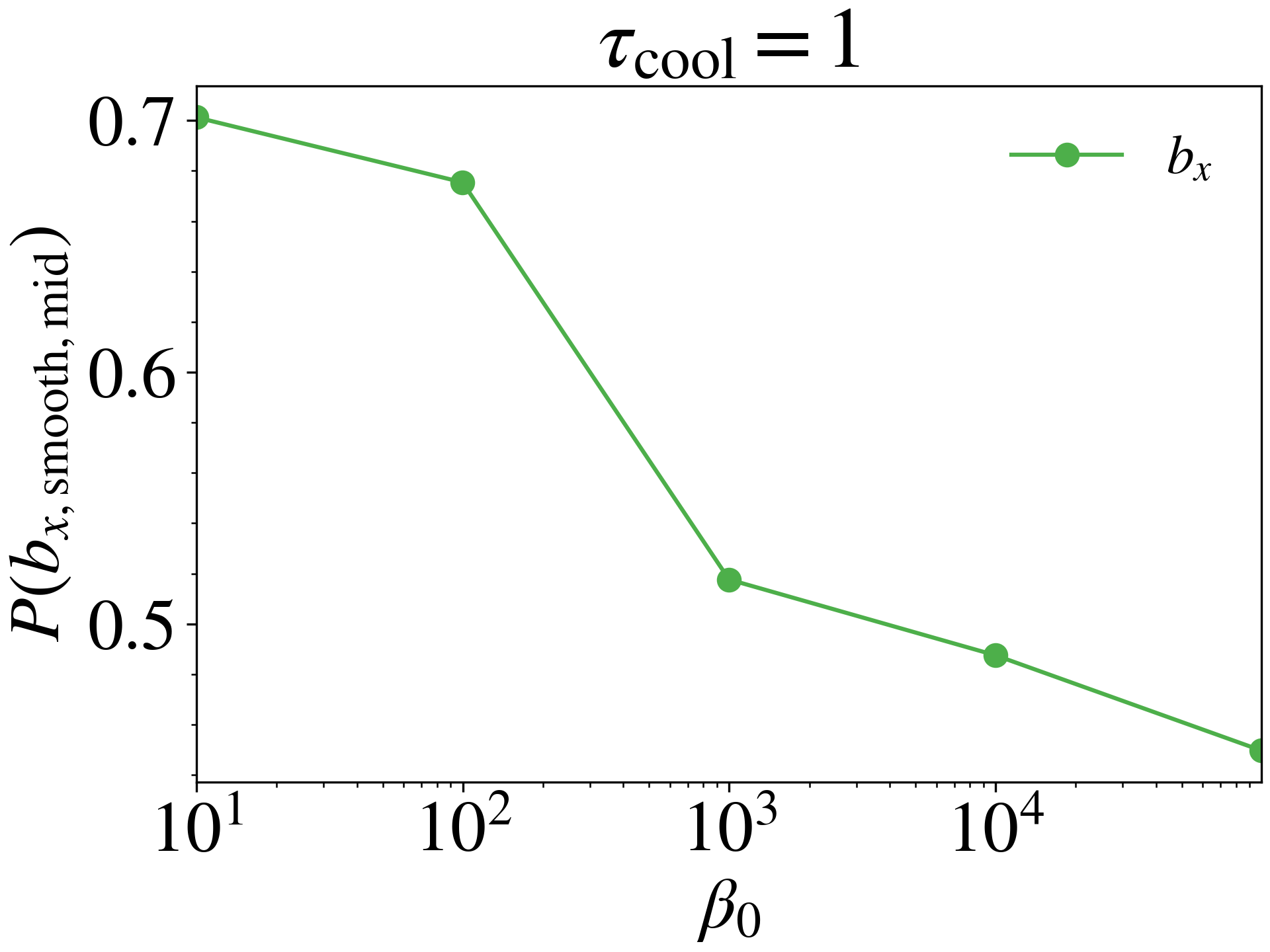} 
    \caption{Top left: Fluctuations of the window-averaged plasma betas $\beta_{x,\mathrm{smooth,mid}}$ (red), $\beta_{y,\mathrm{smooth,mid}}$ (blue) at the mid-plane for different $\beta_0$ cases, measured in terms of the standard deviation of the logarithm (base 10) of the respective window-averaged betas $\beta_i$. Top right: same as the left panel but for the relative radial field contribution $b_x$. Bottom row: Fraction of power contained in temporal frequencies with period longer than $10\Omega^{-1}$ for $\beta_{x,\mathrm{smooth,mid}},\beta_{y,\mathrm{smooth,mid}}$ (left) and $b_{x,\mathrm{smooth,mid}}$ (right).}
    \label{fig:time_variation}
\end{figure}

In Fig.~\ref{fig:timeseries_Bfield_mid} we plot the window-averaged radial field plasma beta $\beta_{x,\mathrm{smooth},\mathrm{mid}} = 2\langle\bar{P}_{g,w,\mathrm{mid}}\rangle_{xy}/\langle\bar{B}^2_{x,w,\mathrm{mid}}\rangle_{xy}$ (left), toroidal field plasma beta $\beta_{x,\mathrm{smooth},\mathrm{mid}} = 2\langle\bar{P}_{g,w,\mathrm{mid}}\rangle_{xy}/\langle\bar{B}^2_{y,w,\mathrm{mid}}\rangle_{xy}$ (middle), and the relative radial field strength $b_{x,\mathrm{smooth},\mathrm{mid}} = [\langle\bar{B}^2_{x,w,\mathrm{mid}}\rangle_{xy}/(\langle\bar{B}^2_{y,w,\mathrm{mid}}\rangle_{xy}+\langle\bar{B}^2_{y,w,\mathrm{mid}}\rangle_{xy})]^{1/2}$ (right) at the mid-plane, using a window length of $w/L_x=1$ (i.e. large scale field), as a function of time. We note that rapid fluctuations in $\beta_x,\beta_y$ are observed for the high $\beta_0=10^4,10^5$ cases. Fluctuations can still be seen in $\beta_x,\beta_y$ for the lower $\beta_0$ cases, but they are less rapid than their high $\beta_0$ counterparts. In the top row of Fig.~\ref{fig:time_variation} we characterize the magnitude of the fluctuations using the quantities $\sigma(\log\langle\beta_i\rangle^\mathrm{smooth}_{xy})$ and $\sigma(\langle b_x\rangle^\mathrm{smooth}_{xy})$, which are the standard deviation of the logarithm (base 10) of $\beta_{i,\mathrm{smooth,mid}}$ ($i=x,y$) (left panel) and that of $ b_{x,\mathrm{smooth,mid}}$ (right). The standard deviations are taken over $\Omega t=100-300$ (for $\beta_0\neq 10^5$) and over $\Omega t=250-450$ for $\beta_0=10^5$. The magnitude of the fluctuations in $\beta_x,\beta_y$ are quite similar across different $\beta_0$ cases, with $\Delta\log\beta_i\sim 0.35-0.45$, except for the $\beta_0=10^3$ case, in which $\beta_y$ is particularly steady. This implies that the large-scale $\beta_x,\beta_y$ at the mid-plane varies typically by 0.35-0.45 dex (i.e. less than an order of magnitude). As observed in the top right panel, $b_x$ typically varies by $\Delta b_x\sim 0.1-0.15$, except for the $\beta_0=10^3$ case, which as a low value of $\sim 0.01$. Depending on the local conditions, such fluctuations could either mean insignificant or appreciable change in the growth rates. As we shall see in \S\ref{subsubsec:midplane} (also Fig.~\ref{fig:growth_midplane}), the resultant fluctuations in the growth rates are modest. 

In the bottom row of Fig.~\ref{fig:time_variation}, we display the fraction of power contained in the low frequency part (with period $>10\Omega^{-1}$) of the temporal spectrum of $\beta_{x,\mathrm{smooth,mid}},\beta_{y,\mathrm{smooth,mid}},b_{x,\mathrm{smooth,mid}}$. We calculate this by 
\begin{equation}
    P(X) \equiv \frac{\int_0^{\omega(T>10\Omega^{-1})}\abs{\tilde{X}}\dd{\omega}}{\int_0^{\infty}\abs{\tilde{X}}\dd{\omega}}, \label{eqn:fft_temporal}
\end{equation}
where $\tilde{X}$ is the temporal FFT of the quantity $X$. The bottom left plot shows that roughly $40\%$ of the power of $\beta_{x,\mathrm{smooth,mid}}$ is contained in the low frequency part, and $40-60\%$ for $\beta_{y,\mathrm{smooth,mid}}$. The radial field contribution $b_x$ is more time-steady, with around $70\%$ contained in the low frequency part for the low $\beta_0$ cases, dwindling to $50\%$ for the high $\beta_0$ cases. These plots imply that while the WKB analysis is, strictly speaking, not valid due to presence of high frequency fluctuations (with fluctuation periods comparable or shorter than the growth time), it remains an insightful tool to understand the growth of GI modes in this complex environment since there is substantial power in the low frequency parts. In particular, $b_x$, which controls whether magnetic field destabilizes axisymmetric CRMG modes (Fig.~\ref{fig:B_transition}), is highly time-steady in the low $\beta_0$ cases.

In summary, there are several noteworthy points regarding the magnetic field structure. First, the radial and vertical fields are multi-scaled, with grid size structures as well as domain-sized mean field, while the toroidal field is mostly large-scaled. Second, the magnetization of the disk in the nonlinear stage generally increases with decreasing $\beta_0$, but not necessarily in a monotonic manner. Specifically, the gas is the most magnetically dominated and toroidally directed for $\beta_0=10^3$, while the radial field contribution increases for $\beta_0<10^3$. Overall, the low $\beta_0$ ($10,10^2,10^3$) cases are considerably more magnetically dominated at the mid-plane than the high $\beta_0$ cases. Third, the large-scale magnetic field at the mid-plane can be considered approximately time-steady, with fluctuations in $\beta_x,\beta_y$ of order $\sim0.35-0.45$ dex --- which only has a modest effect on the growth rates --- and more than  $40\%$ of the power contained in the low frequency part of the temporal spectrum. Taken together, these observations suggest that MRI is able to produce steady radial and toroidal field of appreciable magnitude, particularly in the low $\beta_0$ cases. 

\subsection{Disk fragmentation} \label{subsec:disk_fragmentation}

\begin{figure}
    \centering
    \includegraphics[width=0.4\textwidth]{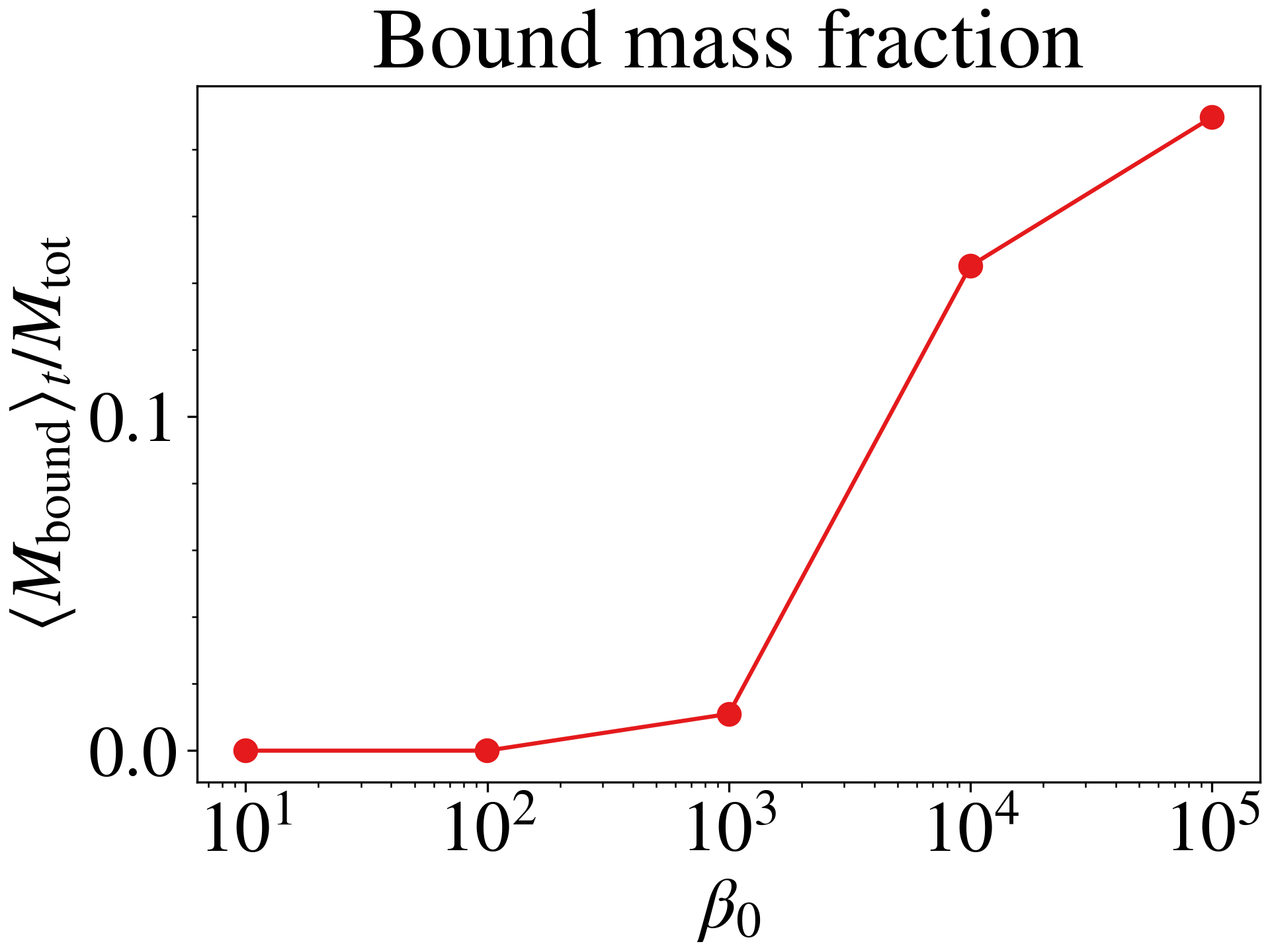}
    \caption{Time-averaged bound mass fraction $\langle M_\mathrm{bound}\rangle_t/M_\mathrm{tot}$, where $\langle M_\mathrm{bound}\rangle_t$ is the time averaged bound mass within the box (i.e., located within TBRs, as defined in eq.~\ref{eqn:frag_condition}), and $M_\mathrm{tot}$ is the total mass within the box, which is kept constant in our setup. 
    This is plotted against the initial mid-plane plasma beta $\beta_0$. Time-average is taken from $\Omega t=100-300$, except for the $\beta_0=10^5$ case, for which time-average is taken from $\Omega t=250-450$.}
    \label{fig:bound_properties}
\end{figure}

Using the clump identification algorithm outlined in \S\ref{subsec:clump_identification}, we identified clumps satisfying eq.~\ref{eqn:frag_condition} within the time period $\Omega t = 100-300$ and collected their properties (e.g. mass, size, location, etc.). As Athena++ is an Eulerian code, we cannot trace the evolution of the clumps as one could with an SPH-based code \citep[e.g. Gizmo, ][]{Hopkins-2015}, as in \citet{Kubli_etal-2023}. Therefore, we focus on the time-averaged properties of the clumps, in which we identify clumps and collect their properties for every output file within $\Omega t = 100-300$ (or $\Omega t=250-450$ for the $\beta_0=10^5$ case), and take the average of these properties over the files we read. To calculate the average fraction of bounded mass in the box, we first sum over the fraction of bounded mass in each output snapshot, then divide this by the number of snapshots we scanned through. 

\begin{figure*}
    \centering
    \includegraphics[width=0.23\textwidth]{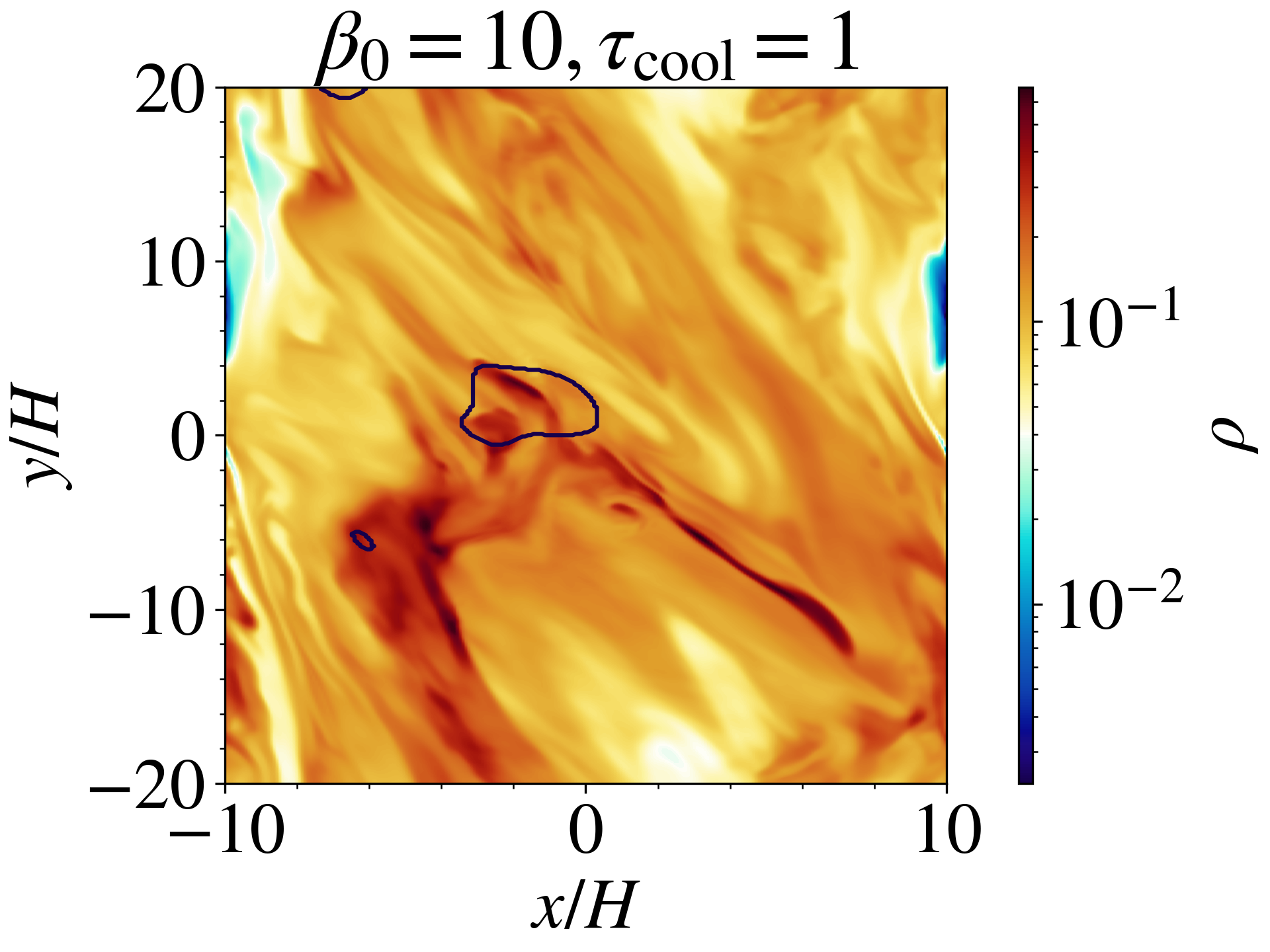}
    \includegraphics[width=0.23\textwidth]{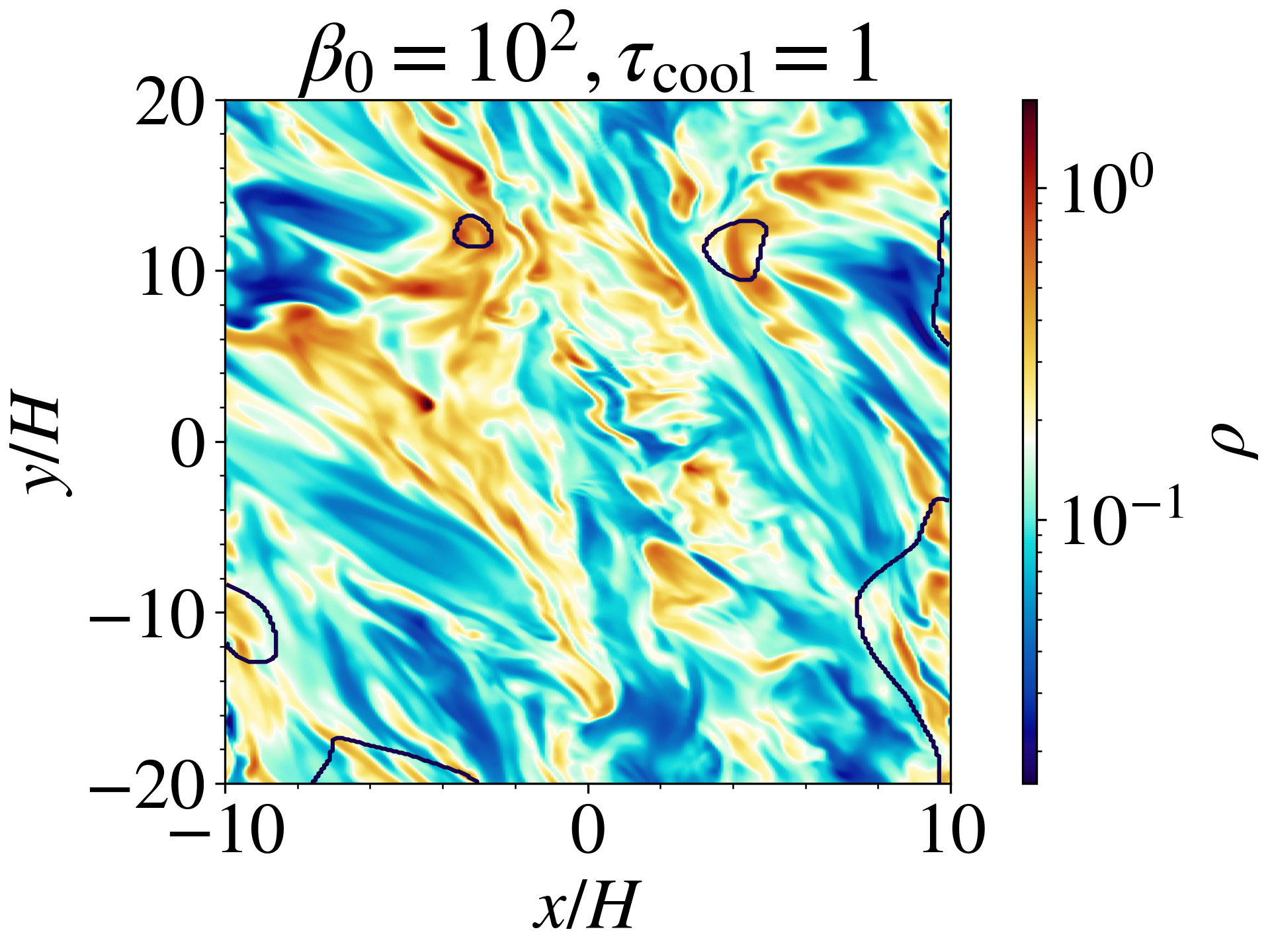}
    \includegraphics[width=0.23\textwidth]{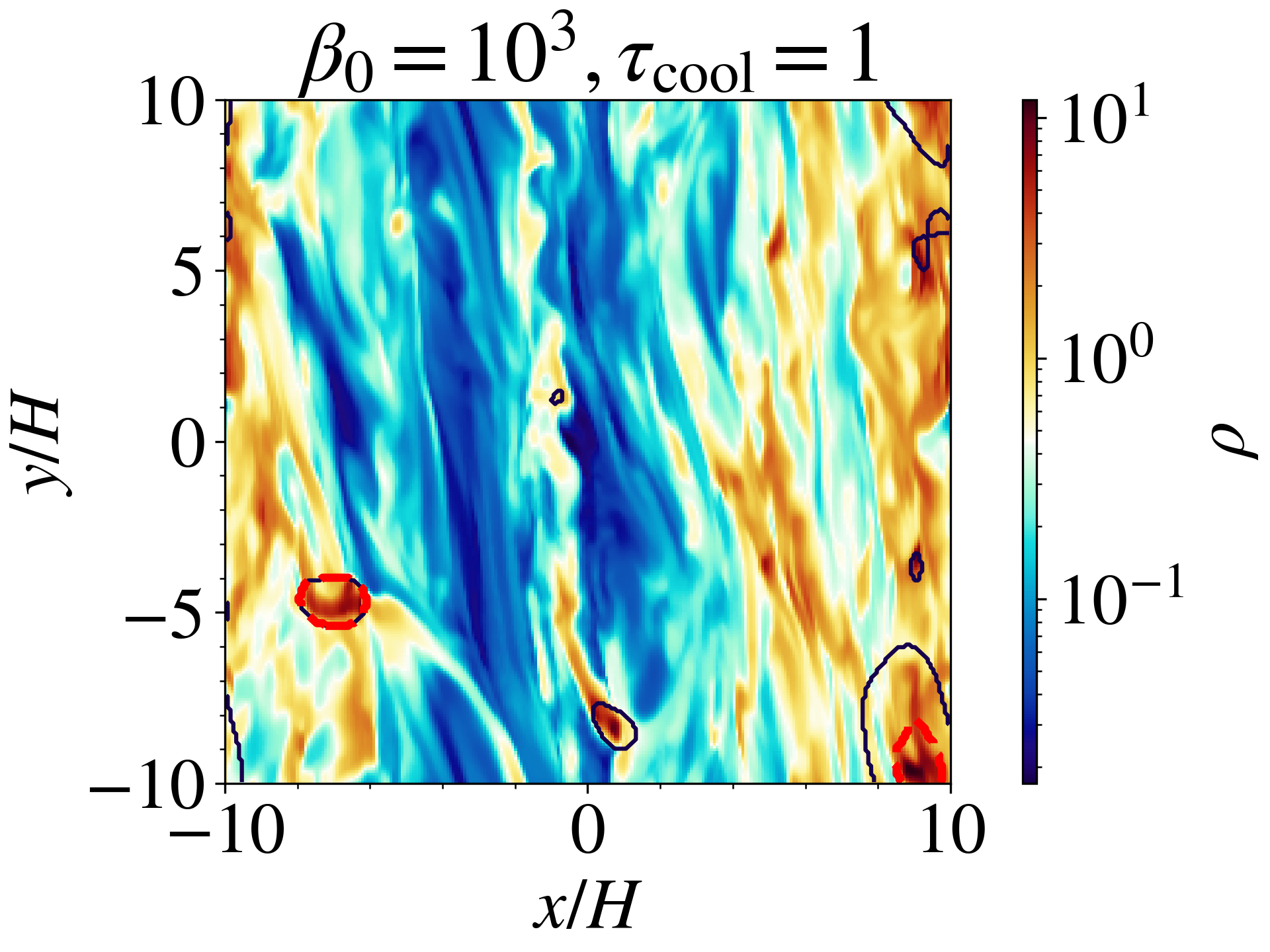}
    \includegraphics[width=0.23\textwidth]{figures/results/sg/production/adi_mhd_beta10000_tc1/rho_midplane.png} \\
    \includegraphics[width=0.23\textwidth]{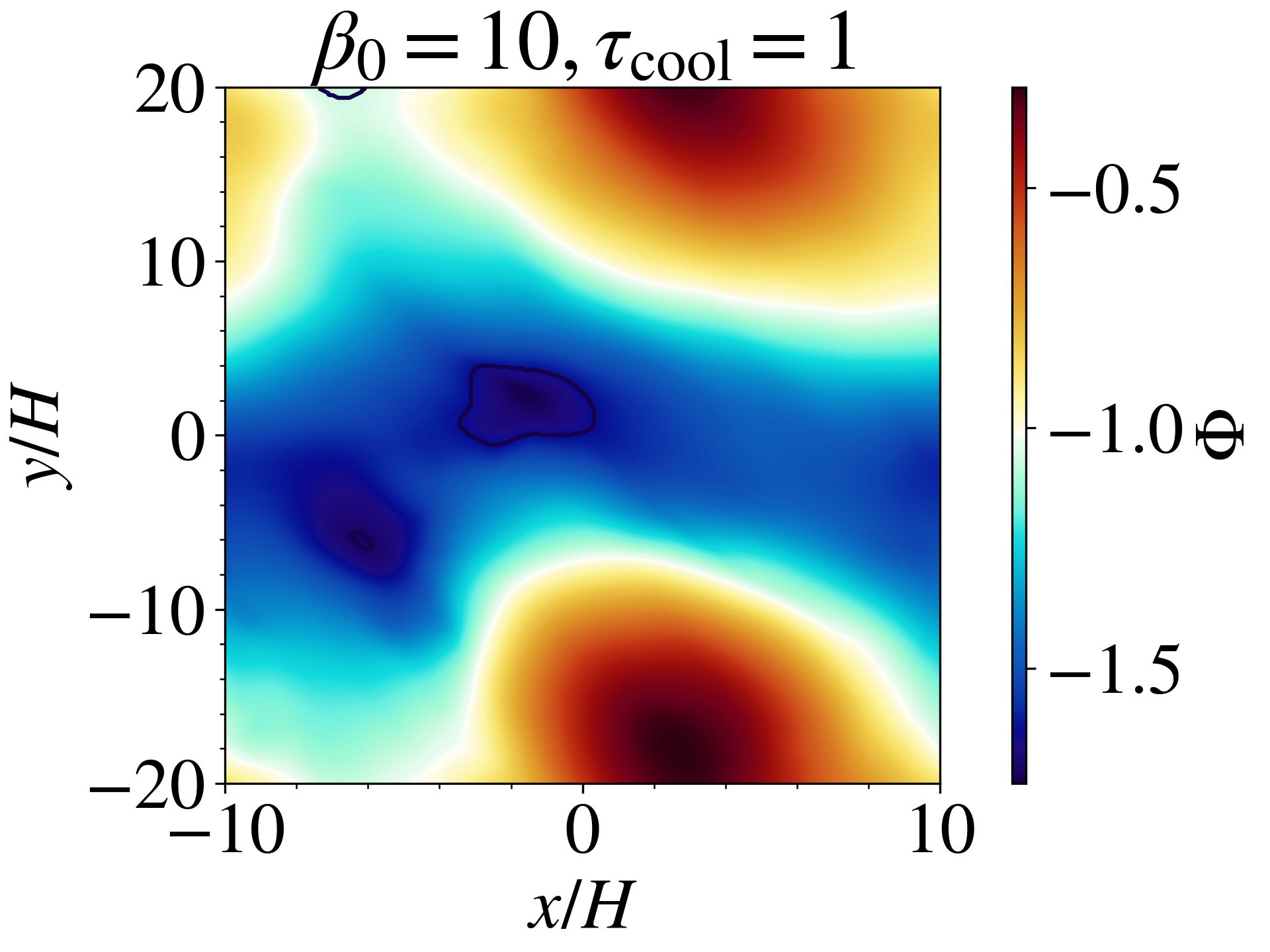}
    \includegraphics[width=0.23\textwidth]{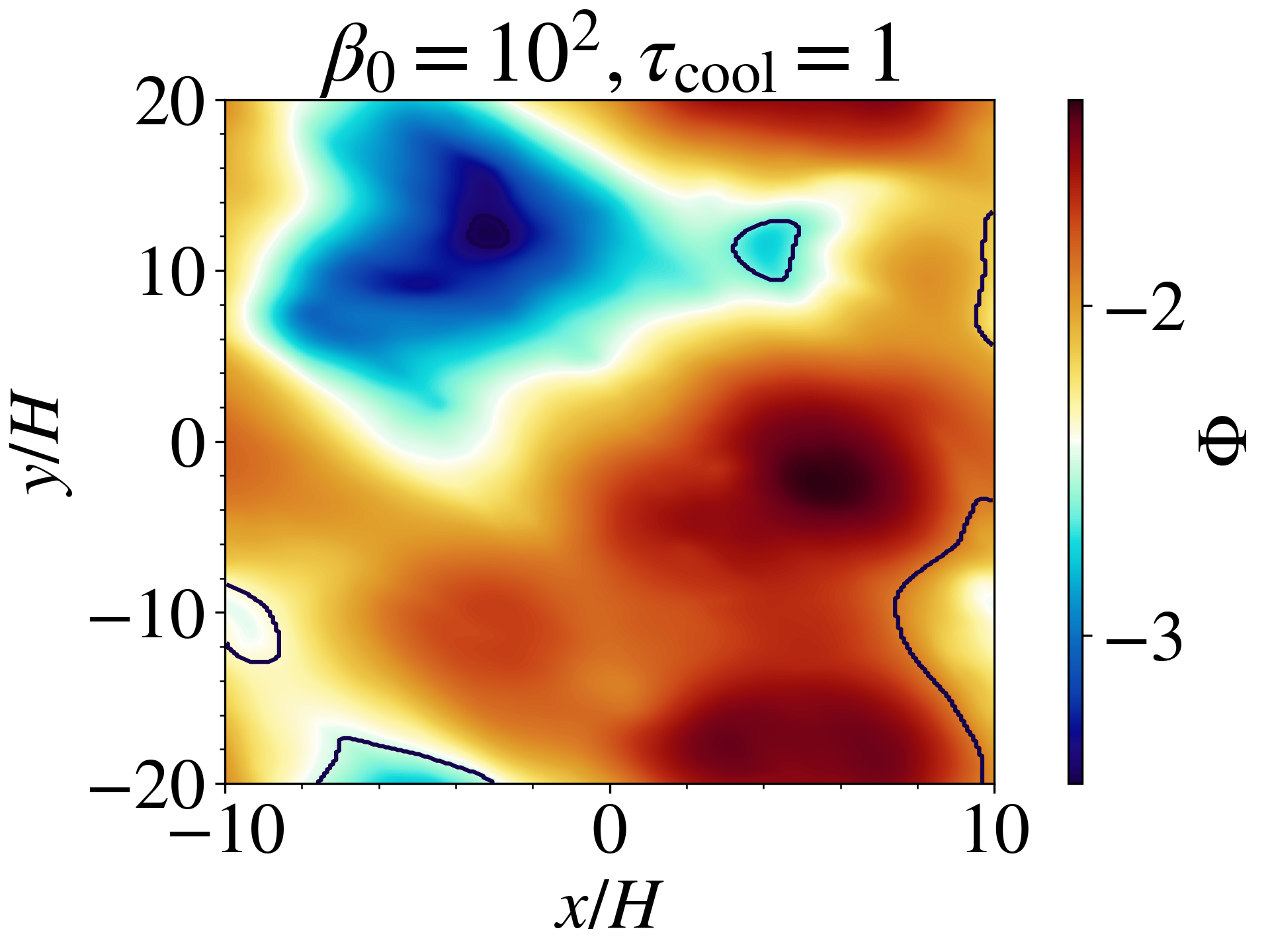}
    \includegraphics[width=0.23\textwidth]{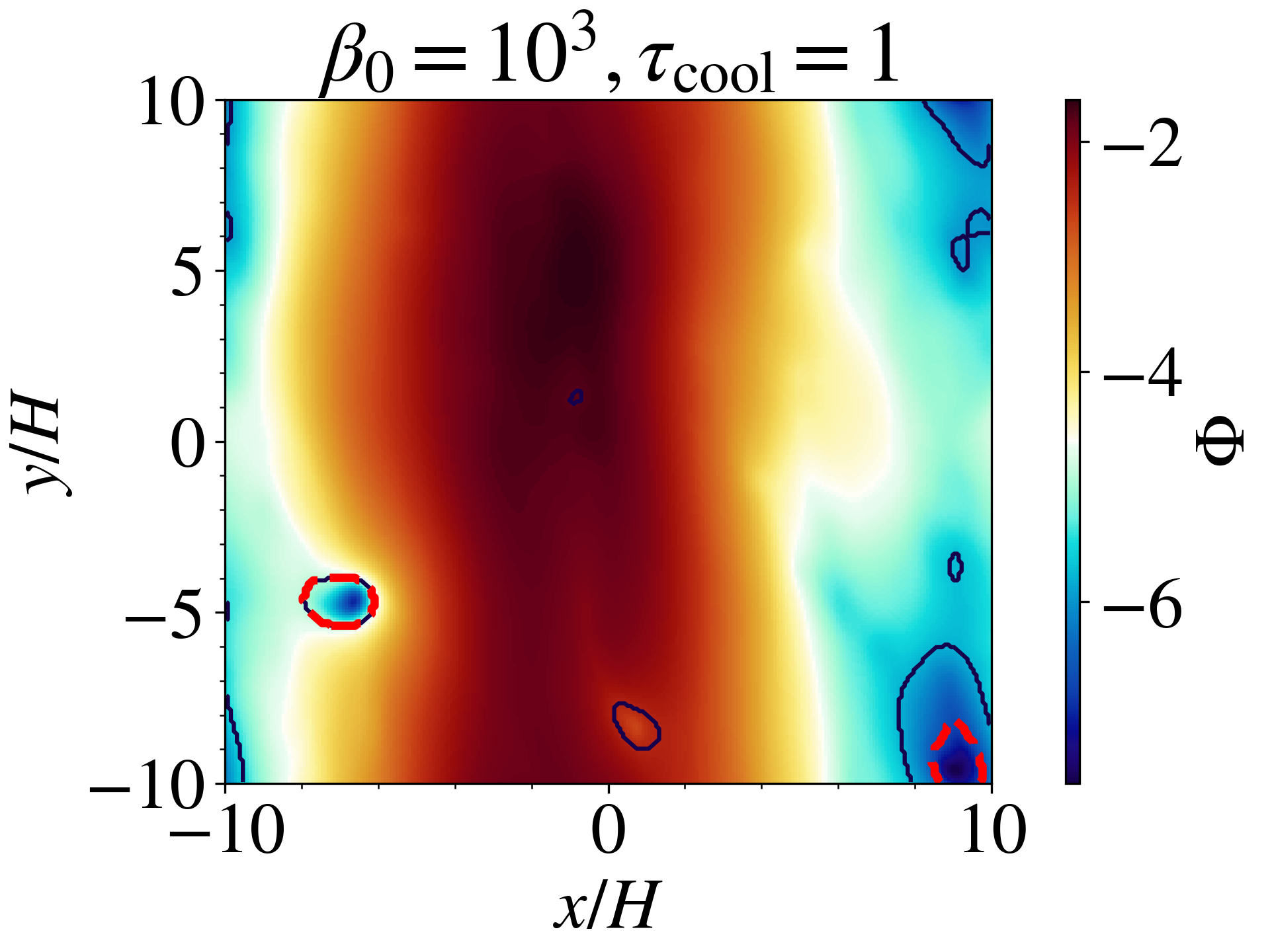}
    \includegraphics[width=0.23\textwidth]{figures/results/sg/production/adi_mhd_beta10000_tc1/phi_midplane.png}
    \caption{Snapshots of the mid-plane $\rho$ (top row) and $\Phi$ (bottom row) taken at $\Omega t = 200$ with contours overlaid to indicate clumps identified. From left to right, $\beta_0 = 10, 10^2,10^3,10^4$, $\tau_\mathrm{cool} = 1$. Black contours: Gravitational Binding Region (GBR, i.e. isolated region that is gravitationally bound); Red dashed contours: Total Binding Region (TBR, i.e. isolated region that is bound when thermal, kinetic and magnetic pressure support are taken in account.)}
    \label{fig:frag_compare}
\end{figure*}

In Fig.~\ref{fig:bound_properties}, we show the averaged bound mass fraction as a function of $\beta_0$. We note that the bound mass fraction decreases for stronger initial fields, dropping roughly a factor of 10 from $\beta_0=10^4$ to $\beta_0=10^3$, and no clumps were identified for $\beta_0=10,10^2$. This strongly suggests that GI is suppressed for low $\beta_0$ cases. 
In Fig.~\ref{fig:frag_compare} we display selected snapshots of the mid-plane density and gravitational potential with GBR (black) and TBR (red) contours overlaid. Clearly, the number of clumps decreases as $\beta_0$ decreases. 

\subsubsection{Midplane properties and magnetic elevation} \label{subsubsec:midplane}

Despite appreciable radial field contribution $b_x$ in some low $\beta_0$ cases and the decrease in plasma beta $\beta$ at the mid-plane going from $\beta_0=10^5$ to $10$ (Fig.~\ref{fig:cool0.5_field_cases}), we observe no uptick in fragmentation (in terms of the bound mass fraction), which seems to contradict expectation from linear theory. This is because the CRMG instability growth rate depends not just on $b_x$ and $\beta$, but also on other parameters which may adversely impact the growth rate when a strong magnetic field is present. In this section, we investigate the conditions at the mid-plane to address why fragmentation/GI is suppressed for the low $\beta_0$ ($\leq 10^3$) cases. 

\begin{figure*}
    \centering
    \includegraphics[width=0.4\textwidth]{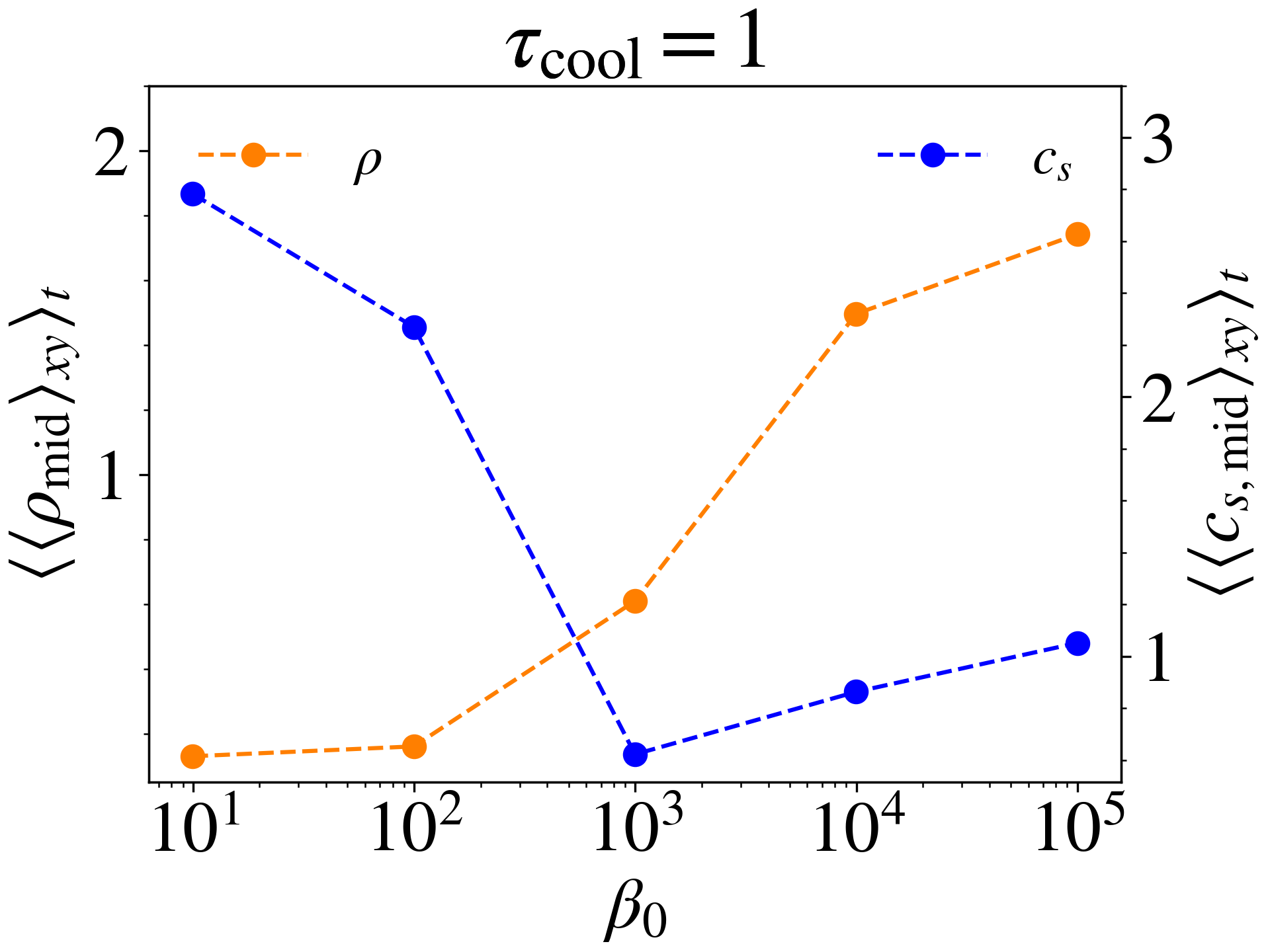} 
    \includegraphics[width=0.4\textwidth]{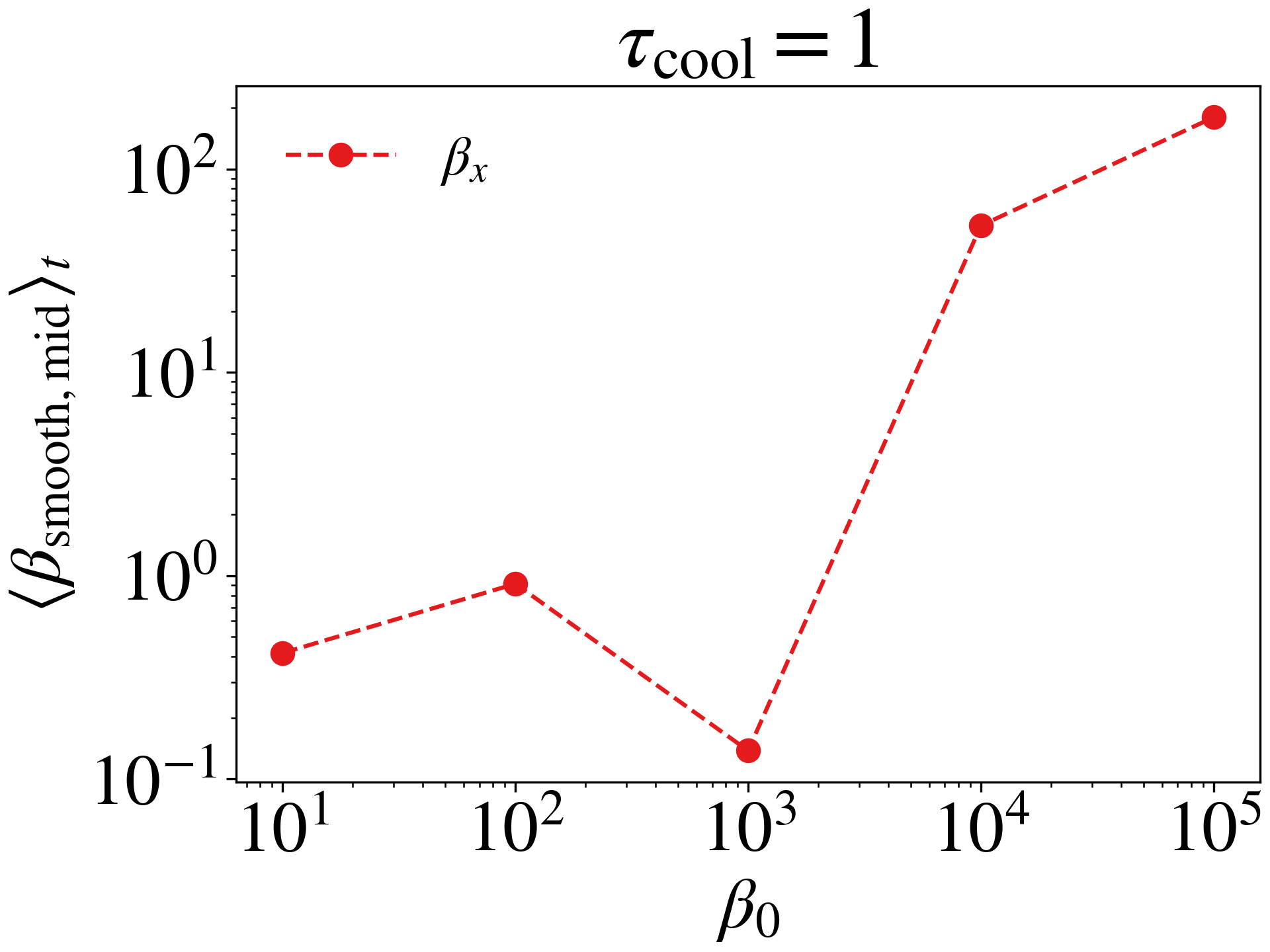} \\
    \includegraphics[width=0.4\textwidth]{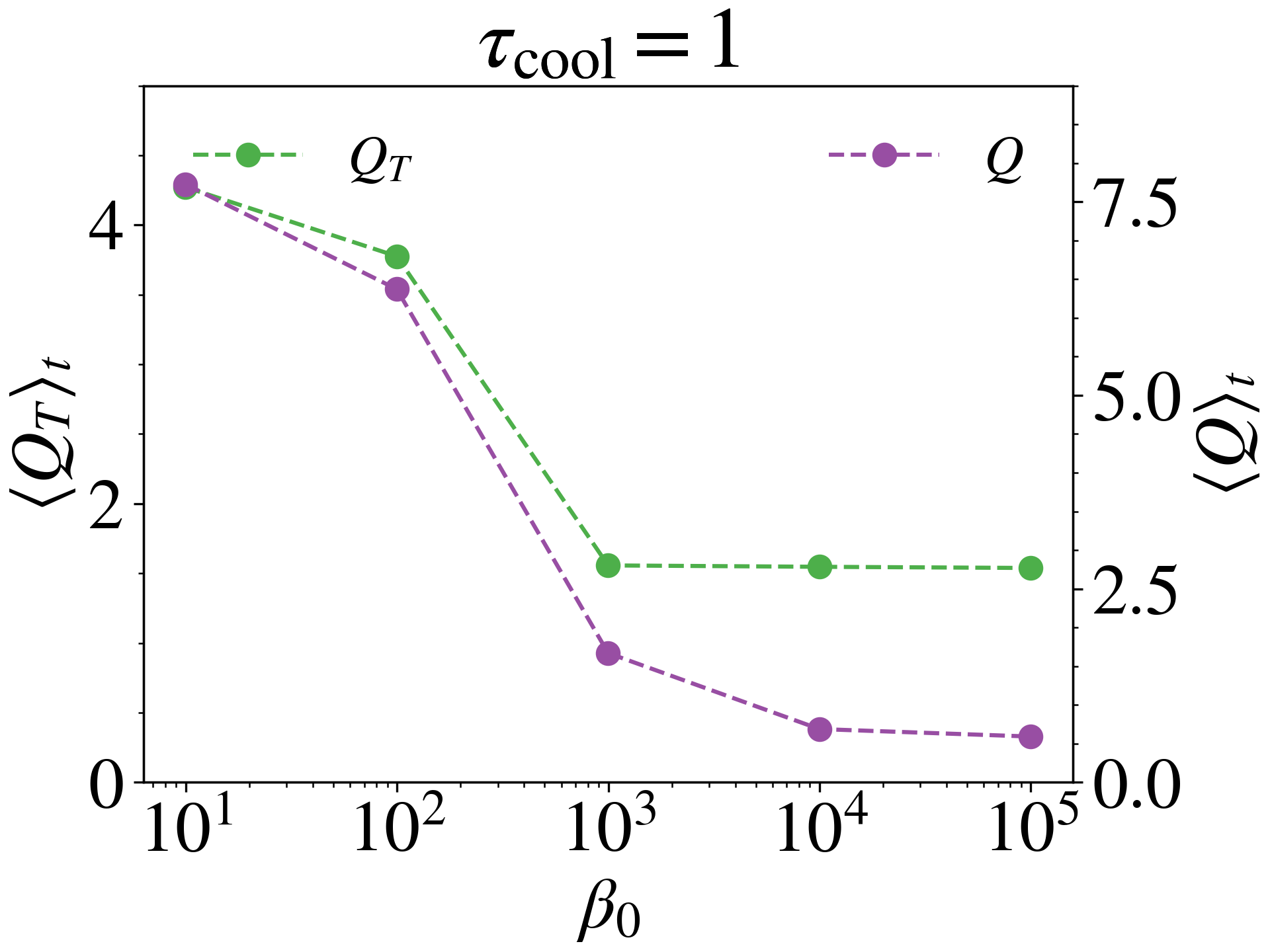} 
    \includegraphics[width=0.4\textwidth]{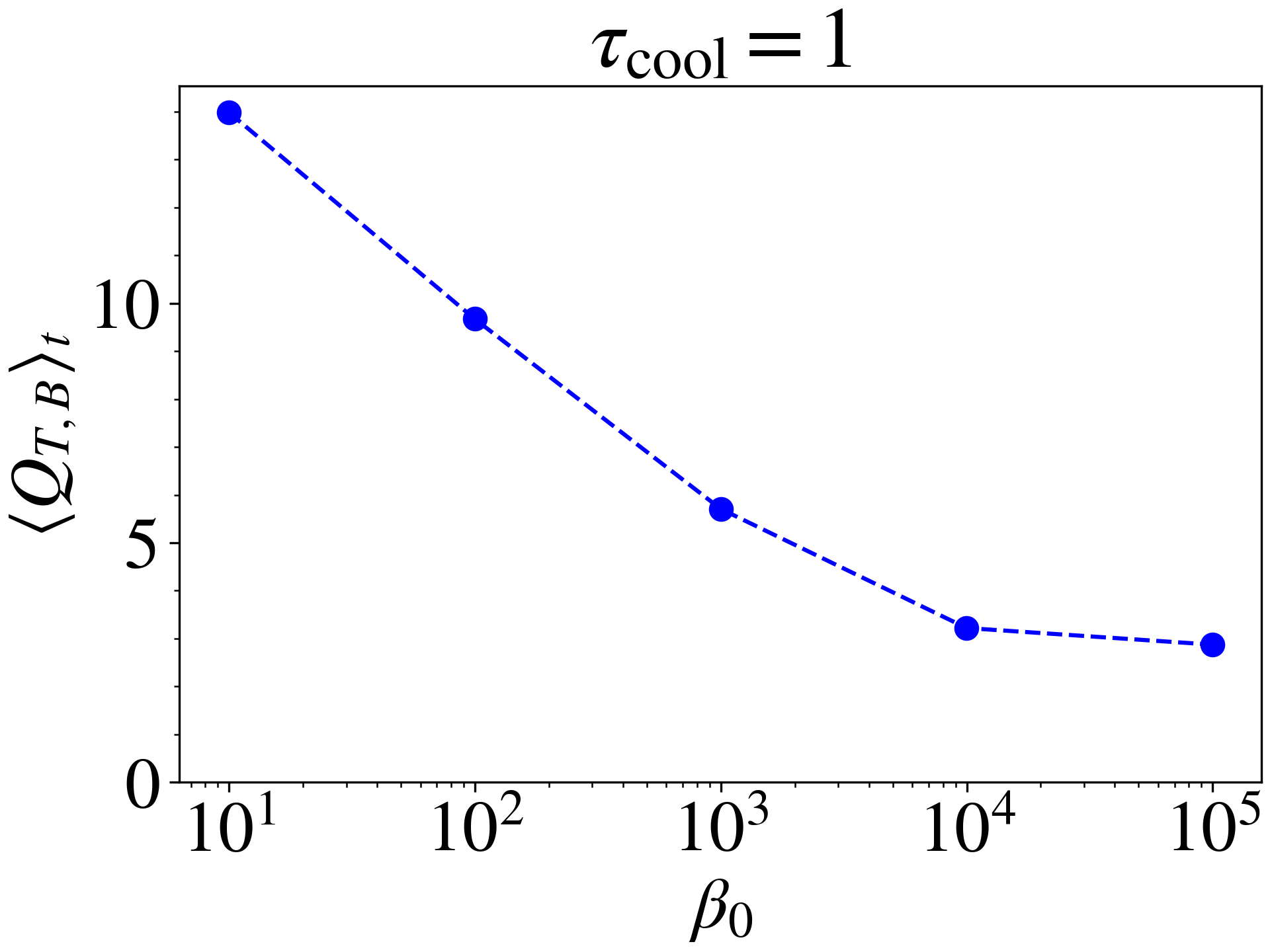} 
    \caption{Horizontally-averaged, time-averaged mid-plane density (top left, orange), thermal sound speed (top left, blue), window-averaged plasma beta (top right, with a smoothing length of $L_x$), Toomre parameter $Q_T$ (bottom left, green), proxy Toomre parameter $Q$ (bottom left, purple) and magnetized Toomre parameter (bottom right) for different $\beta_0$. Time-average is taken from $\Omega t=100-300$, except for the $\beta_0=10^5$ case, which is taken from $\Omega t=250-450$.}
    \label{fig:midplane_elevation}
\end{figure*}

In the top row of Fig.~\ref{fig:midplane_elevation} we plot the horizontally-averaged, time-averaged mid-plane density, thermal sound speed (top left) and plasma beta (top right) as a function of $\beta_0$. There are two note-worthy points in this figure. First, the density at the mid-plane decreases by over an order of magnitude from $\beta_0=10^5$ to $\beta_0=10$. Second, the mid-plane magnetic field becomes dominant for $\beta_0\leq 10^3$. These two points indicate clearly that the disk is magnetically elevated for $\beta_0\leq 10^3$ \citep{Begelman_Silk-2017}. We also observe an increase in the sound speed going to lower $\beta_0$, indicating an enhanced thermal scaleheight at the mid-plane. To quantify the extent to which the increase in the vertical scale height is due to magnetic elevation versus that due to increase in gas temperature, we list for different $\beta_0$ cases in Table \ref{tab:midplane} the scaleheight measured from the time and horizontal-averaged density profile $\hat{H}$, defined by the height at which the density drops by an e-fold ($\rho_\mathrm{max} e^{-1}$), and also the time and horizontal-averaged mid-plane gas sound speed $\langle c_{s,\mathrm{mid}}\rangle_t$. Note that $\hat{H}$ is defined differently from the usual definition of thermal scale-height $c_s/\Omega$, thus we need to calibrate $\hat{H}$ using a reference point to interpret our numbers. Focusing on the $\beta_0=10^5$ case, for which magnetic elevation can be considered absent, $\hat{H}=0.352H$ (where $H$ is the initial mid-plane thermal scale-height). The time-averaged mid-plane thermal scale-height $\langle c_{s,\mathrm{mid}}\rangle_t/\Omega = 1.05H$. Going from $\beta_0=10^5$ to $\beta_0=10$, $\hat{H}$ increases by roughly 16 times while $\langle c_{s\mathrm{mid}}\rangle_t/\Omega$ increases only by 2.8 times. We can reasonably infer that while the increase in gas temperature contributes an increase of 2.8 to the vertical scaleheight, the combined thermal plus magnetic pressure led to an increase of 16 times, indicating the dominant effect of magnetic elevation\footnote{We did not define a magnetic scale height $\langle v_{A,\mathrm{mid}}\rangle_t/\Omega$ in the same manner as thermal scaleheight because the magnetic pressure at the mid-plane is substantially weaker than above and below the disk, so it is unclear how this number would reflect the actual increase in vertical scale-height due to magnetic fields.}

\begin{figure}
    \centering
    \includegraphics[width=0.23\textwidth]{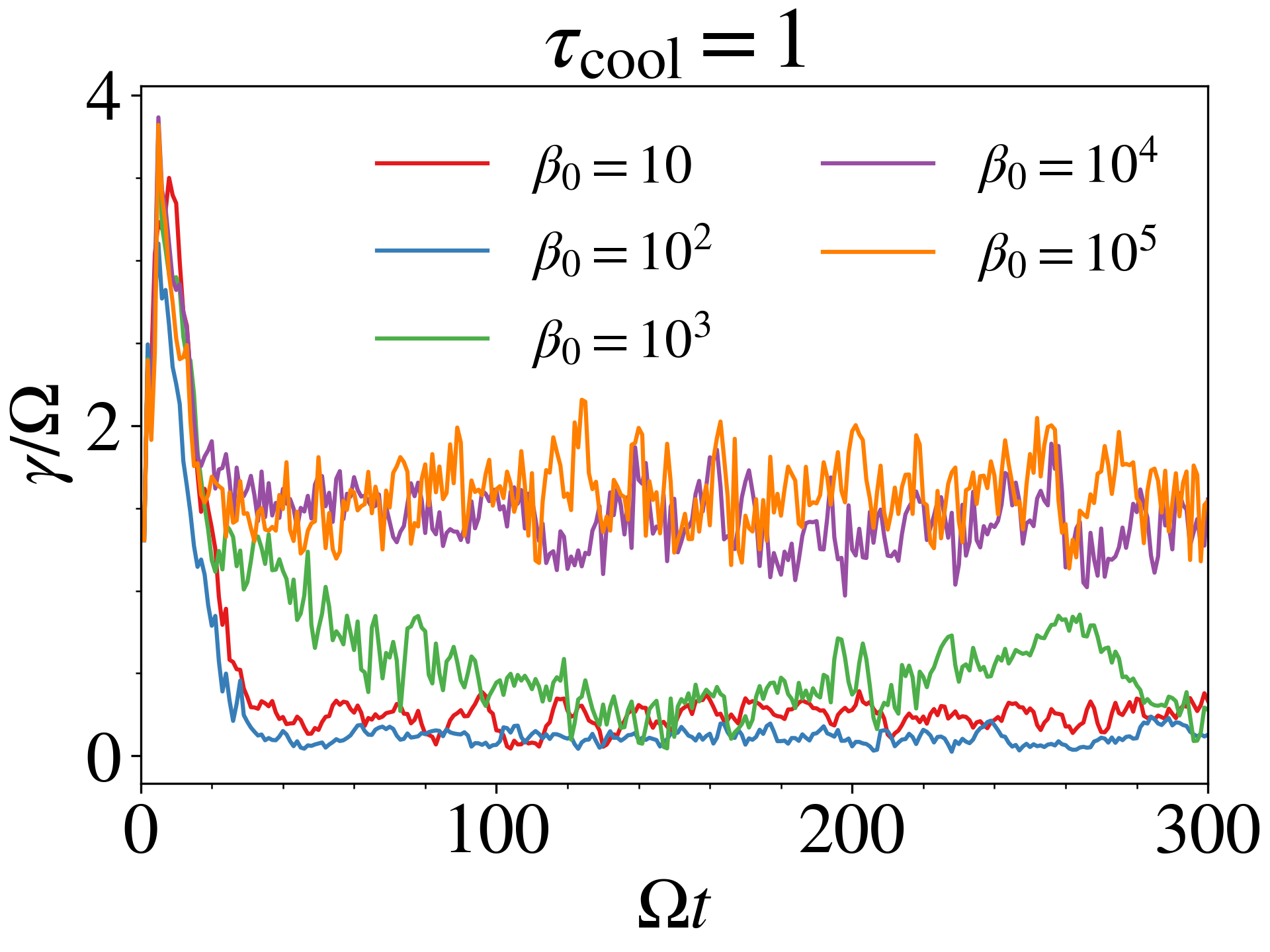}
    \includegraphics[width=0.23\textwidth]{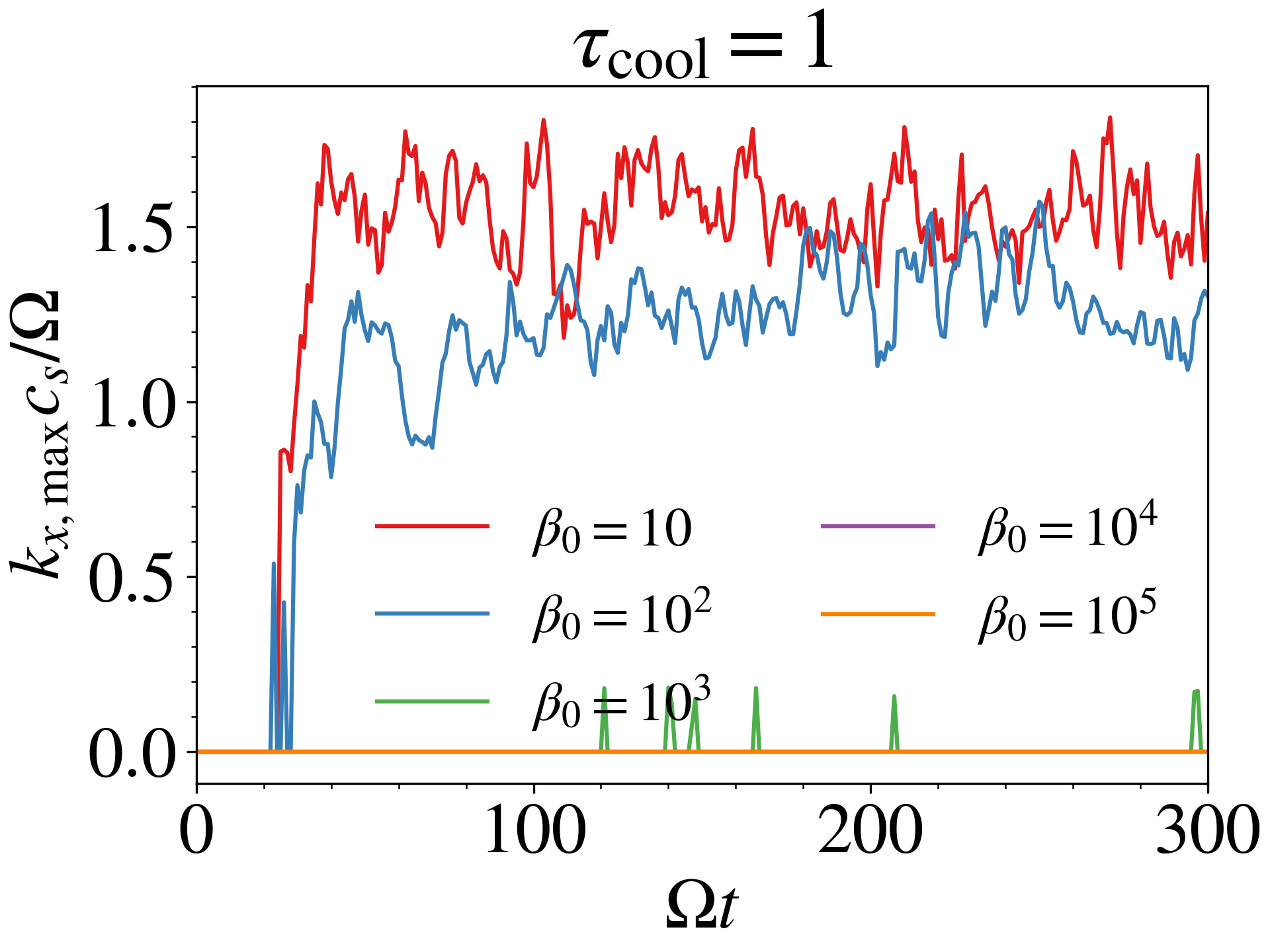} \\
    \includegraphics[width=0.4\textwidth]{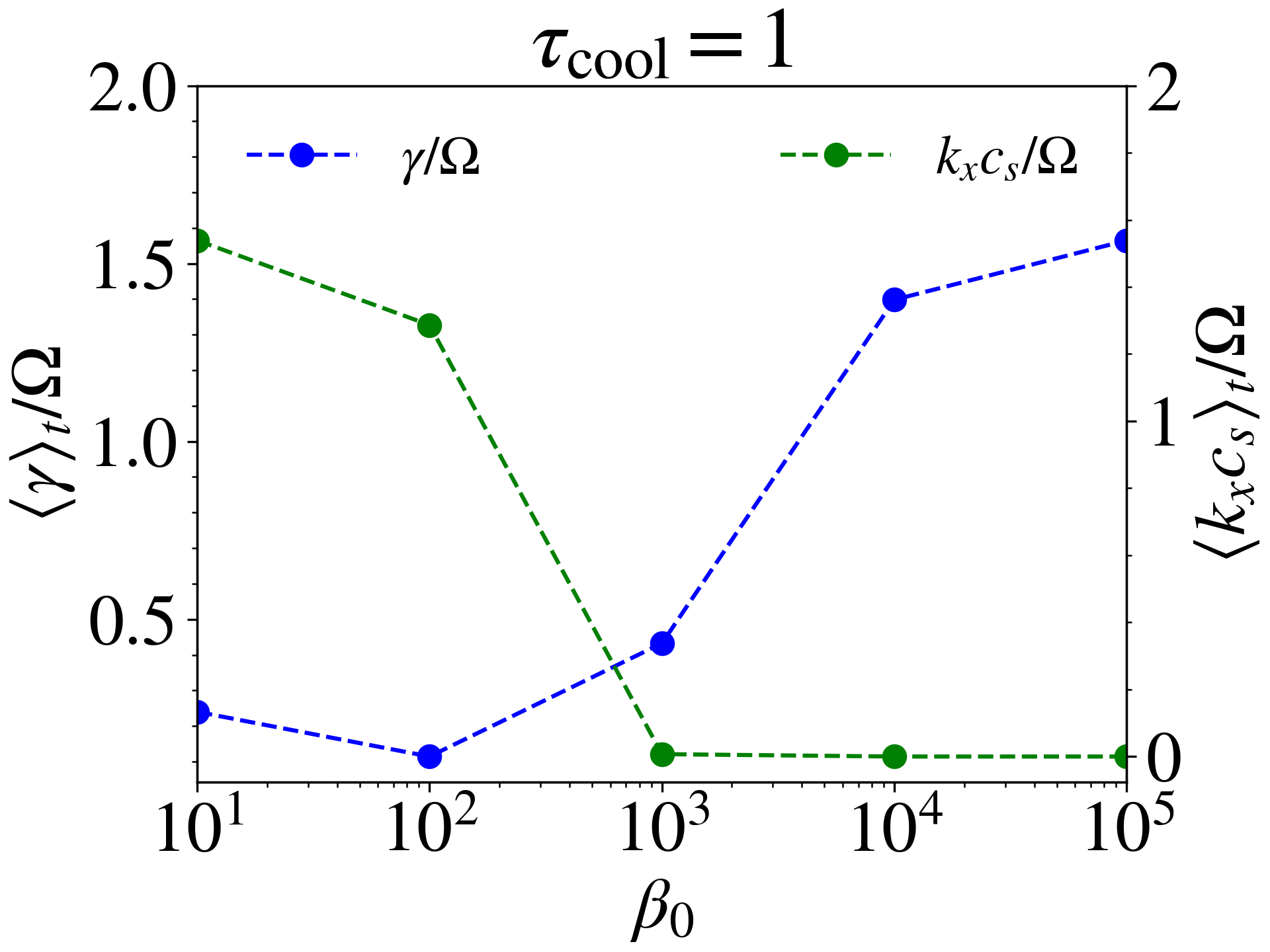} 
    \caption{The CRMG growth rate (top left) and the corresponding wavenumber (top right) of axisymmetric modes against time for various $\beta_0$, calculated using the horizontally averaged mid-plane properties. Bottom: the growth rates and the associated wavenumbers calculated by taking the time-average of the plots in the top two panels. Time-average is taken from $\Omega t=100-300$, except for the $\beta_0=10^5$ case, which is taken from $\Omega t =250-450$.}
    \label{fig:growth_midplane}
\end{figure}

The gravitational instability growth rate depends sensitively on the Toomre parameter $Q_T$ (discussed in \S\ref{subsec:magneto_gravito}). The bottom row of Fig.~\ref{fig:midplane_elevation} shows that the change in mid-plane properties result in larger $Q_T$, which tends to lower the instability growth rate. The proxy Toomre parameter $Q$ is also significantly increased. Similarly, the magnetized Toomre parameter $Q_{T,B}$ increases for lower $\beta_0$, rising way above 1, the stability threshold for magneto-Jeans mode. In Table \ref{tab:midplane} we list some time and horizontally averaged mid-plane properties. Using the mid-plane quantities extracted from each output snapshot, we calculate the most unstable axisymmetric CRMG growth rate and the corresponding wavenumber at the mid-plane for all the cases using the dispersion relation eq.~\ref{eqn:crmg_dispersion_uniform}\footnote{Note that we are using the axisymmetric CRMG dispersion relation assuming a uniform background for this calculation as it is a more appropriate approximation for an elevated disk at the mid-plane.}, plotting them as a function of time in the top panels of Fig.~\ref{fig:growth_midplane}. We also plot, in the same figure, the time-averaged growth rate and the associated wavenumber as a function of $\beta_0$. This figure shows that the axisymmetric CRMG growth rate decreases by close to an order of magnitude as $\beta_0$ decreases from $10^5$ to $10$, indicating suppression of GI due to the different mid-plane properties in the low $\beta_0$ cases, consistent with the lack of fragmentation in those cases. The most unstable wavelengths are also much shorter in the low $\beta_0< 10^3$ cases ($k_{x,\mathrm{max}}\sim 1.5\Omega/c_s$) than the high $\beta_0$ cases ($k_{x,\mathrm{max}}\approx 0$), however, we are unable to confirm this as as we did not capture any clumps in the $\beta_0=10,10^2$ cases and only a few in the $\beta_0=10^3$ case, insufficient for a statistical analysis. In the low $\beta_0$ ($10,10^2$) cases, the simple WKB linear analysis predicts a growth rate of $\gamma\sim 0.1-0.2\Omega$, corresponding to an e-folding time of $\sim 30-60\Omega^{-1}$. Such an increase in growth time results in a drastic decrease in fragmentation and further lowers the validity of the WKB analysis in that scenario. It is unlikely that perturbations would be able to sustain exponential growth over such a long e-folding time; we suspect that only weak transient growth is possible before background conditions check further growth or even damp it.

\begin{table}
    \centering
    \begin{tabular}{c|c|c|c|c|c}
        $\beta_0$ & $\langle\beta_\mathrm{mid}^\mathrm{smooth}\rangle_t$ & $\langle c_{s,\mathrm{mid}}\rangle_t$ & $\langle Q\rangle_t$ & $\langle b_{x,\mathrm{mid}}^\mathrm{smooth}\rangle_t$ & $\hat{H}/H$\\
        \hline
        $10$ & 0.437 & 2.78 & 7.73 & 0.324 & 5.66 \\
        $10^2$ & 0.913 & 2.27 & 6.37 & 0.196 & 3.91 \\
        $10^3$ & 0.131 & 0.622 & 1.67 & 0.041 & 0.773\\
        $10^4$ & 52.679 & 0.863 & 0.688 & 0.198 & 0.305 \\
        $10^5$ & 179.584 & 1.05 & 0.591 & 0.273 & 0.352 
    \end{tabular}
    \caption{Midplane quantities for the different $\beta_0$ cases. Note that we have moved the `smooth' subscript to the superscript position for $\beta, b_x$ to avoid overstretching the table. We are not introducing new variables. $\hat{H}$ in the rightmost column is the scaleheight (in units of initial mid-plane scaleheight $H$) measured from the time-averaged density profile (defined by the height at which the density drops by an e-fold $\rho_\mathrm{max} e^{-1}$).}
    \label{tab:midplane}
\end{table}

Thus, the potential increase in growth rate due to increased magnetic field (decreased $\beta$) is negated by magnetic elevation in the low $\beta_0$ cases. The disk mid-plane becomes magnetic pressure-dominated, making the gas more rarefied (i.e. lower density) and contributing to increased Toomre parameter which drastically lowers the CRMG growth rate. 

\subsection{Fragmentation in low shear environment} \label{subsec:low_shear}

\begin{figure*}
    \centering
    \includegraphics[width=0.4\textwidth]{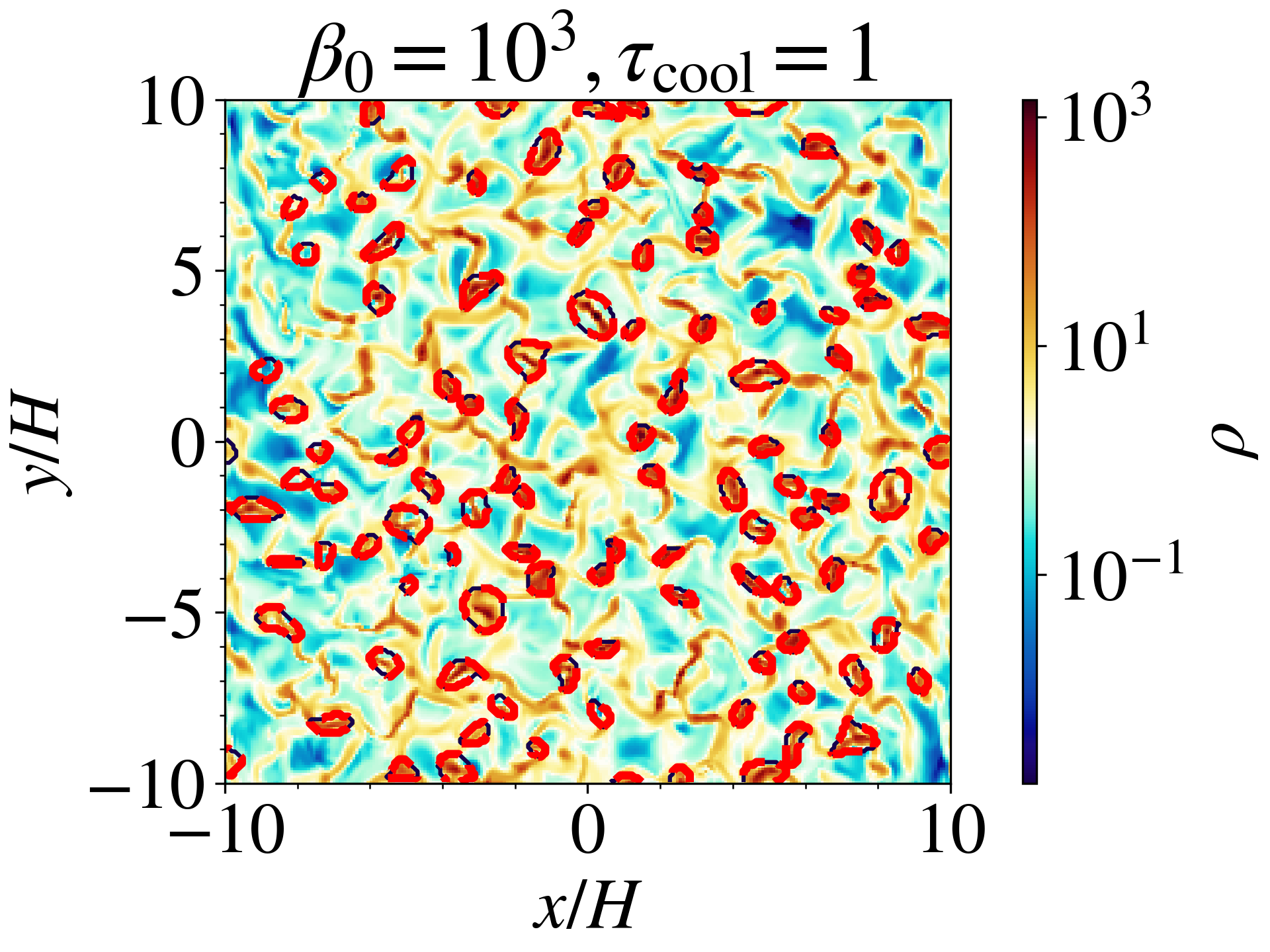}
    \includegraphics[width=0.4\textwidth]{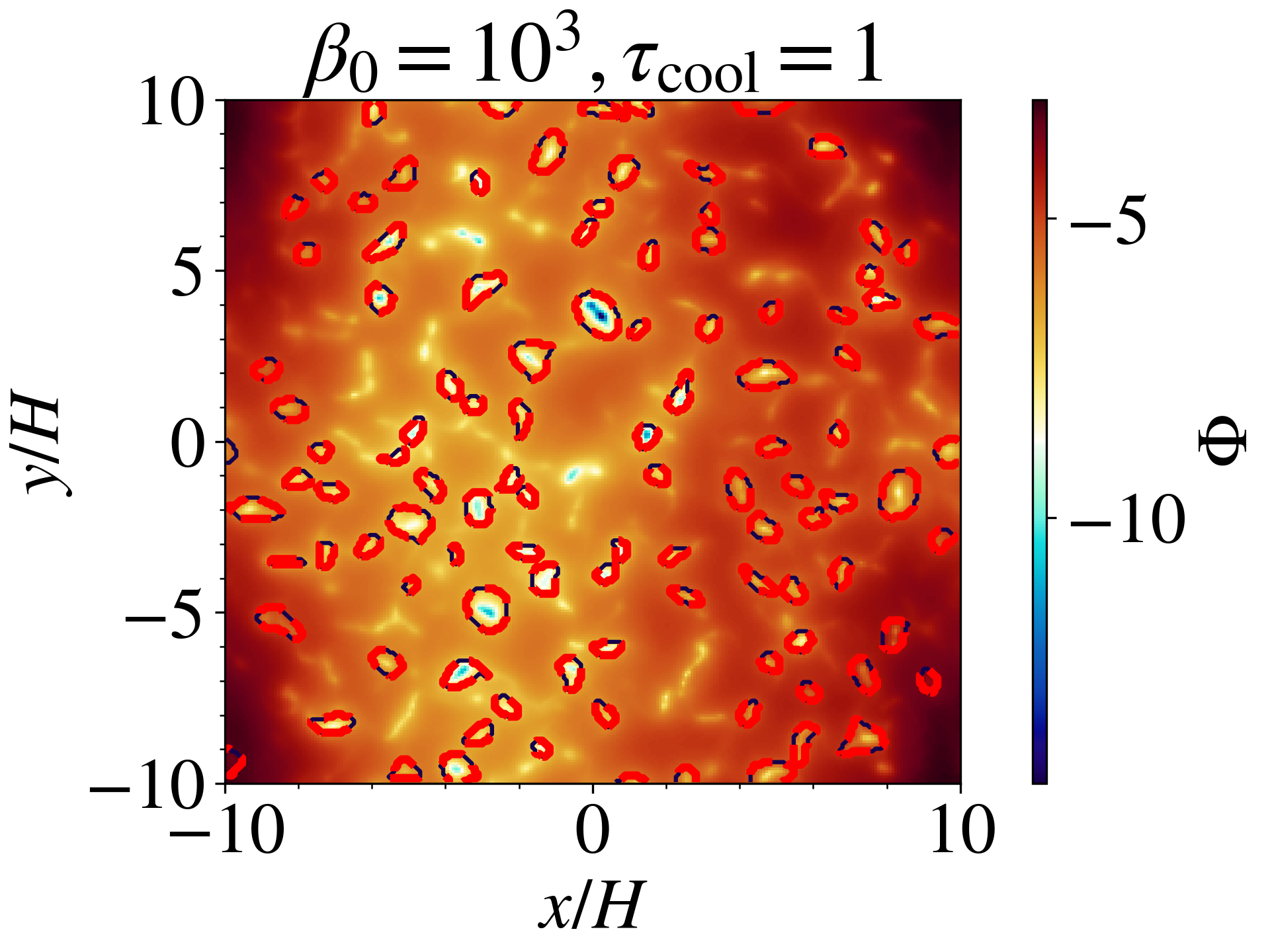}
    \caption{Snapshots of the mid-plane $\rho$ (left) and $\Phi$ (right) taken at $\Omega t = 200$ with contours overlaid to indicate clumps identified, taken from simulation with $\beta_0=10^3, \tau_\mathrm{cool}=1$ and low shearing parameter $q=0.1$.}
    \label{fig:lowq_fragmentation}
\end{figure*}

It was mentioned in \citet{Kubli_etal-2023} that a number of clumps were formed in the low shear $q$ locations of their simulation. Radial fields in low shear environments take much longer to be sheared into the toroidal direction, potentially alleviating the issue proposed by \citet{Gammie-1996} (and discussed in \S\ref{subsec:magneto_gravito}). We repeated one of our simulations ($\beta_0=10^3,\tau_\mathrm{cool}=1$) with very weak shear $q=0.1$, finding drastically enhanced fragmentation (Fig.~\ref{fig:lowq_fragmentation}) and a much larger bound mass fraction of 0.428. However, MRI is very inefficient in low shear environments, and the in-disk field generated in our simulation is weak. Thus, it is unlikely that the magnetic field would have an effect on gravitational instability in our low $q$ simulation; the increased fragmentation seen in this simulation is likely due to stronger growth rates with weaker Coriolis stabilization \citep{Kim_Ostriker-2001}.

\section{Discussion and conclusions} \label{sec:discussion_conclusion}

In this study we performed local shearing box simulations incorporating a net vertical flux and cooling using a $\tau_\mathrm{cool}$ prescription (i.e. setting the cooling time to be some constant multiple of $\Omega^{-1}$) \citep{Gammie-2001}. By fixing the cooling time to a small value ensuring fragmentation in the weakly magnetized cases, we systematically vary the initial mid-plane plasma beta $\beta_0$, which introduces different magnetic field strengths and structures in the saturated stage. We seek to address whether magnetic field promotes or suppresses fragmentation in AGN disk environments. Theoretically, while magnetic pressure tends to suppress GI (as manifested in magneto-Jeans type modes), magnetic tension in a shearing disk context can enhance GI. We call this the Coriolis-Restricted-Magneto-Gravitational (CRMG) instability, and performed WKB stability analysis of the axisymmetric CRMG modes. In this analysis, we find that the instability occurs even for $Q_T, Q_{T,B} > 1$ (compared to magneto-Jeans instability which is unstable only for $Q_{T,B}<1$), but requires a moderate radial contribution ($b_x$) to the in-disk magnetic field in order for the magnetic field to have a substantial destabilizing influence (see Fig.~\ref{fig:B_transition} and discussion in \S\ref{subsec:magneto_gravito}). This may not be achievable in setups with weak vertical flux but is possible with strong vertical flux. We observe in strong vertical flux simulations that the large-scale radial field can achieve $\beta_x\sim 3$, $b_x\sim 0.35$, and the MRI-dynamo seems to be successful in producing a relatively time-steady radial field supporting the use of the WKB analysis. These seem to be promising ingredients for magnetic fields to be destabilizing. However, our simulation results suggest otherwise:

\begin{enumerate}
    \item The bound mass fraction drops drastically in the low $\beta_0$ cases, where the in-disk magnetic field, produced by the MRI-dynamo, is strong. 
    \item Despite MRI being able to produce sustained in-disk magnetic field with appreciable radial contribution, magnetic elevation due to strong fields produced by the MRI-dynamo causes the mid-plane to be evacuated, increasing the Toomre parameter and lowering the growth rates of potential CRMG modes. This washes out potential increase in growth rates due to increased magnetic field.
\end{enumerate}

Therefore, our conclusion is that magnetic field seems to suppress rather than promote fragmentation in AGN disk environments.

The conclusions drawn in this study rely very much on the horizontally ($x,y$) averaged quantities. However, local variations across the disk could introduce subtle effects. For example, spiral density waves tend to align the local magnetic field in the direction of the wavefront \citep{Riols-Latter-2019}, increasing or decreasing the local radial field contribution $b_x$ depending on the orientation of the waves. Since these waves could harbor density seeds for gravitational instability, this alignment could promote or suppress fragmentation through axisymmetric CRMG growth. However, spiral density waves typically require a moderately low Toomre $Q_T\lesssim 1-2$ to accentuate the amplitude \citep{Goldreich_Lynden-Bell-1965}, thus it is hardly a relevant issue in our low $\beta_0$ simulations, for which $Q_T\gtrsim 3$ as a result of magnetic elevation. Indeed, we hardly observe any spiral waves in our low $\beta_0$ test cases.

A recent study by \citet{Kubli_etal-2023}, using global simulations of a protoplanetary disk, reached different conclusions than ours, that magnetic field is conducive rather than oppressive to clump formation. In particular, they discovered clumps of smaller mass and size in their simulations with magnetic field compared to the case without, indicative of the CRMG instability. This seemingly contradictory result to ours can be resolved by noting that 1) most of their clumps are formed at regions of low shear (i.e. small $q$), where non-axisymmetric wavevectors and radially pointing magnetic field can maintain their directionality much longer, conducive to the development of the CRMG instability. 2) The magnetic field is primarily generated through the GI-dynamo, not MRI. 3) There is strong resistivity in their simulations. 4) The magnetic field is sub-thermal at the locations of the clumps, implying absence of magnetic elevation. There are no explicit resistivity effects in our simulations and magnetic field is primarily generated through MRI, not GI, as indicated by the hugely subdominant gravitational stress. In our simulation with a smaller shear parameter $q=0.1$, we find that the lower shear renders a slower MRI-dynamo and weaker in-disk fields, thus the effect of magnetic field is not as apparent. The presence of resistivity and the use of a global geometry could also change this picture, as one can imagine strong magnetic field being generated in high shear regions, which diffuses to the low shear regions, providing sufficient field strength to trigger the CRMG instability. Our local, ideal MHD setup cannot address these issues, which we relegate to future work.

\section*{Acknowledgements}

We acknowledge support from NASA Astrophysics Theory Program grants 80NSSC22K0828 and 80NSSC24K0940, and NASA FINESST Fellowship 80NSSC22K1753 (HGD). The simulations presented in this study were performed using Frontera at TACC under the LRAC computing allocation AST-21007. We also thank the anonymous reviewer for the suggestions, which helped improve this manuscript.

\section*{Data Availability}

The data used for this study is available upon reasonable request to the authors.



\bibliographystyle{mnras}
\bibliography{main} 




\appendix

\section{Initial equilibrium profile} \label{app:initial}

\begin{figure}
    \centering
    \includegraphics[width=0.23\textwidth] {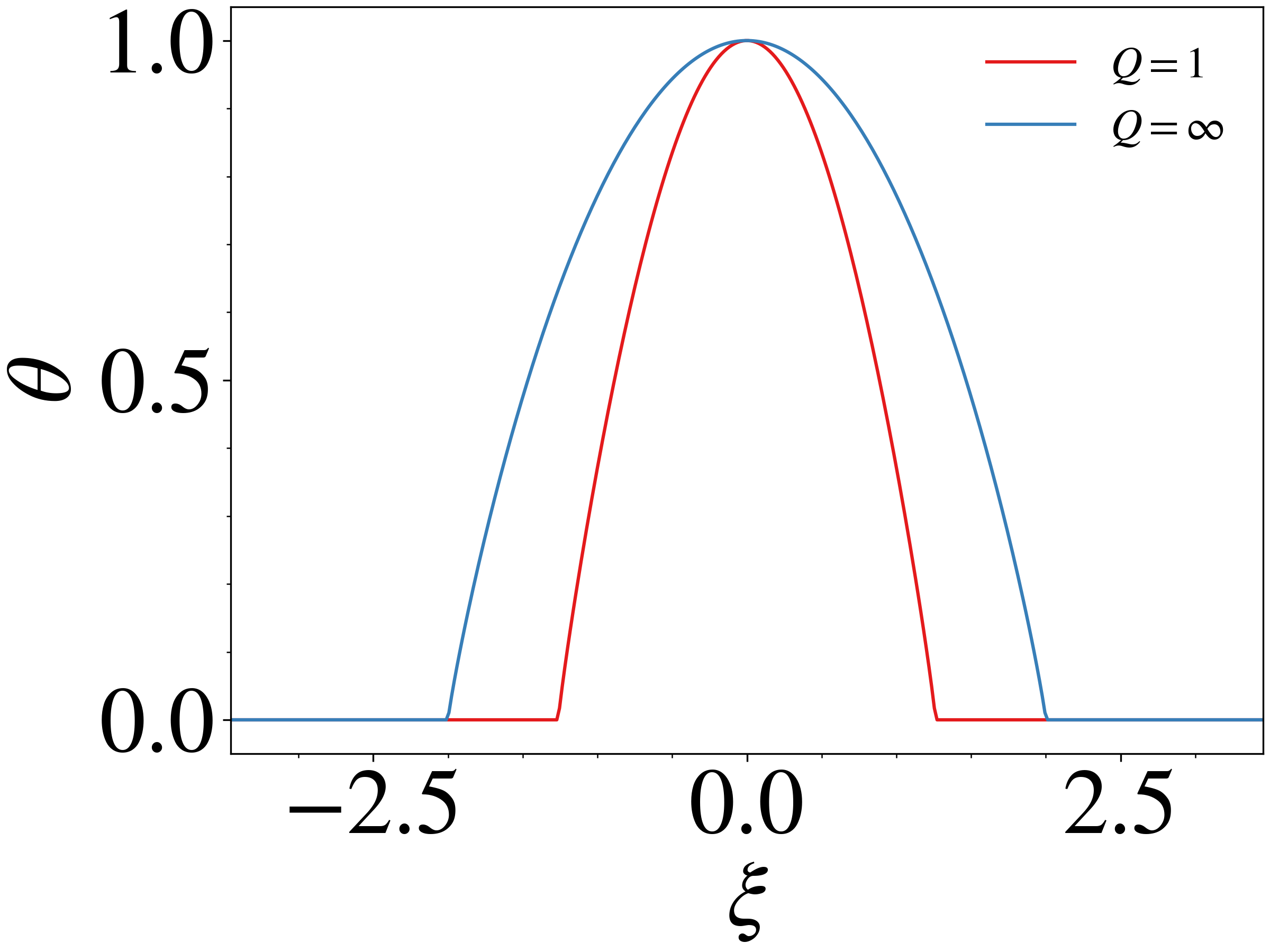} \\
    \includegraphics[width=0.23\textwidth] {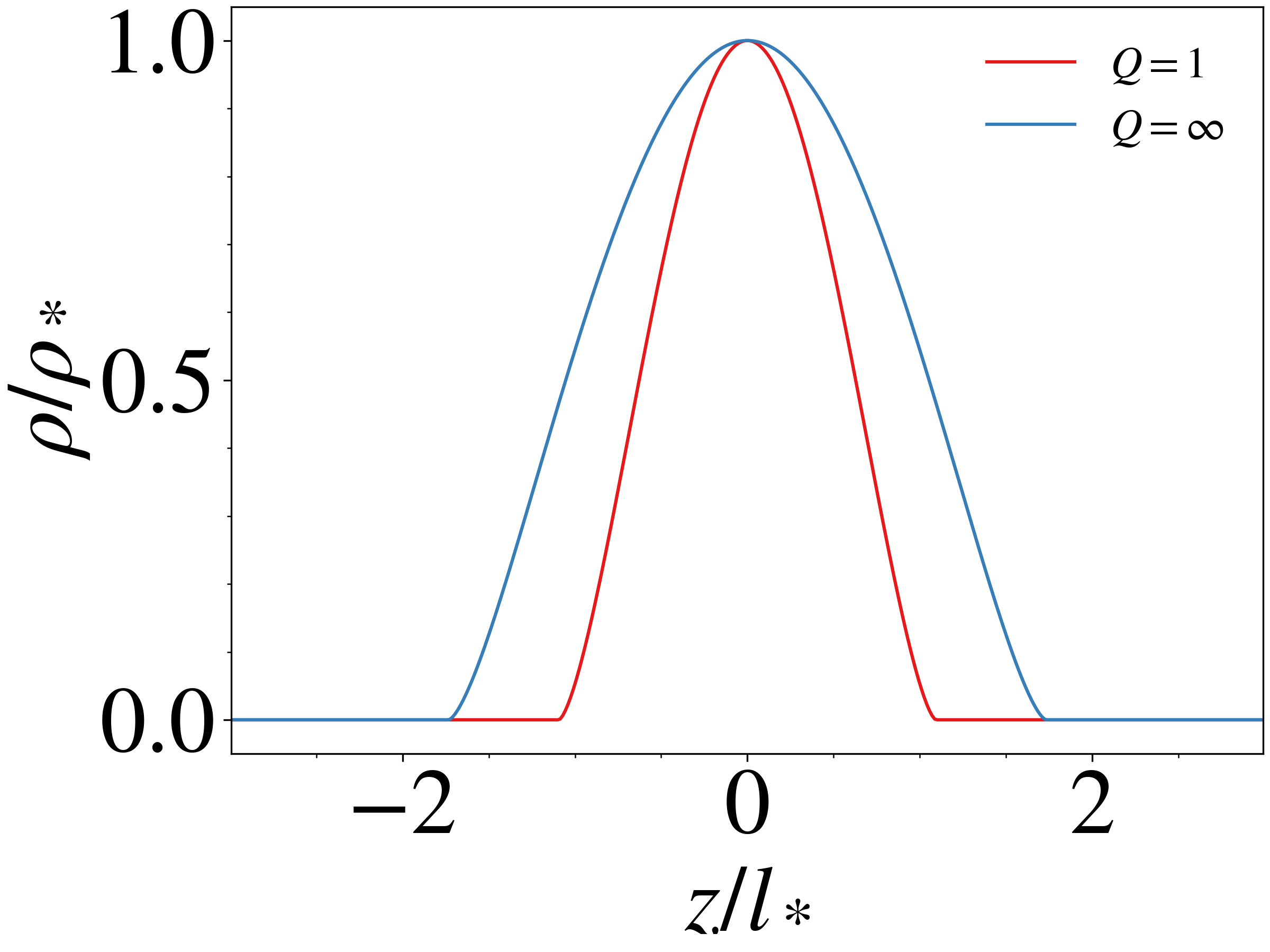} 
    \includegraphics[width=0.23\textwidth] {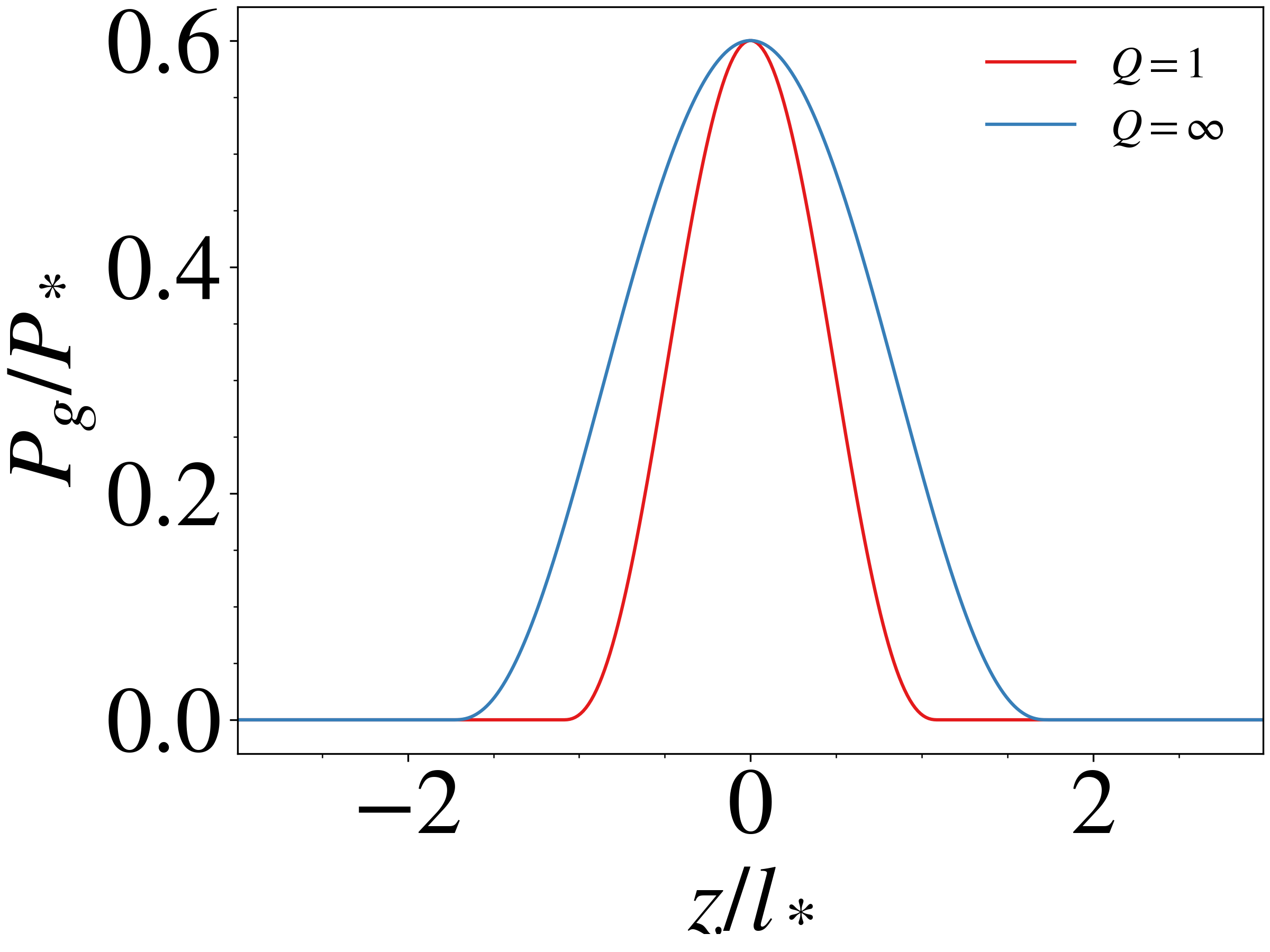} 
    \caption{Normalized profiles of $\theta$ (top), $\rho/\rho_*$ (bottom left) and $P_g/P_*$ (bottom right) for $Q = 1$. Profiles for the non-self-gravitating case, $Q = \infty$, are shown for comparison.}
    \label{fig:compare_theta}
\end{figure}

Under the assumption of a uniform magnetic field, no $x,z$ flow and that $\rho, P_g$ depend only on $z$, the $x$-component of eq.~\ref{eqn:momentum} gives the equilibrium velocity profile $\vb{v} = -q\Omega x\vu{y}$ and the $z$-component gives the following equation:
\begin{equation}
    \frac{1}{\rho}\dv{P_g}{z} = -\Omega^2 z - 4\pi G\int^{z}_0\rho\dd{z'}. \label{eqn:momentum_balance}
\end{equation}
Eq.~\ref{eqn:momentum_balance} can be solved using the method described in \citet{Jiang_Goodman-2011,Chen_etal-2023}. Assuming a polytropic EOS of $P_g\propto\rho^{5/3}$, we can simplify eq.~\ref{eqn:momentum_balance} by transforming into Lane-Emden variables $\rho = \rho_0\theta^{3/2}$, $P_g = P_{g0}\theta^{5/2}$, $\xi = z/h$, where $\rho_0,P_{g0}$ are the initial mid-plane density and gas pressure, giving
\begin{equation}
    \dv[2]{\theta}{\xi} + \theta^{3/2} = -\frac{Q}{2}. \label{eqn:Lane-Emden}
\end{equation}
In doing so, we have defined $h^2 = 5P_{g0}/8\pi G\rho_0^2, Q=\Omega^2/2\pi G\rho_0$. Eq.~\ref{eqn:Lane-Emden} can be solved numerically with the conditions $\theta(0) = 1, \theta'(0) = 0$. In Fig.~\ref{fig:compare_theta}, we see that self-gravity confines the disk profile. Eq.~\ref{eqn:Lane-Emden} shows that the density and gas pressure profiles, which are encoded in $\theta$, depend only on $Q$.

To calculate $\rho, P_g$, we need to know $\rho_0, P_{g0}$. The initial profile is completely specified when the proxy Toomre parameter $Q$, the surface density $\Sigma$ and the angular velocity $\Omega$ are specified. Once the $\theta$ profile is determined from eq.~\ref{eqn:Lane-Emden}, $\rho_0, P_{g0}$ can be calculated using the fact that $\Sigma = 2\rho_0 h\int^\infty_0\theta^{3/2}\dd{\xi}$ and $Q = \Omega^2/2\pi G\rho_0$, which gives
\begin{equation}
    \rho_0 = \frac{\Omega^2}{2\pi G Q},\quad P_{g0} = \frac{2\Sigma \Omega^2}{5\Psi_{3/2} Q},\quad T_0 = \qty(\frac{\mu m_p}{k_B})\frac{4\pi G\Sigma}{5\Psi_{3/2}}, \label{eqn:app_midplane}
\end{equation}
where $\Psi_{3/2}(Q) = \int_0^\infty\theta^{3/2}\dd{\xi}$ is a dimensionless measure of the disk's thickness. For $Q\ll 1$, self-gravity dominates and $\Psi_{3/2}$ will be small, while for $Q\gg 1$ the tidal potential dominates, eq.~\ref{eqn:Lane-Emden} becomes independent of $Q$ and $\Psi_{3/2}$ approaches a constant.

One can observe that $Q$ and the Toomre parameter $Q_T = \langle c_s^2\rangle^{1/2}_\rho\Omega/\pi G\Sigma$ are related in the initial profile. Note that
\begin{gather}
    \langle c_s^2\rangle_\rho = \frac{\gamma\int P_g\dd{V}}{\int\rho\dd{V}} = \frac{c_{s0}^2\int^\infty_0\theta^{5/2}\dd{\xi}}{\int^\infty_0\theta^{3/2}\dd{\xi}} = c_{s0}^2\frac{\Psi_{5/2}\qty(Q)}{\Psi_{3/2}\qty(Q)}, \label{eqn:initial_Q_QT} \\
    \Sigma = \int\rho\dd{z} = 2\rho_0 h\int^\infty_0\theta^{3/2}\dd{\xi} = 2\rho_0 h \Psi_{3/2}\qty(Q), \label{eqn:surface_density}
\end{gather}
where $c_{s0}=\sqrt{\gamma P_{g0}/\rho_0}$ is the initial mid-plane sound-speed and $\Psi_{5/2}(Q) = \int^\infty_0\theta^{5/2}\dd{\xi}$ is another measure of the disk thickness. Then we have:
\begin{equation}
    Q_T = \frac{c_{s0}\Omega}{2\pi G \rho_0 h}\frac{\qty(\Psi_{5/2})^{1/2}}{\qty(\Psi_{3/2})^{3/2}} = \qty(\frac{8\gamma}{5}Q)^{1/2}\frac{\qty(\Psi_{5/2})^{1/2}}{\qty(\Psi_{3/2})^{3/2}}. \label{eqn:QT_Q_relation}
\end{equation}
We note that $\Psi_{5/2},\Psi_{3/2}$ depend only on $Q$ and $(\Psi_{5/2})^{1/2}/(\Psi_{3/2})^{3/2}$ is of order unity generally, so $Q_T$ is a function of $Q$ alone and scales roughly as $Q^{1/2}$. For $Q=1$, $Q_T=0.89$. In addition, with our choice of units, the initial density and pressure profiles in code units are completely specified by $Q$.

\section{Rationale for our choice of box size} \label{app:box_rationale}

Our choice of box size is unprecedentedly large in local shearing box simulations. Nevertheless, it is numerically justified. First of all, gravitational instability is a long-range effect, so a horizontally extended box is warranted. A common measure of the length-scale required for gravitational effects to be important is the Toomre-length $L_T = \pi^2 G\Sigma/\Omega^2$ \citep{Chen_etal-2023}, corresponding to the 2D Jeans length $c_s^2/G\Sigma$. In our initial setup the ratio $L_T/H$ is given by
\begin{equation}
    \frac{L_T}{H} = \frac{\pi}{Q_T}\qty(\frac{\Psi_{5/2}}{\Psi_{3/2}})^{1/2}, \label{eqn:Ltoomre}
\end{equation}
where $\Psi_{5/2}$ and $\Psi_{3/2}$ are defined in Appendix \ref{app:initial} and $H$ is the initial mid-plane scale height $c_{s0}/\Omega$. For $Q_T=0.89$, $L_T/H=3.2$, which sets a box size of $20H$ roughly to $6.25 L_T$. The horizontally averaged Toomre length is unchanged throughout the simulation as $\langle\Sigma\rangle_{xy} = \int\rho\dd{V}/L_x L_y$ is constant since the total mass is fixed. Thus, our horizontal box size should be sufficient to encompass gravitational effects throughout the simulations. We note that it is typical to use a horizontal box length of $20H$ in existing studies \citep[e.g.][]{Lohnert_Peeters-2023,Riols-Latter-2019}.

A large vertical box height is necessary to encompass a sufficient number of scale heights, especially in the strongly magnetized regime where magnetic elevation is active. In the most strongly magnetized case, $\beta_0=10$, we find that the mid-plane scale height $c_{s0}/\Omega$ is elevated by roughly 4 times compared to the initial value, so an initial vertical extent of $24H$ covers roughly 6 scale heights in the nonlinear stage for that case. We find that simulations that are not sufficiently extended vertically tend to have unphysically dominant Reynolds stresses $\alpha_R$.

\begin{figure}
    \centering
    \includegraphics[width=0.23\textwidth]{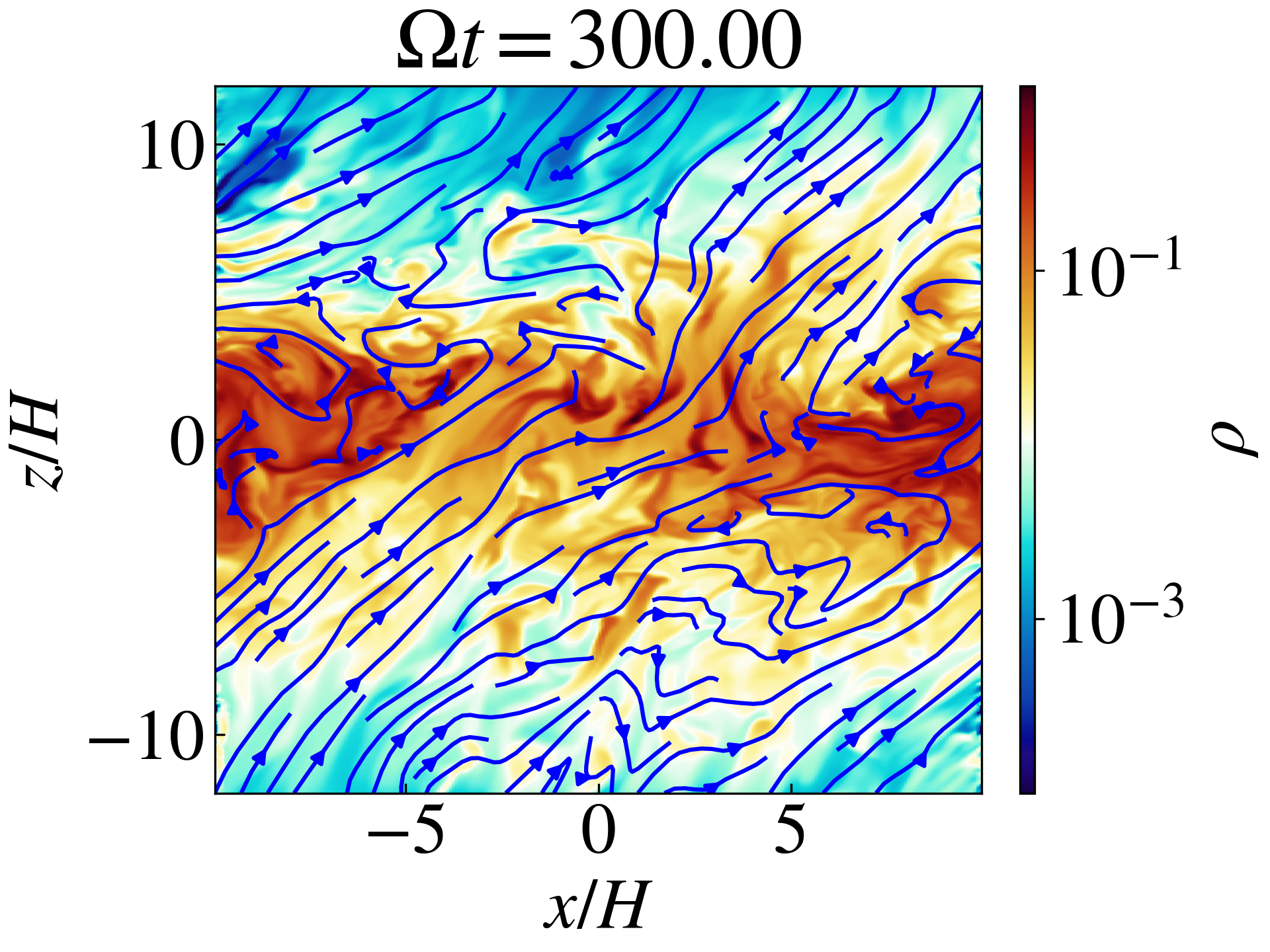}
    \includegraphics[width=0.23\textwidth]{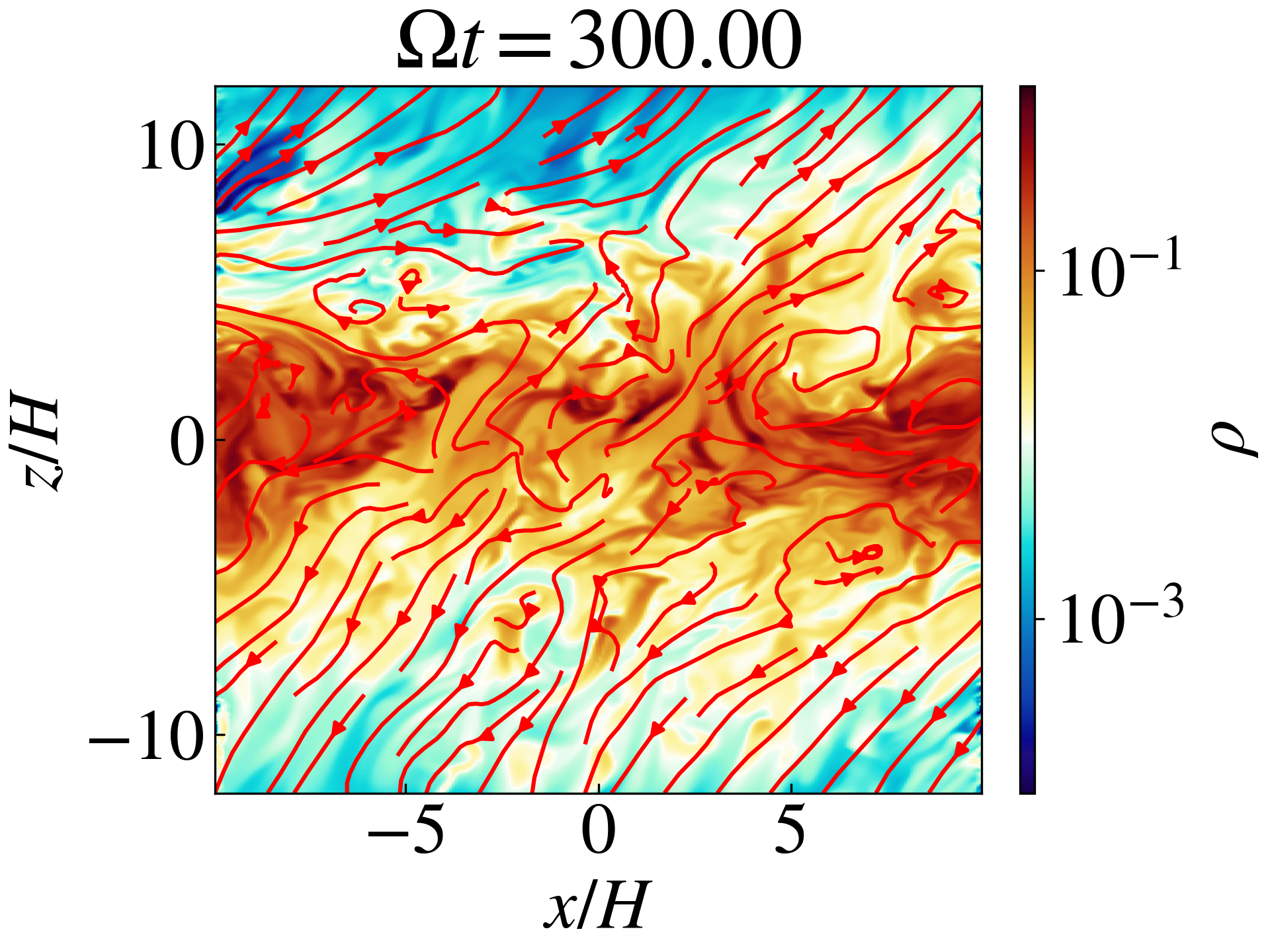} \\
    \includegraphics[width=0.48\textwidth]{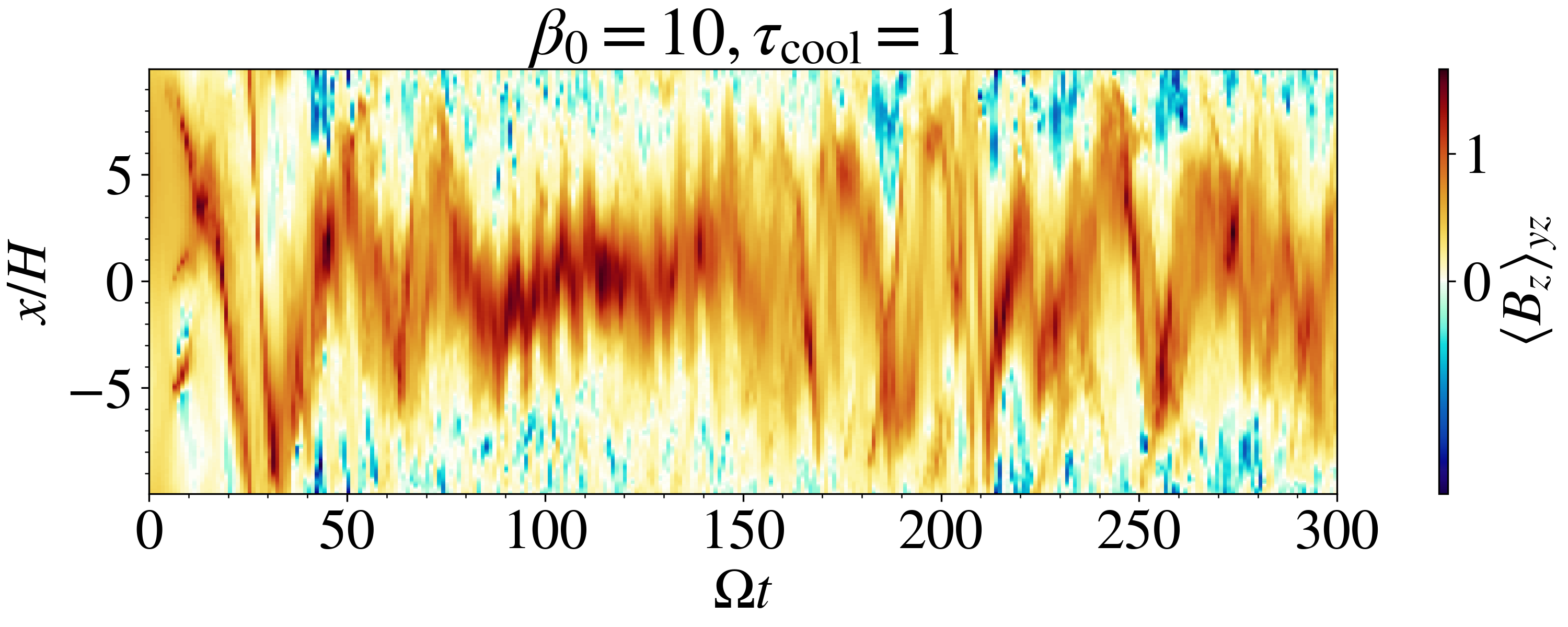} \\
    \includegraphics[width=0.48\textwidth]{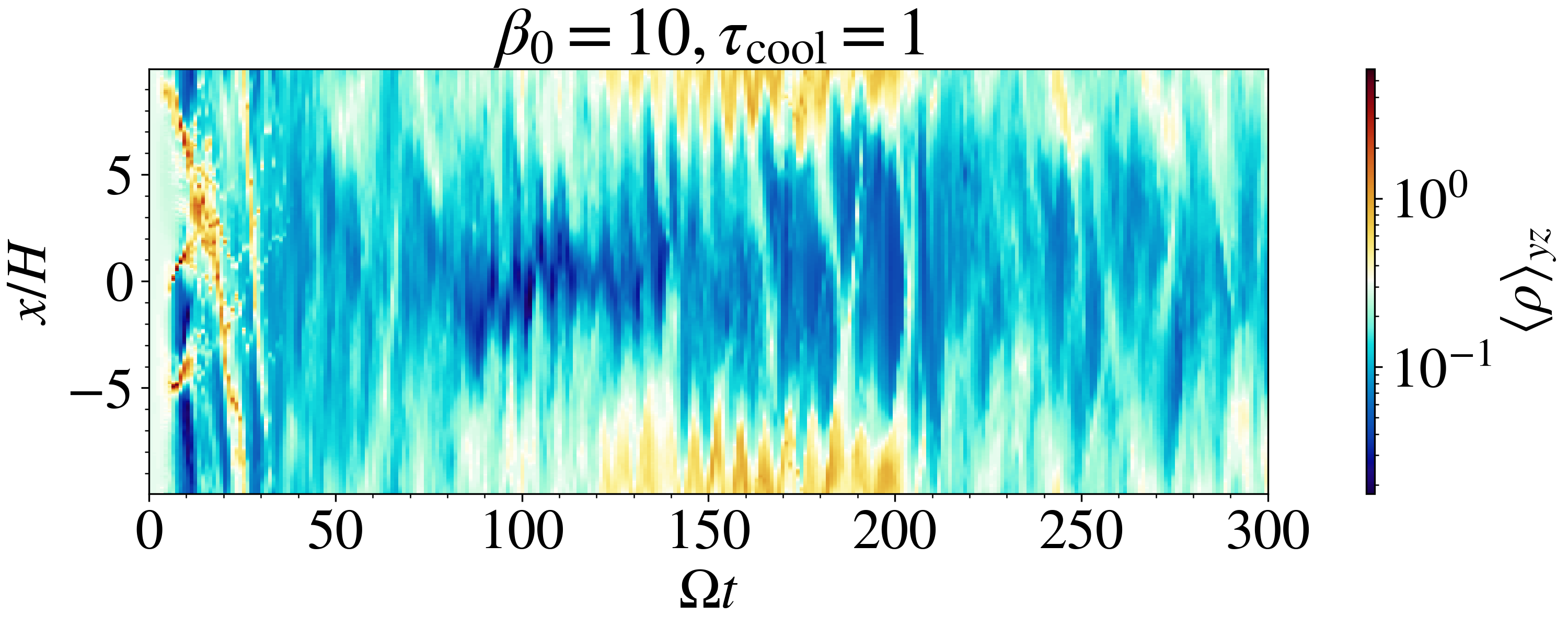} \\
    \includegraphics[width=0.48\textwidth]{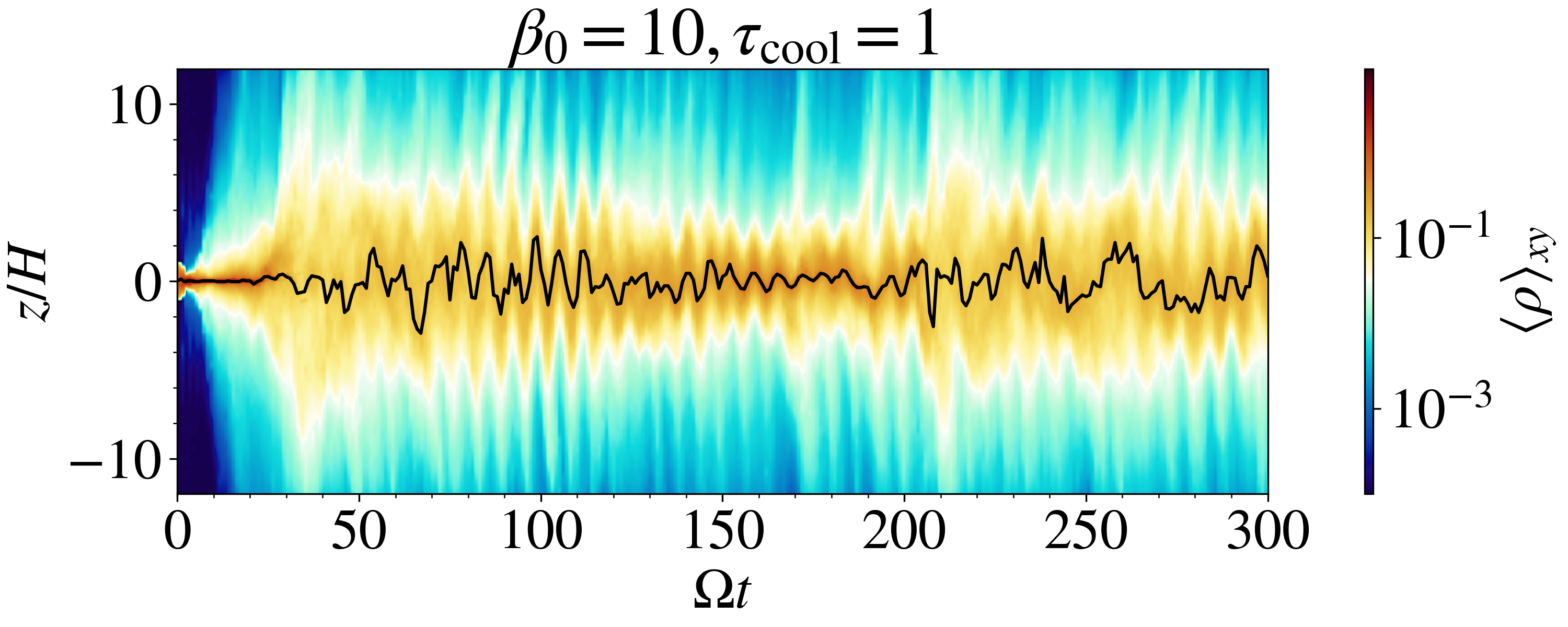}
    \caption{Strong magnetic flux channels develop when a restrictive box ($20H\times 20H\times24H$) is used. Top row: $x-z$ density slice of the $\beta_0 = 10, \tau_\mathrm{cool}=1$ case at $\Omega t = 300$, overplotted with magnetic field (top left, blue) and velocity field streamlines (top right, red). Remaining panels: Time-series diagrams of $\langle B_z\rangle_{yz}$ (second row), $\langle\rho\rangle_{yz}$ (third row), $\langle\rho\rangle_{xy}$ (bottom). The black solid line in the bottom panel denotes the density maximum, which oscillates in position. In the second and third row, $y-z$ average $\langle\cdot\rangle_{yz}$ is taken across the whole azimuthal ($-10H<y<10H$) extent, and only through the disk portion in the vertical ($-2H<z<2H$) extent. The $x-y$ average $\langle\cdot\rangle_{xy}$ is taken across the full horizontal ($-10H<x,y<10H$) extent. Note that the direction of variation in the second and third rows are in the $x$-direction whereas it is in the $z$-direction for the bottom row.}
    \label{fig:cubic_slices}
\end{figure}

When running our low $\beta_0=10, 10^2$ simulations using a reduced box domain of $20H\times20H\times24H$ (resolution of $256\times256\times512$), we observed signatures of zonal flow described in \citet{Riols_Lesur-2019}, in which a flux channel stretching diagonally across the $x-z$ plane is formed (Fig.~\ref{fig:cubic_slices}). This channel is magnetically enhanced and lower in density than the mid-plane disk material on the two sides, as shown by the slice plots at the top of the figure and the time-series plots of $B_z,\rho$, averaged over the $y-z$ directions. Over the course of the simulation, we observe disk material on the two sides attempting to replenish the channel through turbulent diffusion, occasionally closing it. But it would soon be reopened again by the magnetically driven wind. When a channel is formed, the disk also undulates, as shown by the time-series plot of the $x-y$ averaged density in the bottom panel, where the undulations are marked by the black solid line. \citet{Riols_Lesur-2019} explained the origin of such zonal flows as being due to the magnetically driven wind drawing out mass faster than it can be replenished by turbulent diffusion, leading to an instability. The eigenmode of this instability is a diagonal channel, just like the ones we see. Overall, our observations of the zonal structures agree with this explanation. Curiously, when we extend the box domain azimuthally to $20H\times40H\times24H$ (resolution of $256\times512\times384$), the zonal structure is subdued. This could be due to the extra azimuthal freedom giving the turbulence more room to act, closing the channel whenever it begins to form.

\begin{figure}
    \centering
    \includegraphics[width=0.40\textwidth]{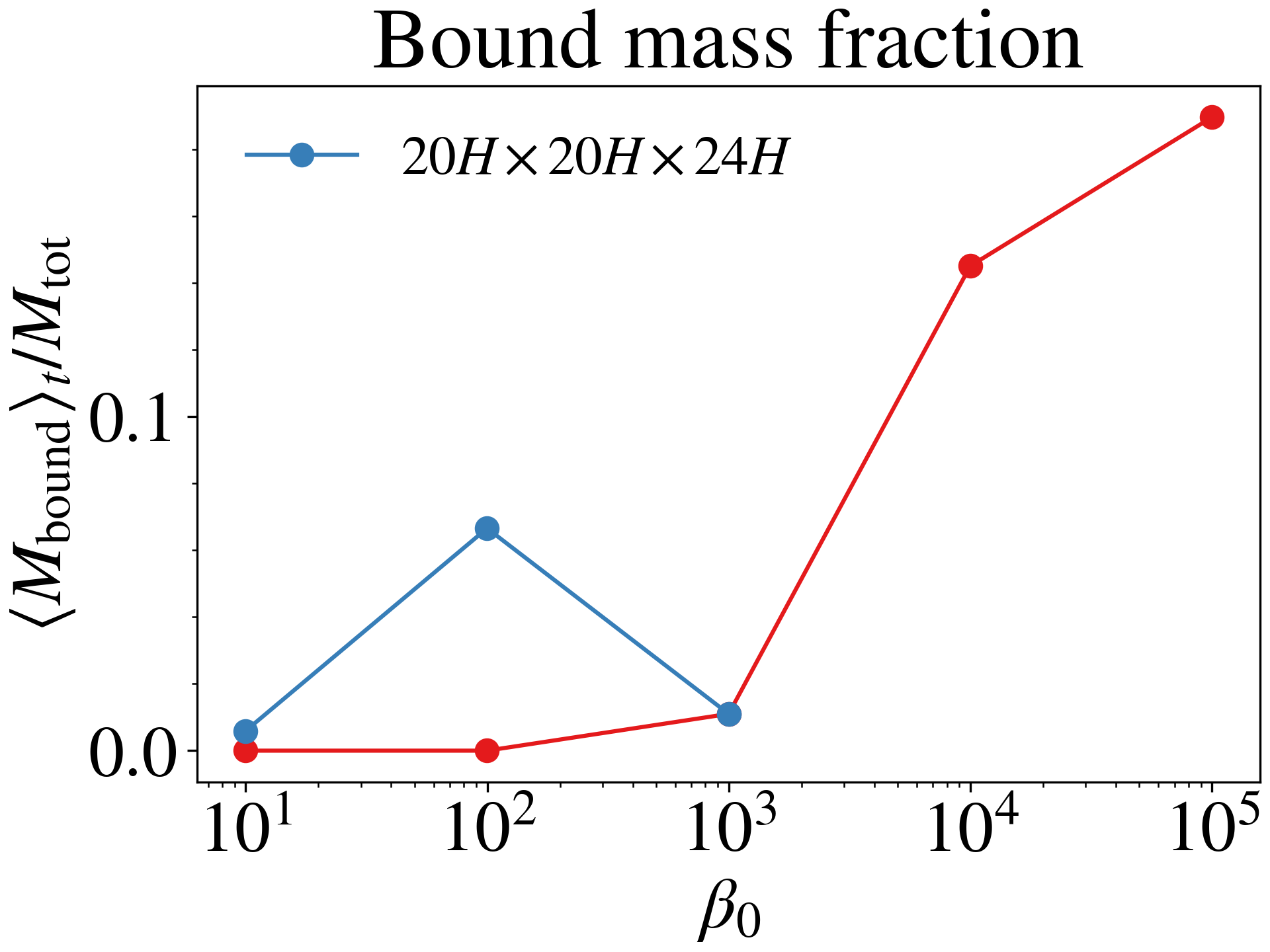} \\
    \includegraphics[width=0.23\textwidth]{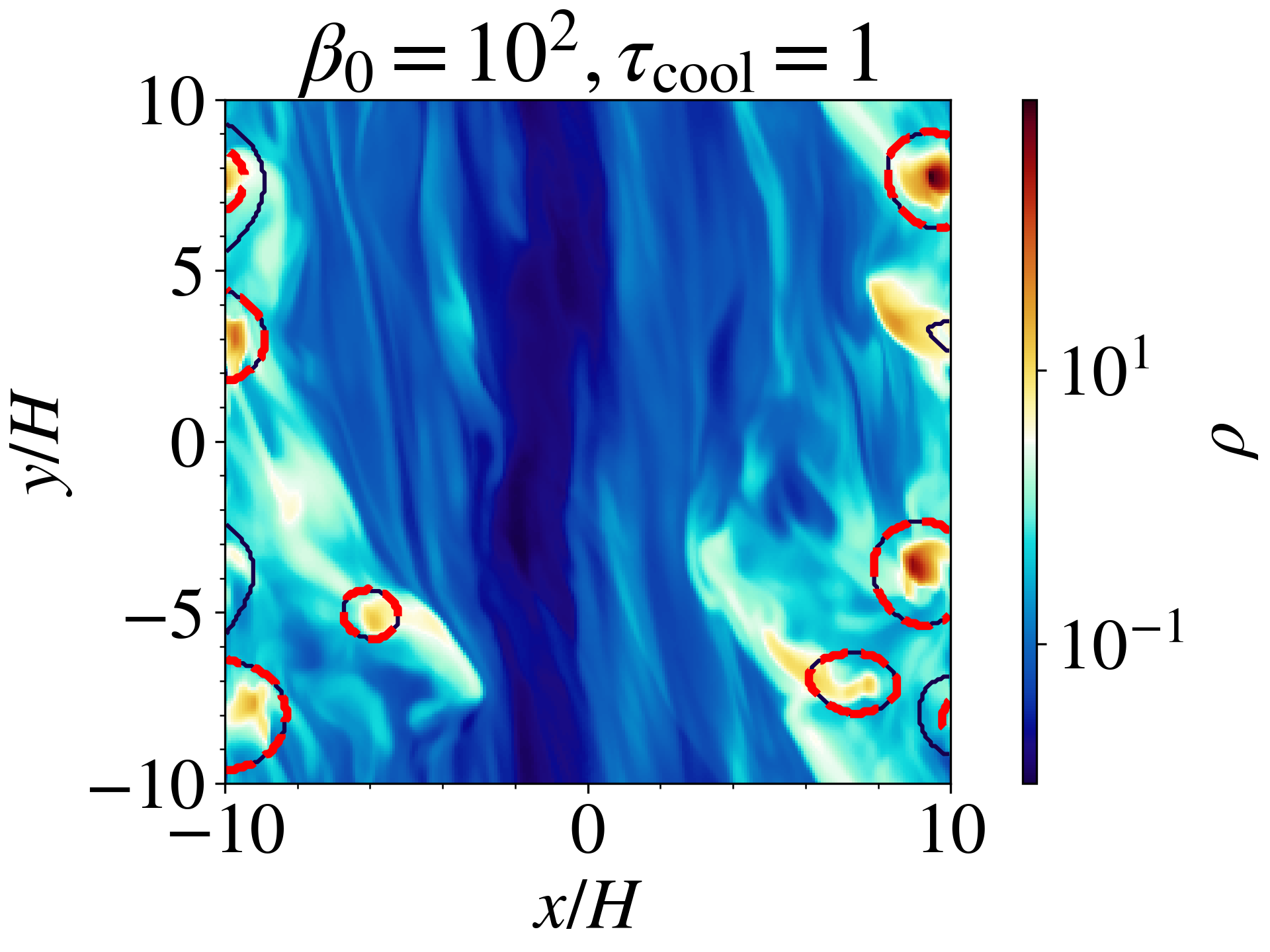}
    \includegraphics[width=0.23\textwidth]{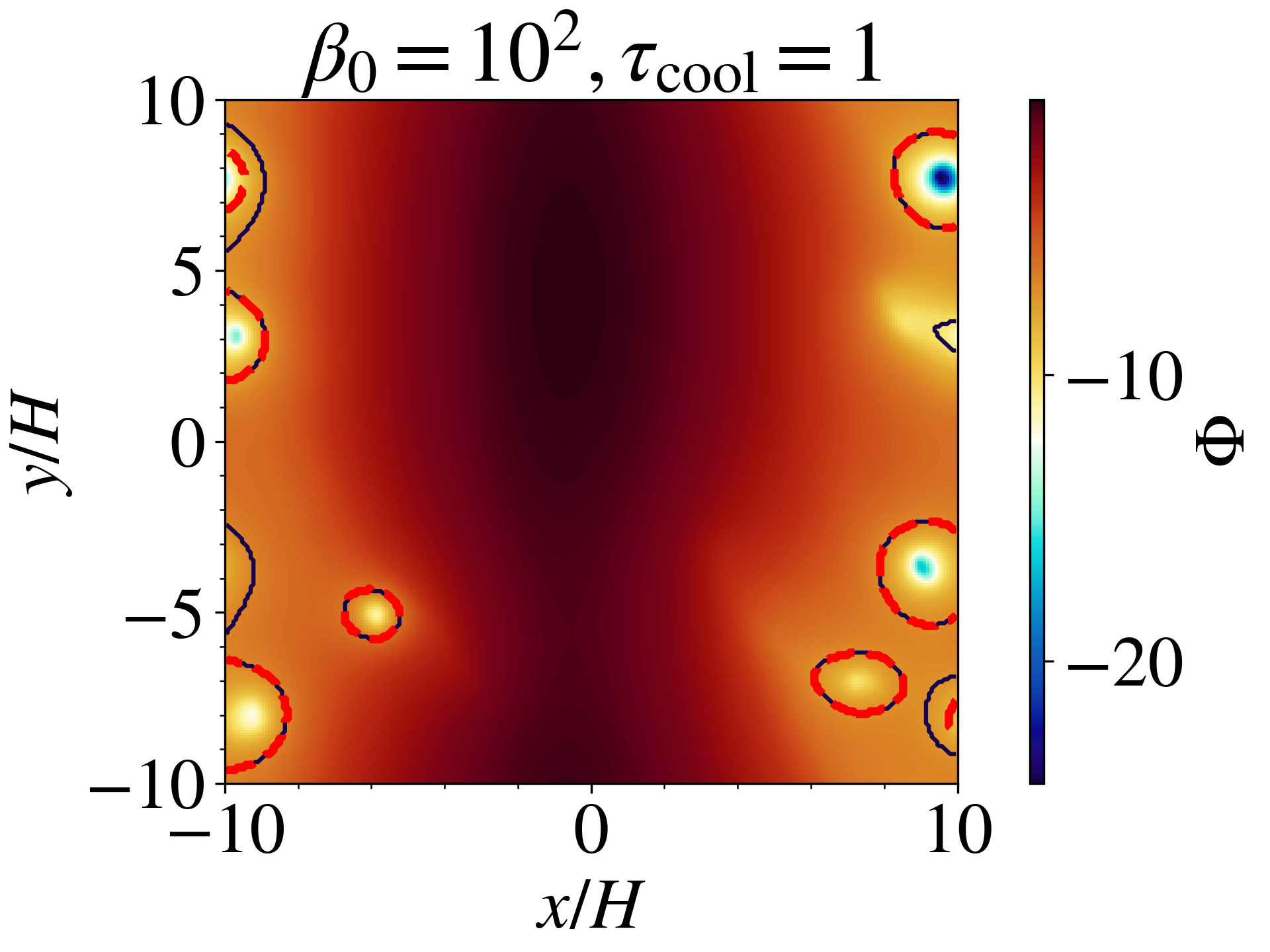} \\
    \includegraphics[width=0.40\textwidth]{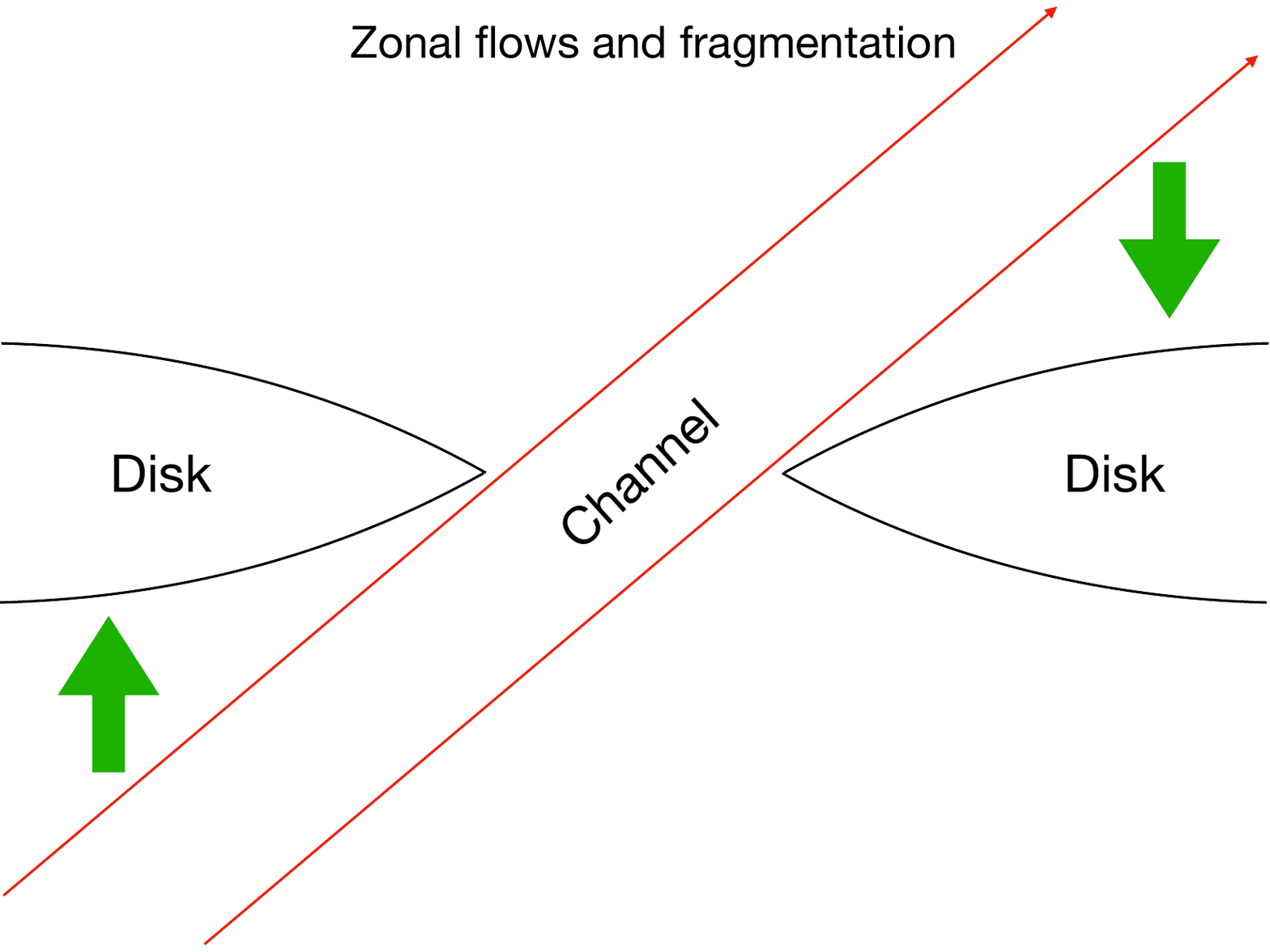}
    \caption{Top: Bound mass fraction for various $\beta_0$ using the fiducial box domain and resolution (red), compared against when a more restrictive box $20H\times 20H\times 24H$ (resolution $256\times256\times512$) is used (blue). Middle two panels: Snapshots of the density and gravitational potential for the case $\beta_0=100,\tau_\mathrm{cool}=1$ when the more restrictive box is used. Bottom: Schematic diagram showing how the ram and magnetic pressure from the channels inhibits gas escape, reinforcing a denser disk, thereby promoting fragmentation.}
    \label{fig:box_rationale_fragmentation}
\end{figure}

Apparently, the channels seen in the zonal flow have an effect on fragmentation. In Fig.~\ref{fig:box_rationale_fragmentation}, we see from the bound mass fraction plot that simulations with a reduced azimuthal extent (where zonal flow is seen), have slightly different fragmentation characteristics (blue line) from the simulations with the fiducial box domain (red line). In particular, the bound mass fraction of the $\beta_0=10^2$ case is greatly enhanced. Looking at a snapshot of this simulation, we observe that most of the fragmentation occurs on the sides, away from the channel. We believe that this enhancement in fragmentation is due to the flux channel forming a ram pressure and magnetic pressure wall preventing escape of material, forcing it back to the disk, and increasing the mid-plane density, which is conducive to more fragmentation (see bottom panel of Fig.~\ref{fig:box_rationale_fragmentation}). While this is an interesting observation from a fragmentation point of view, the properties of these flux channels, e.g. their width, how regularly they appear, etc. in a global, astrophysical context are poorly understood. Thus in this study we choose a box domain in which the zonal flow is subdued, and relegate study of the effect of zonal flows on fragmentation to the future.

\section{Linear analysis of the Coriolis-Restricted-Magneto-Gravitational instability for a razor-thin disk} \label{app:crmg}

In this Appendix we outline the linear stability analysis of a self-gravitating, razor-thin disk subjected to an in-plane magnetic field. We refer the reader to \citet{Elmegreen-1987,Gammie-1996,Kim_Ostriker-2001} for more details of this derivation. The fluid equations governing the dynamics of a razor thin disk $\rho = \Sigma\delta(z)$, written out in component form in shearing box coordinates, are
\begin{gather}
    \pdv{\Sigma}{t} + \qty(v_x\pdv{x} + v_y\pdv{y})\Sigma + \Sigma\qty(\pdv{v_x}{x} + \pdv{v_y}{y}) = 0, \label{eqn:full_continuity} \\
    \pdv{B_x}{t} = B_y\pdv{v_x}{y} - v_x\pdv{B_x}{x} - v_y\pdv{B_x}{y} - B_x\pdv{v_y}{y}, \label{eqn:full_induction_x} \\
    \pdv{B_y}{t} = B_x\pdv{v_y}{x} - v_x\pdv{B_y}{x} - v_y\pdv{B_y}{y} - B_y\pdv{v_x}{x}, \label{eqn:full_induction_y} \\
    \pdv{v_x}{t} + \qty(v_x\pdv{x} + v_y\pdv{y})v_x = -\pdv{x}\qty(h + \Phi) \nonumber \\\quad+ \frac{1}{\Sigma}\qty(B_x\pdv{x} + B_y\pdv{y})B_x - \frac{1}{\Sigma}\pdv{x}\qty(\frac{B^2}{2}) + 2\Omega v_y, \label{eqn:full_momentum_x} \\
    \pdv{v_y}{t} + \qty(v_x\pdv{x} + v_y\pdv{y})v_y = -\pdv{y}\qty(h + \Phi) \nonumber \\\quad+ \frac{1}{\Sigma}\qty(B_x\pdv{x} + B_y\pdv{y})B_y - \frac{1}{\Sigma}\pdv{y}\qty(\frac{B^2}{2}) - 2\Omega v_x, \label{eqn:full_momentum_y} 
\end{gather}
where $P_g = K\Sigma^\gamma$ is the gas pressure at the disk ($K$ being a constant) and $h = \gamma/(\gamma-1)K\Sigma^{\gamma-1}$ is the enthalpy density. Defining the disk sound speed by $c_s^2 = \gamma P_g/\Sigma$, it follows that $\partial_i h = \partial_i P_g/\Sigma$ for $i=x,y$. Eqs.~\ref{eqn:full_continuity}--\ref{eqn:full_momentum_y} can be derived by integrating the hydrodynamic equations eq.~\ref{eqn:continuity}--\ref{eqn:induction} through $z$ assuming a razor thin density distribution $\rho=\Sigma\delta(z)$ and then expanding in component form. $\Sigma$ is the surface density of the disk. These equations are coupled to the Poisson equation
\begin{equation}
    \nabla^2\Phi_\mathrm{3D}\qty(r,\phi,z) = 4\pi G\Sigma\delta\qty(z), \label{eqn:full_poisson}
\end{equation}
where $\nabla^2 = \partial^2/\partial x^2+\partial^2/\partial y^2+\partial^2/\partial z^2$ here. The gravitational potential in eq.~\ref{eqn:full_poisson}, $\Phi_\mathrm{3D}$, is related to that in eq.~\ref{eqn:full_momentum_x},~\ref{eqn:full_momentum_y} by $\Phi = \Phi_\mathrm{3D}(x,y,z=0)$. Consider an unperturbed solution (denoted by subscript 0) where $\Sigma_0 = \mathrm{const.}$, $\vb{v}_0 = -q\Omega x\vu{y}$, $P_{g0} = \mathrm{const.}$ with a spatially uniform magnetic field $\vb{B}_0 = B_{x0}\vu{x} + B_{y0}\vu{y}$. We perturb the fluid variables $Q = Q_0 + Q_1$, where $Q = \Sigma, B_x, B_y, v_x, v_y, h, \Phi$. Keeping only the terms first order in the perturbation, eqs.~\ref{eqn:full_continuity}--\ref{eqn:full_momentum_y} become
\begin{gather}
    \pdv{\Sigma_1}{t} - q\Omega x\pdv{\Sigma_1}{y} + \Sigma_0\qty(\pdv{v_{x1}}{x} + \pdv{v_{y1}}{y}) = 0, \label{eqn:perturb_continuity} \\
    \pdv{B_{x1}}{t} = B_{y0}\pdv{v_{x1}}{y} + q\Omega x\pdv{B_{x1}}{y} - B_{x0}\pdv{v_{y1}}{y}, \label{eqn:perturb_induction_x} \\
    \pdv{B_{y1}}{t} = -q\Omega B_{x1} + B_{x0}\pdv{v_{y1}}{x} + q\Omega x\pdv{B_{y1}}{y} - B_{y0}\pdv{v_{x1}}{x}, \label{eqn:perturb_induction_y} \\
    \pdv{v_{x1}}{t} - q\Omega x\pdv{v_{x1}}{y} = -\pdv{x}\qty(h_1 + \Phi_1) + 2\Omega v_{y1} \nonumber \\\quad+ \frac{1}{\Sigma_0}\qty(B_{x0}\pdv{B_{x1}}{x} + B_{y0}\pdv{B_{x1}}{y}) \nonumber \\\quad- \frac{1}{\Sigma_0}\pdv{x}\qty(B_{x0} B_{x1} + B_{y0} B_{y1}), \label{eqn:perturb_momentum_x} \\
    \pdv{v_{y1}}{t} - q\Omega v_{x1} - q\Omega x\pdv{v_{y1}}{y} = -\pdv{y}\qty(h_1 + \Phi_1) - 2\Omega v_{x1} \nonumber \\\quad+ \frac{1}{\Sigma_0}\qty(B_{x0}\pdv{B_{y1}}{x} + B_{y0}\pdv{B_{y1}}{y}) \nonumber \\\quad- \frac{1}{\Sigma_0}\pdv{y}\qty(B_{x0} B_{x1} + B_{y0} B_{y1}). \label{eqn:perturb_momentum_y}
\end{gather}
The perturbed Poisson equation is
\begin{equation}
    \nabla^2\Phi_{\mathrm{3D},1} = 4\pi G\Sigma_1\delta\qty(z). \label{eqn:perturb_poisson}
\end{equation}

\subsection{Axisymmetric perturbations} 

We first consider axisymmetric perturbations $(k_y = 0)$. Assuming perturbations have the form $Q_1 \propto \exp(ik_x x - i\omega t)$, eqs.~\ref{eqn:perturb_continuity}--\ref{eqn:perturb_momentum_y} become
\begin{gather}
    \omega \Sigma_1 = k_x\Sigma_0 v_{x1}, \label{eqn:wave_continuity} \\
    B_{x1} = 0, \label{eqn:wave_induction_x} \\
    \omega B_{y1} = k_x B_{y0} v_{x1} - k_x B_{x0} v_{y1}, \label{eqn:wave_induction_y} \\
    \qty(\omega - k_x^2 v^2_{Ay}/\omega)v_{x1} = k_x\qty(h_1 + \Phi_1) - \frac{k_x^2 v_{Ax} v_{Ay}}{\omega} v_{y1} + 2i\Omega v_{y1} \label{eqn:wave_momentum_x} \\
    \omega v_{y1} = -\frac{k_x^2 v_{Ax} v_{Ay}}{\omega} v_{x1} + \frac{k_x^2 v_{Ax}^2}{\omega} v_{y1} + i\qty(q - 2) \Omega v_{x1}, \label{eqn:wave_momentum_y}
\end{gather}
where $c^2_s = \gamma P_g/\Sigma_0, v^2_A = B_0^2/\Sigma_0$ are the sound speed and Alfv\'en speed, respectively. The perturbed Poisson equation becomes
\begin{equation}
    \pdv[2]{\Phi_{\mathrm{3D},1}}{z} = k_x^2\Phi_{\mathrm{3D},1} + 4\pi G\Sigma_1\delta\qty(z). \label{eqn:wave_poisson}
\end{equation}
The solution to the perturbed Poisson eq.~\ref{eqn:wave_poisson} is 
\begin{equation}
    \Phi_{\mathrm{3D},1} = -\frac{2\pi G\Sigma_1}{\abs{k}}e^{-\abs{kz}}, \label{eqn:perturb_poisson_sol}
\end{equation}
which gives $\Phi_1 = \Phi_{\mathrm{3D},1}(x,y,z=0) = -2\pi G\Sigma_1/\abs{k}$. The perturbed enthalpy density is $h_1 = c_s^2\Sigma_1/\Sigma_0$. Substituting into eqs.~\ref{eqn:wave_continuity}--\ref{eqn:wave_momentum_y} and rearranging gives the following dispersion relation
\begin{gather}
    \omega^4 + \qty[2\qty(q-2)\Omega^2 - k_x^2\qty(c_s^2 + v_A^2) + 2\pi G\Sigma_0\abs{k_x}]\omega^2 \nonumber \\\quad+ iq\Omega\omega k_x^2 v_{Ax} v_{Ay} \nonumber \\\quad+ k_x^2 v_{Ax}^2\qty(k_x^2 c_s^2 - 2\pi G\Sigma_0\abs{k_x}) = 0.\quad \text{(Razor-thin disk)} \label{eqn:crmg_dispersion_razor_thin}
\end{gather}
If the background density is uniform with value $\rho_0$ instead of razor-thin, one simply needs to replace the $2\pi G\Sigma_0\abs{k_x}$ terms in eq.~\ref{eqn:crmg_dispersion_razor_thin} by $4\pi G \rho_0$, i.e.
\begin{gather}
    \omega^4 + \qty[2\qty(q-2)\Omega^2 - k_x^2\qty(c_s^2 + v_A^2) + 4\pi G\rho_0]\omega^2 \nonumber \\\quad+ iq\Omega\omega k_x^2 v_{Ax} v_{Ay} \nonumber \\\quad+ k_x^2 v_{Ax}^2\qty(k_x^2 c_s^2 - 4\pi G\rho_0) = 0.\quad \text{(Uniform background)} \label{eqn:crmg_dispersion_uniform}
\end{gather}

\section{Wavelength dependence of the axisymmetric CRMG instability with respect to the field strength } \label{app:crmg_wavelength}

\begin{figure}
    \centering
    \includegraphics[width=0.23\textwidth]{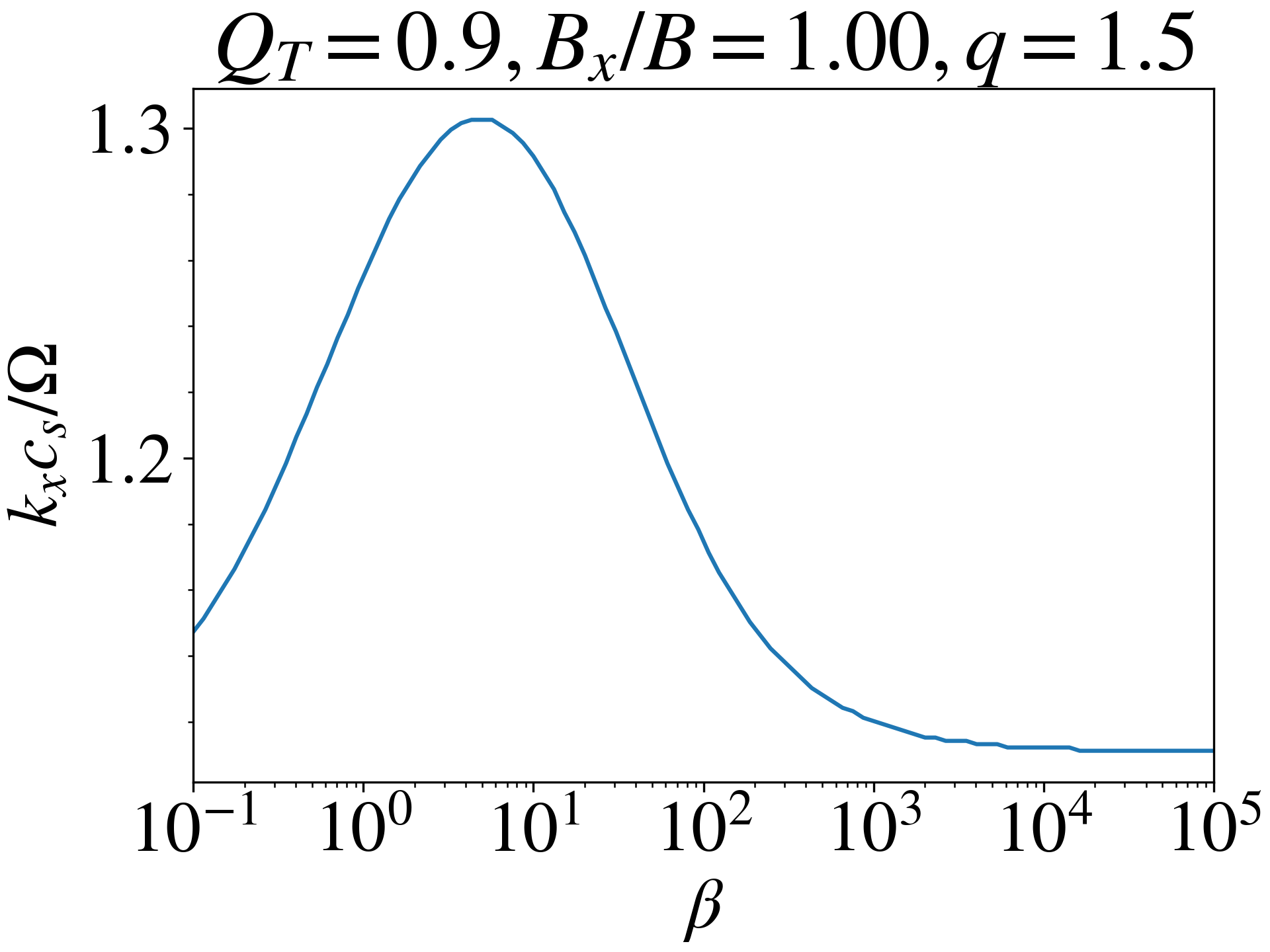}
    \includegraphics[width=0.23\textwidth]{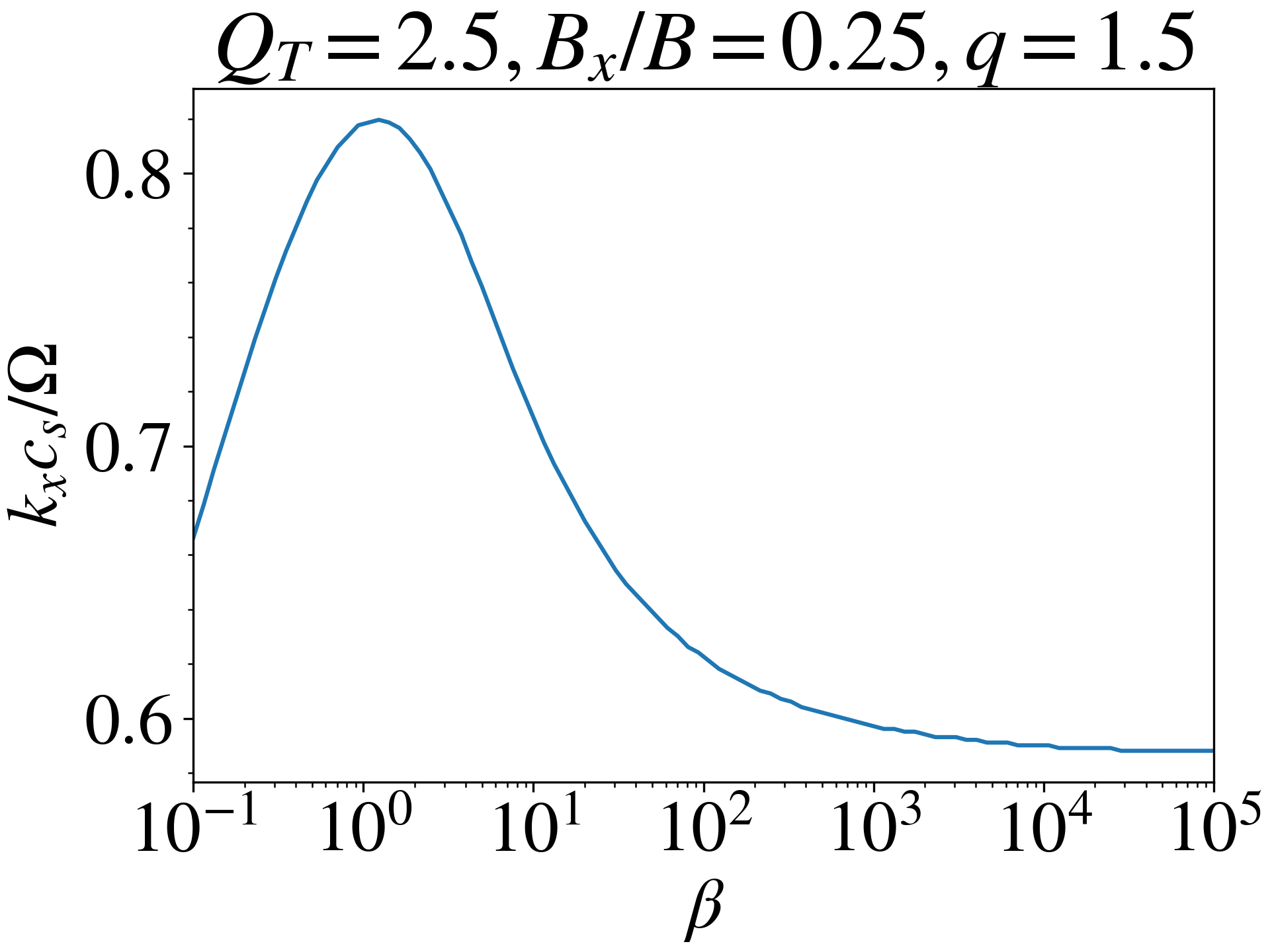}
    \caption{Most unstable axisymmetric wavenumber (expressed as $k_x c_s/\Omega$) for as a function of plasma beta $\beta$ for two sets of $(Q_T, B_x/B, q)$ parameters.}
    \label{fig:axisymmetric_wavelength}
\end{figure}

In Fig.~\ref{fig:axisymmetric_wavelength} we show, for two sets of $(Q_T, B_x/B, q)$ parameters, that the wavelength of the most unstable axisymmetric CRMG mode becomes shorter when the magnetic field becomes stronger. The typical most unstable wavelength is roughly a thermal scaleheight ($c_s/\Omega$). The most unstable wavelength is the shortest at $\beta\approx 1$, and increasing the field strength further below $\beta\approx 1$ yields a longer wavelength. \citet{Kubli_etal-2023} found that this explains the smaller clumps found in their simulations when magnetic field is introduced.

\section{Comparison of the axisymmetric CRMG and Magneto-Jeans type modes} \label{app:crmg_jeans_compare}

\begin{figure}
    \centering
    \includegraphics[width=0.23\textwidth]{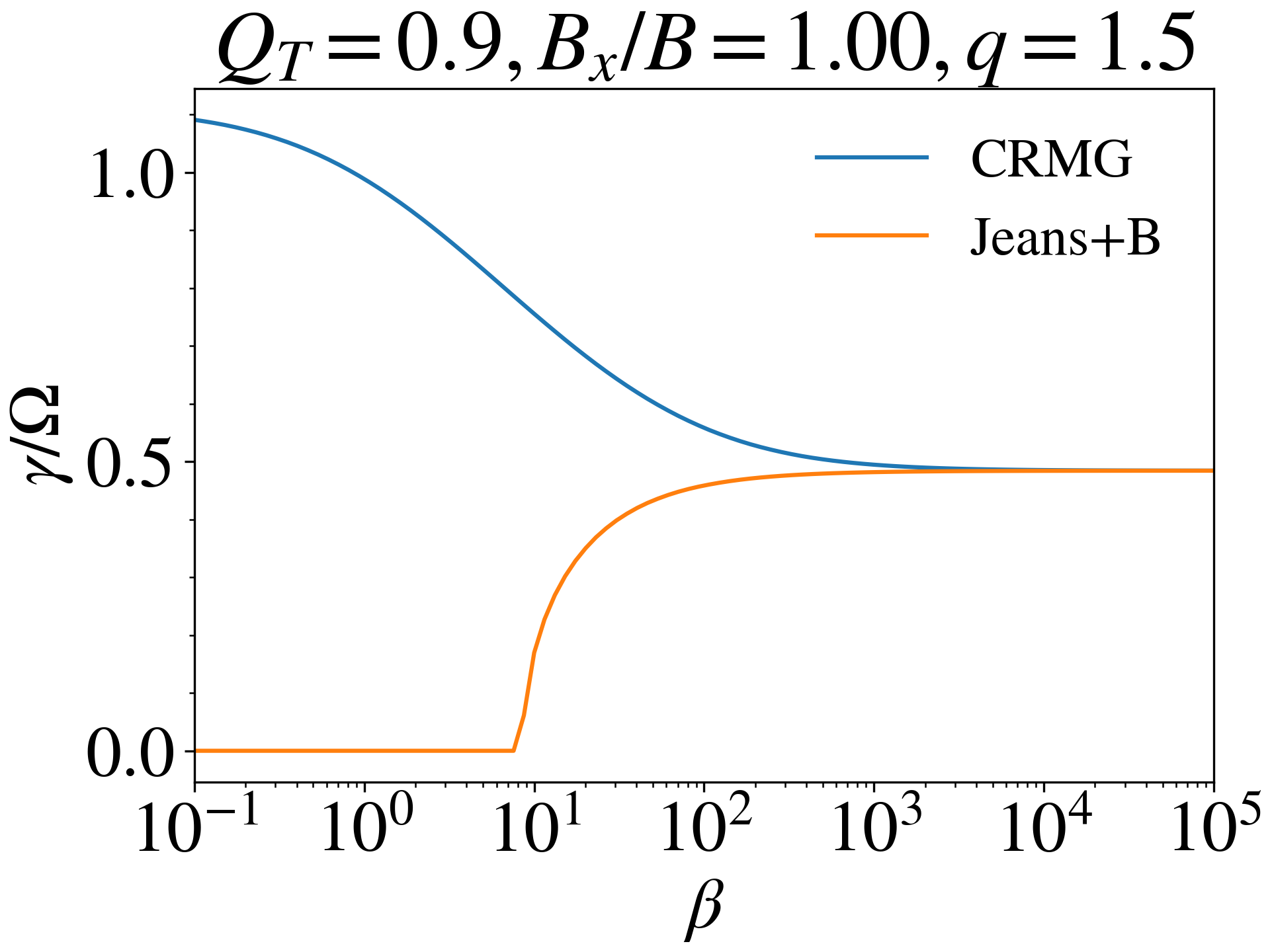}
    \includegraphics[width=0.23\textwidth]{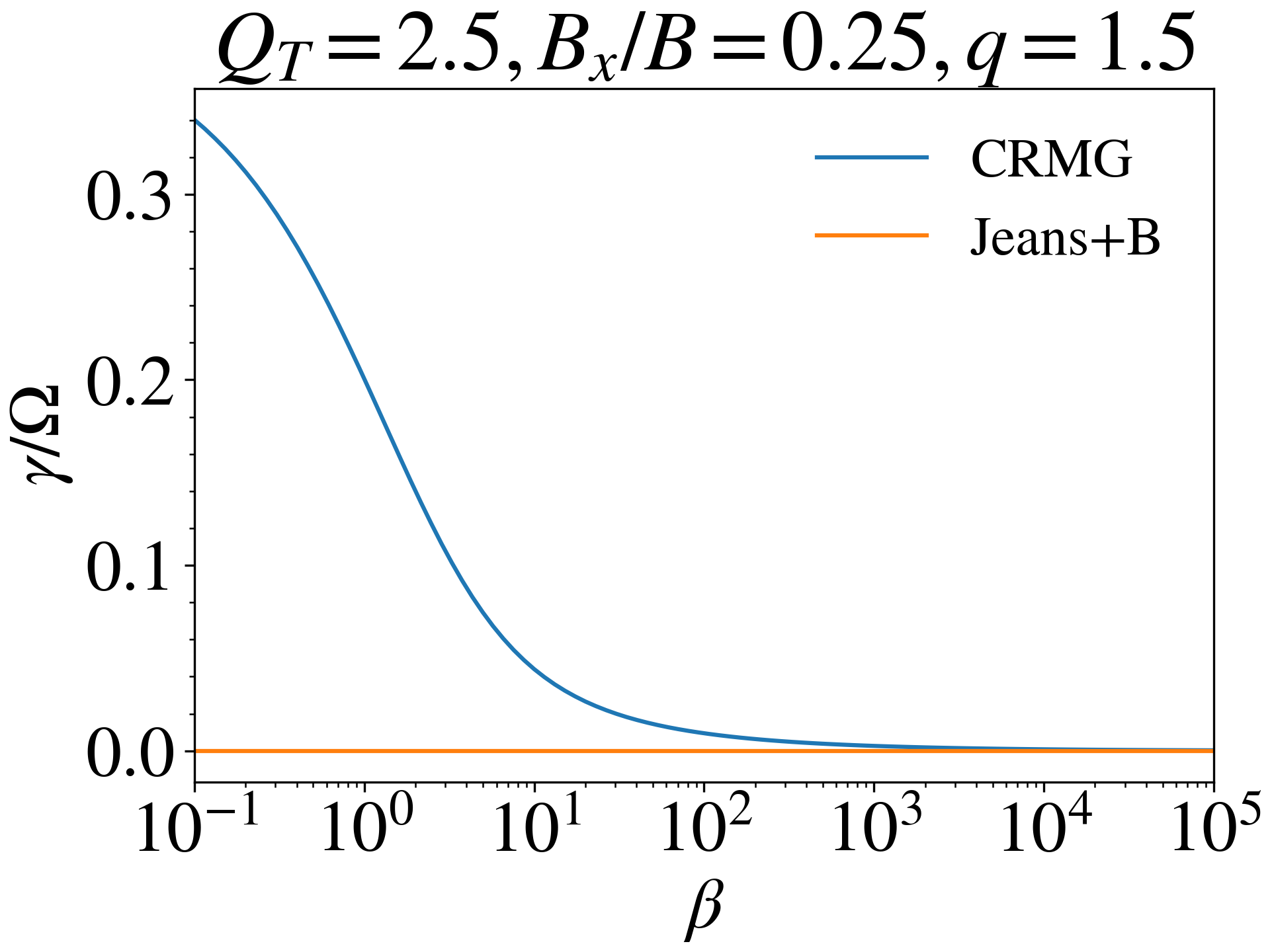}
    \caption{Comparison of growth rates against plasma beta $\beta$ of the CRMG and magneto-Jeans type unstable modes for two sets of $(Q_T, B_x/B, q)$ parameters. For each $\beta$, the the growth rate is calculated over a range of $k_x v_A/\Omega$ and the maximal one is selected and displayed.}
    \label{fig:axisymmetric_compare}
\end{figure}

The CRMG mode described above is fundamentally different from the magneto-Jeans type mode in the sense that it is destabilized for stronger magnetic field while the latter is stabilized in that limit. We illustrate this difference in Fig.~\ref{fig:axisymmetric_compare} by comparing the fastest growth rate of the two modes over a range of plasma beta, with the growth rate of the CRMG mode determined by eq.~\ref{eqn:crmg_dispersion} and that for the magneto-Jeans type mode determined by $\omega^2 = k_x^2(c_s^2 + v_A^2) + 2(2-q)^2\Omega^2 - 2\pi G \abs{k_x}\Sigma_0$. We performed this exercise for two sets of $(Q_T, B_x/B, q)$ parameters. For each $\beta$, the growth rate is calculated over a range of wavenumbers and the maximal one is selected and displayed. At high $\beta$, the two modes converge, with both of them stable when $Q_{T,B}>1$ and unstable otherwise. The growth rate of CRMG mode begins to deviate at $\beta\sim 10^2$, becoming stronger at lower $\beta$ while Jeans type mode dwindles due to $Q_{T,B}>1$ at low $\beta$.

In addition, for CRMG modes, the most unstable wavelength varies non-monotonically with respect to the magnetic field strength while it becomes larger with increasing field strength for magneto-Jeans mode\footnote{Using the dispersion for magneto-Jeans modes: $\omega^2 = k_x^2(c_s^2 + v_A^2) + 2(2-q)^2\Omega^2 - 2\pi G \abs{k_x}\Sigma_0$, the most unstable mode has wavenumber $k_{x,\mathrm{max}}=\pi G\Sigma_0/(c_s^2+v_A^2)$, which decreases (shifts to longer wavelength) when magnetic pressure dominates. If the background density were uniform instead of razor-thin, the most unstable wavenumber would be $k_{x,\mathrm{max}}=0$ instead, i.e. the largest scale mode would be the most unstable.}. In particular, in the regime $\beta>1$, the most unstable wavelength becomes shorter as $\beta$ decreases. 

\section{Numerical verification of the axisymmetric CRMG growth rate in 2D} \label{app:numerical_crmg}

\begin{figure}
    \centering
    \includegraphics[width=0.23\textwidth]{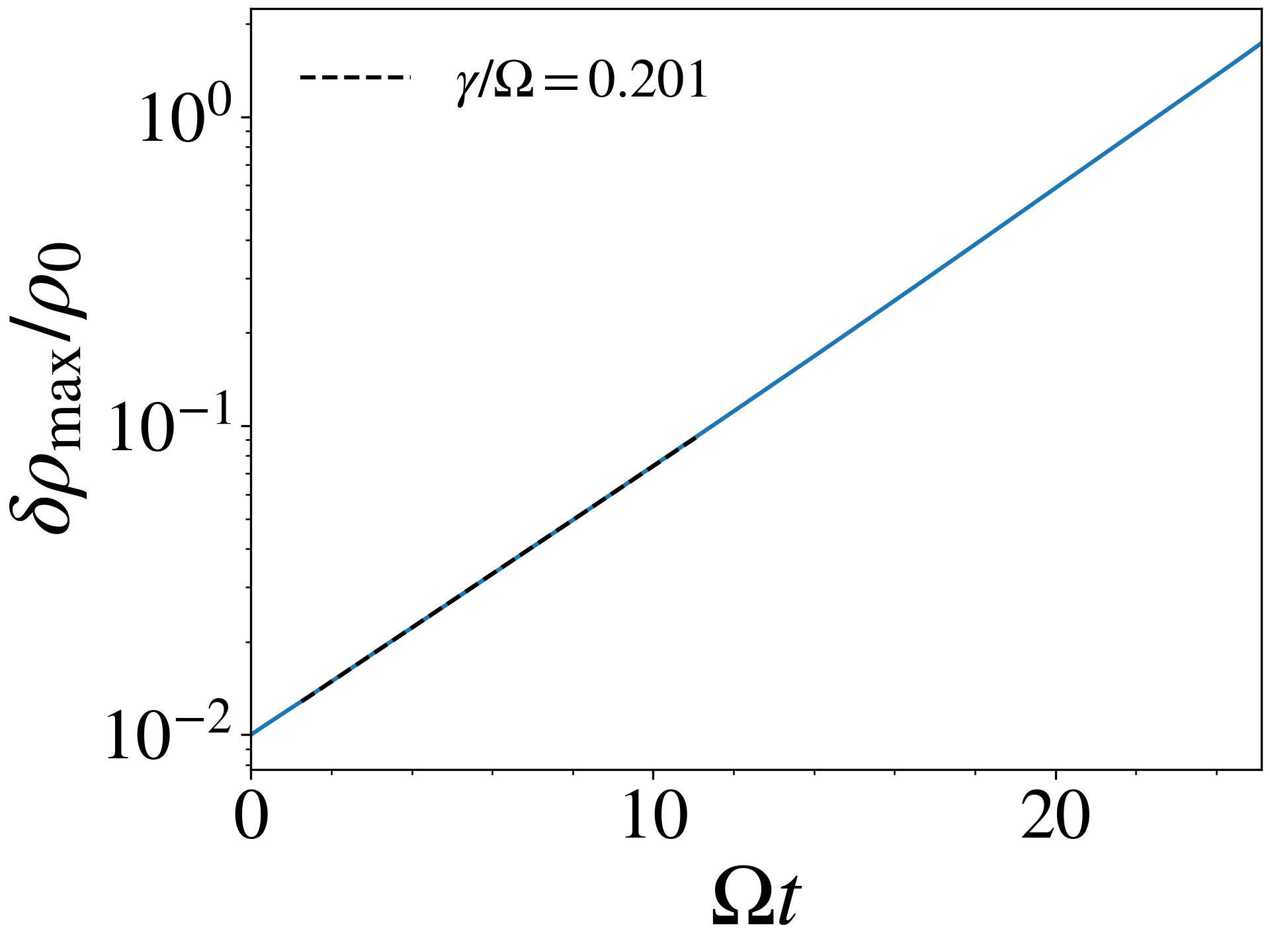} 
    \includegraphics[width=0.23\textwidth]{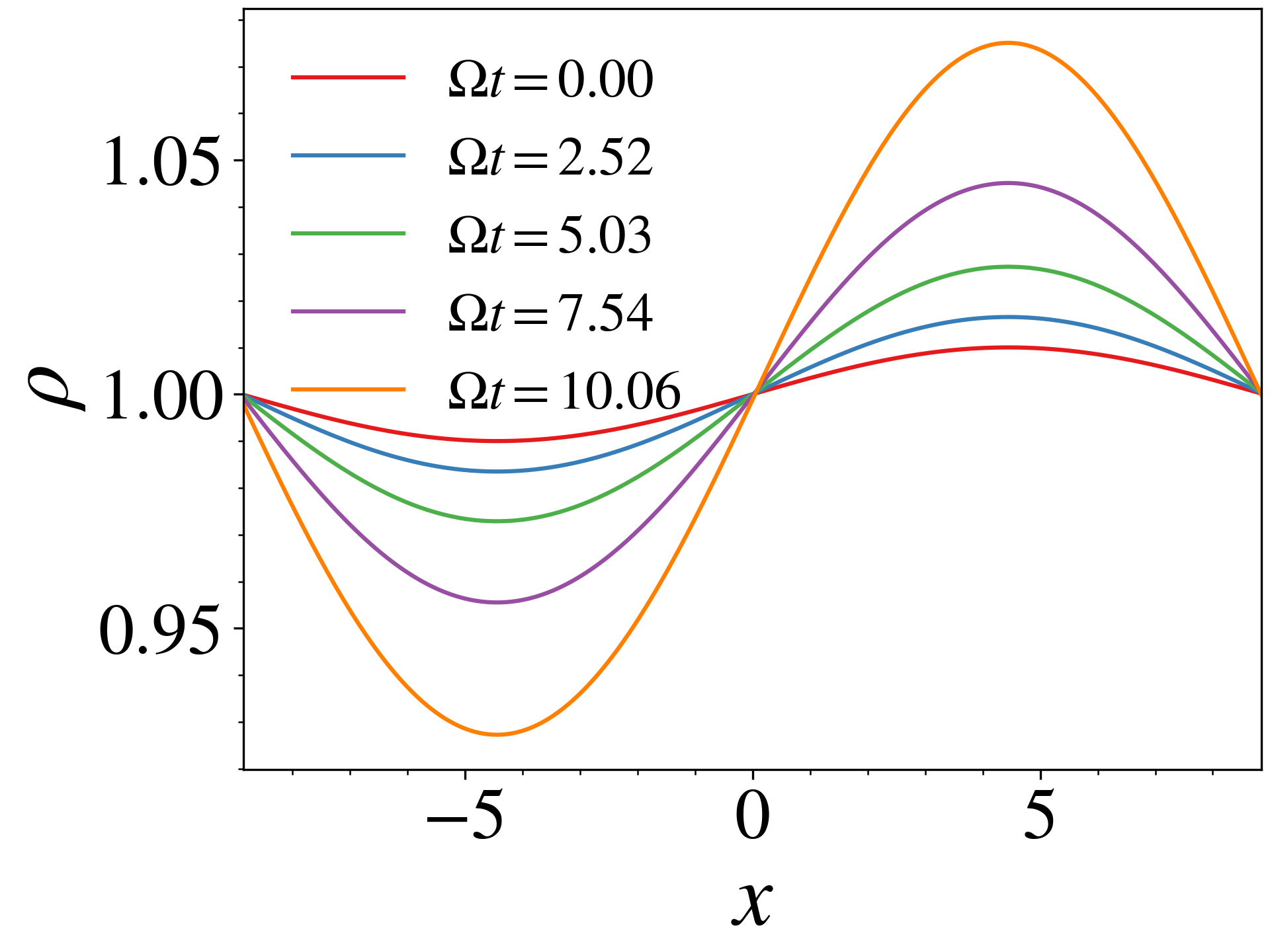} \\
    \includegraphics[width=0.23\textwidth]{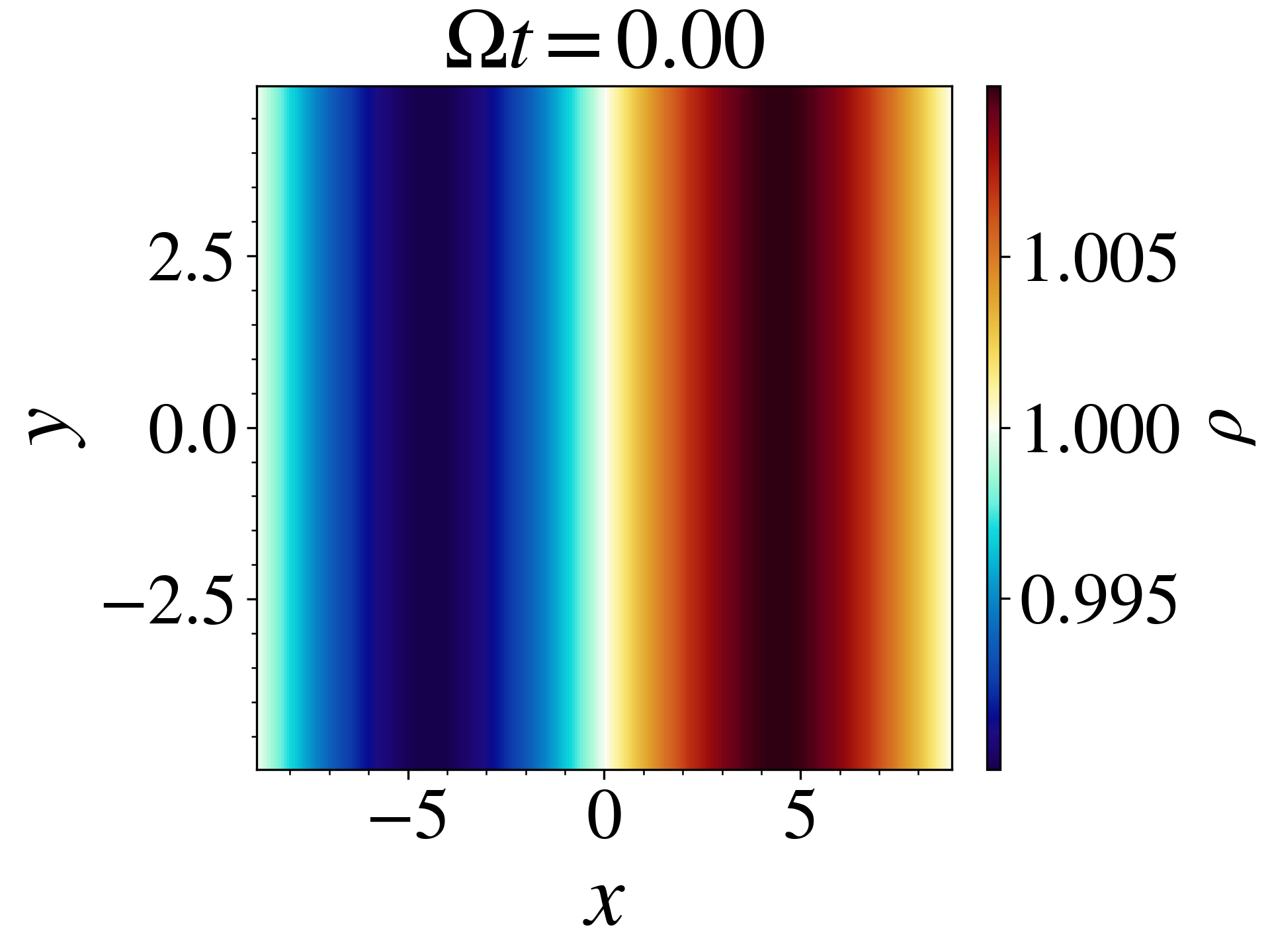} 
    \includegraphics[width=0.23\textwidth]{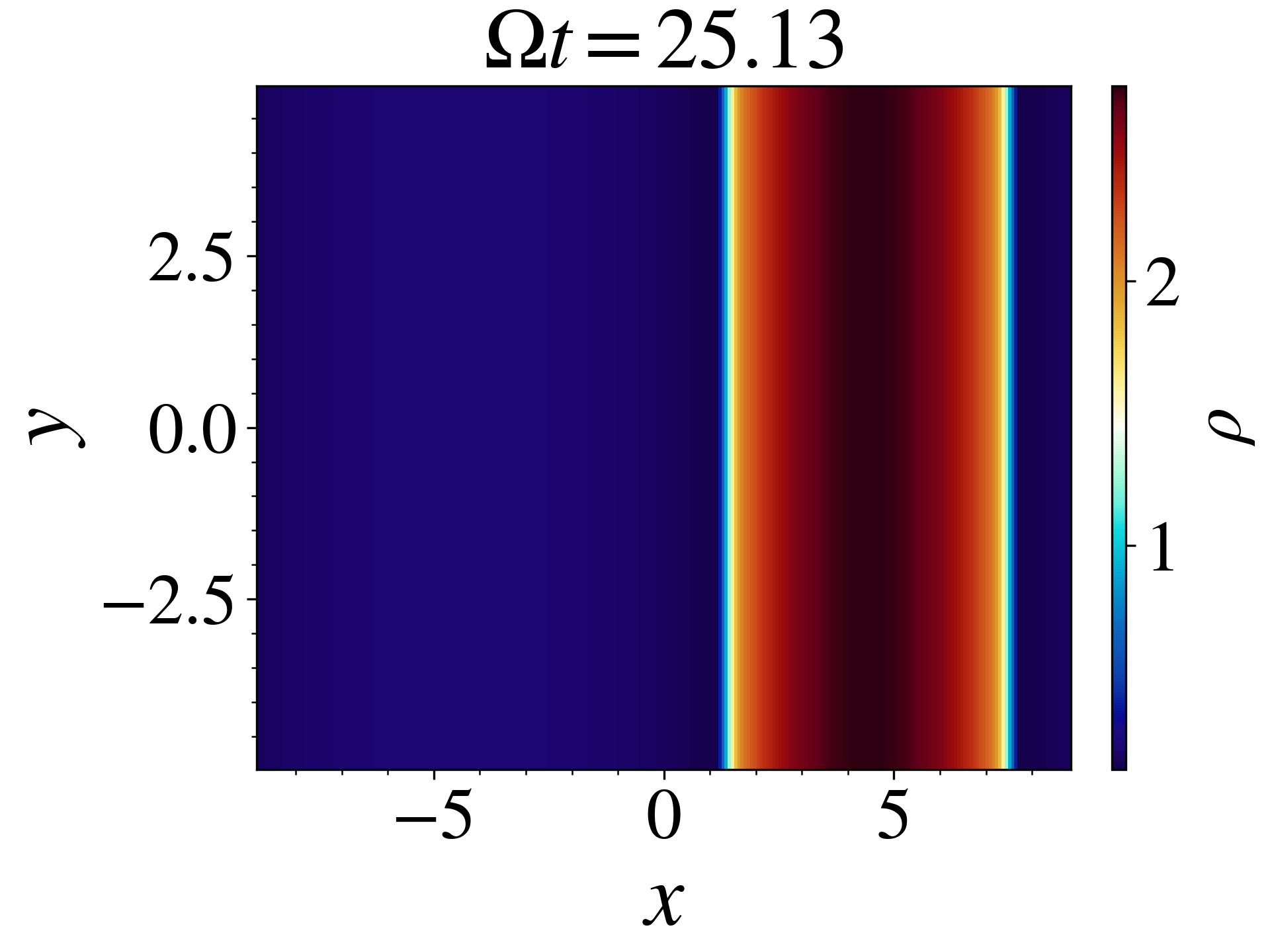}
    \caption{Simulation of a selected CRMG unstable mode with $\Omega^2/4\pi G\rho_0 = 1, k_x v_{A0}/\Omega = 0.5, \beta=2P_{g0}/P_{B0}=1, B_{x0}/B_0=1, q=0$, where the subscript $0$ denotes the background value. Top left: The growth curve of the mode, measured by $\delta\rho_\mathrm{max}/\rho_0\equiv (\rho_\mathrm{max} - \rho_0)/\rho_0$, where $\rho_\mathrm{max}$ is the maximum density. The fitted growth rate of $\gamma/\Omega = 0.201$ (black dashed line) is very close to the expected growth rate of $0.199$. Top right: snapshots of the density profile in the $x$-direction during linear growth, showing the perturbed profile growing in-situ. Bottom left and right: The initial (left) and $\Omega t=25.13$ (right) 2D density snapshots, showing how the imposed density perturbation grows to nonlinear amplitudes.}
    \label{fig:CRMG_numerics}
\end{figure}

We verify that the CRMG instability can be captured by simulation. Using a 2D setup (without cooling) with background density $\rho_0$, sound speed $c_{s0}$ and angular frequency $\Omega$ set to 1, we inserted an axisymmetric mode with density amplitude of $\delta \rho_\mathrm{max}/\rho_0=0.01$ into the background and let it evolve. As it is impossible to set up a razor-thin disk in 2D, we redid the linear analysis using a uniform density background. The dispersion relation remains the same as eq.~\ref{eqn:crmg_dispersion}  except $2\pi G\Sigma_0\abs{k_x}$ is replaced by $4\pi G\rho_0$ (eq.~\ref{eqn:crmg_dispersion_uniform}). The unstable mode we selected for display is characterized by $\Omega^2/4\pi G\rho_0 = 1, k_x v_{A0}/\Omega = 0.5, \beta = 2 P_{g0}/P_{B0}=1, B_{x0}/B_0 = 1, q=0$. We use a shearing parameter $q=0$ for this demonstration to keep the background radial field $B_x$ static (a non-zero shearing parameter $q$ would cause $B_x$ to be sheared into the azimuthal direction on a timescale of $(q\Omega)^{-1}$). As we observe in Fig.~\ref{fig:CRMG_numerics}, the perturbed density profile grew exponentially with a growth rate of $\gamma/\Omega = 0.201$, matching the expected growth rate of $0.199$. We followed the growth into the nonlinear regime, which is characterized by a single overdense peak.

\section{Comparison with lower resolution simulations} \label{app:resolution}

\begin{table}
    \centering
    \begin{tabular}{c|c|c|c|c}
       $\beta_0$ & $\langle Q_y\rangle_t$ (HR) & $\langle Q_y\rangle_t$ (LR) & $\langle Q_z\rangle_t$ (HR) & $\langle Q_z\rangle_t$ (LR) \\
       \hline 
       10 & 482 & 189 & 146 & 73.3 \\
       $10^2$ & 391 & 146 & 88.2 & 29.5 \\
       $10^3$ & 207 & 122 & 35.8 & 11,6 \\
       $10^4$ & 78.6 & 68.2 & 24.8 & 6.14 \\
       $10^5$ & 82.5 & 29.2 & 28.7 & 6.10
    \end{tabular}
    \caption{Comparison of the time-averaged quality factors $\langle Q_y\rangle_t$, $\langle Q_z\rangle_t$ for the fiducial cases (HR) and the reduced resolution cases (LR).}
    \label{tab:quality_factors_resolution}
\end{table}

\begin{figure}
    \centering
    \includegraphics[width=0.23\textwidth]{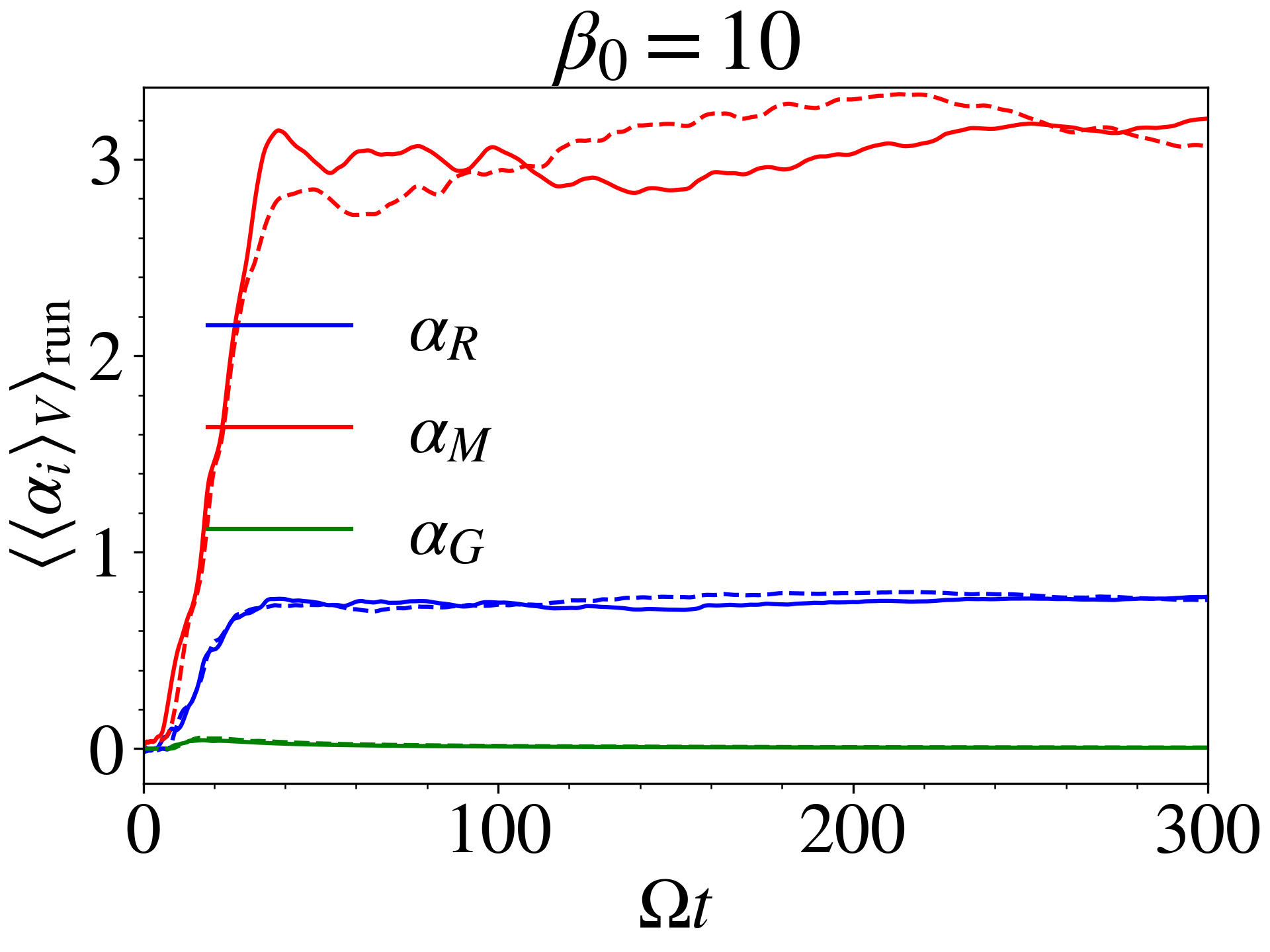}
    \includegraphics[width=0.23\textwidth]{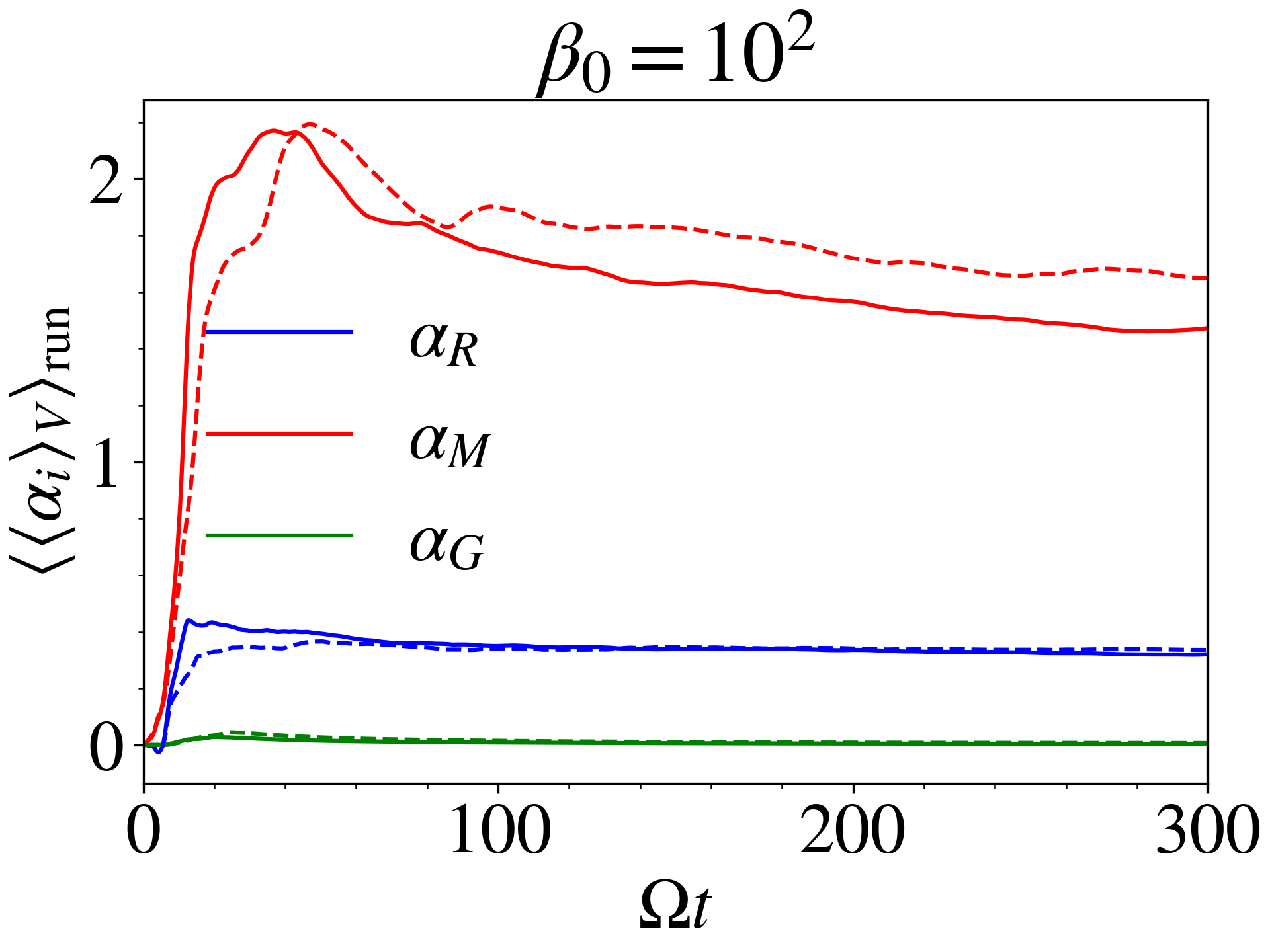} \\
    \includegraphics[width=0.23\textwidth]{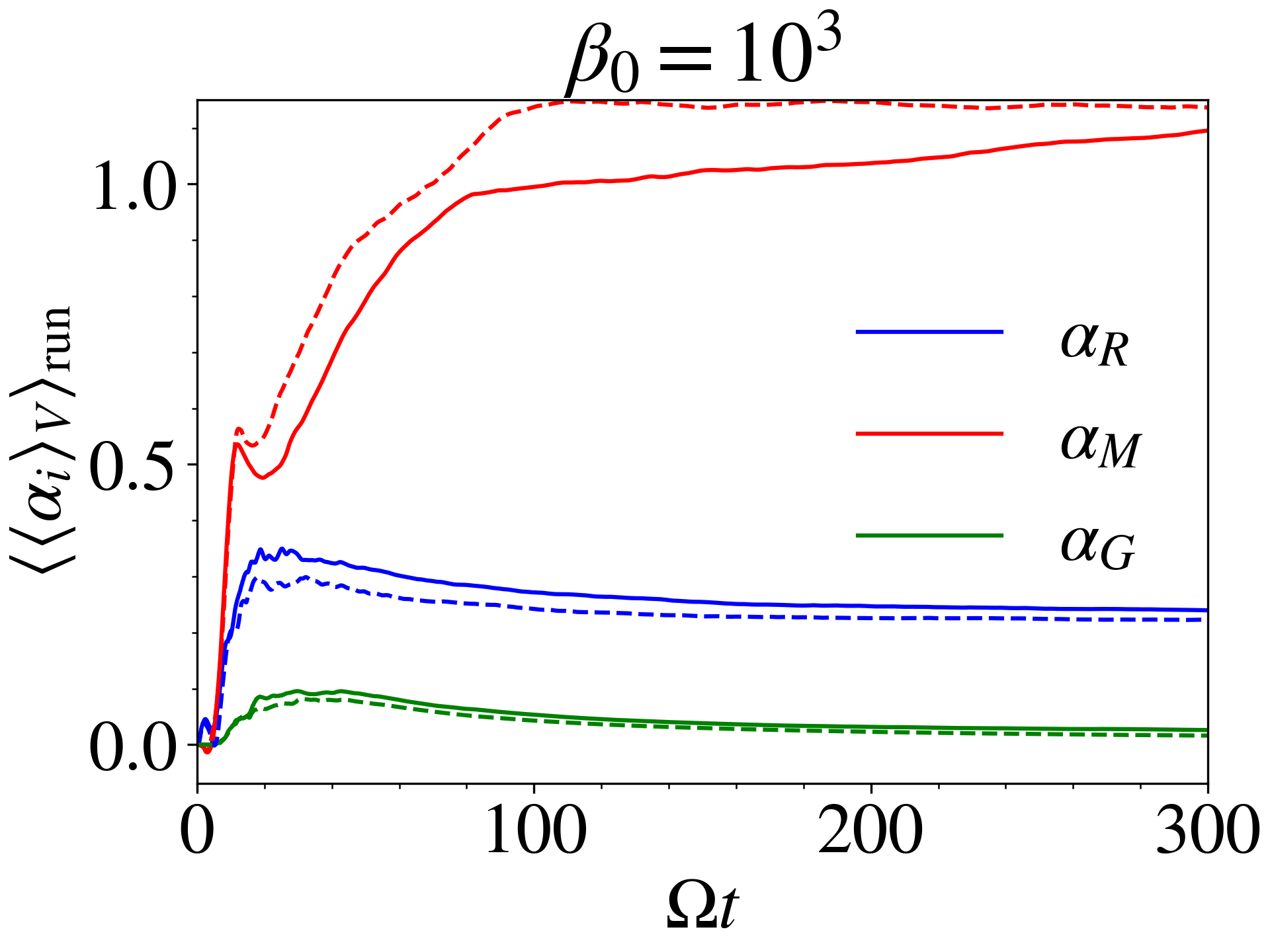}
    \includegraphics[width=0.23\textwidth]{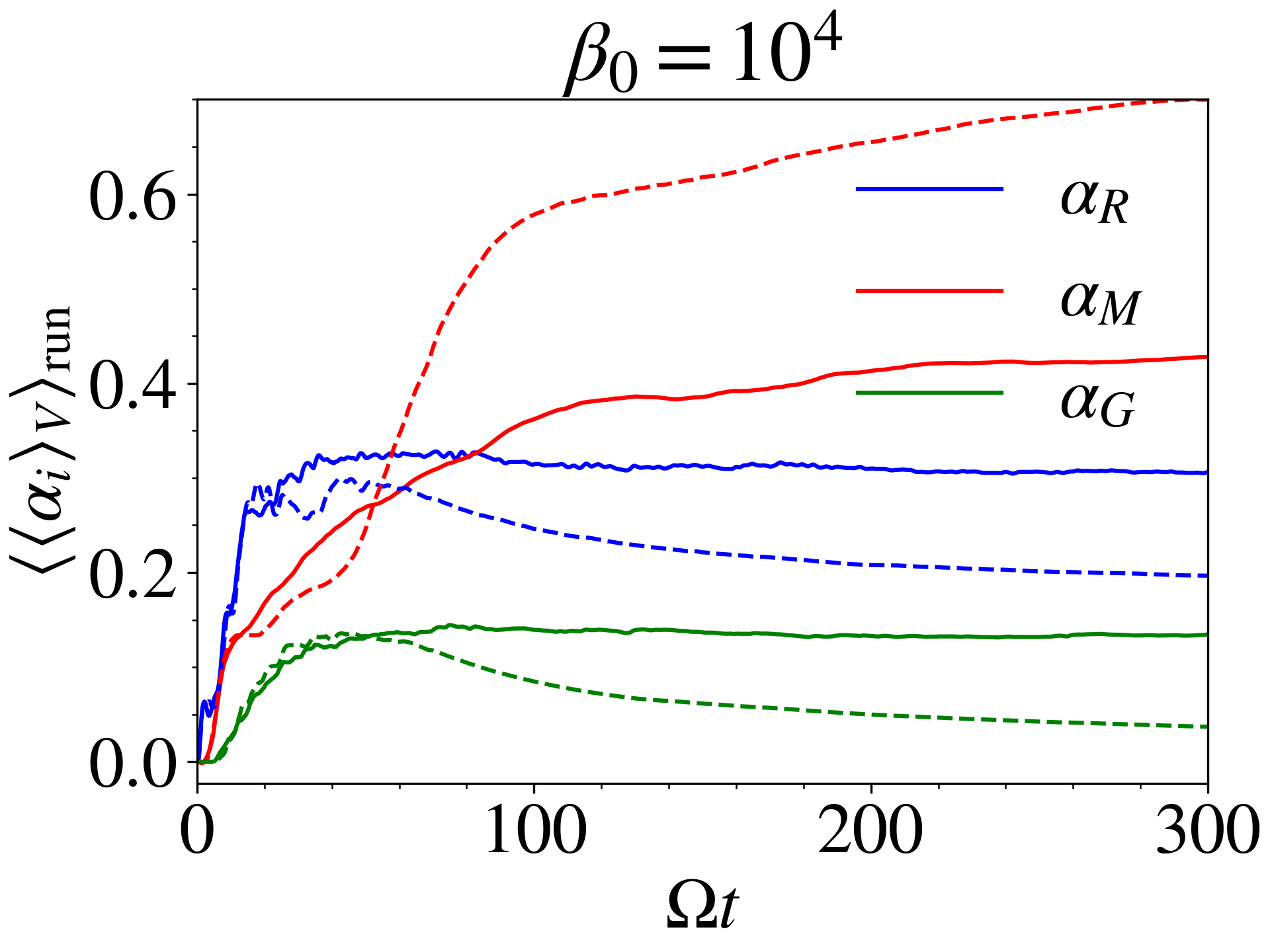} \\
    \includegraphics[width=0.23\textwidth]{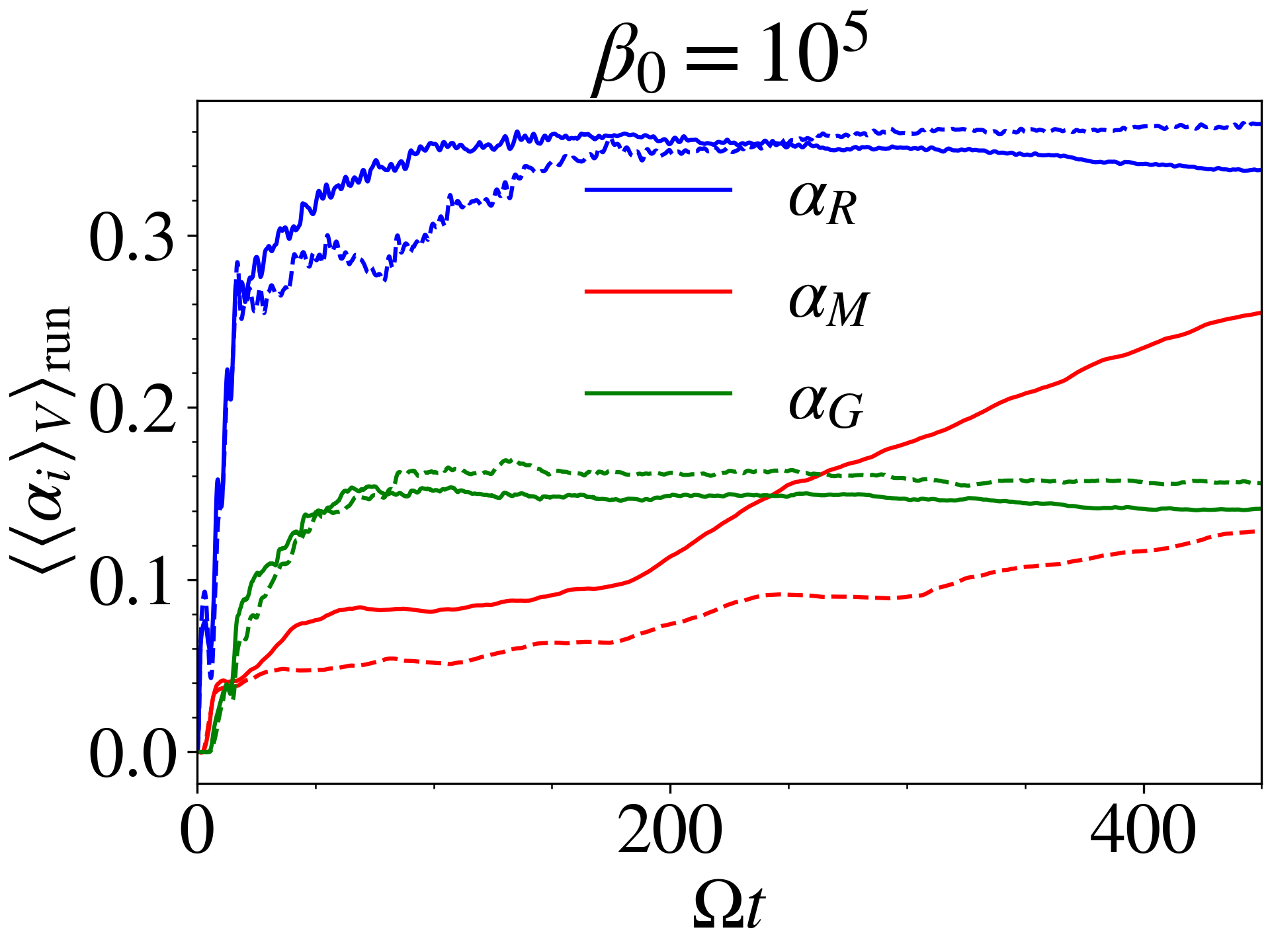}
    \caption{Comparison of the running-averaged stresses for the two resolutions explored. Solid line denotes the fiducial resolution (HR), while dashed lines denotes the reduced resolution (LR). The title in each panel denotes $\beta_0$ of the test case.}
    \label{fig:running_stress_resolution}
\end{figure}

\begin{figure}
    \centering
    \includegraphics[width=0.23\textwidth]{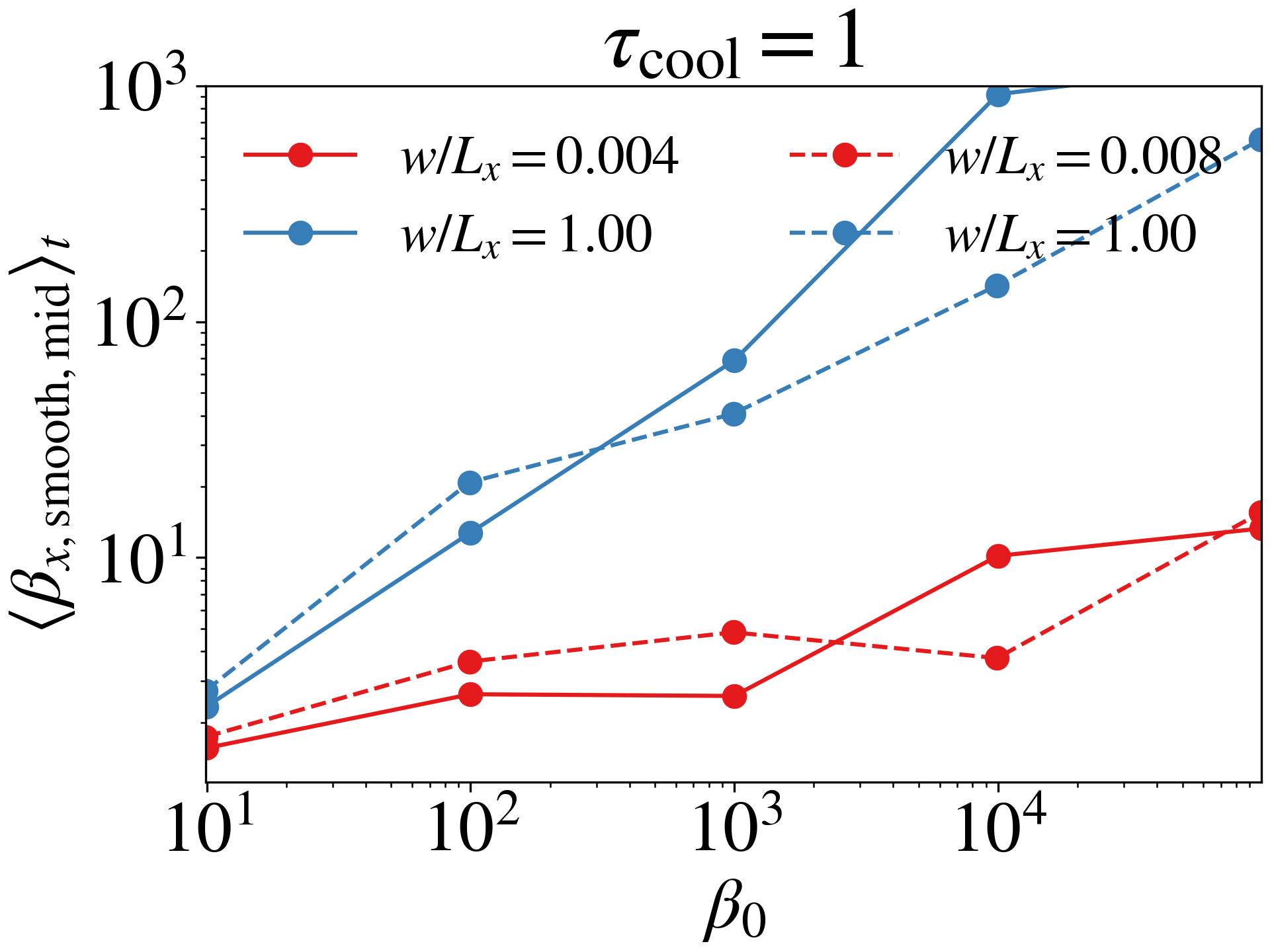}
    \includegraphics[width=0.23\textwidth]{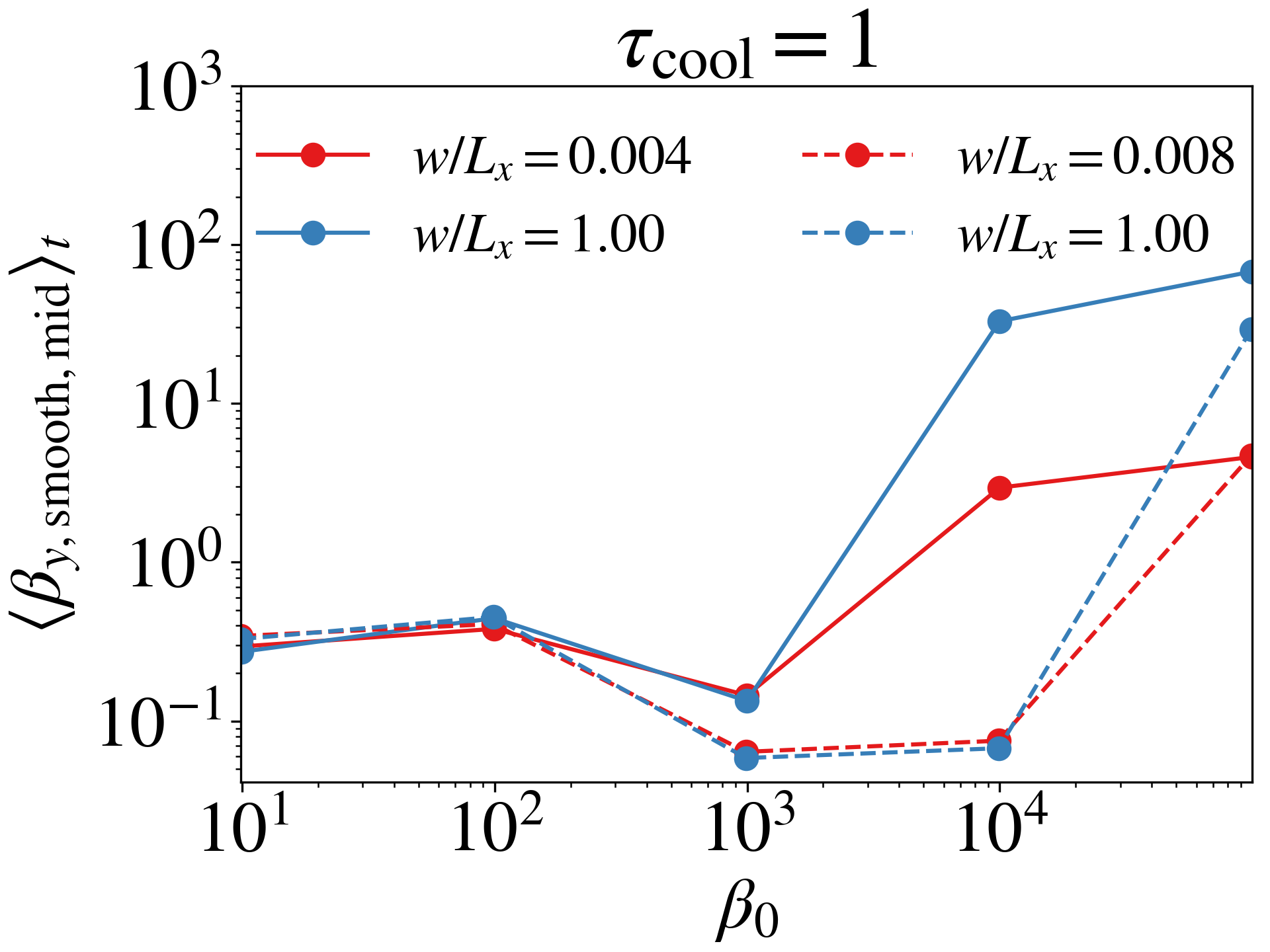} \\
    \includegraphics[width=0.23\textwidth]{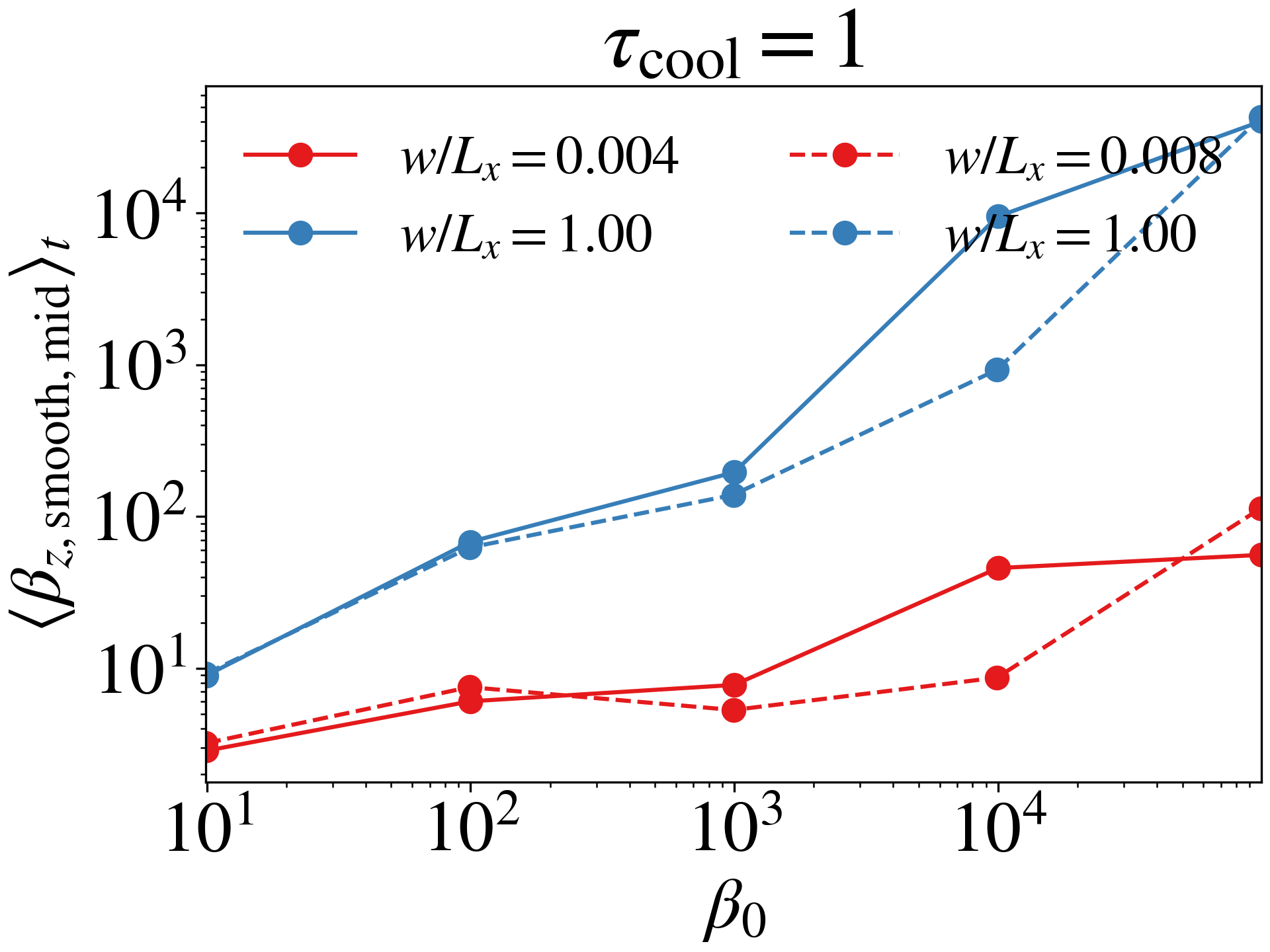}
    \includegraphics[width=0.23\textwidth]{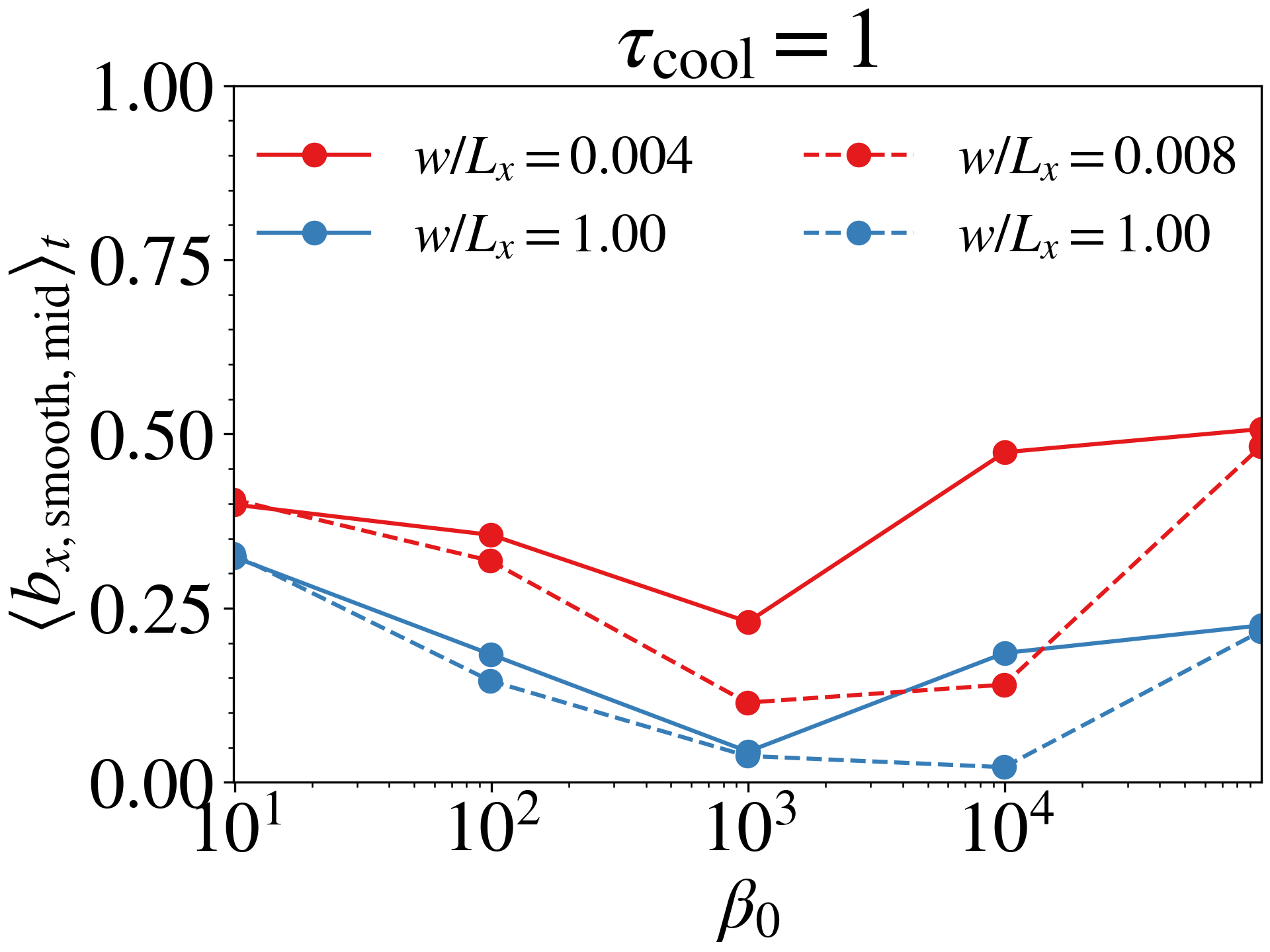} 
    \caption{Mid-plane window-averaged, time-averaged radial field plasma beta $\langle\beta_{x,\mathrm{smooth}}\rangle_t$ (top left), toroidal plasma beta $\langle\beta_{y,\mathrm{smooth}}\rangle_t$ (top right), vertical plasma beta $\langle\beta_{z,\mathrm{smooth}}\rangle_t$, and the relative radial field strength $\langle b_{x,\mathrm{smooth}}\rangle_t$. Solid line denotes HR resolution, dashed line denotes LR resolution.}
    \label{fig:timeavg1D_bfield_resolution}
\end{figure}

\begin{figure}
    \centering
    \includegraphics[width=0.45\textwidth]{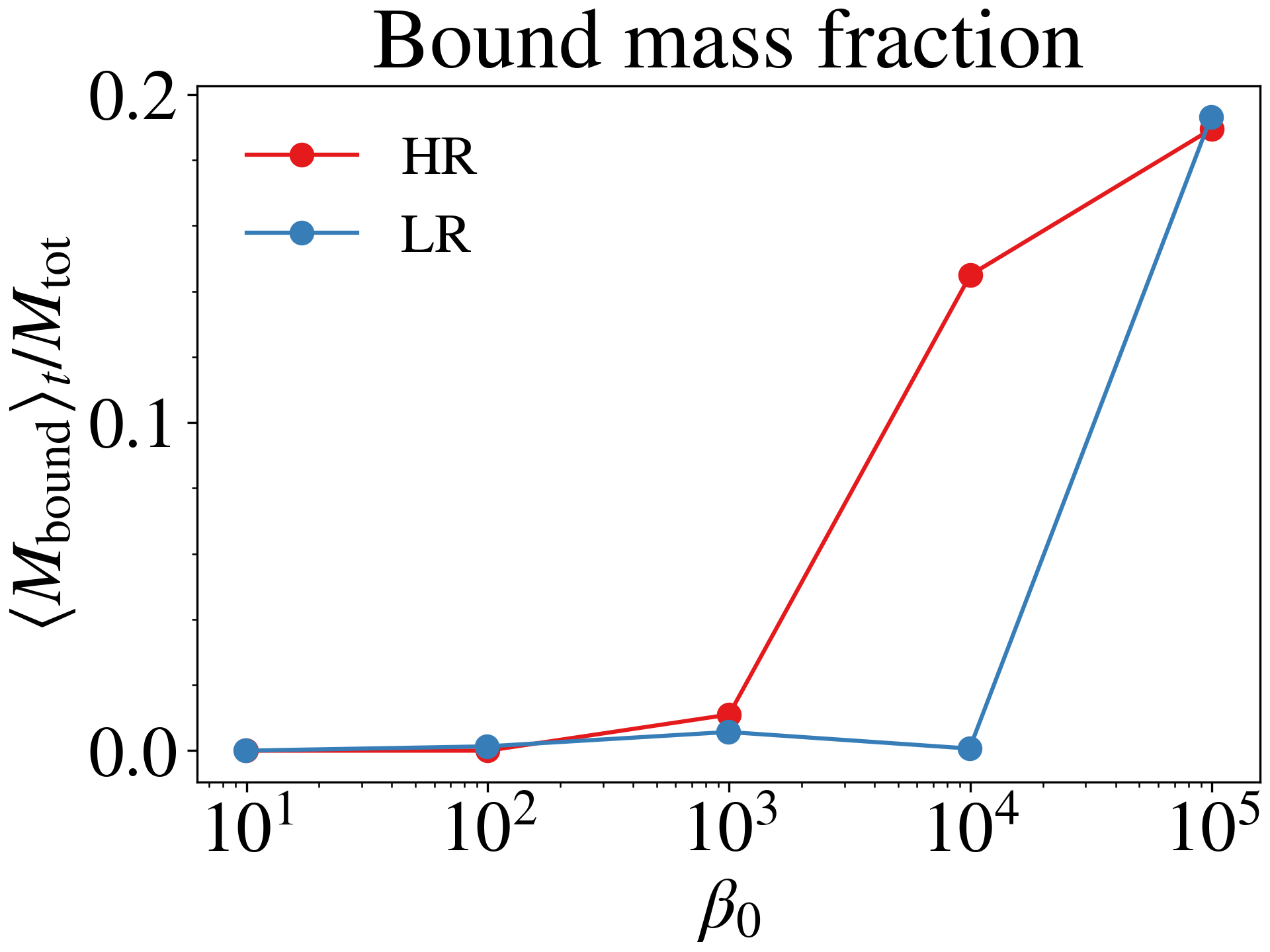}
    \caption{Comarison of the bound mass fraction found in the two resolutions, with the red solid line denoting the fiducial resolution (HR), and the blue solid line denoting the reduced resolution cases (LR).}
    \label{fig:bound_mass_resolution}
\end{figure}

We repeated our simulations at a reduced resolution of $128\times 128\times192$ (for $\beta_0\geq 10^3$) and $128\times 256\times 192$ (for $\beta_0< 10^3$), keeping the box dimensions the same. We present a comparison of the stresses, mid-plane magnetic field, and fragmentation fraction for the two resolutions (HR denoting the fiducial resolution cases, LR denoting the reduced resolution cases). Due to limited computational resources, we could not re-simulate the test cases at a higher resolution. In Table \ref{tab:quality_factors_resolution} we list the time-averaged quality factors in the $y$ ($\langle Q_y\rangle_t$) and $z$ ($\langle Q_z\rangle_t$) directions at the mid-plane. The quality factors of the lower resolution cases are generally a factor of a few lower than the fiducial cases. For $\beta_0\geq10^4$, $\langle Q_z\rangle_t$ are less than 10 for the LR cases, which according to \citet{Hawley-etal-2011} means MRI may be under-resolved for those cases.

From Fig.~\ref{fig:running_stress_resolution}-\ref{fig:bound_mass_resolution}, we see that the stresses, mid-plane magnetic field, and the bound mass fraction in the LR cases are quite close to the HR cases for $\beta_0\leq 10^3$. Surprisingly, there are good agreements for the $\beta_0=10^5$ case as well except for the Maxwell stress $\alpha_M$, which is a factor of 2 lower for the LR case, reflecting under-resolved MRI modes.

\begin{figure}
    \centering
    \includegraphics[width=0.45\textwidth]{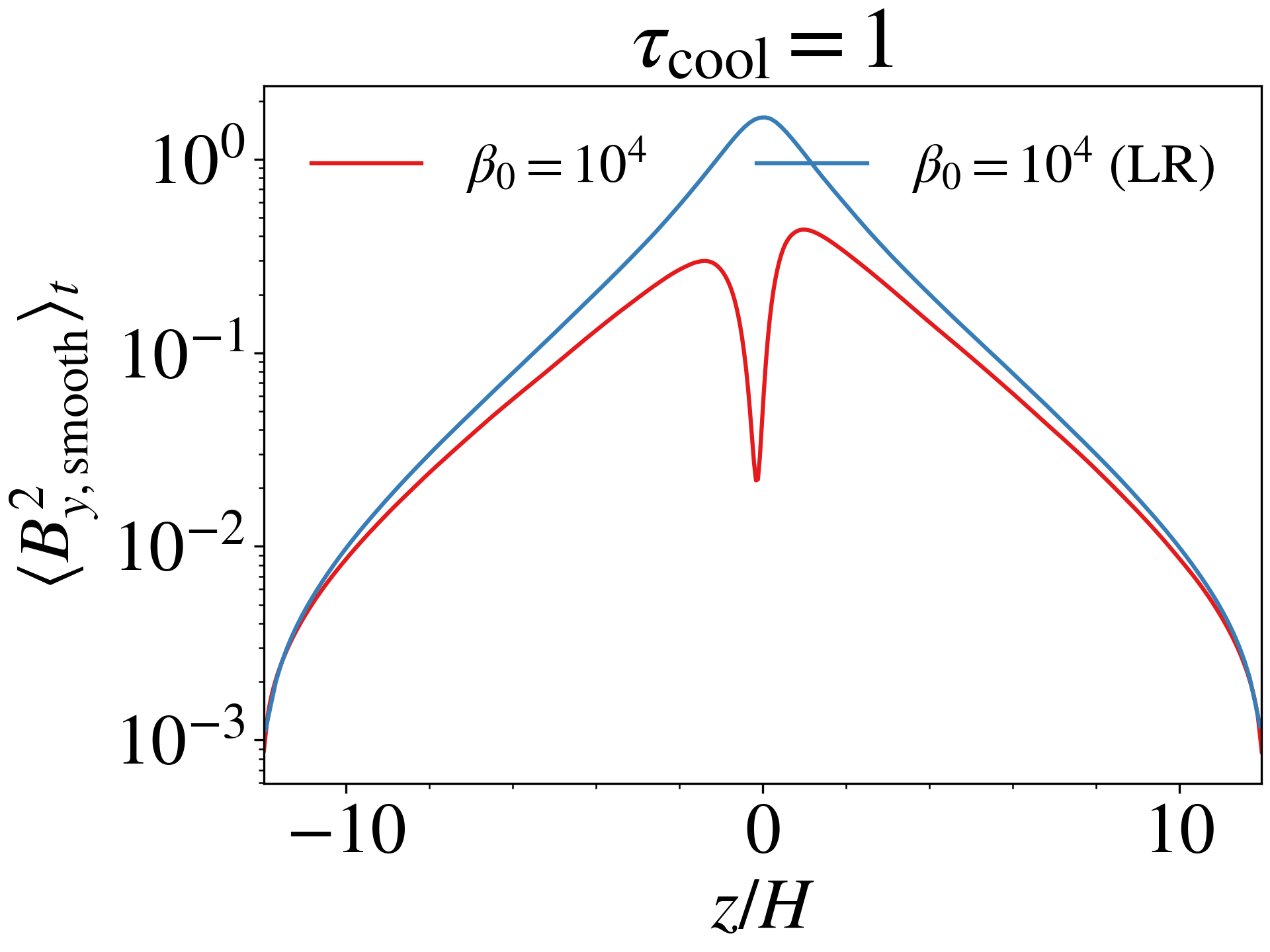} \\
    \includegraphics[width=0.45\textwidth]{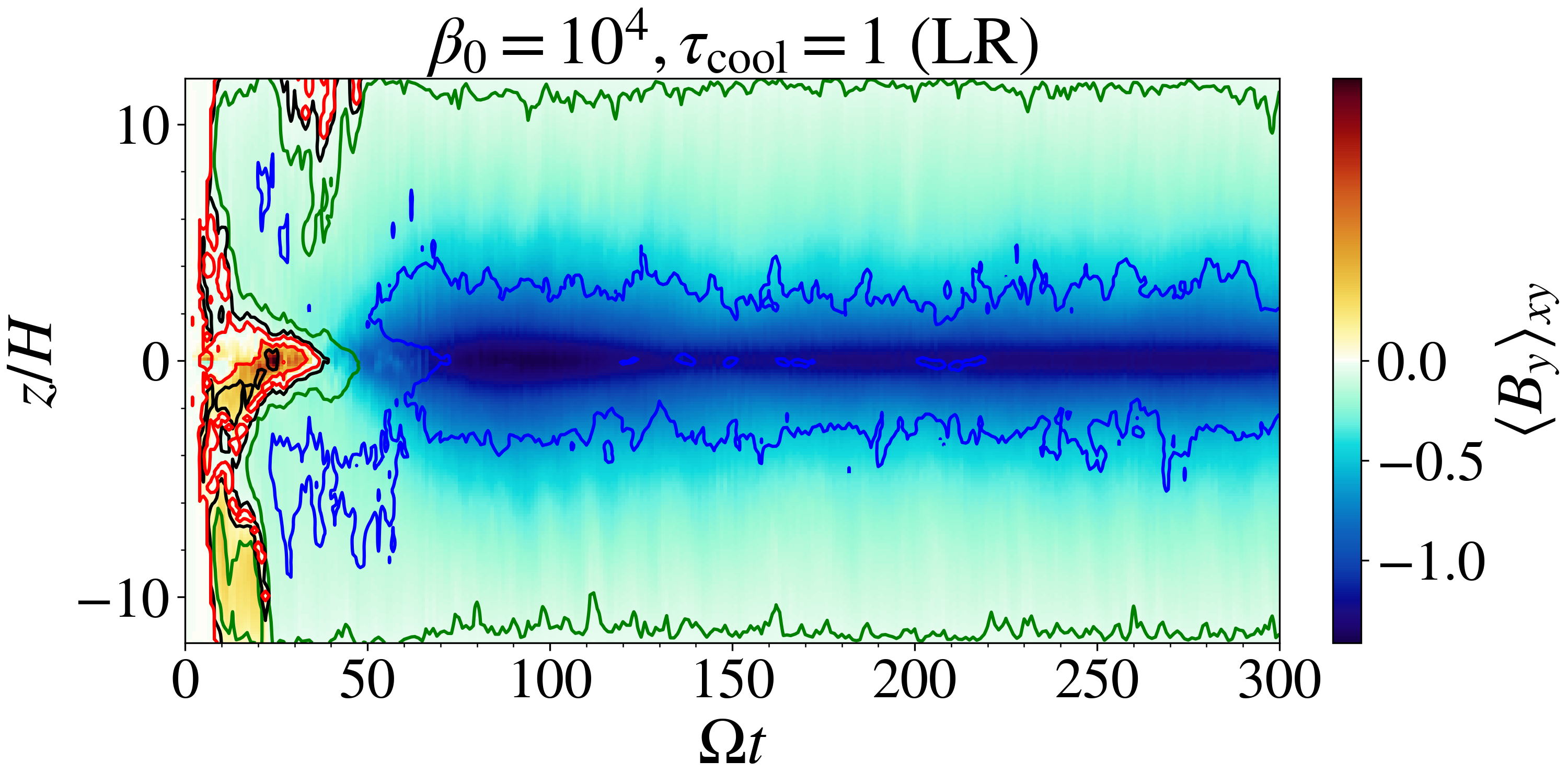} \\
    \includegraphics[width=0.45\textwidth]{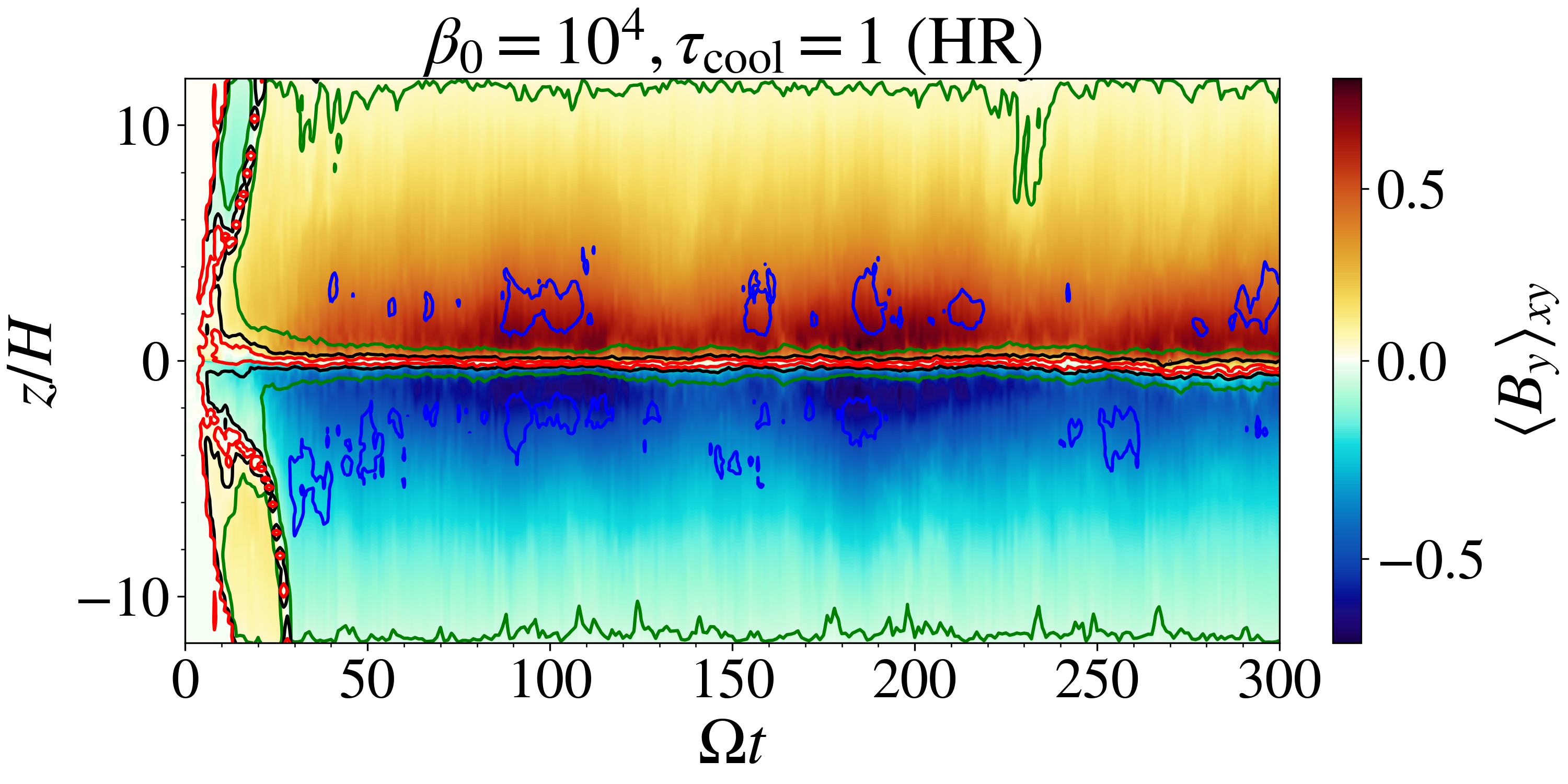}
    \caption{Top: window-averaged, time-averaged $B_y^2$ profiles for the HR (red) and LR (blue) cases. Bottom: Timeseries diagrams of $B_y$ for the LR, HR cases ($\beta_0=10^4$).}
    \label{fig:magnetic_field_resolution}
\end{figure}

The behavior of the $\beta_0=10^4$ case (LR) is rather odd, as the toroidal plasma beta and Maxwell stress are markedly stronger than in the HR case on average, and the bound mass fraction is exceptionally low. This is contrary to the expectation that under-resolved MRI modes should give rise to weaker Maxwell stress and field amplification, although we note that for $\Omega t\leq 50$ the Maxwell stress of the LR case is smaller. At $\Omega t\sim 50$ the magnetic field increased greatly for the LR case. In Fig.~\ref{fig:magnetic_field_resolution} we show the vertical profiles of the toroidal magnetic field (as time-averaged and timeseries plots) of the $\beta_0=10^4$ case for the LR and HR resolutions. We note that the toroidal field in the LR, $\beta_0=10^4$ case is particularly strong at the mid-plane, and does not exhibit the antisymmetric profile observed in the HR case. It is unclear to us how lower resolution would trigger the sudden and overwhelming increase in the toroidal field, and why it happens only to the $\beta_0=10^4$ case at low resolution. This comparison reinforces our fiducial choice of a higher resolution. We note, that the other $\beta_0$ cases appear to be converged with resolution.

\bsp	
\label{lastpage}
\end{CJK*}
\end{document}